\documentclass{article}
\usepackage{graphicx}  % standard LaTeX graphics tool;
\usepackage{amsmath}   % For getting proper math eqns
\usepackage{amssymb}   % Used for getting various symbols eg. gtrsim, lesssim etc.
\usepackage{bm}

\usepackage[compress]{cite}

\usepackage{color}

\def\eq#1{{Eq.~(\ref{#1})}}

\def\frab#1#2{\left(\frac{#1}{#2}\right)}      

\def\ket#1{|#1\rangle}                    %%%%   ket 
\def\bk#1#2#3{{\langle #1|#2|#3\rangle}}  %%%%   bracket
\def\amp#1#2{\langle #1 | #2\rangle}      %%%%   amplitude

 \def\b#1{\bm{#1}}
\def\bkp{{\bm{k}_\perp}}

\def\bp#1{{\bm{#1}_\perp}}

\def\cc{{cosmological\ constant}}

\def\g{{\sqrt{-g}}}

\def\oms{{\Omega_m(t_*)}}

\def\ors{{\Omega_R(t_*)}}

\newcommand{\LL}{Lanczos-Lovelock }

\newcommand{\df}{\delta}

\def\dqg{density of states of the quantum spacetime}

\def\md{microscopic degrees of freedom of the spacetime}

  \title{Gravity and Quantum Theory: Domains of Conflict and Contact}
  \author{T. Padmanabhan\\
  IUCAA, Pune University Campus,\\
  Ganeshkhind, Pune- 411 007.\\
  {\small {email: paddy@iucaa.in}}
  }

  \date{}  %% This command  will supress printing the date.   If date is required, comment out this line.

\begin{document}
  
  \maketitle
  
  \begin{abstract}
There are two strong clues  about the quantum structure of spacetime and the gravitational dynamics, which are almost universally ignored in the conventional approaches to quantize gravity. The first clue is that null surfaces exhibit (observer dependent) thermal properties and possess a heat density. This suggests that spacetime, like matter, has microscopic degrees  of freedom and its long wavelength limit should be described in thermodynamic language and not in a geometric language. Second clue is related to the existence of the cosmological constant. Its understanding from first principles will require the dynamical principles of the theory to be invariant under the shift $T^a_b \to T^a_b + \text{(constant)} \delta^a_b$. This puts strong constraints on the nature of gravitational dynamics and excludes metric tensor as a fundamental dynamical variable. In fact, these two clues are closely related to each other. When the dynamical principles are recast, respecting the symmetry $T^a_b \to T^a_b + \text{(constant)} \delta^a_b$, they automatically acquire a thermodynamic interpretation related to the first clue. The first part of this review provides a pedagogical introduction to thermal properties of the horizons, including some novel derivations. The second part describes some aspects of cosmological constant problem and the last part provides a perspective on gravity which takes into account these principles. 
\end{abstract}
 \tableofcontents

 \vfill\eject

 \section{Two Clues about the Quantum Spacetime (And Two Historical Calamities)}\label{sec:intro}

 \begin{flushright}
  \obeylines{
\textit{
 ``While there has been a lot of very  interesting  
   and imaginative work done .........  it is  safe 
   to say that nothing which could definitely be
   called progress has been accomplished.''\phantom{.......}
 }
 }
 
 \vskip 0.1in
 --- Lee Smolin \cite{leesmolin}

 \end{flushright}
 
\noindent The above statement of Lee Smolin \cite{leesmolin} in 1979, about the status of research in quantum gravity,  remains even more valid today! Given the fact that decades of attempts to combine quantum theory and gravity have not led to any concrete progress, it is  worthwhile to take a step back and ask: \textit{Where have we gone wrong in our approaches?} I will argue that, in the conventional approaches,  we have failed to take cognizance, up-front, of two important clues about the quantum structure of spacetime which Nature has provided us \cite{tpcourses}.
 
 The quantum microstructure of normal matter --- viz., that matter has microscopic degrees of freedom in the form of atoms and molecules, which can store energy --- can be \textit{deduced}  without ever probing the matter at angstrom scales. This is precisely what Boltzmann did through the guiding principle: ``If  it is hot, it must have microstructure''. The  fact that matter can be hot  gives a strong clue about the existence of the quantum microstructure of matter. Boltzmann's vision interpreted heat as microscopic motion, thereby overthrowing a large number  of convoluted ideas regarding heat. 
 
 The situation is identical as regards the microstructure of spacetime though, due to a historical accident, it has not been recognized as such.  Just by using special relativity and quantum field theory in \textit{inertial} coordinates,
 one can deduce that the null surfaces in the spacetime can acquire  a temperature \cite{davies-unruh}  given by: 
 \begin{equation}
  k_BT = \frab{\hbar}{c}\, \frab{\kappa}{2\pi}
  \label{clue1}
 \end{equation} 
 where $\kappa$ is related to the properties of the null surface (see Sec. \ref{sec:horterm}). This result implies that spacetime, just like normal matter, can be hot; so invoking the Boltzmann principle, we can conclude that there must exist \md. It also implies that all thermodynamic notions --- like those of entropy, temperature etc. --- are observer/foliation dependent. 
 We  learn three lessons from these facts: 
 
 \begin{enumerate}
  \item 
 Spacetime, like matter, is made of large number of microscopic degrees of freedom. The description of the dynamics of spacetime must recognize this and --- at macroscopic scales --- the description should  be presented in a purely thermodynamic language (rather than in geometrical language) just as in the case of normal matter. 
 \item
 This description will necessarily be observer/foliation dependent --- i.e., it will involve, say, vector fields related to the observer or the foliation --- if we have  to express geometrical quantities in terms of thermodynamic quantities. Any insistence that the description should be in a  geometrical,   generally covariant, language is totally misplaced and will take us in the wrong direction. 
 \item
 The general covariance will reappear in  --- and only in\footnote{This leads to \textit{new} consequences. The equivalence of the descriptions  in thermodynamic and geometrical languages  will hold only at scales where thermodynamic description itself is valid. That will break down near Planck scales, which is to be expected. But it can also break down at very large cosmic scales where the \md\ might not have reached thermodynamic equilibrium. So general covariance can break down at very large cosmic scales, thereby selecting a preferred frame in which CMBR is isotropic  (see Sec. \ref{sec:altpara}).} --- the context of thermodynamic equilibrium of \md. This arises when we demand that the basic principles must hold for all observers/foliations;  but, the principles themselves, will depend on the foliation. For example, an equation $A_j = B_j$, which relates two generally covariant, geometrical variables can be equivalently stated as the following principle: $A_j u^j = B_j u^j$ for \textit{all} observers with four-velocities $u^j$. In general, quantities like $A_ju^j, B_j u^j$ will have observer dependent interpretation in thermodynamic language which translates into generally covariant geometrical statements like $A_j = B_j$.
 \end{enumerate}

 Let me present a dramatic expression of this point of view. I would claim that the natural (yes, natural!) description  of the ground state of quantum gravity --- made of approximately flat spacetime  in the long wavelength limit, and the inertial vacuum state for the quantum fields ---  is  in terms of a \textit{finite temperature} spacetime metric (in the Euclidean sector) given by:
 \begin{equation}
  ds^2 = (1 + 2\pi T x)^2 d\theta^2 + dx^2 + d\bp{x}^2
  \label{drama}
 \end{equation} 
 where $\theta$ is an angular variable with period $(1/T)$. If you analytically continue this metric to the Lorentzian sector and do QFT on it, you will find that the field quanta are thermally distributed at temperature $T$. This is consistent with the notion that the spacetime (in general) is endowed with a temperature $T$ just like, say, a solid at finite temperature. Of course, one can take the $T\to 0 $ limit to get the conventional description of spacetime but that would be as special as the study of zero temperature solids.
 
 The correct point of view is to think of horizon temperature as the \textit{temperature of the spacetime itself}, rather than as the temperature of the quantum fields residing in that spacetime.
  Though this is partially semantics, there are strong reasons to think of spacetime itself being hot:
(a) The temperature of the horizon is independent of any property of the quantum fields in the spacetime. It is therefore more natural to think of fields
acquiring the temperature of spacetime, just as material kept in a microwave oven acquires oven's temperature.
(b) Gravitons  (even in flat spacetime) will also exhibit thermal properties with the same temperature; so you don't really need to
invoke non-gravitational physics to \textit{define} the temperature.
(c) We will see later on (see Sec \ref{sec:obdeent}) that the surface term in gravitational action, when evaluated on a null surface will
give its heat density; this provides an even deeper link between spacetime dynamics and horizon thermality. 
 
 I stress that the result in \eq{clue1}  could have been discovered by just using special relativistic, inertial frame QFT;  one does not really need notions like accelerated observers, Rindler quantization, Rindler vacuum etc. to obtain this result.
 (I will demonstrate this in Sec. \ref{sec:tdlee} and Sec. \ref{sec:inpro}.). If only this result has been discovered \textit{before} the development of general relativity, one would have realized that \eq{clue1} is telling us something about the microscopic nature of spacetime rather than it having something to do with QFT in curved spacetime. One would have then  tried to develop a thermodynamic description of the evolution of large number of \md --- rather than use a  geometric description for the evolution of the metric tensor. Just as the Boltzmann principle is more fundamental than any specific equation of state describing a particular gas, the \eq{clue1} is more fundamental than any field equation of gravity.
 
 But, alas, a historical accident prevented physicists from moving in that direction. General relativity, black hole solutions and --- most importantly --- the thermodynamics of black hole were discovered first; as a consequence, the thermodynamic description was added as a superstructure to the underlying geometric description of spacetime. People started asking the \textit{\textbf{wrong}} question:
 
 \medskip
 \textit{``How come spacetime geometries describing objects like black holes  exhibit thermodynamic features?''}, 
 \medskip
 
 \noindent instead of asking the \textbf{\textit{right}} question: 
 
 \medskip
 \textit{``How come the thermodynamic limit of \md\ lends itself to a description in terms of  pseudo-Riemannian geometry?''. }
 \medskip
 
\noindent As a consequence of this historical accident, all conventional approaches to quantum gravity ignores the fact that the result in \eq{clue1} provides a fundamental clue about the nature of spacetime; most of the approaches concentrates on \textit{special cases} of \eq{clue1} --- in the context of black holes/de Sitter spacetimes etc--- and tries to interpret these  in a quantum language. It is no wonder that all these approaches to quantum gravity --- which are as convoluted as the calorific theory of heat --- remain resounding failures. I believe progress can only occur when \eq{clue1} is recognized as a fundamental statement about the \md\ and one shifts from a geometric to thermodynamic language for the description of the  macroscopic spacetime.

  The fact that gravitational dynamics can indeed be expressed entirely in thermodynamic language is extremely important for the point of view I am advocating here regarding \eq{clue1} and \eq{drama}. Most people think of
 the result in \eq{clue1} as an intriguing  curiosity about QFT in the presence of horizons, because this result can be derived (see Sec. \ref{sec:horterm}) directly from QFT without ever introducing \md. So the idea of linking a result, obtainable from QFT in \textit{continuum} spacetimes with some unknown \md\ --- and writing flat spacetime as a finite temperature system --- might appear rather far-fetched. 
 
 This objection would have been true except for the fact that all of gravitational dynamics renders itself to a description in thermodynamic language involving temperature, entropy, equipartition of microscopic degrees of freedom etc. \cite{A19}. It is this feature which suggests that, in the correct analysis, one must interpret \eq{clue1} as a fundamental statement about \md. In this sense, the situation is very similar to what happened in the context of normal matter; temperature can be defined and used in the continuum limit but it cannot be understood without discrete microscopic degrees of freedom.
 
 To avoid misunderstanding, let me also stress that  \textit{it is totally insufficient to derive an equation $G_{ab} = \kappa_g T_{ab}$ from some kind of thermodynamic considerations}. In such approaches, we are still describing the spacetime evolution in a geometric language which is incorrect; what we need is a reinterpretation of both sides of the equation,  $G_{ab} = \kappa_g T_{ab}$, entirely in thermodynamic language. The latter sections of this review will show you how this can be achieved.  
 
 Strong support for this point of view comes from the following fact as well. The description of spacetime dynamics in the language of thermodynamics is not a special feature of Einstein gravity. It works seamlessly for all \LL\ models of gravity (\cite{ll}; for a review, see \cite{dktpll}). This is very important because some of the  approaches to establish the connection between thermodynamics and gravity crucially depends upon the entropy being proportional to horizon area (like e.g., models based on entanglement entropy). In \LL\ models of gravity, entropy is \textit{not} proportional to the area; nevertheless the thermodynamic description continues to hold in almost identical fashion. This tells you that the connection is fairly deep, transcends specific field equations and is related to \md. It is similar to the fact that thermodynamic description of matter does not care what kind of matter one is dealing with. (I will not discuss the extension to \LL\ models in this review; but for a sample of references, see Refs. \cite{dktpll,llwork,scbook,ayan}).  
 
 A second strong clue about dynamics, which helps us in arriving at  a correct description,  is provided by the following fact: Observations suggest that the nature is described by \textit{four} fundamental constants, $c, \hbar, A_P, A_\Lambda$. Of these, $c$ has the dimension of velocity, $\hbar$ has the dimension of action and the other two constants have the dimensions of area; these last two are  related to  more familiar Newton's constant $G$ and the cosmological constant $\Lambda$ by (for a review of \cc\ see e.g. \cite{ccreviews,cc1})
 \begin{equation}
  A_P \equiv 4\pi L_P^2 \equiv 4\pi \frac{G\hbar}{c^3}; \qquad A_\Lambda \equiv 4\pi R_\Lambda^2 \equiv \frac{1}{2\Lambda}
 \end{equation} 
 These two areas appear in the 
  standard action principle for gravity, expressed in natural units in dimensionless form:\footnote{The Newton's law of gravity should be presented as $F=(c^3A_P/\hbar)(m_1m_2/r^2).$ Gravity is inherently \cite{inherentqm} quantum mechanical --- this equation blows up if you take $\hbar\to0$ limit --- just as matter is inherently quantum mechanical --- atoms will collapse if you take $\hbar\to0$. Gravity is a macroscopic phenomenon due to large collection of atoms of spacetime.} 
 \begin{equation}
  \mathcal{A}_{\rm grav} = \frac{1}{4 A_P} \int d^4x \, \sqrt{-g}\, \left( R - \frac{1}{A_\Lambda}\right) = \frac{1}{4} \int \frac{\sqrt{-g}d^4x}{A_P^2} \left( A_P R -  \frac{A_P}{A_\Lambda}\right)
  \label{clue2}
 \end{equation} 
 Observations also tell us that the dimensionless parameter $A_P/A_\Lambda$ is non-zero but extremely tiny: 
 \begin{equation}
 \frac{A_P}{A_\Lambda} \approx 10^{-123}.
 \label{ccvalue}
 \end{equation}
 
 This tiny value worries people but the \textit{real issue is more serious}\cite{tpap,A19}. The gravitational action in \eq{clue2} breaks a symmetry present in matter Lagrangian: The equations of motion for matter are invariant under the addition of a constant to the Lagrangian through the transformation $L_m \to L_m - \rho_0$. But,  the gravitational action couples to this $\rho_0$ (through the innocent looking $\sqrt{-g}$ factor in the volume measure) and hence the gravitational field equations are not invariant under this transformation. The change $L_m \to L_m - \rho_0$ changes the energy-momentum tensor of matter by $T^a_b \to T^a_b -\rho_0  \delta^a_b$ and hence a field equation like $G^a_b = \kappa_g T^a_b$ will not remain invariant.
 This implies that the observed value of $A_P/A_\Lambda$ has no invariant meaning; it could be altered by the addition of a constant $\rho_0$ \textit{to the matter sector}. 
 
 It seem natural, therefore, to demand that gravity must respect the symmetry present in the matter sector under the shift $L_m \to L_m - \rho_0$. So,  gravity cannot be correctly described either by an action like the one in \eq{clue2} or by a field equation like $G^a_b = \kappa_g T^a_b$. 
 The correct variational principle, which determines the dynamics of gravity, has to remain invariant under the transformation $T^a_b \to T^a_b + (\text{constant})\delta^a_b$. 
 
 This recognition that gravity breaks a symmetry present in the matter sector and the demand that, in the correct description of spacetime, it should not, constitutes the second most important clue we have about the nature of gravity. The demand that the field equations should remain invariant under the transformation $T^a_b \to T^a_b + (\text{constant})\,  \delta^a_b$ is a very strong clue about the nature of gravitational dynamics. In fact, one can show that such an invariance cannot be achieved in any extremum principle based on a local, generally covariant Lagrangian in which metric is varied as a dynamical variable. We shall see that the correct, thermodynamic, extremum principle has a completely different structure and interpretation. 
 
 As I will argue in Sec. \ref{sec:classgrav}, Einstein would have actually arrived at an alternate description of field equations if he had used this clue as a guiding principle for gravitational dynamics. Instead of demanding $G^a_b = \kappa_g T^a_b$, he would have demanded that the relation $G^a_b \ell_a\ell^b = \kappa_g T^a_b \ell_a\ell^b$ should hold for all null vectors $\ell_a $ in the spacetime.
 We again see that an equation involving additional vector field $\ell_a$ is preferred over the generally covariant equation $G^a_b = \kappa_g T^a_b$ for incorporating this principle. 
 
 Once again, due to a historical accident, the Hilbert action was discovered \textit{before} the introduction of the \cc\ into the fray. Even when Einstein introduced $\Lambda$, it came in and was abandoned, both for  wrong reasons. The fact that the addition of a constant to matter Lagrangian introduces a $\Lambda$, \textit{thereby breaking a symmetry} present in the matter sector  never seems to have been appreciated, due to the manner in which the historical developments took place. 
 
 More importantly, none of the standard approaches to quantum gravity incorporates this second clue about dynamics as a fundamental principle.\footnote{The conventional view is to think of $G$ and $\Lambda$ as two low energy coupling constants in the theory. The RG paradigm, so successful in non-gravitational context, will then tell us that these two coupling constants will run. On the other hand, if one thinks of spacetime as analogous to a fluid/solid, then $L_P$ and $R_\Lambda$ will be like the physical lattice separation and the size of the solid; they do not run in condensed matter physics.}
 In fact, all these approaches treat the metric tensor (or some equivalent geometrical structure) as  a dynamical variable in the theory and hence misses the bus. So, we have been trying to put together the principles of quantum theory and gravity, after ignoring two most useful clues available to us!
 
 The two clues about quantum spacetime I have introduced --- viz., null surfaces are hot and \cc\ `problem' demands a solution --- are not supposed to have any connection with each other in the conventional way of thinking about gravity. So it is remarkable  --- and very encouraging --- to find that these two issues are intimately related. The solution to \cc\ problem demands that the field equation should have the form $G^a_b \ell_a\ell^b = \kappa_g T^a_b \ell_a\ell^b$ (so that it is invariant under the transformation $T^a_b \to T^a_b + (\text{constant})\,  \delta^a_b$.
 As we shall see, the two sides of this equation have the natural thermodynamic interpretation as the heat density contributed by gravity and matter on a null surface which acts as a horizon. We will see later (see Sec. \ref{sec:zero-diss}),  that this equation has a physical interpretation, unlike the usual field equation $G_{ab}=\kappa_g T_{ab}$.
 
 To probe the alternative description deeper, these  two clues need to be supplemented by some procedure for introducing the discrete \md. It is widely believed that such discrete features of spacetime will play a crucial role at scales close to the Planck length $L_P$.
 This is indeed true and if one introduces $L_P$ as the zero-point-length in spacetime --- in a well-defined manner, as described in Sec. \ref{sec:zpl} --- it allows us to count, $\rho_g$, the number of \md. Using an extremum principle based on $\Omega\propto\rho_g \rho_m$ where $\rho_m$ is the corresponding number for matter, one can indeed obtain the gravitational field equation which remain invariant under the shift $T^a_b \to T^a_b + (\text{constant})\,  \delta^a_b$ (see Sec. \ref{sec:grfratoms}). All the principles I have outlined above --- horizon thermality, observer dependent thermodynamics, invariance of field equations under  $T^a_b \to T^a_b + (\text{constant})\,  \delta^a_b$, description of spacetime evolution in a purely thermodynamic language --- get synthesized in this particular approach.
 
 There is something more:
In any such approach, the actual numerical value of the cosmological constant in \eq{ccvalue}  has to come as a relic of quantum structure of spacetime. I will show that this is indeed the case and one can, in fact, determine \cite{tphpplb} the precise numerical value \eq{ccvalue} in this approach, thereby providing a new perspective on what is considered the most intriguing problem in theoretical physics.\footnote{\textit{Notation etc.:} The signature is $(-,+,+,+)$ unless otherwise specified. I use natural units with $c=1, \hbar =1$ and set $\kappa_g = 8\pi G =8\pi L_P^2$ where $L_P$ is the Planck length $(G\hbar/c^3)^{1/2}=G^{1/2}$ in natural units.  In many expressions, we will also set $G=1$ making $L_P^2 =1$. Latin letters $i, j$ etc. range over spacetime indices and the Greek letters $\alpha, \beta$ etc. range over the spatial indices. I will write $x$ for $x^i$, suppressing the index, when no confusion is likely to arise.}

 \section{From the Euclidean Origin to the Lorentzian Horizon}
 
 Since gravity affects the propagation of light rays, it  determines the causal structure of spacetime. In particular, it is usual for  spacetimes to contain  regions  which cannot communicate with  adjacent regions, because they are separated by a null surface $\mathcal{H}$ which acts as  a horizon. This horizon  limits the vision of observers confined to one of the regions, say, $\mathcal{R}_2$, preventing them from accessing information from the region, say, $\mathcal{R}_1$.
 
 This classical feature turns out to be of major significance when quantum effects are brought in. The horizon $\mathcal{H}$ then acquires \textit{observer dependent}\footnote{These thermodynamical features are observer dependent in \textit{any} curved spacetime,  in the sense that it is certainly possible for observers to venture from $\mathcal{R}_2$  to $\mathcal{R}_1$, thereby modifying/eliminating the information blockage and the thermodynamical properties they perceive.}  thermodynamical properties like temperature, entropy etc. vis-a-vis the physical systems (and observers) confined to $\mathcal{R}_2$. 
 A characteristic feature of a class of  spacetimes (like black hole spacetimes, de Sitter universe etc.) which exhibit  horizon thermodynamics is the following: They all possess a timelike Killing vector field $\bm{\xi}_\tau$  in the region $\mathcal{R}_2$ such that the integral curves of this vector field asymptotically approaches the horizon $\mathcal{H}$. In general, these spacetimes will \textit{not} possess any other timelike Killing vector field in the region $\mathcal{R}_2$.
 
  Since horizon thermodynamics is always observer dependent, a natural question to ask would be whether curvature of the spacetime plays any role in this phenomenon. That is, can this effect  arise even in the \textit{flat} spacetime with respect to some specific region $\mathcal{R}_2$ and observers confined to that region? If so, it would be interesting to identify the relevant  timelike Killing vector $\bm{\xi}_\tau$ in such a context. 
  
  In flat spacetime, described in Cartesian coordinates $(t,x, \bp{x})$, where $\bp{x}$ denotes $(D-2)$ transverse coordinates, we can easily list all the Killing vectors describing the symmetries of the $t-x$ plane. The natural time translation symmetry in flat spacetime is along the $t-$direction and is described by the timelike Killing vector field $\bm{\xi}_t = \partial/\partial t=(1,0,\bm{0})$.  Observers traveling along the integral curves of  $\bm{\xi}_t$ (parametrized by $t$ with $(-\infty < t< \infty)$ ) can receive information from all regions of the spacetime when we allow for the full range of $t$ for the integral curves. These observers do not see a  horizon. Of course, we also know from the study of standard QFT in flat spacetime that nothing peculiar happens as regards evolution with respect to this Killing vector.
  
  The $t-x$ plane, however, has another symmetry corresponding to the Lorentz boost along the $x-$axis, described by the Killing vector field $\bm{\xi}_\tau \equiv \partial/\partial \tau  = (x, t, \bm{0})$. This Killing vector is timelike in the region  $|x| > |t|$ and is naturally parametrized by the rapidity $\tau=\tanh^{-1} V_x$, where $V_x$ is the boost velocity.   Varying $\tau$ corresponds to \textit{viewing the same physical system in a sequence of Lorentz frames}, connected to the original one by a series of Lorentz boosts. The set of all Lorentz boosts in $t-x$ plane is covered by the range $(-\infty < \tau< \infty)$ of the rapidity parameter.

  So we have two timelike vector fields, $\bm{\xi}_t$ (corresponding to time translation symmetry) and $\bm{\xi}_\tau$ (corresponding to Lorentz boost symmetry) in this region $|x|>|t|$. One would have thought that, the $t$ coordinate gives the natural time direction for dynamical evolution while the `evolution' along $\tau$ is just a proxy for how the system appears in a sequence of Lorentz transformed frames. But,
  it is easy to see that observers moving along the integral curves of $\bm{\xi}_\tau$ (which are hyperbolas) will not have access to information about the  entire spacetime. For example,  such an observer in the region 
   $x > |t|$ will perceive the null surface $x=t$ as a horizon $\mathcal{H}^+$, limiting the information accessibility. As a result, it turns out that  the surface $\mathcal{H}^+$, which is just a garden-variety null plane in flat spacetime, will  exhibit thermodynamical properties vis-a-vis the region $x>|t|$. (Similar results hold with respect to the region $(-x)>|t|$ with the null plane $x=-t$ acting as the horizon $\mathcal{H}^-$ but we will concentrate on the region $x > |t|$.) Obviously all such null planes in flat spacetime will possess thermodynamical features vis-a-vis a corresponding set of observers. This shows that the horizon thermodynamics does not requires curvature of spacetime and is just an inherent property of the null surfaces.
   
  More surprisingly,  it is the boost Killing vector $\bm{\xi}_\tau$ which generalizes in a natural fashion to a wide class of curved spacetimes which exhibit horizon thermodynamics. In fact, the existence of a Killing vector analogous to  the  boost Killing vector turns out to be a very general feature of the class of curved spacetimes we would be interested in. The integral curves of $\bm{\xi}_\tau$ defines a natural class of observers who attributes thermodynamical properties to the horizons in these spacetimes.   In contrast,  the time translation Killing vector $\bm{\xi}_t$ of the flat spacetime has no natural counterpart in more general spacetimes. 
   Given this background, we will begin by describing the role of boost Killing vector in flat spacetime and its generalization to a  class of curved spacetimes with horizons.

 \subsection{Boost invariance in flat spacetime}
 
 Surprisingly enough, it turns out to be more appropriate (and convenient) to think of Lorentzian geometries as those obtained by analytic continuation from Euclidean geometries. (This is a theme which will run through our entire discussion.) In view of this fact, let us begin by considering an \textit{Euclidean} flat spacetime described in the Cartesian coordinates ($t_E , x_E, \bp x$) as well as  in the polar coordinates ($\theta, \rho, \bp x$) introduced in the 
  $t_E-x_E$ plane.  We will concentrate on this plane for most of our discussion; the transverse coordinates $\bp x$ will just go for a ride.
 The line element is:
 \begin{equation}
  ds^2 =  dt_E^2 + dx_E^2 + d\bp x^2 =  \rho^2 d\theta^2 + d\rho^2 + d\bp x^2
  \label{eq1}
 \end{equation} 
  The Euclidean 2--plane has three obvious symmetries: translation along $t_E$, rotations about the origin corresponding to ``translation'' in the angular parameter $\theta$ and translations along $x_E$. Of these, we will take a closer look at the first two. 
 
 Translation along $t_E$ is described by the Killing vector $\bm{\xi}_t = \partial/\partial t_E$ while the rotation in the plane  is described by the Killing vector $\bm{\xi}_\theta = \partial/\partial \theta$. The integral curves of $\bm{\xi}_t$ are straight-lines parallel to $t_E$ axis while the integral curves of $\bm{\xi}_\theta$ are circles centered at the origin. These circles, having the equation $x_E^2 + t_E^2 = \rho^2$, can be parameterized using an angular variable  $\theta$ in many different ways, of which, we will first concentrate on the following \textit{two}  parameterizations:
 \begin{align}
  &x_E = \rho \cos \theta; \ t_E= \rho \sin\theta; \qquad  x_E = -\rho \cos \theta; \ t_E= -\rho \sin\theta \label{paramet1}
  \end{align}
 The first one is standard; the second one is obtained by replacing $\theta\to\pi+\theta$. Since the polar metric in \eq{eq1} is invariant under translation in $\theta$ both these transformations lead to the same form of polar metric.
 
 Let us now analytically continue from the Euclidean plane to the Lorentzian plane, thereby introducing one coordinate with negative signature in \eq{eq1}. In the Euclidean plane, nothing distinguishes $x_E$ from $t_E$. So we could have replaced $t_E$ by $it$ or $x_E$ by $ix$ and could have obtained a Lorentzian metric with one negative eigenvalue. In the first case, $t_E$ would have become a timelike coordinate in the Lorentzian sector while in the second case $x_E$ would have become a timelike coordinate in the Lorentzian sector. While the notation might appear a bit strange, this is a perfectly valid procedure because, as I said before, the Euclidean plane treats $x_E$ and $t_E$ in an equal footing. 
 
 Let us start with 
  the replacements $t_E \equiv it, x_E=x,  \theta \equiv i\tau$ in \eq{paramet1}. 
  The time translation symmetry generated by $\bm{\xi}_t$ seamlessly translates to the time translation symmetry along the $t-$direction in the Lorentzian plane. (So I have used the same symbol $\bm{\xi}_t$ for the time translation Killing vectors in both cases). On the other hand, the rotational symmetry generated by $\bm{\xi}_\theta$ takes a more complicated avatar in the Lorentzian plane. 
  
  To understand it, we need to recognize \textit{the} most dramatic effect of the analytic continuation: The origin of Euclidean plane (viz., the solution to $x_E^2+t_E^2=0$)  becomes two null planes $x=\pm t$ (viz., the solution to $x^2-t^2=0$) in the Lorentzian sector. 
  The null planes $x=\pm t$  naturally divide   the Lorentzian $t-x$ plane into four wedges which we will denote by R(ight), F(uture), L(eft) and P(ast). Any single integral curve of $\bm{\xi}_\theta$, which was a circle of radius $\rho$ given by $x_E^2 + t_E^2 = \rho^2 $, now becomes \textit{two} hyperbolas $x^2 - t^2 = \rho^2$ in R and L wedges. This is obvious from the analytic continuation of \eq{paramet1} into the  wedges R and L as given below:
 \begin{align}
  (R)\quad x = \rho_R \cosh \tau_R ; \ t= \rho_R  \sinh\tau_R ; \quad  (L)\quad x = -\rho_L \cosh \tau_L; \ t= -\rho_L \sinh\tau_L \label{paramet11}
 \end{align}
 where we have added the subscripts R,L to indicate which of the wedges to which the hyperbolas belong. 
 In the Euclidean sector one could restrict $\theta$ to be in the range $(0\le \theta < 2\pi)$ since the angular variable is  periodic. This particular range will map to the range $(0, 2\pi)$ in $\tau$, which, of course, will cover only part of the hyperbolas in either of the wedges. To get the full integral curves corresponding to  $\bm{\xi}_\tau \equiv \partial/\partial \tau$ we need to use the full range $(-\infty <\tau<\infty)$ in \textit{each} of the wedges. 
 The \eq{paramet11}  also provide two \textit{coordinate transformations} from $(t,x)$ to $(\tau, \rho)$ in R and L wedges.\footnote{It is usual to call the $(\tau-\rho)$ coordinate system as the Rindler coordinate system in R and L. The integral curves of boost Killing vector in R and L corresponds to trajectories of constant acceleration. In our parametrization, the magnitude of the acceleration is unity but this can be set to any value $\kappa$ by rescaling $\tau\to\kappa\tau$ in \eq{paramet11} and \eq{paramet22}. We will introduce $\kappa$, by this scaling,  when required.} 
 In R and L the line element now has the form given by
  \begin{equation}
  ds^2 = - dt^2 + dx^2 + d\bp x^2 = - \rho^2 d\tau^2 + d\rho^2 + d\bp x^2
  \label{le1}
 \end{equation}
  
The fact that a \textit{point} in the Euclidean sector (viz.  $x_E^2+t_E^2=0$) becomes the \textit{surfaces} (viz. $x^2-t^2=0$) is nontrivial --- in spite of its apparent simplicity. One unusual consequence of this result is that,  if  we proceed in the reverse direction --- i.e., from the R wedge in the Lorentzian sector  to Euclidean sector, again using $t_E = it, x_E=x$ --- the null planes \textit{and the region beyond them} `collapse' to the origin of the Euclidean plane. We will see that the same features arise in a large class of curved spacetimes. The correspondence between the Euclidean origin and Lorentzian horizons will be of crucial importance in our discussions. 
 
 The situation is slightly more complicated in F and P wedges. This can be seen from the fact that when you analytically continue $t_E=it, \theta=i\tau$, in \eq{paramet1} to obtain \eq{paramet11} we end up with the constraint  $x^2-t^2>0$; so this analytic continuation works only in R and L. To take care of F and P, let us start with two more parameterizations of the circle in the Euclidean plane, given by:
 \begin{align}
  &x_E = \rho \sin \theta; \ t_E= \rho \cos\theta; \qquad  x_E = -\rho \sin \theta; \ t_E= -\rho \cos\theta \label{paramet2}
 \end{align}
 These are obtained from \eq{paramet1} by replacing $\theta$ by $(\pi/2)-\theta$ which, of course, is another valid parametrization. Comparing with \eq{paramet1} we see that the riles of $t_E$ and $x_E$ are reversed. Of course, in the Euclidean plane, there is nothing that distinguishes `time' from `space' or $t_E$ from  $x_E$.  So, instead of analytically continuing in $t_E$ one could have equally well analytically continues from $x_E$ to $ix$, keeping $t_E=t$. This would have mapped the circle $x_E^2+t_E^2=\rho^2$ to a pair of hyperbolas 
 $x^2-t^2=-\rho^2$ in the F and P regions. In \eq{paramet2}, the analytic continuation $\theta=i\tau, x_E=ix, t_E=t$ now gives the two 
 transformations, similar to \eq{paramet11} with $\sinh$
and $\cosh$ factors interchanged: 
\begin{align}
 (F)\quad x = \rho_F \sinh \tau_F; \ t= \rho_F \cosh\tau_F; \quad  (P)\quad x = -\rho_P \sinh \tau_P; \ t= -\rho_P \cosh\tau_P \label{paramet22}
 \end{align}
 In F and P the corresponding line element has the signs of  of $d\tau^2$ and $\rho^2$ reversed:
 \begin{equation}
  ds^2 = - dt^2 + dx^2 + d\bp x^2 = + \rho^2 d\tau^2 - d\rho^2 + d\bp x^2
  \label{le2}
 \end{equation} 
 Obviously $\tau$ behaves like a time coordinate only in R and L and it is $\rho$ which acts like a time coordinate in F and P.

 We see that the hyperbolas in all the four Lorentzian sectors can be included if we take their equation to be $x^2-t^2=2\xi$ where $\xi$ can be now positive or negative. In R and L, we take $\xi\equiv\rho^2>0$; in F and P we take $\xi\equiv-\rho^2<0$. It is now possible to express the line element in R \textit{and} F together in a different coordinate system $(\tau,\xi,\bp x)$, in the form:
 \begin{equation}
  ds^2 = - dt^2 + dx^2 + d\bp x^2 = -2\xi\ d\tau^2 +\frac{d\xi^2}{2\xi} +d\bp x^2
  \label{le2rf}
 \end{equation} 
 Here $\xi$  varies over the entire real line; $\rho^2=2\xi$ for $\xi>0$ (covering the R wedge) and $\rho^2=-2\xi$ for $\xi<0$ (covering the F wedge).
 The sign flip in $\xi$ takes care of the reversal of the roles of $\tau$ and $\xi$ when we go from R to F.

The line element in all the four wedges is independent of $\tau$ which represents the original rotational symmetry of the Euclidean plane.
 The rotational symmetry in the Euclidean plane manifests in a very non-trivial manner in the Lorentzian plane.
 This symmetry, generated by $\bm{\xi}_\tau$, has a clear physical meaning in the R and L wedges: It corresponds to the Lorentz boost in the $t-x$ plane with rapidity $\tau$. (Recall that the velocity $V_x$ of the Lorentz boost is related to the rapidity $\tau$ by $V_x= \tanh\tau$.) The set of all Lorentz transformations along the $x-$axis `spans' the R and L wedges in the sense that the hyper-plane to which $\bm{\xi}_t$ is orthogonal, rotates from the past light cone $x=-t$ (when $\tau = -\infty$) to the future light cone $x= t$ (when $\tau = +\infty$). For a Lorentz transformation with a given rapidity, the coordinate $\rho$ measures the invariant interval $x^2 - t^2 = \rho^2$  and can be taken as the spatial distance on the $\tau=$ constant hyper-plane.
 
 We thus see that the two basic symmetries of the Euclidean plane --- viz., time translation and rotation about the origin --- maps to translation along  $t$ and translation along $\tau$ with both variables ranging over the entire real line. Given any Killing vector $\xi^a$ and a conserved energy momentum tensor $T^b_a$, one can construct a conserved four-momentum current $P^a \equiv T^a_b \xi^b$. From the local conservation law $\nabla_j P^j =0$ one can also construct a global conserved charge by integrating $P^j d\Sigma_j$ over a space-like hypersurface with volume element $d\Sigma_j$. When the Killing vector $\bm{\xi}$ defining $P^j$ is time-like in a given region, this charge can be interpreted as the Hamiltonian generating translations along the integral curves of $\bm{\xi}$. In our case, we have two Killing vectors $\bm{\xi}_t$ and $\bm{\xi}_\tau$ which are time-like in R (as well as in L but we will concentrate on R for the moment). The corresponding Hamiltonians
 \begin{equation}
  H_t \equiv \int d\Sigma_j \, T^j_k \, \xi^k_{(t)}; \qquad H_\tau \equiv \int d\Sigma_j \, T^j_k \, \xi^k_{(\tau)}
  \label{2H}
 \end{equation} 
 generate evolution along $t$ and $\tau$ through the operators $\exp (-i tH_t)$ and  $\exp (-i \tau H_\tau)$ respectively.
 There are, however, significant differences between the action of these two Hamiltonians. 
 
 First, the action of $H_t$ can take the $t=$ constant hyperplane, containing the physical system, throughout the spacetime when we let $t$ vary over its full range $-\infty<t<+\infty$. On the other hand, when $\tau$ varies over its full range $-\infty<\tau<+\infty$, the hyperplane $\tau=$ constant just rotates from the past light cone $x=-t$ (when $\tau = -\infty$) to the future light cone $x= t$ (when $\tau = +\infty$) and spans only the R and L wedges. So $H_\tau$ can only evolve a system within the R and L wedges. 
 
 Second, the translation along $t$ corresponds to genuine  evolution of the dynamical variables of the system which is generated by the action of $H_t$. On the other hand, translation along $\tau$ (generated by the action of $H_\tau$) merely represents viewing the same physical system from a different Lorentz frame obtained by boosting with velocity $V_x = \tanh\tau$.  For example,  consider the quantum state of a field at time $t=0$. Translation along $t$ direction will tell us how this state is evolving with time.  However, one can also ask how this quantum state appears in a sequence of Lorentz frames parameterized by the rapidity $\tau$. The variation of the state along $t$ is a genuine dynamical evolution, while the variation of the state along $\tau$ is  purely kinematical. 
 
 So, in flat spacetime, the $t$ coordinate gives the natural time direction while the `evolution' along $\tau$ is just a proxy for how the system appears in a sequence of Lorentz transformed frames. However --- somewhat surprisingly --- 
 the study of how different dynamical variables vary with $\tau$, as we view the system from  a sequence of Lorentz frames, reveals an interesting connection between null surfaces and thermodynamics. Because $\tau$ varies along the entire real line, one can think of it as an alternative time coordinate (in R and L) and use it to define a Fock space for the QFT.  This vacuum state and definition of particles in this Fock basis
   turns out to be inequivalent to the corresponding (standard) vacuum state and particle definition in a Fock basis constructed in a QFT based on $t$ coordinate. In particular, the standard vacuum state defined using evolution along $t$ contains a Planckian distribution of particles defined with respect to $\tau $ coordinate.
 
 More importantly, the boost invariance acquires larger significance in the study of a
 large class of other spacetimes describing black holes, de Sitter universe etc., 
 In all these spacetimes,
 a time coordinate analogous to the rapidity variable $\tau$ turns out to be more natural! In these spacetimes --- characterized by, what is known as, a bifurcate Killing horizon --- the vector analogous to $\bm{\xi}_\tau$ remains a Killing vector while the vector analogous to $\bm{\xi}_t$ ceases to be a Killing vector.
 In fact, the physics in these spacetimes turns out to be remarkably similar to the physics in flat spacetime described using $\tau$ coordinate. This provides an additional, strong, motivation for studying the flat spacetime example since virtually every result can be extended to curved spacetime in a straightforward manner. We shall next consider how the notion of Lorentz boost generalizes to curved spacetime.
 
 \subsection{Boost invariance in curved spacetime}
 
 It turns out that one can perform an identical analysis for a wide class spacetimes with a bifurcation horizon. To introduce these concepts in the simplest possible way, let us again start with  a \textit{Euclidean} curved spacetime with the line element of the form:
   \begin{equation}
 ds^2 = F(\rho) \, [dt_E^2 + dx_E^2] + dL_\perp^2; \quad \rho^2 \equiv x_E^2 + t_E^2
     \label{14a}
   \end{equation} 
   where  $dL_\perp^2=h_{AB}(\rho, y^A)dy^Ady^B $ gives the metric in the transverse direction, which, in general need not be flat.
   We will be interested in a class of metrics  which satisfies the following conditions: Near the origin of the $t_E-x_E$ plane (that is when $\rho\to0$), (a) $F(\rho) \to 1$  and (b) the transverse line element $dL_\perp^2 $ reduces to the flat form $d\bp{x}^2$. This implies that the coordinate system ($t_E,x_E, \bp{x}$) becomes a locally flat, Cartesian, coordinate system near the origin.
   
   The main characteristic of the curved geometry described by the line element in \eq{14a} is the following. Since $F$, in general,  depends on $t_E$ we do \textit{not} have translation symmetry in the time coordinate $t_E$ (unlike in the flat spacetime with $F=1$). However, we still have rotational invariance about the origin of the $(x,t_E)$ plane. 
   This ensures the existence of a Killing vector $\bm{\xi}_\theta \equiv \partial / \partial \theta$ where $\theta $ is the standard polar angle. The integral curves of this vector field are circles centered at the origin of $(x_E,t_E)$ plane. In this regard the coordinate system and the metric in \eq{14a} still resembles the flat Euclidean spacetime in Cartesian coordinates. To make the rotational invariance manifest we can introduce the standard transformation $x_E=\rho\cos \theta$, $t_E= \rho \sin \theta$, to the polar coordinates. Then the metric reduces to the form 
   \begin{equation}
  ds^2 = F(\rho)\, [\rho^2 d\theta^2 + d\rho^2] + dL_\perp^2
    \label{14b}
   \end{equation} 
   which is clearly invariant under translations in $\theta$ with the existence of a Killing vector $\bm{\xi}_\theta=\partial/\partial\theta$. One more coordinate transformation will reduce the line element to the form
   \begin{equation}
  ds^2 = f(\xi) \, d\theta^2 + \frac{d\xi^2}{f(\xi)} + dL_\perp^2
    \label{14c}
   \end{equation} 
   where $f(\xi)=\rho^2 F(\rho)$ and the coordinate $\xi$ is related to $\rho$ through the relations: 
   \begin{equation}
    \rho^2 \equiv  e^{2\xi_*}; \qquad \xi_* \equiv   \int \frac{d\xi}{f(\xi)} + \mathrm{const}
    \label{defrstar} 
   \end{equation}
   Alternatively, one can start with a class of geometries in \eq{14c} with the  property that the function $f(\xi)$ behaves as $f(\xi)\approx 2\xi$ near the origin. We then define $\xi_*$ --- called the tortoise coordinate --- through the second equation in \eq{defrstar} (so that $\xi_*\approx(1/2)\ln 2\xi$, with appropriate choice of constants, near $\xi=0$) and the $\rho$ through the first equation in \eq{defrstar} (so that $\rho^2\approx 2\xi$ near the origin). This reduces the metric in \eq{14c} to the form in \eq{14b} with $F=f/\rho^2$; clearly $F\to 1$ as $\xi\to0$. 
   
   As we did in the case of flat spacetime, we will now analytically continue from the Euclidean to the Lorentzian structure.
   Let us begin with the replacement $t_E\to it, x_E =x$ in \eq{14a} thereby producing the Lorentzian spacetime metric of the form:
   \begin{equation}
 ds^2 = F(\rho) \, [-dt^2 + dx^2] + dL_\perp^2 
 \label{14cc}
 \end{equation} 
  This coordinate system ($t,x,\bp{x}$) is analogous to  the global inertial coordinate system of the flat spacetime we introduced in the first part of \eq{le1}.
  The analytic continuation changes $\rho^2 = x_E^2 +t_E^2$ to $\rho^2 = x^2 - t^2$ though we will continue to use the same symbol $\rho $. We assume that $F(\rho)$ remains positive definite under this change so that $t$ retains its time-like character all throughout the manifold. 
   Just as in the case of the flat spacetime, the analytic continuation introduces the null surfaces where $\rho^2 = x^2 - t^2$ vanishes; that is, the origin in the Euclidean plane gets mapped to the null planes, $\mathcal{H}^\pm$ given by $x= \pm t$, in the Lorentzian sector, when we confine our attention to the two-dimensional subspace with $dL_\perp^2=0$.  
   Since the original Euclidean metric was flat near the origin (since $F\to 1$ near the origin), the line element in $(t,x,\bp{x})$ is locally inertial \textit{all along the vicinity of the horizons} $x= \pm t$. 
   Hence the coordinate system ($t,x,\bp{x}$) has the physical meaning as the locally inertial, freely falling frame (FFF), all along the vicinity of the horizons
    This fact will play an important role in our future discussions.
   
   The horizons $x= \pm t$ again divide the $x-t$ plane into four wedges just as in the flat spacetime. 
   The coordinate transformation to the polar coordinates in the Euclidean sector  can now be implemented in four different ways in the four different wedges exactly as in the case of flat spacetime through the equations \eq{paramet11} and \eq{paramet22}.
   In each of the four wedges one can introduce a metric similar to \eq{14c}. But note that, while $F(\rho)>0$ throughout the manifold (keeping $t$ timelike everywhere), the function $f(\xi)=\rho^2 F(\rho)=(x^2-t^2)F(\rho)$ now flips sign when we cross
   $\mathcal{H}^\pm$. 
   For example, in the right wedge (where $\rho^2>0$ so that  $f(\xi)>0$) we will now get 
   \begin{equation}
 ds^2 = F(\rho) \, [-dt^2 + dx^2] + dL_\perp^2 =  -f(\xi) \, d\tau^2 + \frac{d\xi^2}{f(\xi)} + dL_\perp^2 
 \label{14ccc}
 \end{equation} 
 with the $\tau$ retaining a timelike character in R. But in F, where $\rho^2<0$, we will have the same form of the metric but with $f(\xi)<0$. This is completely analogous to the situation described by the metric in \eq{le2rf}; in fact $f(\xi)\approx 2\xi$ near $r=0$, which is the horizon surface, and the metric in \eq{14ccc} reduces to the one in \eq{le2rf}.
 While $\bm{\xi}_\tau=\partial/\partial\tau$ is still a Killing vector, it is no longer timelike in F region. Similar comments apply to L and P wedges and the situation is completely analogous to what we saw in the flat spacetime example.
 
   Once we are back in the Lorentzian sector with light cones, it is convenient to introduce the null coordinates $u\equiv \tau-\xi_*$,  $v \equiv \tau+\xi_*$ (where $\xi_*$ is defined in \eq{defrstar}) in terms of which  the metric in \eq{14ccc} becomes 
   \begin{equation}
   ds^2 = F(\rho) \, \rho^2\, [-du\, dv] \, + dL_\perp^2
   \end{equation} 
   The condition that $F(\rho) \to 1$ near the origin of the Euclidean spacetime  tells us that this metric reduces to the Rindler form of the metric along the null surfaces $\mathcal{H}^\pm$. It is also convenient introduce another null coordinate system $(U,V,\bp{x})$ based on the  line element in \eq{14cc}, with $U, V$ related to $u,v$ by:
   \begin{equation}
   U =t-x= - e^{-u} ; \qquad V=t+x = e^v
    \label{notethree}
   \end{equation}
   which will reduce the line element to the form 
    \begin{equation}
   ds^2 = F(UV) \, [ - dU\, dV] + dL_\perp^2
   \end{equation} 
   Again, since $F \to 1$  on the two horizons $\mathcal{H}^\pm$, tells us that the coordinates ($U,V,\bp{x}$) represent a freely falling frame in the vicinity of $\mathcal{H}^\pm$. 
   Our different coordinate systems can be summarized as follows:
   \begin{equation}
    ds^2 = Fe^{2\xi_*} \left( -dt^2 + d\xi_*^2\right) + d{L_\perp}^2 = -Fe^{v-u}(du\, dv) + d{L_\perp}^2 
    =- FdU dV + d\bp{x}^2
   \end{equation} 
   where we have used the coordinate transformations 
   \begin{equation}
   \rho \equiv e^{\xi_*}; \quad u \equiv \tau - \xi_*; \quad v\equiv \tau + \xi_*;\quad
    U = - e^{-u} ; \quad V = e^v
   \end{equation} 
   One can think of these coordinate  transformations as relating the two forms of   the line elements in \eq{14ccc}: starting from the Rindler-like coordinates ($\tau,\rho, \bp{x} $) and transforming to the freely falling frame in the coordinates ($U,V,\bp{x}$).
   The sequence of coordinate transformations which accomplishes this is given by: 
   \begin{equation}
(\tau,\xi)\to (\tau,\xi^*)\to  (v,u) \to (V,U)\to (t,x)
\end{equation}
 Taken together, this is equivalent to:
 \begin{align}
  &(R)\quad x = e^{ \xi^*} \cosh \tau ; \ t= e^{ \xi^*} \sinh\tau ; \quad  (L)\quad x = -e^{ \xi^*} \cosh \tau; \ t= -e^{ \xi^*} \sinh\tau \label{paramet111}\\
  &(F)\quad x = e^{ \xi^*} \sinh \tau; \ t= e^{ \xi^*} \cosh\tau; \quad  (P)\quad x = -e^{ \xi^*} \sinh \tau; \ t= -e^{ \xi^*} \cosh\tau \label{paramet222}
 \end{align}
 in the four wedges. (These are identical to \eq{paramet11}, \eq{paramet22} except that we have dispensed with separate subscripts R,F etc. for the coordinates)  The full manifold is covered by $(t, x, \bp{x})$ coordinates while the each of the wedges is covered by $(\tau, \xi^*, \bp{x})$ coordinates.  In the case of Schwarzschild metric, for example, the standard Schwarzschild coordinates in the right wedge play the role similar of Rindler coordinates in flat spacetime. The time translation invariance of Schwarzschild metric due to the existence of the Killing vector $\bm{\xi}_\tau \equiv \partial/\partial \tau$ is a relic of the rotational invariance in the Euclidean spacetime encapsulated by   $\bm{\xi}_\theta \equiv \partial/\partial \theta$.

  The thermal features of horizons continue to hold in the curved spacetimes represented by metrics of the form in \eq{14cc}. These spacetimes clearly have a Killing vector $\bm{\xi}_\tau$ which generalizes the boost Killing vector in flat spacetime. 
   The form of the metric in \eq{14cc} can describe a host of spacetimes with a bifurcation horizon like, for e.g. black hole spacetime, de Sitter spacetime etc.  (The Rindler and inertial coordinate systems clearly correspond to the simple choice $F(\rho) =1$ leading to $f(\xi)=2\xi$.) It is now obvious that, translations in $\tau$ coordinate remains a symmetry while translations in $t $  coordinates is not. 
   The best one can do is to introduce a freely falling frame close to the vicinity of $\mathcal{H}^\pm$; in that FFF translations in $t$ coordinate will be an approximate symmetry. 
   
   Of the two Hamiltonians defined in \eq{2H}, the $H_\tau$ remains a valid evolution operator with a conserved (boost) energy as its eigenvalue; on the other hand, since there is no Killing vector analogous to $\bm{\xi}_t$ in the curved spacetime, the Hamiltonian $H_t$ does not correspond to a conserved charge and loses its significance. 
  Thus the  boost invariance of the flat spacetime generalizes to a very wide class of curved spacetimes which are relevant to us. 
   The importance of Lorentz boost and the study of variations with respect to $\tau$ in flat spacetime should be now obvious. We will now turn to this study from different perspectives.

 In proceeding from flat spacetime to curved spacetime we confined ourselves to a special class of metrics of the form in \eq{14a} with specific properties for $F(\rho)$. However, the ideas developed here have a far greater domain of validity and allows us to introduce an important notion of local boost invariance. I will conclude this section describing how this comes about. 
 
 Consider an arbitrary  event $\mathcal{P}$ in an arbitrary --- in general, curved --- \textit{Euclidean} space and choose a coordinate system $S$ such that $\mathcal{P}$ is at the origin of the coordinate system. Transform to  a locally flat 
 coordinate system $S_{\rm flat}$ around the origin with coordinates $(T_E,X, \bp{X})$. Such a coordinate system will be valid in a region of size $\mathcal{O}(L)$ where $L^{-2}$ is the typical magnitude of the background curvature at the origin. 
 The Euclidean plane $(T_E,X)$ can equally well be described in the polar coordinates $(\theta, \rho)$ and in the locally flat region it exhibits rotational invariance around the origin. The vector $\bm{\xi}_\theta = \partial/\partial \theta$ will be an \textit{approximate} Killing vector in the locally flat region. 
 We can now analytically continue from the Euclidean to Lorentzian spacetime exactly as in our previous discussion. The origin of the Euclidean plane will now become the horizons $\mathcal{H}^\pm$ in the Lorentzian sector and 
  $S_{\rm flat}$ will become a FFF  in the vicinity of $\mathcal{H}^\pm$. 
 One can now introduce  Lorentz boosts with different rapidities $\tau$ in the Lorentzian sector such that $\bm{\xi}_\theta$ maps to $\bm{\xi}_\tau$. In the process the \textit{approximate} rotational invariance in the Euclidean sector becomes an \textit{approximate} boost invariance in the Lorentzian sector. Circles arbitrarily close to the Euclidean origin --- where the validity of locally flat coordinate system becomes better and better --- will map to hyperbolas straddling arbitrarily close to the horizons $\mathcal{H}^\pm$. Such a mapping extends the validity of locally flat coordinate system in a \textit{compact} region in the Euclidean sector to a \textit{non-compact} region in the Lorentzian sector. I stress that this construction remains valid around any event in any spacetime. 
 
 With future applications in mind, we will also note the following aspect of this analytic continuation. Instead of considering a circle $X^2 + T_E^2 = \rho^2$ in the Euclidean plane, we can also consider a spherical surface $R^2 + T_E^2 = \rho^2$ (where $R^2 = X^2 + \bp{X}^2$) in the Euclidean sector. This sphere maps to a hyperboloid $R^2 - T^2 = \rho^2$ enveloping the light cone $R= \pm T$. As we reduce the value of $\rho$, the sphere in the Euclidean sector shrinks to the origin while the hyperboloid becomes the light cone! Something interesting  happens to the normals of the sphere and the hyperboloid. When the sphere shrinks to a point, the unit normal to its surface becomes degenerate and could point in any direction at the origin; that is, it becomes a vector of unit norm and indeterminate direction. The \textit{unnormalized} normal $n_a=\partial_a\rho^2=2(T_E,R)$ vanishes in this limit. In the Lorentzian sector, when the hyperboloid collapses to the light cone, the (unnormalized) normal to the hyperboloid $2(-T,R)\propto (-1,1)$ survives and becomes a null  normal to the light cone. In summary:
 
 \begin{itemize}
  \item[$\blacktriangleright$] The origin of a Euclidean sector, in a locally flat coordinate system, maps to null horizons $\mathcal{H}^\pm$ of the freely falling frame in the Lorentzian sector, around \textit{any} event in \textit{any} arbitrary spacetime. 
  \item[$\blacktriangleright$]
  Spherical surfaces  around the Euclidean origin maps to hyperboloids around the light cones with vertex at the  origin of the Lorentzian sector. The (unnormalized) normal to these spherical surfaces in the Euclidean sector, maps to null normals to the light cone when the radius of the sphere shrinks to zero. 
   \item[$\blacktriangleright$]Two events $x_1$ and $x_2$ separated by a large ($\gg L_P$) spacelike or timelike distance in a Lorentzian spacetime will also have a large separation in
the Euclidean sector under analytic continuation. On the other hand, two  widely separated events connected by a light ray (with  affine distance being large compared to Planck length) 
in the
Lorentzian spacetime  will collapse to a single point in the Euclidean sector.

\item[$\blacktriangleright$] If the quantum theory of
spacetime has to be formulated in the Euclidean sector --- and analytically continued to the Lorentzian sector --- then events separated by
a light ray,  can inherit quantum features of spacetime from the Euclidean sector. This is why events separated by
null interval, and null surfaces, will  play a crucial role in our discussions.
\end{itemize}
 This metamorphosis --- of a point in a Riemannian spacetime to a null surface in a pseudo-Riemannian spacetime --- can occur at any event in the Euclidean space (or Lorentzian spacetime) when we use locally flat (or locally inertial) coordinates. It is, of course, true that an arbitrary pseudo-Riemannian metric may not have a global analytic continuation to a Euclidean metric with definite signature. But as long as we are studying local physics we can adopt the above rule. The fact that null surfaces --- which will turn out to be very important in our discussion --- maps locally to a single point in the Euclidean sector, is telling us something significant about the structure of spacetime.
We will have occasion to exploit this general feature of this local analytic continuation  in our later discussions.

%\vfill\eject   
 
 \section{Horizon Thermality: The First Clue} \label{sec:horterm} 

 \begin{flushright}
 \obeylines{
 \textit{
 ``If it is hot, it must have microscopic degrees of freedom''}
 \vskip0.1in
 --- The Boltzmann Principle
 } 
 \end{flushright}

\noindent The null surfaces $X=\pm T$ in flat spacetime --- or in a locally flat coordinate system around any region in a curved spacetime --- act as horizons $\mathcal{H}^\pm$ with respect to the R wedge. These null surfaces are endowed with thermodynamic properties, especially temperature, when viewed from R. Since this is essentially a result about two surfaces $\mathcal{H}^\pm$ in \textit{inertial} coordinate system,  we should be able to derive it, working entirely \textit{in inertial coordinates}. Further, since the geometrical structure of the wedge-like regions arise from analytic continuation from the Euclidean sector, it must also be possible to obtain the same result working entirely in the Euclidean sector --- \textit{in spite of the fact that there are no null surfaces or horizons in the Euclidean sector! }

In the next few sections, I will provide different derivations of horizon thermality, each highlighting a different aspect of this result. In Sec. \ref{sec:tdlee}, I will provide an extremely simple proof, based entirely on Euclidean QFT, using the rotational invariance in the Euclidean plane \cite{leeNL}. Next, in  Sec. \ref{sec:tfromp}, I will show how the result can be obtained directly \cite{tppropagator} from the structure of the \textit{inertial} Feynman propagator. (Neither of these derivations use accelerated observers, Rindler frame quantization, Rindler vacuum etc. which play a central role in the text book derivations. In fact, our approach will lead us \textit{to discover} the Rindler modes.)  Section \ref{sec:boostmodes} provides a more conventional derivation based on Bogoliubov transformation. But even here, I stress the fact that one can obtain the result by simply  rewriting the inertial modes in a different representation, using rapidity $\tau$. Finally in Sec. \ref{sec:horcft},  I discuss the dimensional reduction and CFT, on the horizon, to explain the universality of this effect. Even though I present the derivations in the language of flat spacetime, it should be obvious that they remain valid in curved spacetime in any local Rindler frame around any event. 
 
\subsection{Thermality from the Euclidean field theory}\label{sec:tdlee}

The Euclidean sector has no horizon. All the same, the \textit{simplest}  derivation of the thermality \cite{leeNL} of Lorentzian horizon is based on an Euclidean 
construction. This arises from the result emphasized above, viz that Euclidean origin maps to Lorentzian horizon and the 
translation along $\tau$ corresponds to rotational symmetry in Euclidean plane.   We can show, using Euclidean quantum field theory, that the vacuum state of a quantum field will lead to a thermal density matrix (with temperature $T= 1/2\pi$) as far as the expectation value of operators confined to the R wedge are concerned. 

To obtain this result, let us begin from a standard result in quantum mechanics, which expresses the quantum mechanical path integral in terms of the stationary states of the system:
\begin{equation}
 \int \mathcal{D}q\exp\left( i A [ t_2,q_2; t_1,q_1]\right) = \amp{t_2,q_2}{t_1,q_1} = \sum_n \phi_n (q_2) \phi_n^*(q_1) e^{-i E_n(t_2-t_1)}
\end{equation} 
We next analytically continue to Euclidean time and set $t_1^E =0,  q_2=0, q_1 = q$ and take the limit  $t_2^E \to \infty$. In the infinite time limit, the right hand side will be dominated by the lowest energy eigenvalue which will correspond to the ground state $\phi_0(q)$. We can always add a constant to the Hamiltonian to make the ground state energy zero. In that case, the only term which survives in the right hand side will be $\phi_0(q_2) \phi_0^*(q_1) = \phi_0(0) \phi_0(q) \propto \phi_0(q)$.
Thus the ground state wave function can be expressed as a Euclidean path integral with very specific boundary conditions:
\begin{equation}
 \phi_0(q) \propto \int \mathcal{D}q \, \exp(-A_{E}[\infty, 0; 0,q])
\end{equation} 

This result  directly generalizes to field theory in Schrodinger picture, if we think of $q_1$ and $q_2$ as (spatial) field configurations $q_1(\b x), q_2(\b x)$ and $\phi_0 (q_1)$ as the ground state (`vacuum') wave functional $\Psi_{\rm vac}[\phi(\b x)]$. 
Then we get:
\begin{equation}
 \Psi_{\rm vac} [\phi(\b x)] \propto \int \mathcal{D}\phi \, e^{-A_E(\infty, 0; 0,\phi(\b x))}
\label{20Marhp12}
\end{equation}
In our case, it is convenient to take the  field configuration $\phi(\b x)$ on the $t_E=0$ surface  as being given by two functions $\phi_L(\b x), \phi_R(\b x)$ in the left and right halves $x<0$ and $x>0$ respectively.
Therefore, the ground state wave functional $\Psi_{\rm vac}[\phi(\b x)]$ now becomes a functional of these two functions:
$\Psi_{\rm vac} [\phi(\b x)] = \Psi_{\rm vac} [\phi_L(\b x), \phi_R(\b x)]$. The path integral expression in \eq{20Marhp12} now reads
\begin{equation}
  \Psi_{\rm vac} [\phi_L(\b x), \phi_R(\b x)] \propto \int_{t_E=0;\phi=(\phi_L,\phi_R)}^{t_E=\infty;\phi=(0,0)}
\mathcal{D}\phi e^{-A}\label{euclpath}
\end{equation} 
Here the
path integral on the right hand side is being  evaluated  (in the Euclidean sector)  by analytically continuing the inertial time coordinate $t$ to $t_E$. 

\begin{figure}
\begin{center}
\scalebox{0.8}{\input{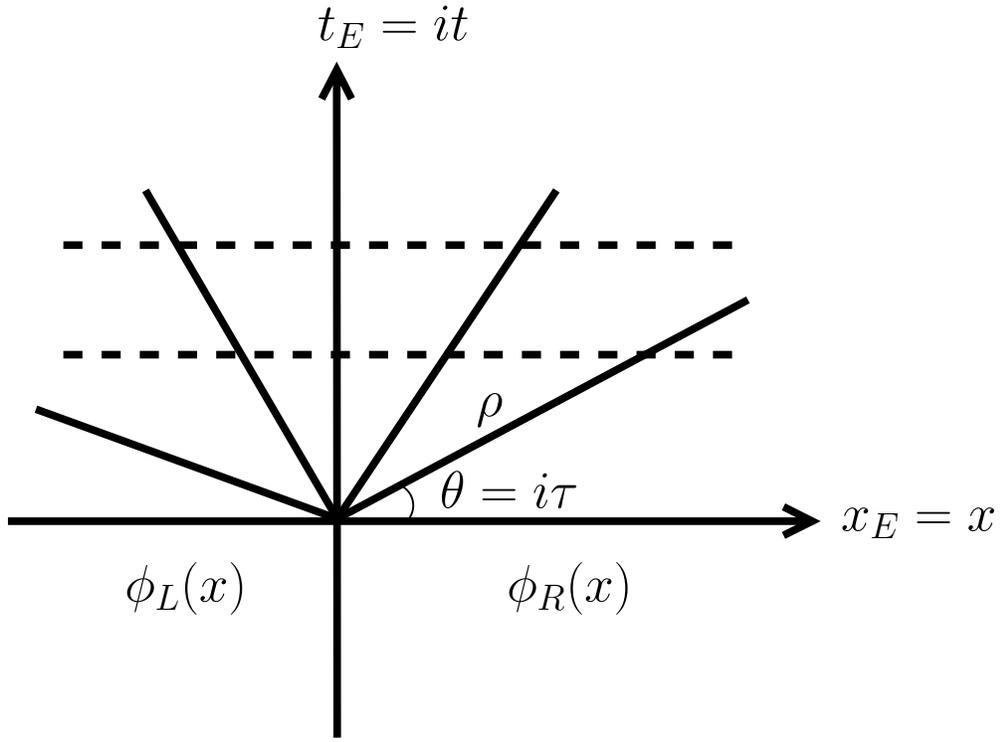}}
\caption{When one uses the path integral to determine the ground
state wave functional on  the $t_E =0$ surface, one needs to integrate over the field configurations
in the upper half ($t_E>0$) with a boundary condition on the field configuration on $t_E=0$. This can be
done either by using a series of hypersurfaces parallel to the horizontal axis (shown
by broken lines) or by using a series of hypersurfaces corresponding to the radial
lines. Comparing the two results, one can show that the ground state in one
coordinate system appears as a thermal state in the other.}
\label{fig:gr141}
\end{center}
\end{figure}

But from Fig.~\ref{fig:gr141}  it is obvious that 
this path integral could also be evaluated by analytically continuing the rapidity parameter $\tau$ to the polar angle $\theta$ by varying the angle
$\theta$ from 0 to $\pi$ and the radial coordinate in the range $0<\rho<\infty$. 
(The two sets of coordinates are related by the usual transformation, 
$
  x_E=\rho \cos \theta;  t_E = \rho \sin \theta
$.)
While the evolution in $t_E$ will take the field configuration from 
 $t_E =0$ to $t_E\to \infty$, the same time evolution gets mapped in terms of 
 $\theta$ into evolving the ``angular'' coordinate $\theta$ from $0$ to $\pi$.  Obviously (see
 Fig.~\ref{fig:gr141}),  the entire upper half-plane
 $t>0$ are being covered in two completely different ways in terms of the evolution
 in $t_E$, compared to the  evolution in $\theta$. 
 In $(t_E,x_E)$ coordinates, we vary $x_E$ in the range $(-\infty,\infty)$ for each $t_E$ 
 and vary $t_E$ in the range $(0,\infty)$. In $(\theta,\rho)$ coordinates, we vary $\rho$ in the range $(0,\infty)$ for each $\theta$ 
 and vary $\theta$ in the range $(0,\pi)$.

When $\theta =0$, the field configuration corresponds to 
  $\phi = \phi_R$ and when $\theta = \pi$ the field configuration corresponds to $\phi = \phi_L$.
  Therefore \eq{euclpath} can also  be written as
  \begin{equation}
\Psi_{\rm vac} [\phi_L(\b x), \phi_R(\b x)] \propto \int_{\theta=0;\phi=\phi_R}^{\theta=\pi;\phi=\phi_L}
\mathcal{D}\phi e^{-A}
\label{rindact}
\end{equation}
   Let $H_\theta$ be the  Hamiltonian  which  describes the evolution in the $\theta$ direction, generated by the Killing vector $\bm{\xi}_\theta=\partial/\partial\theta$. (This is the Euclidean version of $H_\tau$ defined in \eq{2H}.) Then, in the
    Heisenberg picture, rotating from $\theta =0$ to 
    $ \theta =\pi$ is a `time' evolution governed by $H_\theta$. So the path integral \eq{rindact} can also be represented  as a matrix element of the  Hamiltonian, $H_\theta$
    giving us the result:   
    \begin{equation}
 \Psi_{\rm vac} [\phi_L(\b x), \phi_R(\b x)]\propto \int_{\theta=0;\phi=\phi_R}^{\theta=\pi;\phi=\phi_L}
\mathcal{D}\phi e^{-A} 
= \langle \phi_L|e^{-\pi H_\theta}|\phi_R\rangle
\end{equation}
The proportionality constant  $C$, say,  can be fixed by the normalization condition:
   \begin{eqnarray}
 1&=& \int \mathcal{D} \phi_L\, \mathcal{D} \phi_R\, \big| \Psi_{\rm vac} [\phi_L(\b x), \phi_R(\b x)] \big|^2 
\nonumber\\
&=& C^2 \int \mathcal{D} \phi_L\, \mathcal{D} \phi_R\, \langle\phi_L|e^{-\pi H_\theta} | \phi_R \rangle \, \langle \phi_R|e^{-\pi H_\theta} | \phi_L\rangle
= C^2 \, {\rm Tr}\, \left(e^{-2\pi H_\theta}\right)
\end{eqnarray}  
allowing us to write  the normalized vacuum functional  as:
\begin{equation}
\Psi_{\rm vac} [\phi_L(\b x), \phi_R(\b x)] = \frac{\langle \phi_L \vert e^{-\pi H_\theta } \vert \phi_R \rangle}{ \left[ {\rm Tr}(e^{-2 \pi H_\theta })\right]^{1/2} } \label{eqn:fifteen}
\end{equation}
   
Using  this result, we can immediately show that, for operators $\mathcal{O}$ made out of  variables 
having support in $x>0$,   the \textit{vacuum } expectation values 
$\langle{\rm vac} \vert \, \mathcal{O} (\phi_R)\vert {\rm vac}\rangle$
 become
\textit{thermal} expectation values. This arises from straightforward algebra of inserting
a complete set of states appropriately:
\begin{eqnarray}
&&\hskip-2em \langle{\rm vac} \vert \, \mathcal{O} (\phi_R)\vert {\rm vac}\rangle 
 = \sum\limits_{\phi_L} \sum\limits_{\phi^{(1)} _R, \phi^{(2)}_R} 
\Psi_{\rm vac} [\phi_L, \phi^{(1)}_R] 
\langle \phi^{(1)}_R \vert \mathcal{O} (\phi_R)\vert \phi^{(2)}_R \rangle 
\Psi_{\rm vac} [ \phi^{(2)}_R , \phi_L]  
 \nonumber \\
&&\hskip1em = \sum\limits_{\phi_L} \sum\limits_{{\phi^{(1)}_R}, \phi^{(2)}_R} \frac{\langle \phi_L \vert e^{-\pi H_\theta}\vert \phi^{{(1)}}_R \rangle \langle \phi^{{(1)}}_R \vert \mathcal{O} \vert \phi^{(2)}_R \rangle \langle \phi^{(2)}_R \vert e^{-\pi H_\theta } \vert \phi_L \rangle}{Tr(e^{-2\pi H_\theta })}
= \frac{Tr (e^{-2\pi H_\theta } \mathcal{O}) }{Tr(e^{-2\pi H_\theta })}
 \label{eqn:twentytwo}
\end{eqnarray}
Thus, tracing over the field configuration $\phi_L$ in the region $x<0$ (``behind the horizon'') leads to a thermal density
matrix $\rho \propto \exp[-2\pi H]$ for the observables in the region $x>0$.\index{path integral!Davies-Unruh effect}
In particular, the expectation value of the number operator will be a thermal spectrum at the temperature $T=1/2\pi$ in our units.

The role of Euclidean ``time'', represented by the polar angle $\theta$ in the above derivation is crucial.  As we have seen earlier, the Hamiltonian
$H_\tau$ corresponding to the boost Killing vector $\bm{\xi}_\tau$ (see \eq{2H}), defined in Lorentzian spacetime can only evolve the system from the past horizon $\mathcal{H}^-$ to the future horizon $\mathcal{H}^+$.
Therefore, the evolution cannot take the system from positive values of $x$ to negative values of $x$. 
On the other hand, the corresponding \textit{Euclidean} Hamiltonian, $H_\theta$, built from $\bm{\xi}_\theta$ can evolve the system from positive values of $x$ to negative values of $x$.
This is, of course, related to the fact we have mentioned earlier in connection with equations \eq{paramet11} and \eq{paramet22}: viz., $x = \rho \cosh\tau$ is always positive while its Euclidean analog $x = \rho\cos\theta$ can have either sign. A closely related fact is that allowing $\tau$ to have complex values and replacing $\tau \to \tau - i\pi$ changes $x$ to $-x$. This, in fact, allows a very general way of obtaining horizon thermality  by making suitable complex excursions.
 (For a general result, see Sec. 3.3 of  \cite{tpreviews2}; for some special cases, see e.g., \cite{Keski-Vakkuri:1997xp} and references therein.)
 
How come we could get a result about the horizon working in Euclidean sector which has no horizon? The effect of horizon is implicit in \eq{eqn:twentytwo} when we confine our attention to observables which can be completely determined from $\phi_R(x)$ at the initial hypersurface $t_E = 0 = t$. This is  a statement in Lorentzian spacetime using the notions of Lorentzian causality in QFT. For example, an observable in $F$ wedge could very well be influenced by the field configuration $\phi_L$; on the other hand, an observable in $R$ wedge at $t>0$ is entirely determined by the field configuration $\phi_R$. The result in \eq{eqn:twentytwo} 
 is restricted to observables in $R$ which cannot be influenced by $\phi_L$. The fact that we are excluding the propagation of influence from $\phi_L$ is equivalent to assuming the existence of a one-way membrane along $x=t$. This, of course, is equivalent to the notion of a horizon. In this sense, the Euclidean QFT has a ``horizon without horizon''.

\subsection{Thermality from the inertial frame QFT}\label{sec:tfromp}

The above derivation used \textit{Euclidean} QFT to obtain the thermality of Lorentzian horizons. I will next show how the same result can be obtained, again very simply, from the structure of the Feynman propagator, working entirely in the \textit{Lorentzian} sector \cite{tppropagator}.

Such a derivation is important from a conceptual point of view because of the following reason: 
The Feynman propagator $G(x_1,x_2)$ encodes \textit{all} the physics contained in a free  field and, hence, we should be able to discover the thermality of Rindler horizon, just by probing the structure of this propagator. The  thermal nature of the  horizon is indeed contained --- though hidden --- in the standard, inertial, Feynman propagator. 
 We will see that one can unravel this by studying how the standard Feynman propagator in flat spacetime varies with the rapidity $\tau$ describing a sequence of Lorentz frames. 
 
 \subsubsection{Horizon temperature from the Feynman propagator}\label{sec:inpro}

 The path integral representation of the (Feynman) propagator,  given by the sum over paths prescription, involving the (square-root) action for a relativistic, spinless, neutral, particle \cite{tpqft}
 suggests that one can interpret $G(x_1,x_2)$  as an amplitude for a particle/antiparticle to propagate between two events in the spacetime.\footnote{I define the propagator in the momentum space $G(p)=i(p^2-m^2-i\epsilon)^{-1}$  with an $i$ factor, so that  $G(x_1,x_2)=\bk{0}{T[\phi(x_1)\phi(x_2)]}{0}$. In this section, I will use the mostly negative signature (+ - - - .....) since it turns out to be more convenient.}. 
 Since $\bm{\xi}_\tau$ is a Killing vector, the propagator, when expressed in $(\tau, \rho, \bp{x})$ coordinates, will depend only on the difference $\tau \equiv \tau_1 - \tau_2$ and we can write $G(x_1,x_2) = G(\tau, \bm{x}_1, \bm{x}_2)$. 
  The explicit form of $G(\tau)$ in a $D$-dimensional flat spacetime given by (with $m^2$ treated as $m^2-i\epsilon$):
  \begin{equation}
 G(\tau) = i \frab{1}{4\pi i}^{D/2} \int_0^\infty \frac{ds}{s^{D/2}} \ \exp\left[-ism^2 - \frac{i}{4s}\, \sigma^2(\tau)\right]
   \label{rt2}
  \end{equation} 
  where $\sigma^2(\tau) \equiv \sigma^2 (x_1,x_2)$ is the squared line interval between the two events. This expression, of course, is valid for all events in spacetime irrespective of whether the two events are in the same wedge or at different wedges.
  When the two events are in the same wedge, say R, this denotes the amplitude for propagation without crossing a horizon. But when the two events are at different wedges, say in R and F, this amplitude describes the propagation of a particle across  the horizon $\mathcal{H}^+$. 
  
  To study the variation of $G$ with the rapidity $\tau$, now treated as a time coordinate, it is convenient to  introduce the Fourier transform of $G$ with respect to $\tau$ and interpret 
 \begin{equation}
   A(\Omega; \bm{x}_1,\bm{x}_2) = \int_{-\infty}^\infty d\tau\ G(\tau; \bm{x}_1,\bm{x}_2) \ e^{i\Omega \tau};
   \qquad \tau=(\tau_1-\tau_2)
   \label{rt1}
  \end{equation} 
  as the amplitude for a particle to propagate between the events $\bm{x}_1$ and $\bm{x}_2$ with the  (boost) energy $\Omega$, introduced  as the Fourier conjugate to the time coordinate $\tau$. In order to simplify the notation, I will write  $G(\tau) $ for $G(\tau; \bm{x}_1,\bm{x}_2)$ and $A(\Omega)$ for $A(\Omega; \bm{x}_1,\bm{x}_2)$, suppressing the dependencies on the spatial coordinates. While evaluating the amplitude $A(\Omega)$ in \eq{rt1} we will   assume that $\Omega > 0$ and  interpret $A(-\Omega)$ as the expression obtained by replacing $\Omega$ by $-\Omega$ in the resulting  integral in \eq{rt1}. 
  Our interest lies in comparing $A(-\Omega)$ with $A(\Omega)$ when: (i) both events are in the same wedge as well as when (ii) they are on different wedges separated by a horizon. The motivation is the following:
  
   The propagation of a particle with energy $\Omega$ from a spatial location in F to a spatial location in R can be thought of as an emission of a particle by the horizon surface, since an observer confined to R cannot (classically) detect anything beyond the horizon. By the same token, the propagation of a particle with an energy $-\Omega$ can be thought of as the absorption of energy $\Omega$ by the horizon. Therefore, we have $P_e/P_a = |A(\Omega)|^2/|A(-\Omega)|^2$ where $P_e, P_a$ denote the probabilities for emission and absorption.
   On the other hand, if we think of the horizon as a surface with emitters/absorbers in thermal equilibrium, $P_e \propto N_{\rm up}$ and $P_a \propto N_{\rm down}$ where $ N_{\rm up}$ and $ N_{\rm down}$ are the population of the upper and lower levels separated by energy $\Omega$. 
   If the horizon is perceived as a hot surface with temperature $T=1/2\pi$, then we will expect $ N_{\rm up}/ N_{\rm down} = e^{-2\pi \Omega}$. This demands that when the propagation is between two events separated by a horizon we must find $|A(\Omega)|^2/|A(-\Omega)|^2  = e^{-2\pi \Omega}$ for consistency of interpretation. 
 Equally important is the condition that we must have $|A(\Omega)|^2/|A(-\Omega)|^2  =1$ when both events are at the same R wedge. \textit{This is a strong algebraic constraint on the Feynman propagator if horizons have to exhibit thermality!} Let us verify this.
  
   From the form of the integral in \eq{rt1} it is obvious that, if $G(\tau) = G(-\tau)$, then $A(\Omega) = A(-\Omega)$ so that nothing very interesting happens. This is trivially true if we use the standard inertial time coordinate $t$ so that $\sigma^2(t) = t^2 - |\bm{x}_1 - \bm{x}_2|^2$ making  $\sigma^2$ and $G$  even functions of the time difference; we then have  $A(\Omega ) = A(-\Omega)$. 
  Gratifyingly enough, the same result holds  when both events $x_1$ and $x_2$ are on the right Rindler wedge (R) as well. In R the Rindler coordinates $(\tau, \rho)$ can be defined in the usual manner (see \eq{paramet11}) as $t=\rho \sinh \tau$, \ $x= \rho \cosh \tau$.   Then the line interval $\sigma^2_{RR}$ for two events in the right wedge has the form 
  \begin{equation}
 \sigma^2_{RR}(\tau) = - L_1^2 + 2 \rho_1\rho_2 \, \cosh \tau
   \label{rt3}
  \end{equation} 
  where 
  $
 L_1^2 = (\Delta \mathbf{x}_\perp^2 + 2 \xi_1 + 2 \xi_2)
  $,
  with the $\xi$ coordinate defined through the relation  $x^2 - t^2  \equiv 2\xi$. The $\sigma^2_{RR}(\tau)$ is  an even function of $\tau$ which implies  that \textit{when a particle propagates between any two events within the right Rindler wedge R, we have $A(\Omega) = A(-\Omega)$ as expected.}
  
Let us now consider what happens when  one event is in R and the second event is in F where the Rindler-like coordinate system is introduced through $t=\rho \cosh \tau$ and $x=\rho \sinh \tau$; see \eq{paramet22}. (If one uses the $\xi$ coordinate, then the relation $x^2 - t^2 = 2\xi$ allows us to cover the region F  by the range $-\infty < \xi < 0$ and the region R  by the range $0<\xi<\infty$.) The line interval $\sigma^2_{FR} (\tau)$ between an event $(\tau_F, \rho_F) $ in F and an event $(\tau_R, \rho_R) $ in R is  easily computed to be:\footnote{The following point is worth mentioning: The line interval $\sigma^2(\mathcal{P}_1,\mathcal{P}_2)$ between any two events in the spacetime is, of course,  symmetric with respect to the interchange of events; i.e.,
 $\sigma^2(\mathcal{P}_1,\mathcal{P}_2)=\sigma^2(\mathcal{P}_2,\mathcal{P}_1)$. But this symmetry may or may not be apparent in the coordinate labels if we use two different coordinate charts. The symmetry of the line interval is obvious in the inertial coordinates and we do have $\sigma^2(t_F,\mathbf{x}_F;,t_R,\mathbf{x}_R)=\sigma^2(t_R,\mathbf{x}_R;,t_R,\mathbf{x}_F)$
 but not in the Rindler coordinates $\sigma^2(\tau_F,\rho_F,\mathbf{x}_F^\perp;\tau_R,\rho_R, \mathbf{x}_R^\perp)\neq\sigma^2(\tau_R,\rho_R,\mathbf{x}_R^\perp;\tau_F,\rho_F, \mathbf{x}_F^\perp)$.
 This emphasizes the fact that if you introduce arbitrary coordinate labels (in two different coordinate charts) to events in spacetime,  there is no assurance that the interchange of coordinate labels will correspond to the interchange of events.}
  \begin{eqnarray}
 \sigma^2_{FR}(\tau)&\equiv& (t_F-t_R)^2-(x_F-x_R)^2-\Delta \mathbf{x}_\perp^2\\
 &=&\rho_F^2-\rho_R^2-2 \rho_F\rho_R\sinh(\tau_R- \tau_F)-\Delta \mathbf{x}_\perp^2
 \label{vpone}\\
 &\equiv&- L_2^2 - 2 \rho_F\rho_R \, \sinh \tau; \qquad\qquad \tau\equiv(\tau_R-\tau_F)
 \label{tp1}
 \end{eqnarray} 
 with  $L_2^2 \equiv (\Delta \mathbf{x}_\perp^2 + 2 \xi_R + 2| \xi_F|)$. The line interval $\sigma^2$ between the events in R and F only depends on the difference in `time' labels, even though $\tau$ is not  a time variable in F. This is because one can indeed introduce,  a  coordinate system covering both R and F in which the 2-D metric takes the form $ds^2=(2\xi)d\tau^2-(2\xi)^{-1}d\xi^2$ (see \eq{le2rf}). The coordinate $\tau$ retains its Killing character both in R and F, though $\partial/\partial\tau$ is timelike only in R. It is the Killing character which ensures that $\sigma_{FR}^2$ only depends on the difference in the `time' labels.

Let us now compute the Fourier transform in \eq{rt1} with  respect to $\tau\equiv(\tau_R-\tau_F)$. From the sign convention in \eq{rt1} we see that $G$ picks up a contribution $A(\Omega)\exp -i\Omega(\tau_R-\tau_F)$ --- which will correspond to  positive energy with respect to $\tau_R$ when $\Omega>0$ (and negative energy when $\Omega<0)$. The energy is defined with respect to $\tau_R$ which is a valid time coordinate in R. (Therefore we do not have to worry about the fact that $\tau_F$ has no clear meaning as a time coordinate in F; it is an ignorable constant which goes away when I do the integral over the range $-\infty<\tau<\infty$.)   
 The Fourier transform in \eq{rt1} involves the integral:
   \begin{equation}
 I = \int_{-\infty}^\infty d\tau\ e^{i\Omega \tau - \frac{i}{4s} \sigma_{FR}^2 (\tau) }
 = 2\, e^{\frac{iL^2}{4s} } \ e^{-\pi \Omega/2} K_{i\Omega}(2\alpha)
   \label{rt7}
  \end{equation}
  where $\alpha \equiv  (\rho_1\rho_2/2s)$. This result uses  the standard integral representation for the McDonald function:
   \begin{equation}
  \int_0^\infty \frac{dq}{q} \, q^{i\omega} \, e^{i\alpha\left( q - \frac{1}{q}\right)} = 2\, e^{-\pi\omega/2}\ K_{i\omega} (2\alpha); \qquad (\alpha>0)
   \label{rt8}
  \end{equation}
 Substituting \eq{rt7} in \eq{rt1}, we find that the relevant amplitude is given by 
 \begin{equation}
 A(\Omega) = e^{-\pi\Omega/2} \, \int_0^\infty ds\ F(s) K_{i\Omega}(2\alpha)
   \label{rt9}
  \end{equation}
  where 
  \begin{equation}
  F(s) = 2i \frab{1}{4\pi is}^{D/2}\, e^{-im^2 s + \frac{iL_2^2}{4s}}
   \label{rt10}
  \end{equation}
  Since $K_{i\Omega} = K_{-i\Omega}$ is an even function of $\Omega$, we find that: 
  \begin{equation}
  A(-\Omega) = e^{\pi\Omega/2} \, \int_0^\infty ds\ F(s) K_{i\Omega}(2\alpha) = e^{\pi \Omega}\ A(\Omega)
   \label{rt11}
  \end{equation}
  leading correctly to the  Boltzmann factor 
  \begin{equation}
  \frac{|A(\Omega)|^2}{|A(-\Omega)|^2} = e^{-2\pi\Omega}
   \label{rt12}
  \end{equation}
  corresponding to the  temperature $T=g/2\pi = 1/2\pi$ in our units.    
Our result in \eq{rt12}
implies that the effective number of emitters and absorbers on the horizon surface satisfies the Boltzmann distribution corresponding to the temperature $T= 1/2\pi$ with  $ N_{\rm up}/ N_{\rm down} = e^{-2\pi \Omega}$. As we said before, this is what we need for consistency of interpretation.\footnote{A similar analysis leads to corresponding conclusions for other situations when the events are separated by a horizon, like for e.g., between region P and region L. I will concentrate on F and R.} 
 
 It is nice to see that the  thermal behaviour of the Rindler horizon has indeed left a trace in the variation of the inertial propagator with respect to the boost parameter. 
  That is, the propagator can distinguish  clearly between the propagation across the horizon from the propagation within one side of the horizon.\footnote{This fact \textit{prevents} you from `understanding'  \eq{rt12} in a trivial manner: One might think, at first sight, that if we are Fourier transforming $G$ with respect to the \textit{Rindler time} $\tau$ (and define positive/negative energies through $\exp\mp i\Omega\tau$) then it is a foregone conclusion that we will get the thermal factor. \textit{This is  not true.} Recall that, when we do the Fourier transform with respect to Rindler time etc. but for two events within the right wedge R, we get \textit{nothing} interesting. So the usual suspect, viz., $\exp-i\Omega t$ being a superposition of $\exp\mp i\Omega \tau$), is \textit{not} responsible for our result.} There are two other crucial ingredients which have gone into it. First, you need the horizon crossing to break the symmetry between $G(\tau)$ and $G(-\tau)$; this is obtained in \eq{tp1}. Second, it is crucial that the result in \eq{tp1} depends only on the coordinate difference $\tau\equiv(\tau_R -\tau_F)$. So  when we integrate over all $\tau$, we don't have to worry what $\tau_F$ means, since it is not a time coordinate in F. We can stay in R and interpret everything using $\tau_R$. Therefore, it is not just the use of the Rindler time coordinate which leads to the result. \textit{The structure of the inertial propagator is more nontrivial than one would first imagine.}
  
  In obtaining  this result, we have worked entirely in the Lorentzian sector with a well-defined causal structure and the horizons at $x^2-t^2=0$. We have also emphasized the key role played by the horizon in obtaining  this result. On the other hand, we saw in Sec. \ref{sec:tdlee} that one can obtain the same result working in the Euclidean sector. So one  
  may wonder what happens to this analysis if it is done with the inertial  propagator in the \textit{Euclidean} sector. In the conventional approach, the right wedge (with $t=\rho\sinh\tau, x=\rho\cosh\tau$) itself will fill the \textit{entire} Euclidean plane $(t_E,x_E)$ if we take $it=t_E, i\tau=\tau_E, x=x_E$ leading to $t_E=\rho\sin\tau_E, x_E=\rho\cos\tau_E$. The horizons ($x^2-t^2=0$) will then map to the origin ($x_E^2+t_E^2=0$) and the F,P,L wedges will disappear. At first sight, it is not clear how we can recover the information contained in the F,P,L wedges if we start with the Euclidean, inertial, propagator. However, it can be indeed be done  using four different types of analytic continuations to proceed from the Euclidean plane to the four Lorentzian sectors (R, F, L, P). This is described briefly in Appendix \ref{appen:apenA} for the sake of completeness.
  
  While obtaining the above result, we did not compute the final integral in \eq{rt9} because it was unnecessary. However, this can be done in both cases; when the two events are in R or when the  two events are separated by a horizon. The relevant integrals are simpler to present if we first get rid of the transverse coordinates, by Fourier transforming both sides of \eq{rt1} 
  with respect to the transverse coordinate difference  ($\bm{x}^\perp_1-\bm{x}^\perp_2$), thereby introducing the conjugate variable $\bm{k_\perp}$. (As usual, we will write $G^{(RR)}(\tau)$ for $G(\tau; \rho,\rho'; \mathbf{k}_\perp)$ when both events are in R etc.)
  It can be shown  that, when both events are located in R, the  Fourier transform in \eq{rt1} is given by:
  \begin{equation}
 A_{RR}(\Omega) = \int_{-\infty}^\infty d\tau\ G^{(RR)}(\tau) \, e^{i\Omega \tau} = \frac{i}{\pi} \, K_{i\Omega} (\mu \rho_2) \, K_{i\Omega}(-\mu \rho_1);\qquad \tau =(\tau_1 -\tau_2)
   \label{rt13}
  \end{equation}
 with the ordering,  $\rho_1<\rho_2$ and $\mu^2 \equiv k_\perp^2 + m^2$.   But when the events are in F and R the corresponding Fourier transform is (See Appendix \ref{appen:apenA}.):
  \begin{equation}
 A_{FR}(\Omega) = \int_{-\infty}^\infty d\tau\ G^{(FR)}(\tau) \, e^{i\Omega \tau} = \frac{1}{2} \, H_{i\Omega}^{(2)} (\mu \rho_F) \, K_{i\Omega}(\mu \rho_R);\qquad \tau =(\tau_R -\tau_F)
   \label{rt14}
  \end{equation}
  
   The replacement of the McDonald function in \eq{rt13} by  the Hankel function $H_{i\Omega}^{(2)}$ in \eq{rt14}  makes all the difference because --- while McDonald function is even in its index --- Hankel function has the property $H_{i\nu}^{(2)} = e^{-\pi \nu}  H_{-i\nu}^{(2)} $. This  gives 
  \begin{equation}
  \left[\frac{A(\Omega)}{A(-\Omega)}\right]_{FR} =\frac{H_{i\Omega}^{(2)}}{H_{-i\Omega}^{(2)}} = e^{-\pi\Omega}
   \label{rt15}
  \end{equation}
  which is, of course, the same as \eq{rt11}. In contrast, because $K_{i\Omega}= K_{-i\Omega}$ we trivially get $A_{RR}(\Omega) = A_{RR}(-\Omega)$.
  This explicit computation verifies the previous result but  the original approach offers greater generality.

 \subsubsection{The thermal spectrum hiding in the Feynman propagator}\label{sec:hiddenthermal}

  Given the above results, it is worth probing the structure of the inertial propagator a little more closely. 
  What we have obtained above was just the temperature of the horizon. It would be nice if we can discover the  alternative, Rindler frame QFT and the  thermal spectrum of particles as well from the propagator. It is not a priori clear how a thermal spectrum is going to arise from the inertial, Feynman, propagator in \eq{rt2} but it does. We will now see how. 
  
  To do this, let us start with the \textit{Euclidean} version of the inertial propagator for two events in R, which can be written as:
  \begin{equation}
 G_{\rm Eu}^{inertial} (\bm{k}_\perp ; \, \rho_1, \rho_2, \theta-\theta') = \frac{1}{2\pi^2} \int_{-\infty}^\infty d\nu\, e^{\pi \nu} \, K_{i\nu}(\mu \rho_2) \, K_{i\nu}(\mu\rho_1)\ e^{-\nu|\theta-\theta'|}
   \label{rt22new}
  \end{equation}
  As before, we have already Fourier transformed with respect to the transverse coordinate difference  ($\bm{x}^\perp_1-\bm{x}^\perp_2$) thereby introducing the conjugate variable $\bm{k_\perp}$ and defining $\mu^2 = k_\perp^2 + m^2$.
 (This expression, which contains a $|\theta-\theta'|$ is known in literature and is  easy to derive. Appendix \ref{appen:apenA}, contains the derivation as well as its relation with the form in \eq{rt13}, which contains  $(\theta-\theta')$ without the modulus sign; this one is somewhat nontrivial to derive.). Using purely a series of Bessel function identities and \textit{without any  physics input}, this result can be re-expressed in the following form:
 \begin{equation}
  G_{\rm Eu}^{inertial} (\theta-\theta')=\sum_{n=-\infty}^{\infty} G_{\rm Eu}^{Rindler}[\theta-\theta'+2\pi n]
  \label{more1}
 \end{equation} 
 where the function $G_{\rm Eu}^{Rindler}$ is given by:
 \begin{equation}
G_{\rm Eu}^{Rindler} \equiv \frac{1}{\pi^2} \int_0^\infty d\omega\ (\sinh \pi \omega) \, K_{i\omega} (\mu \rho)\, K_{i\omega}(\mu\rho')\, e^{-\omega|\theta -\theta'|}
\label{more2}
\end{equation} 

This result expresses the Euclidean version of the  inertial propagator  as an infinite, periodic, sum in the (Euclideanised) Rindler time. Of course, the fact that the  inertial propagator is  periodic in (Euclideanised) Rindler time is obvious from the fact that  the $\sigma_{RR}^2$ in  \eq{rt3} is periodic in $i\tau$. But \eq{more1} and \eq{more2} tell us more because they  explicitly express $G_{\rm Eu}^{inertial}$ as an infinite periodic sum of \textit{another specific function}  $G_{\rm Eu}^{Rindler}$, thereby identifying the latter. More importantly, from the product structure of $G_{\rm Eu}^{Rindler}$ in \eq{more2}, we see that, when analytically continued back to Lorentzian sector,    
$G^{Rindler}$can be thought of as a propagator built from a complete set of mode functions:  
 \begin{equation}
  \phi_\nu (\tau,\rho) = \frac{1}{\pi} \left( \sinh \pi \nu\right)^{1/2} \, K_{i\nu}(\mu \rho)e^{-i\nu \tau}
   \label{rt17}
  \end{equation}
 in the standard fashion with time ordering with respect to $\tau$. This allows us to actually discover the Rindler mode functions, Rindler quantization, Rindler vacuum etc., just from analyzing the inertial propagator and rewriting it as in \eq{more1} and \eq{more2}. (As can be easily verified, the modes in \eq{rt17} satisfy the Klein-Gordon equation and are properly normalized.) So just by staring at the inertial propagator, we can discover the Rindler modes and the Rindler quantization.
 
 There is another, closely related, feature which is worth mentioning. To do this, I
   will introduce a \textit{reflected} wave function $\phi_\nu^{(r)}$ by the definition 
  \begin{equation}
  \phi_\nu^{(r)} (\rho,\tau) = \phi_\nu (-\rho,\tau-i\pi) = \phi_\nu(\rho^r, \tau^r)
   \label{rt19}
  \end{equation}
  The adjective ``reflected''  for $\phi_\nu^{(r)}$ is justified by two facts: (i) The coordinates $\rho$ and $-\rho$ are obtained by a reflection through the origin of the $t-x$ plane and (ii) the replacement of $\tau$ by $\tau-i\pi$  in the Rindler coordinate transformation in \eq{paramet11} takes  you from R to L. (If you do both --- i.e, replace $\rho$ by $-\rho$ and \textit{also replace} $\tau$ by $\tau-i\pi$ --- in the coordinate transformations ($x=\rho\cosh\tau, t=\rho\sinh\tau$), you will get back  to the same event in $R$. But the $\phi_\nu^{(r)} (\rho,\tau)\neq\phi_\nu (\rho,\tau)$, so that the reflected wave function is different from the original one.) 
  The  propagator for two events \textit{within the right wedge} can now be expressed in a very suggestive form as:\footnote{The proofs for all these results are sketched in Appendix \ref{appen:apenA}.}  
  \begin{equation}
 G^{(RR)} = \int_0^\infty d\nu\, \left[ (n_\nu + 1) \, \phi_\nu\, \phi_\nu^{(r)} + n_\nu \phi_\nu^*\, \phi_\nu^{(r)*}\right]
   \label{rt20}
  \end{equation}
  where $n_\nu$ is the thermal population: 
  \begin{equation}
  n_\nu = \frac{1}{e^{2\pi \nu} - 1}
   \label{rt21}
  \end{equation}
  The second term in \eq{rt20} suggests  an absorption process weighted by $n_\nu$  while the first term  represents an emission with the factor $n_\nu +1$ coming from a combination of stimulated emission and spontaneous emission.  If we think of $\phi_\nu$  and $\phi_\nu^r$  as the wave functions for a fictitious particle, then this structure of \eq{rt20} again encodes the usual 
  thermality.\footnote{The factors multiplying $(1+n)$ and $n$ can also be related to the Bremsstrahlung by an accelerating source; recall that  both terms will correspond to emission when viewed in the inertial frame \cite{matsasrev}.}
  
  Since the Rindler frame is just a coordinate transformation of the inertial frame and the propagator $G(x_1,x_2)$ transforms as a bi-scalar under coordinate transformation,  representing it  in the Rindler coordinates is easy. Having done that --- because $G(x_1,x_2)$ encodes \textit{all} the physics contained in a free  field --- we should be able to discover the horizon thermality just by probing $G(x_1,x_2)$. In other words, it should not be necessary for us to quantize the field in Rindler coordinates, identify positive frequency modes, construct Rindler vacuum and particles etc etc. Everything should flow out of $G(x_1,x_2)$ expressed in Rindler coordinates including the alternative, Rindler, quantization. This is what we have demonstrated in the above discussion.

 \subsubsection{Generalization to curved spacetime}
  
  The approach, and the result, have obvious generalizations to specific curved spacetimes with bifurcate Killing horizons. To begin with, the result can be extended to de Sitter spacetime in a straight forward manner because the dependence of the propagator on the geodesic distance (see, for e.g., \cite{kkvp}) is known and allows the same derivation to go through. Further, in the case of curved spacetimes with horizons, like in Schwarzschild, Reissner-Nordstrom etc. in $D=2$, we get the same result by explicit computation. In $D>2$, we do not usually have closed expressions for $G(x,x')$, but one can again compute it close to the horizon. This is because, close to the horizon, we again get a 2D CFT (see Sec. \ref{sec:horcft}) and one can compute approximate form of the modes --- and through them --- the propagator $G(x,x')$. A large class of spacetimes with horizons can be embedded in a higher dimensional \textit{flat} spacetime; one can then use the form of the propagator in higher dimensions to perform a similar analysis \cite{sc-trg}.
  All these will lead to the same result as above.
 
 More generally, one can use this approach to attribute thermality to any \textit{local Rindler horizon} in any spacetime, along the following lines: 
 In an arbitrary spacetime, let us pick an event $\mathcal{P}$ and introduce the Riemann normal coordinates around $\mathcal{P}$. These coordinates will be valid within a region, $\mathcal{V}$, of size $L$ where the typical background curvature is of the order of $L^{-2}$. We now introduce  a local Rindler coordinate system by boosting with an acceleration $\kappa$ with respect to the local inertial frame, defined in $\mathcal{V}$. If we now confine our attention to events  $(x_1, x_2)$ within $\mathcal{V}$, then the standard Schwinger-DeWitt expansion of the propagator tells us that the form in \eq{rt2} will be (approximately) valid. The Fourier integral in \eq{rt1} can still be defined formally, though the range of $\tau$ outside the domain $\mathcal{V}$ is not physically meaningful. To bypass this issue, we have to arrange matters such that most of the contribution to the integral in \eq{rt1} comes from the range  $\tau \lesssim L$. This will indeed be the case for  high frequencies with $\Omega \gg L^{-1}$. In this (high frequency) limit our analysis will go through as before and one will obtain the local Rindler temperature to be $T = \kappa/2\pi$. For consistency, it is necessary to ensure that $\kappa L \gg 1$ which, of course, can be satisfied around any event with finite $L$. In fact, this approach provides a procedure for obtaining the curvature corrections to the temperature systematically, using the Schwinger-DeWitt expansion. I stress that --- in this very general context  of a bifurcate Killing horizon,   introduced into a local inertial frame --- this approach gets you whatever you could reasonably expect; after all, in a curved spacetime, one can expect  thermality (with an approximately constant temperature)  only when the modes do not probe the curvature scale. This is what is ensured by concentrating on the Feynman propagator at two events which are localized within $\mathcal{V}$. 
 
  \subsection{Thermality from the boost modes}\label{sec:boostmodes}

I will next discuss an approach for  obtaining the thermodynamics of Rindler horizon using  a relatively straightforward  procedure. The purpose of presenting this --- admittedly rather unimaginative --- approach is to connect up horizon thermality directly with the nature of Lorentz boosts. This essentially involves  standard quantum field theory re-expressed in terms of Lorentz boost variables.

 In standard QFT, one expands the scalar field (operator) as $\phi(x)$ as $\phi(x) = \mathcal{A}(x) + \mathcal{A}^\dagger (x)$ where 
\begin{equation}
  \mathcal{A} 
  = \int\frac{d^3 k}{(2\pi)^{3/2}} \, a_{\bm{k}}\, \frac{e^{-i\omega_{\bm{k}} t + i \bm{k} \cdot \bm{x}}}{\sqrt{2\omega_{\bm{k}}}}
  \equiv \int \frac{d^2 k_\perp}{(2\pi)} \frac{dk_x}{\sqrt{2\pi}} \frac{1}{\sqrt{2\omega_{\bm{k}}}}\ a_{\bm{k}} f_{\bm{k}}
  \label{eleven}
\end{equation}
Here $\omega_{\bm{k}}^2\equiv m^2+\mathbf{k}^2$ and, in the last expression,  we have separated the $d^3k$ integration into $dk_x$ and transverse integrations. 
This expansion uses the standard mode functions $f_{\bm{k}}$ parameterized by the three components of the momentum $\bm{k}$. Since we are interested in using rapidity $\tau$ as a time variable, it is convenient to use a different representation for the momentum vector $\bm k$ by writing                                                                                                                                                    
$ 
\omega_{\bm k} \equiv \mu \cosh \theta;  k_x \equiv \mu \sinh \theta
$
 with $\mu^2 = m^2 + \bkp^2$. This allows us to use the parameterizations in terms of the variables $(\bm{k}_\perp, \theta)$ instead of $(\bkp,  k_x)$. 
We will also express  the original mode functions $f_{\bm k}(t, x, \bm x_\perp)$ in terms of the coordinates $(\tau, \rho, \bm x_\perp)$
using \eq{paramet11} to get the modes  $f_{\bkp \theta} (\tau, \rho, \bm x_\perp)$. (For the moment we will concentrate on the R wedge; we will comment about the other wedges towards the end.)
 One can introduce corresponding annihilation operators $a_{\bkp} \theta$ and write \eq{eleven} with $a_{\bkp} \theta f_{\bkp} \theta $ replacing $a_{\bm{k}} f_{\bm{k}}$.
 The combination $(k_x x - \omega_k t)$ occurring in the original modes  $f_{\bm k}$ transforms to $\mu \rho \sinh (\theta - \tau)$
so that $f_{\bkp \theta} (\tau, \rho, \bm x_\perp)$ will acquire a dependence on $\tau$ through the factor [$\exp{i\mu \rho \sinh (\theta - \tau)}$].
We see that a translation in $\tau$ --- which involves boosting to a different Lorentz frame --- corresponds to a translation in $\theta$ which takes into account the variation of the vector $k^a$ under such a boost. 
 
 The original plane wave modes  have a simple behaviour under the
  translation in the time coordinate $t$, viz. $t\to t+q$;  they just get multiplied by the phase factor   $\exp(-i\omega_{\bkp} q)$. However, the $\sinh (\theta - \tau)$ factor  tells us that the same plane wave modes have  a much more complicated variation under translation in $\tau$. It would be nice if we could redefine our mode functions such that the translation in the $\tau$ coordinate multiplies it by a pure phase. This can be achieved by using --- what is known as --- Minkowski-Bessel modes \cite{gerlach}. To introduce them, we will express the modes $f_{\bkp \theta}$ as a Fourier transform in the variable $\theta$ through the equation 
\begin{equation}
  f_{\bkp \theta} (\tau,\rho,\bp{x}) \equiv \frac{1}{\sqrt{2}} \,e^{i\bkp \cdot \bm{x}_\perp}\ e^{i\mu \rho \sinh (\theta - \tau)} \equiv \int_{-\infty}^\infty \frac{d\omega}{\sqrt{2\pi}} \, e^{+i\omega \theta} f_{\bkp \omega}(\tau,\rho,\bp{x})
\end{equation} 
 These modes $f_{\bkp \omega}$ are given by the inverse Fourier transform\footnote{We stress that this result is obtained in R wedge; for completeness, we will have to work out the form of similar modes in the L wedge as well but we will not need them for our purpose.}
 \begin{align}
  f_{\bkp \omega}(\tau,\rho,\bp{x}) =  \int_{-\infty}^\infty \frac{d\theta}{\sqrt{2\pi}} \,  f_{\bkp \theta}  \ e^{-i\omega \theta} 
    = \frac{1}{\sqrt{\pi}} K_{i\omega}(\mu\rho) \, e^{\pi\omega/2} \, e^{-i\omega \tau + i \bkp \cdot \bm{x}_\perp}
  \label{MB1}
\end{align} 
 where $K_{i\omega}(z)$ is the modified Bessel function.
 Fortunately, we will not need any property of $K_{i\omega}(z)$ except that it is real for real arguments; that is, $K_{i\omega}(x) = K_{-i\omega}(x)$. We see from \eq{MB1} that the modes $f_{\bkp \omega}$ have a simple behaviour with respect to translation in $\tau$ coordinate; viz. they get multiplied by a pure phase. In terms of these modes we can write the mode expansion of the scalar field as $\phi=\mathcal{A}+\mathcal{A}^\dagger$ with:
 \begin{align}
  \mathcal{A}(\tau,\rho,\bp{x}) 
  \equiv  \int_{-\infty}^\infty \frac{d^2 k_\perp d\omega}{(2\pi)^{3/2}} a_{\bkp \omega} \,f_{\bkp \omega};\quad
  a_{\bkp \omega} \equiv  \int_{-\infty}^\infty \frac{d\theta}{\sqrt{2\pi}} \, e^{i\omega \theta}\ a_{\bkp \theta}
 \label{three}
\end{align} 
 where all the integrals range over the entire real line and we have defined a new set of annihilation operators by a simple Fourier transform. (Some more details of the calculation are given in Appendix \ref{appen:boost}.)
 
 As it stands, \eq{three} involves integration over $\omega$ in the range $(-\infty < \omega < +\infty)$. We would next like to re-express this result with the integration range limited to positive frequencies $(0<\omega<\infty)$ with respect to the $\tau$ coordinate. This can be easily done using the fact that $K_{i\omega}(x) = K_{-i\omega}(x)$ and
 leads to the expression:
  \begin{align}
  \mathcal{A}(\tau,\rho,\bp{x}) =  \int \frac{d^2 k_\perp}{(2\pi^2)}\int_0^\infty \frac{d\omega}{\sqrt{2}} \, K_{i\omega}(\mu\rho)
    \left\{  a_{\bkp \omega} e^{\pi\omega/2}  e^{-i\omega \tau + i\bkp \cdot\bm{x}_\perp} + a_{-\bkp-\omega}\ e^{-\pi\omega/2}\, e^{i\omega \tau - i\bkp \cdot\bm{x}_\perp} \right\}
 \label{six}
\end{align} 
where we have re-labeled $\bkp$ as $-\bkp$ in the second term
to obtain a nicer form.
 Adding up $\mathcal{A}$ and its Hermitian conjugate, we find that the scalar field $\phi(x) =\mathcal{A} +  \mathcal{A}^\dagger$ has the expansion 
 \begin{align}
 \phi(x) = 
    \int \frac{d^2 k_\perp}{(2\pi^2)}\int_0^\infty \frac{d\omega}{\sqrt{2}} \, K_{i\omega}(\mu\rho) \left\{e^{-i\omega \tau + i \bkp \cdot \bm{x}_\perp} \left[  a_{\bkp \omega} e^{\pi\omega/2} +  a^\dagger_{-\bkp-\omega}\ e^{-\pi\omega/2}\right] + \text{h.c.}\right\}
 \label{seven}
\end{align}
 I stress that up to this point we have just performed straight forward algebraic manipulations thereby rewriting the standard mode expansion in \eq{eleven} in the form in \eq{seven}. This  mode expansion is motivated by our desire to study how different physical variables, in particular the scalar field, varies with the $\tau $ coordinate. 
 
 At this stage we note that the combination of creation and annihilation operators appearing \eq{seven} suggests defining a new operator 
 $  C_{\bkp \omega} = a_{\bkp \omega} e^{\pi\omega/2} +  a^\dagger_{-\bkp-\omega}\ e^{-\pi\omega/2}$
 and its Hermitian conjugate $C^\dagger_{\bkp \omega}$. One can immediately see that the commutator between $C_{\bkp \omega}$ and its Hermitian conjugate suggests defining a new annihilation operator given by:
 \begin{equation}
 a^R_{\bkp \omega} \equiv \frac{C_{\bkp \omega}}{\sqrt{2\sinh \pi \omega}} = \frac{1}{\sqrt{2\sinh \pi \omega}} \, \left[a_{\bkp \omega} e^{\pi\omega/2} +  a^\dagger_{-\bkp-\omega}\ e^{-\pi\omega/2}\right]
 \label{btrans1}
 \end{equation} 
 This operator $a^R_{\bkp \omega}$ and its Hermitian conjugate obey the standard commutation rules for the creation and annihilation operators. In terms of these operators, the scalar field can be expanded as
\begin{equation}
 \phi (x) = \int \frac{d^2 k_\perp}{(2\pi^2)}\int_0^\infty d\omega \ \sqrt{\sinh \pi \omega}\ K_{i\omega} (\mu \rho) \left( a^R_{\bkp \omega}\ e^{-i\omega \tau + i \bkp \cdot \bm{x}_\perp} + \text{h.c}\right)
 \end{equation}
 This expansion allows an alternative quantization of the scalar field in terms of the Fock basis built from the set of operators $( a^R_{\bkp \omega}$, $a^{R\dagger}_{\bkp \omega} )$.
 The Bogoliubov transformation in \eq{btrans1} between this set of operators $(a^R_{\bkp \omega}$, $a^{R\dagger}_{\bkp \omega})$ and the original set of creation and annihilation operators $(a_{\bkp \omega}, a^\dagger_{\bkp \omega})$ tells us that the two Fock basis are distinct.  
 An elementary calculation now shows that the number of particles defined in terms of the new creation/annihilation operators, in the standard vacuum state $\ket{M}$ (defined in terms of the original creation and annihilation operators) has a thermal distribution:
 \begin{equation}
 \bk{M}{a^{R\dagger}_\omega a_\omega^R}{M} = \frac{e^{-\pi\omega}}{2\sinh \pi \omega} = \frac{1}{e^{2\pi \omega} -1}
 \end{equation} 
 We see that the thermality arises from a Bogoliubov transformation between the standard plane wave modes and the boost modes.

\subsection{Horizon CFT and thermality}\label{sec:horcft}

As we have mentioned before, the  thermal behaviour exhibited by the horizon $\mathcal{H}^+$ vis-a-vis the R wedge turns out to be a very general feature. It persists in a large class of curved spacetimes with line elements in the form 
\eq{14cc}.  The algebraic reason for such a thermal behaviour is quite simple and can be understood along the following lines.

Consider a spacetime with a bifurcate Killing horizon expressed in two coordinate systems ($u,v,\bp{x}$) and ($U,V,\bp{x}$) with the null coordinates related by \eq{notethree}. 
  The reason for the universal behaviour of spacetimes with bifurcate Killing horizon is closely related to the fact that,  near $\mathcal{H}^\pm$, a scalar field theory undergoes dimensional reduction (see e.g., section 2.5 of \cite{tpreviews2}) and behaves (essentially) as a two-dimensional conformal field theory (CFT).  Such a CFT has several  universal features which have a bearing on our discussions. 
  
  To see how this comes about, consider the action for a scalar field with a potential $V(\phi)$ in the ($\tau, \xi, \bp{x}$) coordinates defined in \eq{14cc}. The action reduces to the form
  \begin{equation}
 \mathcal{A} = \int d^Dx \, \sqrt{-g}\, \left[ -\partial_a \phi\, \partial^a \phi - V(\phi)\right] = \int d^D x\, \sqrt{\sigma}\, \left\{ \frac{\dot \phi^2}{f} - f \phi_\xi^2 - (\partial_\perp \phi)^2 - V\right\}
  \end{equation} 
  where $\sqrt{\sigma} $  is the determinant of the metric in the $(D-2)$ dimensional transverse space,
  $\dot\phi\equiv \partial \phi/\partial \tau$,  $\phi_\xi \equiv \partial\phi/\partial \xi$ and $\partial_\perp \phi $ denotes the derivatives in the transverse space. Changing coordinates to $\xi_*$ with $d\xi = f  d\xi_*$, this simplifies to 
\begin{equation}
  \mathcal{A} = \int d\tau \, d\xi_* \, d^\perp x\sqrt{\sigma} \left\{ -\dot \phi^2 + \phi'^2 + f\left( (\partial_\perp \phi)^2 - V\right)\right\}
   \label{twentyone} 
  \end{equation}
  where $\phi' \equiv \partial \phi/\partial \xi_*$. 
  It is clear that, near $\mathcal{H}^\pm$ where $f\to 0$, the transverse degrees of freedom as well as the effects due to the potential are suppressed in the last term of the Lagrangian in \eq{twentyone}; so we are dealing with an effective 2-dimensional field theory governed by the reduced Lagrangian  $L_{\rm cft} \propto (-\dot \phi^2 + \phi'^2)$. This, in turn, implies that the relevant modes near $\mathcal{H}^\pm$ can be taken to be the set
\begin{equation}
 \phi_\omega \propto \left( e^{-i\omega u}, \, e^{-i\omega v}\right) = \left( U^{i\omega}, V^{-i\omega }\right)
   \label{twotwo} 
  \end{equation}
Expanding the field in terms of these  modes we can define the annihilation and creation operators, $a_\omega$ and $a_\omega^\dagger$. The corresponding Rindler-like vacuum, $\ket{R}$, is defined through the relation $a_\omega \ket{R} =0$. 

 Close to the horizon $\mathcal{H}^+$ one can also introduce the freely-falling-frame (FFF) coordinates $(U,V,\bp{x})$ as discussed around \eq{notethree}. Within the domain of validity of the FFF, which straddles the horizon $\mathcal{H}^+$, we can also introduce  the inertial modes $(e^{-i\Omega U}, e^{-i\Omega V})$.  
These modes, in turn, allow us to define the corresponding annihilation and creation operators, $A_\Omega$ and $A_\Omega^\dagger$. The Minkowski-like FFF vacuum, $\ket{M}$, satisfies $A_\Omega \ket{M} =0$. 
Given the two QFT structures defined with $(a_\omega, a_\omega^\dagger)$ and $(A_\Omega,A_\Omega^\dagger)$ --- near the entire vicinity of the horizon --- one can easily show that
 the FFF vacuum $\ket{M}$ will appear to be thermally populated with the Rindler-type particles; that is, $\bk{M}{a_\omega^\dagger a_\omega}{M}$ has a thermal distribution. 
 
 The algebraic reason for this result can be understood, in fair amount of generality through the following steps: Consider the Bogoliubov transformation between the two kinds of modes we have introduced:
\begin{equation}
\frac{1}{\sqrt{\Omega}} \, e^{-i\Omega U} = \int_0^\infty \frac{d\omega}{\sqrt{\omega}} \left( \alpha_{\omega\Omega}\, e^{-i\omega u} - \beta^*_{\omega\Omega}\, e^{i\omega u}\right)
\end{equation} 
 These coefficients can be evaluated by a simple Fourier transform and we obtain (see Appendix \ref{appen:planck} for the evaluation of the integral):
\begin{align}
\label{alphaft}
\left\{ 
\begin{array}{l}
\alpha_{\omega\Omega}\\
\beta_{\omega\Omega}
\end{array}
\right\}
&=  \frac{1}{2\pi \kappa} \sqrt{\frac{\omega}{\Omega}}\, e^{\pm \frac{\pi \omega}{2\kappa}} \exp\left(\pm \frac{i\omega}{\kappa} \ln \frac{\Omega}{\kappa}\right) \Gamma\left(\mp\frac{i\omega}{\kappa}\right)
\end{align}
This implies that:
 \begin{align}
 \bk{M}{a^{(R)\dagger}_{\omega'}a^{(R)}_{\omega}}{M} &= \frac{\sqrt{\omega\omega'}}{(2\pi \kappa)^2}\ \Gamma \left(\frac{i\omega'}{\kappa}\right)\Gamma \left(\frac{-i\omega}{\kappa}\right)e^{-\pi (\omega+\omega')/2\kappa}
  \int_0^\infty \frac{d\Omega}{\Omega} \frab{\Omega}{\kappa}^{i(\omega-\omega')/\kappa}\\
 &= \frac{\omega}{2\pi \kappa} \left|\Gamma \left(\frac{i\omega}{\kappa}\right)\right|^2\, e^{-\pi \omega /\kappa} \delta_D(\omega - \omega') = \frac{\delta_D(\omega - \omega')}{e^{2\pi \omega/\kappa} - 1} 
\end{align}
 which represents a thermal number \textit{density}. (If you set $\omega=\omega'$ you get a $\delta(0)$ in the frequency space which can be interpreted as density). So the main result arises from the nature of the Bogoliubov transformations\footnote{You can, of course, obtain the same result by expanding the mode $U^{i\omega}$ in \eq{twotwo} in terms of $\exp\pm i\Omega U$.} between the modes $\exp\pm i \Omega U$ and $\exp\pm i \omega u$ when the null coordinates are related by \eq{notethree}. 
 
 One may worry about the fact that the FFF is defined only close to the horizon. However, a more sophisticated analysis, in which the Bogoliubov coefficients are computed  using the Klein-Gordon inner product on a hypersurface close to $\mathcal{H}^+$ justifies this procedure. I will briefly describe how it comes about. In the space of solutions to KG equation one can define the inner product:
 \begin{equation}
   (\phi_1, \phi_2) =
-i\int_{\Sigma}d\Sigma^{m}(\phi_1\partial_{m}\phi_2^{*}-\phi_2^{*}\partial_{m}\phi_1)
 \end{equation} 
 It is straightforward to show that the value of the scalar product is independent of the spacelike hypersurface over which the integral is evaluated. Consider two complete sets of orthonormal modes labeled $f_j(x)$ and $F^{}_i(x)$ connected by the Bogoliubov coefficients:
 \begin{equation}
  f_j(x)=\sum_i (\alpha _{ji}F^{}_i(x)+\beta_{ji}{F^{*}_i(x)}) 
 \end{equation}
 The KG scalar product allows us to express the Bogoliubov coefficients in the form:
 \begin{equation} 
\alpha_{ij}=(f_i, F^{}_j) \qquad \beta_{ij}=-(f_i, F^{*}_j)
  \label{a6}
 \end{equation}
The crucial point is the following: Since the value of the scalar product is independent of the spacelike hypersurface over which the integral is evaluated, we can choose any convenient hypersurface on which the mode functions have a reasonably simple form. In the case of spacetimes with horizons, a spacelike hypersurface closely straddling the horizon (and becoming the null surface $\mathcal{H}^+$ in the limiting sense) is very convenient for this purpose. On such a surface, the Rindler-like modes $(f_j(x))$ reduce to those in \eq{twotwo}; further one can introduce a FFF on this hypersurface with corresponding inertial-like modes $(F_k(x))$ being $(e^{-i\Omega U}, e^{-i\Omega V})$. It can be easily shown that the scalar products in \eq{a6} then reduces to the Fourier transforms like the one in 
 \eq{alphaft}, giving the latter a generally covariant interpretation. 
 
 In fact, carrying this analysis a bit further, one can express the number density of Rindler-like particles in the inertial-like vacuum in a formal manner as follows ( see e.g,, \cite{agullo}). We begin with the expression:
 \begin{eqnarray} 
   \bk{M}{n_{pq}}{M}
   &=&\sum_k
\beta_{pk}\beta_{qk}^*=-\sum_k
(f^{}_{p}, F^{*}_k)(f^{*}_{q}, F^{}_k)\nonumber \\
&=&\sum_k\left(\int_{\Sigma}d\Sigma^{m}_1f^{}_{p}(x_1)
{\buildrel\leftrightarrow\over{\partial}}_m F^{}_k(x_1)\right)
\left(\int_{\Sigma}d\Sigma^{n}_2f^{*}_{q}(x_2){\buildrel\leftrightarrow\over{\partial}}_n F^{*}_k(x_2)\right)
  \label{a26}
 \end{eqnarray}
 and rewrite it in terms of the Wightman function:
 \begin{equation}
 G^+_M(x_1,x_2)\equiv \bk{M}{\phi (x_1)\phi (x_2)}{M}
  = \sum_k F_k^{}(x_1){F_k^{}}^*(x_2)
  \label{a27}
 \end{equation}
 thereby obtaining the result:
 \begin{equation}
  \bk{M}{n_{pq}}{M}
  =   \int_\Sigma d\Sigma_1 ^m d\Sigma_2 ^n
[f^{}_{p}(x_1){\buildrel\leftrightarrow\over{\partial}}_m
][f^{*}_{q}(x_2){\buildrel\leftrightarrow\over{\partial}}_n
]G^+_M(x_1,x_2)
  \label{a28}
 \end{equation}
To avoid any divergences in this expression one needs to define $G^+$ with the usual $i\epsilon$ prescription. Alternatively, one can simply subtract from $G^+_M(x_1,x_2)$ the corresponding Wightman function defined in the Rindler-like vacuum, $G^+_R(x_1,x_2)$ because $\bk{R}{n_{pq}}{R}$ is identically zero. So we can also write the above expression as:
  \begin{equation}
  \bk{M}{n_{pq}}{M}
  =   \int_\Sigma d\Sigma_1 ^m d\Sigma_2 ^n
[f_{p}(x_1){\buildrel\leftrightarrow\over{\partial}}_m
][f^{*}_{q}(x_2){\buildrel\leftrightarrow\over{\partial}}_n
][G^+_M(x_1,x_2)-G^+_R(x_1,x_2)]
  \label{a28tp}
 \end{equation}
These results relate to what we saw in Sec. \ref{sec:hiddenthermal} in a simple manner. In general, the Rindler-type modes will evolve as $f_{k,\omega}(\mathbf{x})\exp(-i\omega\tau)$ and it is convenient to  choose $f_{k,\omega}(\mathbf{x})$ to be real (`standing waves') which can always be done without loss of generality. We will normalize them by the relation:
\begin{equation}
 \int dV_{\bf{x}}\ f_{k,\omega}(\mathbf{x})f_{k',\omega'}(\mathbf{x})
 =\frac{1}{2\omega}\delta(\omega-\omega')\delta(k-k')
 \label{norm}
\end{equation} 
where $dV_{\bf{x}}$ is the appropriate measure on the spacelike hypersurface. We can then easily show that \eq{rt20} becomes:
\begin{equation}
 G_M=\sum_k\int_0^\infty d\omega\ f_{k,\omega}(\mathbf{x_>})f_{k,\omega}^*(\mathbf{x_<})
 \left[
 n_\omega e^{i\omega(\tau_>-\tau_<)} +(n_\omega+1)e^{-i\omega(\tau_>-\tau_<)}
 \right]
\end{equation} 
On the other hand, $G_R$ is given by an identical expression with $n_\omega$ set to zero. Therefore, $G_M-G_R$ has the structure:
\begin{equation}
 G_M-G_R=\sum_k\int_0^\infty d\omega\ f_{k,\omega}(\mathbf{x_>})f_{k,\omega}^*(\mathbf{x_<})\; 2n_\omega\cos [\omega(\tau_>-\tau_<)]
 \label{gmmgr}
\end{equation} 
This specific structure of $G_M-G_R$ which we saw earlier in  Sec. \ref{sec:hiddenthermal} (which can be easily generalized for the Wightman function as well) is the key reason for the \eq{a28tp} to work.
When the expression in \eq{gmmgr}  is used in \eq{a28tp}, the normalization condition in \eq{norm} ensures that the integral over the spacelike hypersurface picks out the $n_\omega$ on the right hand side of \eq{gmmgr}.

There is a more elegant --- though somewhat opaque --- way of obtaining these results using the properties of CFT near $\mathcal{H}^\pm$  which I will now outline. The  Wightman function of the  standard QFT in flat Minkowski coordinates is defined as   $G_M(x_1,x_2)\equiv \bk{M}{\phi(x_1)\, \phi(x_2)}{M}$. Consider its behaviour on a null surface $U = 0$ (in a limiting sense to be made precise below) between two event, separated in $V$ coordinates by $V\equiv V_2 - V_1$ and in the transverse coordinates by $\Delta \bp{x} \equiv \bp{x}_2 - \bp{x}_1$. It is straightforward to show that the \textit{second derivative} of the Wightman function 
  \begin{equation}
  \partial_{V_1}  \partial_{V_2} \, G_M (x_1,x_2) = \bk{M}{ \partial_{V_1}\phi(x_1)\, \partial_{V_2}\phi(x_2)}{M} \equiv Q_M(x_1,x_2)
   \label{twentyfive}
  \end{equation} 
  has a universal behaviour on the null surface and is given by 
  \begin{equation}
 Q_M = -\frac{1}{4\pi}\frac{ \delta_D(\Delta \bp{x})}{(V_2 - V_1)^2}
   \label{threethreea}
  \end{equation} 
  (This result arises because of the conformal dimension of the fields $\partial_V(\phi)$ in CFT; but, of course, it can be derived by direct computation; see Appendix \ref{appen:wightman})
  We can exploit this universal behaviour to derive horizon thermality by the following procedure. We will first express $G_M$ and $Q_M$  in terms of the CFT modes given in \eq{twotwo}. This, in turn, will bring in the expectation values of the type 
 $\bk{M}{a_\alpha^\dagger\, a_\beta}{M}$ and $\bk{M}{a_\alpha\, a_\beta^\dagger}{M}\equiv \delta_{\alpha\beta} + \bk{M}{a_\alpha^\dagger\, a_\beta}{M}$.  comparing this result with the form of $Q_M$ in \eq{threethreea}, we can determine $\bk{M}{a_\alpha^\dagger\, a_\beta}{M}$ and show that it is thermal. 
 
 Let us start with the mode expansion of $\phi(x)$, near $\mathcal{H}^+$, in the CFT coordinates ($u,v,\bp{x}$) which is given by:
 \begin{equation}
 \phi = \sum_\alpha \left( a_\alpha f_\alpha(x)+ \text{h.c.}\right)
   \label{twothree}
  \end{equation} 
 with  
 \begin{equation}
 f_\alpha \equiv f_{\bkp \omega} = \left[ \frac{1}{(2\pi)^{d/2}} \frac{e^{i\Gamma}}{\sqrt{4\pi\omega}} \ e^{i\bkp\cdot \bp{x} - i \omega v } + f_1(u)\right]
  \end{equation} 
 where $ f_1(u) $ is the part of mode function which is independent of $v$ (and hence is irrelevant to the derivative $\partial_V\phi$) and $\Gamma (\bkp, \omega)$ is an unimportant phase. This phase cannot be determined from CFT considerations alone but, as to be expected, this is also irrelevant for our calculation. The 
  $V$ derivative of the mode function   $g_\alpha (x) \equiv \partial f_\alpha/\partial V=e^{-v}(\partial f_\alpha/\partial v)$  is given by
  \begin{equation}
 g_\alpha (x) = g_{\omega\bkp} \equiv \frac{e^{i\Gamma}\, e^{i\bkp\cdot \bp{x} }}{(2\pi)^{d/2} \, \sqrt{4\pi\omega}} \ e^{-v-i\omega v} \, (-i\omega)
   \label{threeone}
  \end{equation} 
  A simple computation shows that $Q_M$ in \eq{twentyfive} can be expressed in the form
  \begin{equation}
  Q_M (x_1,x_2) = \sum_{\alpha\beta} \left[ \bk{M} {a_\alpha\, a_\beta^\dagger}{M}\ g_\alpha (x_1) \, g_\beta^* (x_2) 
   + \bk{M}{a_\alpha^\dagger \, a_\beta}{M}\ g_\alpha^* (x_1) \, g_\beta (x_2)\right]
   \label{twoseven}
  \end{equation} 
  The expectation values, $\bk{M}{a_\alpha\, a_\beta}{M}$ and $\bk{M}{a_\alpha^\dagger \, a_\beta^\dagger}{M}$, vanish because of time translation invariance with respect to $\tau$. In Heisenberg picture $a_\alpha \equiv a_{\bkp \omega}$ will evolve in time as $e^{-i\omega \tau}$ so that $\bk{M}{a_\alpha\, a_\beta}{M}$ will evolve to $\bk{M}{a_\alpha\, a_\beta}{M} \, \exp [ - i\tau (\omega_\alpha + \omega_\beta)]$ which can be independent of $\tau$ only if $\bk{M}{a_\alpha\, a_\beta}{M} =0$, because the $\omega_\alpha$s are positive. Comparing the right hand side of \eq{twoseven} with the expected form  in \eq{threethreea}, we find that we must have $\bk{M}{a_\alpha^\dagger \, a_\beta}{M} = n_\alpha \delta_{\alpha\beta}$. More explicitly, we must have 
  \begin{equation}
   \bk{M}{a_\alpha^\dagger \, a_\beta}{M} \equiv \bk{M}{a_{\omega \bkp}^\dagger \, a_{\nu \bp{p}}}{M} =  n_\omega \delta(\omega -\nu) \delta(\bkp - \bp{p})
   \label{twoeight}
  \end{equation} 
 In particular, the expectation value of the number operator $n_\omega$ must be independent of transverse momentum variable in order to reproduce the Dirac delta function in transverse coordinates in \eq{threethreea}. Using \eq{twoeight} in \eq{twoseven} we can express $Q_M$ in the form 
 \begin{equation}
 Q_M = \sum_\alpha \left[ \left(n_\alpha + 1\right)  g_\alpha (x_1) g_\alpha^*(x_2) + n_\alpha  g_\alpha^* (x_1) g_\alpha(x_2)\right]
 = \sum_\alpha g_\alpha(x_1) g_\alpha^*(x_2) + \sum_\alpha 2 n_\alpha \left( \text{Re}\left[g_\alpha (x_1)  g_\alpha^*(x_2)\right] \right)
 \end{equation} 
 We note that the first term is exactly what we would have obtained if we have been working with the Wightman function in the Rindler like vacuum $\ket{R}$. Denoting this by $Q_R$ and using the explicit form of $g_\alpha(x)$ in \eq{threeone} we can write 
 \begin{equation}
  Q_M - Q_R = 2 \sum_\alpha n_\alpha \text{Re} \left[  g_\alpha (x_1) g_\alpha^*(x_2)\right] = \delta(\bp{x}^1-\bp{x}^2) e^{-(v_1+v_2)} \int_0^\infty \frac{d\omega}{2\pi} \, \omega \, n(\omega) \, \cos\omega(v_2-v_1)
  \label{threetwo}
 \end{equation} 
 To proceed further it is convenient to rewrite the left hand side in terms of CFT coordinates. Using the transformation $V = e^v$, we can rewrite \eq{threethreea} as:
 \begin{equation}
Q_M = -\frac{\delta_D(\Delta \bp{x} )}{4\pi} \frac{e^{-(v_1+v_2)}}{4\sinh^2 (v/2)}
  \label{threefoura}
 \end{equation} 
 Similarly, the $Q_R$ can be computed from the explicit form of $g_\alpha$ (or could be guessed!) to be 
 \begin{equation}
 Q_R = -\frac{\delta_D(\Delta \bp{x} )}{4\pi} \frac{e^{-(v_1+v_2)}}{v^2}
  \label{threethreeb}
 \end{equation} 
 Comparing the form of $Q_M - Q_R$ arising from \eq{threefoura} and \eq{threethreeb} with \eq{threetwo}, we find that 
 \begin{equation}
-\frac{1}{4\pi}\left[\frac{1}{4 \sinh^2(v/2)} - \frac{1}{v^2}\right] = \frac{1}{2\pi} \int_0^\infty d\omega \, \omega\, n(\omega) \cos\omega v
\label{finres1}
 \end{equation} 
 This equation determines $n(\omega)$ as an inverse cosine transformation of the function in the left hand side. This inverse cosine transformation can be obtained from the standard result 
 \begin{equation}
\frac{1}{2} \left[ \frac{1}{y^2} - \frac{\pi^2}{\alpha^2}\frac{1}{\sinh[(\pi/\alpha)y]}\right]=\int_0^\infty \frac{x\cos xy}{(e^{\alpha x}-1)} \, dx 
\label{ict}
 \end{equation} 
 with the parameters, on the left hand side of \eq{ict} set to $\alpha = \pi$ and $y= v/2$. This gives us: 
 \begin{equation}
  \left[ \frac{1}{4 \sinh^2(v/2)}- \frac{1}{v^2}\right] = -\int_0^\infty d\omega \, \frac{2\omega}{e^{2\pi\omega }-1}\, \cos \omega v
 \end{equation} 
 Comparing this expression with \eq{finres1} we can read off the expectation value of the number operator to be 
 \begin{equation}
  n(\omega) = \frac{1}{e^{2\pi \omega} - 1}
 \end{equation} 
 which is a Planck spectrum with temperature $T= 1/2\pi$.
 
 \subsection{Cautionary Notes on two similar phenomena}
 
 There are two phenomena,  widely discussed in the literature, which are similar to the thermal ambience of horizons described in the previous sections. The first is the response of detectors moving on the integral curve of the boost Killing vector and the the second is the radiation emitted by collapsing matter, especially when it forms a black hole. While these two situations are similar to the ones we have discussed there are some important conceptual differences which needs to be stressed.
 
 Let me start with the response of detectors \cite{takagi,vacprobe,detector}. It is possible to construct a simple model for a particle detector which, when traveling along the integral curve of the boost Killing vector, in $(1+3)$ dimensions, will respond as though it is at rest in a thermal radiation field at temperature $T=1/2\pi$. 
 This result is often misinterpreted  as providing an operational procedure to demonstrate the ``reality'' of the Rindler (type) particles in the Minkowski (type) vacuum. The purpose of the next, short, section is to stress that --- while the study of such particle detectors is a rewarding enterprise by itself --- it is misleading to use it to demonstrate the ``reality'' of Rindler type particles in Minkowski vacuum. I explain this fact in Sec. \ref{sec:detres} below.
 
The second issue has to do with the radiation emitted by a collapsing body \cite{aseem-tp-bh,brout}.  The thermal phenomena, described in the  previous sections, arise in manifolds with metrics (e.g., \eq{14cc}) which are time reversal invariant under $t\to -t$ where $t $ is a \textit{global} time coordinate covering the entire manifold. The thermodynamical phenomena which arise will then correspond to \textit{equilibrium}, time-reversal invariant, thermodynamics. There will be a thermal ambience, felt for e.g., in the R wedge, due to the horizon but \textit{no} flux of radiation will arise in the spacetime. A completely different conceptual situation arises when, for e.g., the spacetime manifold hosts collapsing matter. Such a collapse breaks time reversal invariance and irreversible phenomena can arise even with  natural boundary/initial conditions for the quantum field theory. In the case of black hole spacetimes, the resulting flux of radiation is approximately thermal (at late times) corresponding to the temperature $T=1/8\pi M$. This is also the temperature of $\mathcal{H}^+$ as perceived from, say, the R wedge in the case of an eternal, time-symmetric, black hole manifold. In spite of this fact, these two phenomena --- thermal ambience of $\mathcal{H}^+$ and flux of radiation emitted to spatial infinity --- are completely different conceptually, which is a distinction worth emphasizing. 
 
 \subsubsection{Response of particle detectors}\label{sec:detres}
 
 It is possible to construct physical systems which could act as idealized  `particle detectors' and study their response when they travel along different trajectories. In particular one could ask whether they will `click' (i.e., detect particles) in the inertial vacuum when they travel along non-inertial trajectories \cite{detector}. When the detector is moving along an arbitrary trajectory, its response (for e.g., its rate of clicking) will be time dependent. But if it is moving along an integral curve of  some timelike Killing vector field then the response will be stationary and it will click in a steady rate. Recall that the 3+1 flat  spacetime has 10 Killing vector fields corresponding to the Poincare group of symmetries. It is possible to construct (nontrivial) linear combinations of these Killing vector fields so that they remain time-like in some regions of the spacetime. If a particle detector follows the integral curve of any such time-like Killing vector, $\bm{\xi_{\bar\tau}} \equiv \partial/\partial \bar\tau$ (where $\bar\tau$ parametrizes the curve) it will  register --- in general --- a stationary rate of detection with specific spectral signature. So the response will be as though the detector is at rest  in a bath of particles with some spectral characteristics.
 
 Given the timelike Killing vector $\bm{\xi}_{\bar\tau}$ one can also construct a coordinate system with $\bar\tau$ as the time coordinate; in this coordinate system $\bm{\xi_{\bar\tau}}$ will have components $(1,\bf{0})$. One can then quantize the field in this coordinate system  using modes which vary as $\exp(\pm i\omega \bar\tau)$. The corresponding  annihilation operator, $\bar a_\omega$ will define a vacuum state and the excitations above this vacuum state can be used to construct the standard Fock basis. One can then compute  the number density $\bk{M}{\bar{a}_\omega^\dagger \bar{a}_\omega}{M}$ of particles defined using $\bar n_\omega=\bar a_\omega^\dagger \bar a_\omega$ in the \textit{inertial} vacuum $\ket{M}$.  A natural question is whether $\bk{M}{ \bar a_\omega^\dagger \bar a_\omega}{M}$, computed using QFT and Bogoliubov coefficients etc., will match with the response registered by the `particle detector'. If they are identical, then particle detectors can be thought of giving an operational meaning to the particles defined as excitations of the quantum field.
 
 The simplest timelike Killing vector which can be used for this purpose, of course, is the one corresponding to Lorentz boosts. We have seen earlier that, in this case, alternative quantization leads to a thermal spectrum for $\bk{M}{ \bar a_\omega^\dagger \bar a_\omega}{M}$ in any $(1+D)$ spacetime. 
 It turns out that, \textit{in the $(1+3)$ spacetime,}  a detector traveling along the integral curve of the boost Killing vector in the R wedge will \textit{also} register a  constant rate of transitions  corresponding to a thermal spectrum of particles. However, this is more of an accident than a general result.
 An interpretation of the response of particle detectors as providing an operational meaning to the `reality' of these particles can be misleading. To understand why,  we only have to consider two important facts:
 
 (a) First, the result does not extend to more general class of time like trajectories corresponding to other Killing vector fields which can be constructed. In general, the detector response will not match with $\bk{M}{ \bar a_\omega^\dagger \bar a_\omega}{M}$ computed using QFT and Bogoliubov coefficients. These  examples  show that particles defined 
  as excitations of underlying quantum field, using  a particular scheme of quantization, are not the same as the `particles' defined by detector response. From the detailed analysis of the response of the particle detectors, it is obvious that these detectors actually measure the spectral pattern of the vacuum fluctuations of the quantum field. These fluctuations do receive contributions from  particle-like excitations but they also receive other contributions; in general, there is no one-to-one correspondence between  particle excitations and the pattern of vacuum fluctuations. (For a review and earlier references, see \cite{vacprobe}).

(b) A more dramatic discrepancy between particle detectors and QFT arises when we consider arbitrary spatial dimension $D$. A particle detector, \textit{even while traveling along the integral curve of the boost Killing vector}, will not register the correct thermal spectrum  in odd spatial dimensions (see e..g \cite{takagi}). In general, the detector will respond as though it is immersed in radiation with the spectral function $F(\Omega)$ where (see Appendix \ref{appen:detectorresp} for a simple derivation)
 \begin{equation}
 F(\Omega) = \frac{2^{3-2D}\, \pi^{1-D/2}}{\Gamma(D/2)} \ \frac{Q_D( \Omega)}{e^{2\pi  \Omega} + (-1)^D}
 \end{equation} 
 where $D$ is the \textit{spatial} dimension and $Q_D(\Omega)$ is a specific polynomial function defined by 
 \begin{equation}
 Q_D(\Omega) \equiv -e^{i\pi D/2}\, \lim_{z\to 0}\frac{d^{D-2}}{dz^{D-2}} \,\frac{z^{D-1} e^{-2i\Omega z}}{\sinh^{D-1}z}
 \end{equation} 
 The QFT of a \textit{bosonic} scalar field in $1+D$ dimension, on the other hand,  tells us that the inertial vacuum is populated by a bosonic thermal distribution of Rindler type particles for all $D$; but the detector response will coincide with this spectrum only in odd dimensions. (Most popular dimensions in the literature are $D=1$ and $D=3$, where this discrepancy will be missed!).

 \subsubsection{Black hole formation and evaporation}
 
 All the thermal phenomena discussed in the previous sections arise in manifolds with metrics which are time reversal invariant under $t\to -t$ where $t $ is a global time coordinate covering the entire manifold; see e.g., \eq{14a} or \eq{14cc}. The metric, in general, \textit{depends} on this time coordinate $t$ and hence $\bm{\xi}_t \equiv \partial/\partial t $ is not a Killing vector (except in the special case of $F \equiv 1$ corresponding to flat spacetime). But this time dependence is through $t^2$ which preserves invariance under time reversal.  Any quantum field dynamics, in such a manifold with time reversal invariance, will also respect this symmetry (unless, of course, you specifically break the symmetry by the choice of boundary conditions). This, in turn, implies that the thermodynamical phenomena which arise will correspond to \textit{equilibrium}, time-reversal invariant, thermodynamics. There will be a thermal ambience due to temperature but without any flux of radiation in the spacetime. No irreversible thermodynamic feature will arise unless --- as mentioned before --- we artificially introduce it by peculiar boundary conditions. The thermal ambience is experienced in, for e.g., the R and L wedges wherein we have a time-like Killing vector, $\xi_\tau\equiv \partial/\partial \tau$. 
 
 Collapsing matter in a spacetime, on the other hand,  will lead to genuine particle production because of the time-dependent gravitational field and a flux of radiation can be emitted towards spatial infinity. By and large, the nature of this radiation will depend on the details of the collapse. However, in the context of a collapse leading to a black hole with a bifurcate Killing horizon, the radiation detected at spatial infinity, at late times, will be approximately thermal with a temperature equal to the horizon temperature obtained earlier.  The fact that these two phenomena --- viz., thermal radiation from collapsing matter and thermal ambience in a time-reversal invariant manifold --- are characterized by the same temperature should not mislead one into ignoring the major conceptual differences between the two phenomena. Given the importance of this issue, let me briefly point out this key differences by the following example \cite{aseem-tp-bh, brout}:
 
 Consider a spherically symmetric configuration of matter, of mass $M$, collapsing under self-gravity with its surface radius $R(\tau)$ decreasing in a specified manner as measured with respect to the Killing time coordinate $\tau$ in the outside Schwarzschild metric. Let this spacetime host a scalar field which was initially in a natural vacuum state in the static geometry. It is possible to solve for the dynamics of the scalar field for simplified models both analytically (in 1+1 spacetime) and numerically (in 1+3 spacetime). One can then compute the expectation value of the (suitably regularized) energy-momentum tensor of the scalar field and investigate the nature of radiation emitted towards spatial infinity by the collapsing matter. One will then find the following results \cite{aseem-tp-bh, brout}: 
 
 (a) Let the spherical body start from an initial radius $R_0\gg 2M$ and collapse towards $R\to 2M$. Initially there will be very little radiation; but as the body collapses towards $R\gtrsim  2M$ there will be a flux of radiation towards spatial infinity which is very close to thermal with a temperature $T=1/(8\pi M)$. 
 
 (b) If the body continues to collapse and form a black hole, then the radiation received at  spatial infinity as $\tau \to \infty$ will approach  thermal radiation more and more closely. (In (1+3) dimension there will be a correction, called grey-body factor, which arises due to the transmission of the radiation in the outside Schwarzschild metric.)  
 
 (c) A more interesting situation arises when the collapse asymptotically approaches the radius $R\equiv 2M(1 -\epsilon^2)^{-1}$ and a \textit{black hole is never formed}. One will the find that there is still emission of (approximately) thermal radiation with the temperature $T= 1/(8\pi M)$ (which is the same as our `horizon' temperature, even though no horizon has been formed), for an arbitrarily large but finite duration of time. More precisely, this radiation will last for a time interval
 \begin{equation}
  M\lesssim \tau \lesssim M\ln (1/\epsilon^2)
 \end{equation} 
 which can be an arbitrarily large --- but finite --- interval of time when $\epsilon $ becomes arbitrarily small but remains non-zero. This happens even without a black hole or horizon formation.
 After this time interval, for $\tau\gg M\ln (1/\epsilon^2)$ the flux of radiation towards spatial infinity decays to zero exponentially. Asymptotically, the system will remain static and time invariant. 
 
 These results clearly tell us that the black hole radiation in situation (b) is best interpreted as the limiting case of (c) when $\epsilon \to 0$. By keeping $\epsilon$ finite but arbitrarily small, we can prevent the formation of a black hole and the horizon but we will still encounter nearly thermal radiation propagating to spatial infinity for arbitrarily large but finite interval of time (with the interval varying as $M\ln (1/\epsilon^2$). In other words, black hole radiation with a temperature $T=1/(8\pi M)$ arises due to the \textit{time dependence} of the collapsing metric. on the other hand, the thermal ambience discussed in the previous sections is a phenomenon closely related to the existence of a null surface acting as a one way membrane.
 
 How come the two temperatures in these two, conceptually different, phenomena are the same? The physical reason is that the temperature obviously has to be the horizon temperature once the horizon is formed, that is, when $\epsilon =0$. Continuity in the parameter $\epsilon$ ensures that the temperature is close to the horizon temperature when $\epsilon \ll 1$ but non-zero. Mathematically, the thermal spectrum arises due to a simple algebraic fact: The power spectrum of complex plane wave undergoing exponential redshift is Planckian (see Appendix \ref{appen:planck}). In fact, this algebraic feature lies at the heart of the computation of Bogoliubov coefficients in the context of the horizon temperature as well. Same integrals lead to same physics. 
 
 \subsection{Observer dependent entropy of null surfaces}\label{sec:obdeent}
 
  The mathematical formulation leading to the association of temperature with any horizon is fairly universal and it does not distinguish between different horizons, like, for example, Rindler horizon in flat space or a Schwarzschild black hole event horizon or a de Sitter horizon. But since horizon blocks information, they are also endowed with an entropy \textit{with respect to} the observers who perceive the horizon as blocking their information access. Since we expect temperature and entropy to arise for fundamentally the same reason, it would be natural  to associate entropy with \textit{all} the  horizons. I will now comment on several aspects of this in a more general context,  as well as in the specific context of black hole entropy.
  
  The main result is that horizon entropy --- like temperature --- should be treated as an observer dependent concept. It is true that the event horizon of a black hole can be given a purely geometrical definition while  the Rindler horizon as well as the de Sitter horizon is observer dependent.
This  fact, however, is irrelevant for the purpose of associating an entropy with the horizon.
An observer moving into a black hole will have access to different amount of information (and  will attribute different thermodynamic properties to the black hole) compared to an observer who is remaining stationary outside the horizon. This situation is similar to what happens in the case of Rindler frame as well; an observer whose trajectory crosses the horizon will certainly have access to different regions of spacetime compared to the  observer confined to R wedge. In both the cases the physical effect of horizon in blocking information depends on the set of world lines one is considering and, in this sense, all horizon entropies are observer dependent.
  
 If you want to relate entropy to underlying degrees of freedom, \textit{\`{a} la} Boltzmann, you have to accept that these degrees of freedom are also observer dependent. For example, the often asked (and sometimes even answered!) question, viz., ``what are the degrees of freedom responsible for black hole entropy?'' is ill-posed until you specify the observer --- or more generally --- the spacetime foliation. As it stands, this question is as meaningless as asking what is the energy of a given photon without specifying the four velocity of the observer who is measuring it.  
 
 This result also brings vacuum fluctuations and thermal fluctuations closer to each other. As we have seen in Sec. \ref{sec:tdlee}, vacuum expectation values of  observables in R wedge exhibit standard thermodynamic properties like thermal fluctuations, because their physics is governed by a thermal density matrix. These thermodynamical features arise because the  vacuum state, which is a \textit{pure} quantum state, leads to  a thermal density matrix when we integrate out the unobservable modes.
In this sense,  these thermal effects are intrinsically quantum mechanical,   suggesting that the distinction between quantum fluctuations and thermal fluctuations  could be artificial (like e.g., the distinction between energy and momentum of a particle in non-relativistic mechanics) and might fade away in the correct description of spacetime, when one properly takes into account the fresh observer dependence induced by the existence of  horizons.   

For a  concrete computation of the entropy, let us consider an excited state of a quantum field with energy  $\delta E$ above the ground state in an inertial spacetime. Once we integrate out  the unobservable modes  we will get a density matrix $\rho_1$ for this state (as viewed in R)
  and the corresponding entropy will be $S_1 = - {\rm Tr}\ (\rho_1 \ln \rho_1)$.
  On the other hand, the inertial vacuum state itself has the density matrix $\rho_0$ and the entropy $S_0 = - {\rm Tr}\ (\rho_0 \ln \rho_0)$ in the R wedge. The difference $\delta S = S_1 - S_0$ is finite and 
  represents the entropy attributed to this (excited) state by the observers confined to R. (This difference is
  finite though $S_1$ and $S_0$ can be divergent.) In the limit of $\kappa \to \infty$,
  which  corresponds to  a n observer who is very close to the horizon,
   we can easily compute it and show that 
  \begin{equation}
\delta S = \beta \delta E = \frac{2\pi}{\kappa} \delta E
\label{delS}
\end{equation}
To prove this result, note that if we write $\rho_1=\rho_0 + \delta \rho$, then in the limit of
$\kappa \to \infty$ we can concentrate on states for which $\delta \rho/\rho_0\ll 1$.
Then we obtain:
\begin{eqnarray}
-\delta S &=& {\rm Tr}\ (\rho_1 \ln \rho_1) - {\rm Tr}\ (\rho_0 \ln \rho_0) 
\simeq {\rm Tr}\ (\delta \rho \ln \rho_0)\nonumber\\
&=& {\rm Tr}\ (\delta \rho (-\beta H_R)) = - \beta {\rm Tr}\ \left((\rho_1 -\rho_0)H_R\right) \equiv -\beta \delta E
\end{eqnarray} 
where we have used the results Tr $\delta \rho \approx 0$ and
$\rho_0 =Z^{-1}\exp(-\beta H_R)$ where $H_R$ is the Hamiltonian for the system in the 
Rindler frame. The last equality defines the $\delta E$ in terms of the difference in 
the expectation values of the Hamiltonian in the two states. 
This is the amount of entropy which will be lost, from the perspective of a Rindler observer close to the horizon, when the matter
disappears into the horizon.

The above result is valid in spite of the fact that, formally, matter takes  an infinite amount of 
coordinate time to cross the horizon as far as the outside observer is concerned.
This is  because
 quantum gravitational effects will smear the location of the horizon by $\mathcal{O} (L_P)$ effects \cite{D2c,D2d,tplimitations}.
 So one cannot really talk about the location of the event horizon ignoring spacetime fluctuations of this order.\footnote{We have emphasized that the horizon in Lorentzian sector $X^2-T^2=0$ corresponds to the origin $X^2+T_E^2=0$ in the Euclidean sector. This will require specifying a point with infinite accuracy in the Euclidean space. If we assume that quantum fluctuations require $X^2+T_E^2\gtrsim L_P^2$ then the horizons will satisfy the condition $X^2-T^2\gtrsim L_P^2$ in the Lorentzian sector.}
 So, from the operational point of view, we only need to consider matter reaching within few Planck lengths of the horizon to talk about entropy loss.
Naive reasoning would suggest that the expression for entropy of matter
crossing the horizon 
should consist of its energy $\delta E$ and \textit{its own} temperature $T_{\rm matter}$
rather than the horizon temperature. 
But the correct expression is $\delta S = \delta E/T_{\rm horizon}$ with the horizon temperature replacing the matter temperature;\footnote{In fact, physical processes very close to the horizon must play an important role
in order to provide a complete picture of these issues. There is 
already some evidence \cite{Padmanabhan:1998jp,Padmanabhan:1998vr} that the infinite redshift induced by the horizon plays a crucial
role in these phenomena.}
  it is as though the  horizon acts as a system with some internal degrees of freedom and temperature $T_{\rm horizon}$ \textit{as far as Rindler observer is concerned} so that when one adds an energy $\delta E$ to it, the entropy change is $\delta S = (\delta E/ T_{\rm horizon})$.
All these are not \textit{independent} features but are only the consequence of the basic
result that 
a Rindler observer attributes a non-zero temperature to inertial vacuum. This temperature, in turn,  influences every other thermodynamic variable. 

\subsubsection{Surface term in the Hilbert action and the  entropy}

So far we have been working in the context of a non-dynamical spacetime, either flat or curved. A quantum field  in such a spacetime is seen to exhibit thermal features when the spacetime has a horizon. But, as I argued in Sec. \ref{sec:intro}, this thermality is actually a clue to the microscopic structure of spacetime. Such a point of view necessarily demands that there should exist deeper connections between gravitational \textit{dynamics} and the thermodynamic concepts like temperature and entropy which we have attributed to the null surfaces in the  spacetime. Conventionally, dynamics of gravity is obtained from the Hilbert action. So if the thermodynamic concepts have deeper relations to the dynamics of gravity, it \textit{must} reflect in the structure of the action principle \cite{ayan,inherentqm,tpreviews2,tpreviews3,horts,KBP}.  

At first sight, this  appears impossible  due to the following reason: We have seen that thermal features are exhibited by the null surface $X=\pm T$ even in \textit{flat} spacetime. One would have thought that the flat spacetime cannot host any gravitational dynamics and, in fact, the Hilbert Lagrangian, proportional to Ricci scalar $R$, vanishes in flat spacetime. So if the horizon thermality is linked to the dynamics of gravity, expressed through the action principle, how can we explain the existence of thermal phenomena in \textit{flat} spacetime? 

The answer is very beautiful and provides us with a deep insight into the structure of  gravitational action itself. It is related to the fact that the correct Lagrangian  for gravity is \textit{not} just $R$; such a Lagrangian has no well-defined variational derivative because of the existence of second derivatives of the metric tensor in $R$. To get the correct action, which has a  well-defined variational principle, one can proceed in two different ways: 
(a) We can add an extra \textit{surface} term to the Hilbert Lagrangian (usually the integral over the extrinsic curvature at the boundary, though the choice is not unique) such that its variation cancels that arising from the second derivatives of the metric, thereby leading to the correct equations of motion.\footnote{In fact, by examining  the structure of $R$, one can discover the surface term that needs to be added; see \cite{krooth}.} 
(b) In the Ricci scalar, $R$, having the structure $R= \partial \Gamma + \Gamma^2$, we can simply discard a total derivative term $\partial \Gamma$ and use the $\Gamma^2$ term as the Lagrangian. 

\textit{In either procedure we can obtain a non-zero gravitational action even in flat spacetime!} In approach (a) the surface term which we have added  remains non-zero even in flat spacetime with $R=0$. In the approach (b), we get the Lagrangian $\Gamma^2$ to be a surface term $\Gamma^2 = -\partial \Gamma$ when $R=0$.  In other words, the gravitational action in flat spacetime reduces to a pure surface term in non-inertial coordinates. We will see that this surface term actually gives the heat density $H_{\rm sur} = Ts$ when evaluated on any null surface \cite{horts,KBP,36a}. 

This result provides the first link between horizon thermality and \md. When we obtained the temperature and entropy of $\mathcal{H}^+$ using standard QFT in Sec. \ref{sec:horterm}, we did not introduce any feature about gravitational dynamics or gravitational action. The fact that the   gravitational action in flat spacetime is closely related to what we found using quantum field theory, demands an explanation. If we take the point of view that horizon thermality has nothing to do with \md, it is extremely difficult to understand why the surface term in gravitational action should have such a simple thermodynamic interpretation. 
 
We will see now how this nontrivial connection comes about. 
The gravitational action, separated into bulk (quadratic) term and surface term, can be expressed in the form:
\begin{align}\label{bulksur1}
\sqrt{-g}R=\frac{1}{2}N^{c}_{ab}\partial _{c}f^{ab}-\partial _{c}\left(f^{ab}N^{c}_{ab}\right)
\equiv L_{\rm quad} -L_{\rm sur}
\end{align}
where 
\begin{align}\label{Paper06_Sec_01_Eq01}
f^{ab}\equiv\sqrt{-g}g^{ab};\qquad N^{c}_{ab}\equiv-\Gamma ^{c}_{ab}+\frac{1}{2}\left(\delta ^{c}_{a}\Gamma ^{d}_{db}+\delta ^{c}_{b}\Gamma ^{d}_{ad}\right)
\end{align}
The set $(f^{ab},N^{c}_{ab})$, of course, contains the same amount of information as $(g_{ab},\Gamma^{c}_{ab})$ but has more direct thermodynamic interpretation \cite{KBP}. These two are also dynamically conjugate variables in the sense that
\begin{equation}
 N^c_{ik}=\frac{\partial  L_{quad}}{\partial (\partial_cf^{ik})}
\end{equation}
It turns out that the surface term in the action, obtained by integrating $L_{sur}$, has a direct thermodynamic interpretation when evaluated on a null surface \textit{in any spacetime}.
Let $\mathcal{H}$ be  a null surface which is perceived as a horizon by the local Rindler observers. Let them attribute to it a temperature $T$ and entropy density $s=\sqrt{\sigma}/4$ where $\sigma$ is the metric determinant of the transverse metric with $\sqrt{\sigma}\, d^{D-2} x$ being the transverse area element.
Then, one can show that \cite{KBP}:
\begin{itemize}
 \item The combination  $ N^{c}_{ab} f^{ab}$, when integrated  over $\mathcal{H}$ with the usual measure $d^3 \Sigma_c=\ell_c\sqrt{\sigma}d^2xd\lambda$ (where $\ell_c$ is the null normal to $\mathcal{H}$) gives its
 heat content; that is:
 \begin{equation}
  \frac{1}{16 \pi L_P^2}\int d^3 \Sigma_c (N^{c}_{ab} f^{ab}) =\int d\lambda\ d^2x\ T  s
  \label{surf1}
 \end{equation} 
 
 \item Consider next the metric variations $\delta f$ which preserve the null surface.  Remarkably enough, the combinations   $f\delta N$ and $N\delta f$ will then correspond to the variations  $s\delta T$ and $T\delta s$, when integrated over the null surface. That is, we can show:
 \begin{eqnarray}
\frac{1}{16 \pi L_P^2}\int  d^3 \Sigma_c(N^{c}_{ab}\df f^{ab})&=&  \int d\lambda\ d^2x\ T \df s; 
\label{stsdt0}\\
\frac{1}{16 \pi L_P^2}\int  d^3 \Sigma_c (f^{ab}\df N^{c}_{ab})&=&  \int d\lambda\ d^2x\ s \df T
\label{stSdT}
\end{eqnarray}
So the variations ($N\delta f, f\delta N$) exhibit \textit{thermodynamic conjugacy} very similar to that seen in the corresponding   variations $(T\delta s, s\delta T)$. 
\end{itemize}

These results also hold in flat spacetime. The fact that neither the surface term nor the bulk term is individually covariant is crucial for this result; it implies that even in flat spacetime, these two terms can be individually non-zero, though the sum, being proportional to $R$, vanishes in flat spacetime. This allows us  to obtain the thermality of a null surface in flat spacetime by using non-inertial coordinates. 

Considering its conceptual importance, let us examine the flat spacetime case a bit more closely \cite{horts}. This is best done by rewriting the surface term in a different, equivalent, form (with $L_P^2 =1$):
\begin{equation}
 L_{\rm sur}=\frac{1}{16\pi}\partial_c(\sqrt{-g}V^c); \qquad V^c \equiv -\frac{1}{g} \partial_b(gg^{bc})
\end{equation}  
(see e.g., eq (6.15) of \cite{gravitation}). The surface action $\mathcal{A}_{\rm sur}$ is defined as the integral over $L_{sur}$. 
The heat energy $H_{\rm sur}$ can be computed \cite{tpstructure,tpnoeeng} as the Hamiltonian associated with the surface term of the Hilbert action  through $H_{\rm sur} = -(\partial \mathcal{A}_{\rm sur}/\partial \tau )$.
Because the near horizon metric can be approximated as a Rindler metric (with $-g_{00}=1/g_{xx}=N^2=2\kappa x$ we can evaluate the surface term in the action on the $N=$const surface. We get:
\begin{equation}
\mathcal{A}_{\rm sur}= \frac{1}{16\pi}\int_{x}d\tau  d^2x_\perp V^x
=\pm \tau \left(\frac{\kappa A_\perp}{8\pi}\right)
\label{surfaceH}
\end{equation} 
where $A_\perp$ is the transverse area.\footnote{This result is in the Lorentzian sector. Since the horizon maps to the origin of the Euclidean plane, the corresponding calculation in the Euclidean sector proceeds as follows: We evaluate the Euclidean surface term on a cylinder with a circular base $X^2+T_E^2=\epsilon^2$ in the $XT_E$ plane and extending in transverse directions. The Euclidean `time' integration covers the range $(0,2\pi/\kappa)$. When we take $\epsilon\to 0$ limit --- going to the  Euclidean origin, which maps to the Lorentzian horizon --- we get the surface term to be $A_\perp/4$, i.e equal to the entropy. This is easily seen by setting $\tau =2\pi/\kappa$  in \eq{surfaceH}.}
(The overall sign depends on the convention chosen for the outward normal and whether the contribution of the integral is taken at the inner or outer boundaries; see e.g., the discussion in \cite{surH}. I will choose the negative sign in \eq{surfaceH}.) 
More generally, for any  static, near-horizon, geometry the integrals in \eq{surfaceH}  leads to the same result.
From $A_{sur}$, we get the surface Hamiltonian to be: 
\begin{equation}
 H_{sur}\equiv -\frac{\partial\mathcal{A}_{sur} }{\partial \tau }
 =\frac{1}{16\pi}\int_{x} d^2x_\perp V^x
=\left(\frac{\kappa A_\perp}{8\pi}\right)=TS
\label{horsurfham}
\end{equation} 
with suitable choice of sign.

\subsubsection{Degrees of freedom of null surfaces}

There is another aspect to the observer dependence of the degrees of freedom which contribute to the entropy of a null surface.
In conventional physics, we are accustomed
to thinking of degrees of freedom of a system as absolute, i.e, independent of observer or the coordinate foliation
used by the observer. Then the entropy, 
related to the logarithm of the degrees of freedom,
will also be absolute and independent of the observer. On the other hand,
we  now know that horizon entropy, horizon temperature
etc. \textit{must be} treated as observer dependent notions; e.g.,
a freely falling observer through a black hole
horizon and a static observer outside the black hole will
attribute different thermodynamic properties to the horizon.
It  follows that any microscopic degrees of freedom
which leads to horizon entropy must also be necessarily
observer (foliation) dependent. Let us see how this comes about \cite{horts,bibhas-tp}. 

The usual description of gravity is invariant under the set $\mathcal{C}$ of \textit{all} possible diffeomorphisms. These diffeomorphisms allow us, in principle, to remove all the gauge degrees of freedom in the description of spacetime,  retaining only the diffeomorphism invariant physical degrees of freedom. Consider next   the physical context in which a particular null surface is treated as special and we restrict ourselves 
to only those diffeomorphisms in the set $\mathcal{C}'$, which retain the horizon structure of this null surface. (This could be prescribed, for example, in terms of specific boundary behaviour of metric on and near the null surface.) While it is a difficult (and largely unsolved) problem to  quantify what this restricted class $\mathcal{C}'$ is, it is reasonable to assume that $\mathcal{C}'$ will be a proper subset of  $\mathcal{C}$. This, in turn, implies that using only the diffeomorphisms available in $\mathcal{C}'$,  we cannot remove all the redundant gravitational degrees of freedom --- which we could have originally removed using the full set $\mathcal{C}$. Therefore, certain degrees of freedom which would have been treated as pure gauge (when the theory was invariant under $\mathcal{C}$) now gets `upgraded' to physical degrees of freedom (when the theory is invariant under 
$\mathcal{C}'$). The entropy attributed to the null surface, by the observers who perceive it as a horizon, arises from these degrees of freedom. 

To make progress towards a more concrete realization of the above ideas, let us investigate the following issue \cite{horts}: In a flat spacetime, the inertial observers do not attribute  thermal properties to any null surface.  Consider  an \textit{infinitesimal} coordinate transformation $x^a\to x^a + q^a(x)$ from the inertial coordinate system to the Rindler coordinate system and ask: Is it  possible to obtain the thermal properties, associated with the null surface, $\mathcal{H}^\pm$ in flat spacetime, in terms of the vector field $q^a$? 

It turns out that this is indeed possible. In this sense we can consider the infinitesimal transformations described by the vector field $q^a$ as having upgraded some of the gauge degrees of freedom into physical degrees of freedom. The same results hold in  much wider class of spacetimes with horizons, when we consider the infinitesimal coordinate transformations between the freely falling frame near the horizon and the frame of the static observers (like, for e.g., in terms of the infinitesimal coordinate transformation from Kruskal to Schwarzschild coordinates) and even in a more general contexts \cite{horts}. We will stick to flat spacetime because of its simplicity and the rather intriguing nature of the result.

Let us start  with a flat spacetime described in the inertial coordinates ($T,X,\mathbf{X}_\perp$) and make a coordinate transformation from the inertial coordinates ($T,X,\mathbf{X}_\perp$) to the Rindler coordinates $(\tau,x,\mathbf{X}_\perp)$ by: 
\begin{eqnarray}
 T = \kappa^{-1}(1+2\kappa x)^{1/2} \sinh (\kappa \tau); \quad
X  = \kappa^{-1}(1+2\kappa x)^{1/2} \cosh (\kappa \tau)-\kappa^{-1}.
\label{1.03}
\end{eqnarray}
This leads to the metric:
\begin{eqnarray}
ds^2 = - (1+2\kappa x)d\tau^2 + \frac{dx^2}{1+2\kappa x} + dL_{\perp}^2~.
\label{1.02}
\end{eqnarray}
These transformations  reduce to identity when the acceleration  $\kappa=0$ making the metric in \eq{1.02} reduce to the  flat spacetime metric. 
The infinitesimal form of the coordinate transformations in \eq{1.03} are obtained by retaining only terms up to linear\footnote{Since $\kappa$ is a dimension-full parameter, the `smallness'  of $\kappa$ is somewhat ill-defined. This can be handled by replacing $\kappa$ by $\epsilon \kappa$ in all expressions where $\epsilon$ is a \textit{dimensionless} infinitesimal parameter and perform the Taylor series expansion in $\epsilon$. We will not bother to do this since it ultimately leads to the same results.} order in $\kappa$ in \eq{1.03}. 
In general, such an  infinitesimal transformation between two coordinate systems is implemented by a vector field $q^a(x)$ in the spacetime by the shift $x^a \to \bar x^a = x^a + q^a(x)$ .  In our case,
the vector field $q^a \equiv x^a-X^a$ has the components in the inertial frame given by:
\begin{eqnarray}
&&q^{T} = \tau- T = - \kappa XT;
\nonumber
\\
&&q^X = x - X = -\frac{1}{2}\kappa T^2  + \frac{1}{2}\kappa X^2~.
\label{1.08}
\end{eqnarray}

How does the surface $X=T$ acquire an interpretation in terms of a heat content $H_{\rm sur}$ when we made a coordinate transformation from inertial to Rindler frame? This is related to the  computation of $H_{\rm sur}$,  based on surface term in the action, discussed previously.
 In the inertial coordinates all $\Gamma$s vanish making $A_{\rm sur} = 0 = A_{\rm bulk}$ in the Hilbert action and hence $H_{\rm sur} =0$; so we cannot attribute any heat content to the null surface $T=X$ in inertial coordinates. But since $A_{\rm sur}$ is not generally covariant, it acquires non-zero value under the infinitesimal coordinate transformations $x^a \to x^a + q^a(x)$. The connections, generated to the lowest order, by this infinitesimal transformation are  given by
$
 \Gamma^a_{bc} = -\partial_b \partial_c q^a.
$
For the infinitesimal transformation between inertial and Rindler coordinates in \eq{1.08}, this leads to the non-zero components (in the Rindler frame) to be
$\Gamma^\tau_{\tau x}  = \Gamma^x_{\tau \tau} = \Gamma^x_{xx}= \kappa$.
Correspondingly, the $V^c$ now has the component
$ V^x = -2\kappa$. 
Therefore, we generate a  non-zero surface term in the action, and corresponding surface Hamiltonian, given by
\begin{equation}
\mathcal{A}_{\rm sur}=- \tau\left(\frac{\kappa A_\perp}{8\pi}\right);\qquad
H_{\rm sur} = - \frac{\partial A_{\rm sur}}{\partial \tau} =   \left(\frac{\kappa A_\perp}{8\pi}\right)
\label{surfaceH1}
\end{equation} 
Of course,  we get $H_{\rm sur} \ne 0$ because $A_{\rm sur}$ is not generally covariant.
What \textit{is} rather surprising is that the \textit{infinitesimal} coordinate transformations are capable of leading to the exact result! The vector field $q^a$ was  obtained by taking the linear limit of the full transformations in   \eq{1.03} in $\kappa$.
The $A_{\rm sur}$ was also computed  only to linear order in $\kappa$ because we used the linearized expressions for the connection in its calculation. In spite of this approximation, the result in \eq{surfaceH1} matches with the exact result in \eq{surfaceH}.

 \subsubsection{Aside: Entropy of black hole and de Sitter horizons}

 Finally, let me comment on some results specifically related to black hole (and de Sitter) entropy. In the case of black hole, we take $E = M$ and $T = 1/(8\pi M)$. The standard party line is to integrate the relation $dS = dE/T(E)$ and obtain 
 \begin{equation}
 dS=\frac{dE}{T}\to S=\int (8\pi M)dM=4\pi M^2=\frac{A_H}{4}  
 \end{equation}
 This result is widely discussed in literature. However,
 there exist two more relations connecting these variables which have not received the attention they deserve. To begin with, the expression for $E,T$ and $S$ tells us that the bulk energy which is responsible for gravity can be related to degrees of freedom on the surface of the horizon through the relation \cite{A19,tpcqg}
 \begin{equation}
 E=2TS=\frac{1}{2}N_{sur}(k_BT); \qquad N_{sur}=\frac{A_H}{L_P^2}
  \label{eqns11}
 \end{equation}
 This relation associates one degree of freedom with one Planck area $L_P^2$ at the surface of the event horizon. This, in turn, allows us to compute the equipartition energy of the horizon surface as $(1/2) N_{\rm sur} (k_BT)$; this computation is purely local and uses the temperature and the degrees of freedom on the horizon. Equation (\ref{eqns11}) then tells us that \textit{equipartition energy of the horizon surface is equal to the bulk gravitating energy} in the spacetime. 
 If we define the bulk degrees of freedom in a spatial region $\mathcal{V}$ as $N_{\rm bulk} \equiv E_{\rm bulk}/[(1/2)k_BT]$ where $T$ is the  temperature of the surface $\partial \mathcal{V}$ enclosing the bulk region $\mathcal{V}$,  then, in this specific context, \eq{eqns11} is equivalent to the statement: $N_{\rm bulk}= N_{\rm sur}$ which one might call holographic. We will see later (see Sec. \ref{sec:geotherm}) that this result is very general and holds in all static spacetimes, when reformulated properly. Further, when the spacetime is not static, one can show that   the evolution of the geometry is sourced by the difference ($N_{\rm bulk} - N_{\rm sur}$).
  
The attribution of  $N_{\rm sur}$ degrees of freedom to the surface also allows us to write down the energy balance in the context of black holes in a different manner. Because $E= 2 TS$ for the black hole, we have $\delta E = 2 T \delta S + 2 S \delta T$. Using the relation $S \propto M^2 \propto T^{-2}$ we find that 
  \begin{equation}
\delta E= T\delta S=-2S\delta T=-\frac{1}{2}N_{sur}(k_B\delta T)
  \label{eqns12}
 \end{equation}
 This suggests that the change in the energy of the black hole can be related to change in the temperature with $N_{\rm sur}$ playing the role of effective degrees of freedom. This is exactly what you will find, say, for a gaseous system except for two crucial differences: (a) When you heat up the gas in a container the temperature changes but not the number of gas molecules. Here, since $N_{\rm sur}\propto S \propto T^{-2}$, change in temperature also changes $N_{\rm sur}$ even though it is not transparent in \eq{eqns12}. (b) There is a crucial minus sign in the right hand side of \eq{eqns12} suggesting that the effective specific heat is negative --- which is a characteristic feature of gravitating systems.
 
 I will conclude this section mentioning another intriguing feature of horizon entropy which has not at all received any serious attention in the literature. We know that, in the case of Schwarzschild metric, the limit $M\to 0$ leads to global flat spacetime in inertial coordinates. Therefore we expect horizon thermodynamics to disappear in this limit. The entropy $S\propto M^2$ does vanish when $M=0$; but the temperature $T=1/(8\pi M)$ of the horizon blows up in the same limit. So flat spacetime, treated as a limit of a series of Schwarzschild spacetimes of decreasing mass, has \textit{infinite temperature} and zero entropy! More intriguingly, one finds that different spacetimes with horizons behave differently when we take the flat spacetime limit. For example, the de Sitter spacetime reduces to flat spacetime when we take the limit $H\to 0$. However,  de Sitter spacetime has the temperature $T=H/2\pi$ and entropy $S=\pi / H^2$. Therefore, when we take the $H\to 0$ limit, the de Sitter spacetime reduces to a flat spacetime  now endowed with zero temperature and \textit{infinite entropy} --- which is exactly the opposite of what we find in the case of Schwarzschild spacetime! Naively speaking, the temperature and entropy of flat spacetime, when treated as a limiting case of a curved geometry, seems to depend on the manner in which this limit is taken. 
 
 While the actual dependence of entropy and temperature on the parameters do not show any universal behaviour, it turns out that the result $E=2TS$ always holds \cite{tpcqg}. If we set $dS/dE=1/T$, this relation integrates to give $S(E)\propto E^2$ for all horizons. (For example, this holds for both de Sitter and Schwarzchild horizons.) This feature is telling us something very important about the microstructure of spacetime but nobody knows what exactly it is. We will discuss this a little more in the next section. 
 
 \subsubsection{Density of states in spacetimes with horizon}
 
 In normal statistical mechanics, one can obtain the micro-canonical entropy $S(E)$ as a logarithm of the density of states $g(E)$:
 \begin{equation}
  g(E) \equiv \int d\Gamma\ \delta[H(q,p) - E]; \qquad S(E) = \ln g(E)
  \label{gofe}
 \end{equation} 
where $d\Gamma$ is the measure of integration in the phase space. There are couple of interesting issues which come up when we try to apply this relation to spacetimes with horizon, say, black holes \cite{Padmanabhan:1998jp,Padmanabhan:1998vr}. 

The first one is related to, what could be called, the transmutation of the Hamiltonian.  To understand this, consider a black hole formed by the collapse of some kind of matter, for e.g. a ball of dust or a  neutron star. This matter, made of large number of atoms, originally will have an energy $E=M$ and a density of states $g(E)$; it will also have a corresponding entropy $S(E) = \ln g(E)$ which will depend on the detailed composition of the matter. But once the material collapses to form a black hole, it has a universal form for $S(E)$ given by
 $S(E) = 4\pi (E/ E_{P})^2$ so that all the thermodynamic features of the black hole are correctly reproduced. This, in turn, corresponds to a universal density of states for the system:
 \begin{equation}
 g(E) \approx \exp[4\pi ( E/E_P)^2+\mathcal{O}(\ln(E/E_P) \cdots]
  \label{pap1two}
 \end{equation} 
  where we have indicated the possibility that there could be subleading corrections to both entropy and density of states. (Note that even though the behaviour of temperature and entropy show serious differences between the black hole and de Sitter spacetime, both satisfy the relation $S(E) \propto E^2$; see \cite{tpcqg}.) Since a specific Hamiltonian can only lead to a specific $g(E)$ through \eq{gofe}, this implies that the Hamiltonian of a material system loses its significance when the material collapses to form a black hole! This universal transmutation of Hamiltonian has important consequences \cite{Padmanabhan:1998vr} which I will not discuss here.
  
  Second, one can ask what kind of physical system will exhibit the density of states in \eq{pap1two}, with a more conventional density of states when $E_{\rm Pl} \to \infty$. It turns out that such a behaviour will require, at least in the context of field theory, non local interactions at Planck scales \cite{Padmanabhan:1998jp,Padmanabhan:1998vr}. A simple toy model is provided by a Euclidean field theory with the Lagrangian
  \begin{equation}
L = {1\over 2} \int d^D{\bf x} \, \dot \phi^2 - {1\over 2} \int d^D{\bf x} d^D{\bf y}\, \phi({\bf x}) F({\bf x} - {\bf y}) \phi({\bf y}) = \int {d^D{\bf k}\over (2\pi)^D} {1\over2} \left[ |\dot Q_{\bf k}|^2 - \omega_{\bf k}^2 |Q_{\bf k}|^2\right]
   \label{pap1three}
  \end{equation} 
  The dispersion relation for particle like excitations of this system is given by $\omega^2 (\bm{k}) = F(\bm{k}) $
  where $F(\bm{k}) $ is the Fourier transform of $F(\bm{r})$ occurring in the Lagrangian. The partition function for such an excitation is given by
  \begin{equation}
 Z(\beta) \cong \int {d^D{\bm k}\over (2\pi)^D} \exp[-\beta\omega({\bm k})] = \int dE\,  g(E) e^{-\beta E}
  \end{equation} 
  where the density of states is given by the Jacobian $g(E)=|d^D(\bm{k})/dE|$. A simple example of the dispersion relation $\omega^2(\bm{k})$ which  interpolates between $\omega^2 = k^2$ when $E_{P} \to \infty$ and has a logarithmic dependence when $k \gg E_{P}$, is given by the function
  \begin{equation}
 \omega^2(k) = {E_P^2\over 8\pi} \ln \left( 1 + {8\pi k^2\over E_P^2}\right)
   \label{pap1seven}
  \end{equation} 
  For simplicity, I have taken $D=1$ but the same ideas can be extended for any $D$. 
  One can easily verify \cite{Padmanabhan:1998jp} that this will lead to a density of states given in \eq{pap1two}. The interaction term in \eq{pap1three} is now governed by the function 
  \begin{equation}
 F(x) = {E_P^2\over 8\pi}\int^\infty_{-\infty} {dk\over 2\pi} e^{-ikx} \ln \left( 1 + {8\pi k^2\over E_P^2}\right) = -{E_P^3\over 8\pi} \left( {L_P\over |x|}\right) \exp \left( - {|x|\over \sqrt{8\pi} L_P}\right)
  \end{equation}
  This is a non local interaction at Planck scales. When $L_P \to 0$, $F(x)$ becomes proportional to the second derivative of the Dirac delta function and thus will lead to a correct free field theory in \eq{pap1three}.
  
  This toy model suggests that, if one wants to model the density of states of black hole horizon, one may have to introduce non local  interactions at Planck scales. In this particular field theoretic model we have just constructed, the one particle excitations which have the correct asymptotic density of states as that of a black hole. That is, the black hole is being modeled --- as far as statistical properties are concerned --- as an excitation of this underlying Euclidean field theory.  In a more realistic picture, one would expect the dynamics of \md\ to be described by an effective field theory such that an interpolation like the one in \eq{pap1seven} arises in a natural fashion.

 \section{The Cosmological Constant: The Second Clue}

\hfill {\textit{``It's not that they can't see the solution,}}

\hfill {\textit{it's that they can't see the problem'' }}

\hfill {G.K. Chesterton,}
 
\hfill {{Scandal of Father Brown} (1935)}

\vskip0.2in

\subsection{Cosmic constants and their problems}\label{sec:cosmicconstants}
 
 Observations suggest our universe can be described in terms of three distinct evolutionary phases: (i) A quantum  gravitational phase possibly described by some pre-geometric concepts; (ii) a radiation dominated phase with the (dominant) equation of state $p\approx (1/3) \rho$ followed by a matter dominated phase with $p\approx 0$. (iii) An accelerated phase with the equation of state $p\approx-\rho$ which I will take to be dominated by the cosmological constant.\footnote{One can postulate and inflationary phase between (i) and (ii) but, as we will see, it is not necessary. I will follow a minimalistic approach. True believers in inflation can think of $a_{\rm QG}$ as the epoch at the end of reheating.} Observational astronomers and cosmologists describe the  evolutionary history of such a model through the equation: 
  \begin{equation}
H^2(t) = \frac{\dot a^2}{a^2} =
H_0^2 \left[ ( 1 - \Omega_R - \Omega_m)+ \Omega_R a_0^4/a^4 + \Omega_m a_0^3/a^3\right]\qquad (a>a_{\rm QG})
 \label{eqn4.8}
  \end{equation} 
where $a_{\rm QG}$ is the epoch of transition from quantum gravitation phase to radiation dominated phase. (I  set $k=0$ for simplicity.) 

While these parameters $H_0,\Omega_R,\Omega_m, ....$ etc. are  convenient to compare observations with theory, they are totally unsuited for describing the universe as a physical system. The description in terms of these parameters is not epoch-invariant, in the following sense: Consider, for example, cosmologists living in a star system located in a galaxy at $z=8$; if they will use \eq{eqn4.8} with the corresponding parameters evaluated at $z=8$ which, of course, will differ numerically from the ones we earthlings use. In other words, the parameters used in \eq{eqn4.8} have no epoch-invariant significance and are specific to a an epoch at which the CMB temperature is 2.73 K. This is, of course, unsatisfactory when we want to think of the universe as a physical system described by certain cosmic constants which will act as unique signatures characterizing our universe. It is important to describe the evolution of the universe using  parameters which will have epoch-independent significance \cite{hptp1,hptp2}.
  
  This  is easy to do and, in fact, one can do it in an infinite number of ways. One convenient set of parameters we can use are the following: (i) We denote the end of  the quantum gravitational phase by a constant density $\rho_{\rm QG}$ so that the classical evolution is valid for $\rho\lesssim \rho_{\rm QG}$. (ii) Similarly, we introduce another constant density $\rho_\Lambda $ corresponding to the cosmological constant, so that at  late times we have a  de Sitter universe with  $H^2(t) \approx (8\pi G\rho_\Lambda)/3$. (iii) To describe the in between (radiation and matter dominated) phase, it is convenient to introduce another \textit{constant} density 
  \begin{equation}
  \rho_{\rm eq}\equiv\frac{\rho_m^4(a)}{\rho_R^3(a)}=\sigma T_{eq}^4
   \label{eqn4.9}
  \end{equation}
  where $\sigma$ is the Stefan-Boltzmann constant and the second equality defines the temperature $T_{\rm eq}$.  We  also introduce the parameter $a_{\rm eq} $ by the epoch-independent definition 
  \begin{equation}
 a_{\rm eq}\equiv \frac{a \rho_R(a)}{\rho_m(a)}
   \label{eqn4.10}
  \end{equation}
  and work with the variable $x\equiv (a/a_{\rm eq})$.  Equation (\ref{eqn4.8}) can now be rewritten in the form
  \begin{equation}
  \left(\frac{\dot x}{x}\right)^2=
  \frac{8\pi G}{3}\ 
   \left[ \rho_\Lambda + \rho_{\rm eq} \left( x^{-3} + x^{-4}\right)\right]\qquad \qquad (\text{for}\  \rho < \rho_{\rm QG})
   \label{eqn4.11}
  \end{equation}
  in terms of the three densities $(\rho_{\rm QG}, \rho_{\rm eq}, \rho_\Lambda)$. This is a  more meaningful way of describing our universe than by using the parameterization in \eq{eqn4.8}. In particular, our cosmologist friend who lived in the $z=8$ galaxy would have written exactly the same equation, \eq{eqn4.11}, with exactly the same numerical values\footnote{More precisely, if we measure these densities in terms of the Planck density $\rho_{\rm Pl} = c^5/G^2\hbar$, we will get three dimensionless numbers which will be the same as those used by the $z=8$ cosmologist. So it does not matter that we are using the CGS system which might not have existed at $z=8$!} for $(\rho_{\rm QG}, \rho_{\rm eq}, \rho_\Lambda)$. These three numbers are the signatures of our universe.\footnote{We could also easily convey to the $z=8$ cosmologist our normalization convention for $a(t)$: 
  We tell her to set $a(t)=1$ at the epoch when the CMB temperature was equal to $T_{\rm eq}$, which is again an invariant statement characterizing the description of our universe. Incidentally, there are certain constants in the universe the numerical value of which we cannot determine uniquely. For example, consider  the combinations  $aT(a)$ or $a\rho_R(a)/\rho_M(a)$. These  are constants, independent of the epoch $a$ at which they are measured, as the universe evolves. But their numerical value depends on the numerical value you attribute to $a_0$ which, however, is not determined by theory for a flat universe. Recall that, when $k=0$,  Einstein's equation only fixes $H(t)$ and not the expansion factor $a(t)$. So while we know that $aT(a) = a_0T_0 = a_{\rm eq} T_{\rm eq}$, we will never be able to determine its numerical value without making an additional assumptions.  This is why it is more meaningful to use the scaling freedom and set $a=1$ at the epoch when the radiation temperature was equal to $T_{\rm eq}$. This is a normalization which is independent of the `current' epoch so that, it is something with which our cosmologist at $z=8$ will agree.}
  Thus, \eq{eqn4.11} describes the universe as a physical system (like, for e.g. an elastic solid) determined by certain constants (like, for e.g. the Young's modulus  etc. for a solid).

 It is also convenient to use the variable $H_\Lambda^2=(8\pi G/3)\rho_\Lambda$ and introduce the epoch-independent ratio:
\begin{equation}
 \sigma^4 \equiv\frac{\rho_\Lambda}{\rho_{\rm eq}}=  
 \frac{\rho_R^3}{\rho_m^4} \, \rho_\Lambda  = \frac{\Omega_R^3(t_*)}{\Omega_m^4(t_*)}\, [1-\oms-\ors]
 \label{a7}
\end{equation} 
(which could be computed using the parameters measured at any time $t=t_*$) 
 in terms of which the Friedmann equation  becomes:
 \begin{equation}
 \left(\frac{\dot x}{x}\right)^2 = H_\Lambda^2\left[ 1 + \frac{1}{\sigma^4}\, \left( \frac{1}{x^3} + \frac{1}{x^4}\right)\right]
 \label{a6tp}
\end{equation}
where $x(t) \equiv a(t)/a_{\rm eq}$ and $x(t) = a(t)$ in the epoch invariant normalization\footnote{It is obvious that any epoch of equality, at which the energy densities of two different components are equal, will allow us to introduce an epoch invariant normalization. But since $\rho_R$ and $\rho_m$ (describing radiation and matter) are better understood theoretically compared to $\rho_\Lambda$, it is prudent to use the equality epoch of matter and radiation densities for our normalization.} $a_{\rm eq} =1$. 
Equation (\ref{a6tp})   describes the cosmic evolution in terms of epoch-independent parameters. 

So the evolution of the universe can be described in terms of three densities $[\rho_{\rm QG}, \rho_{\rm eq}, \rho_\Lambda]$; thus the set of all universes forms a three parameter family of evolutionary histories. Of these, our universe is selected out by specific values for these three parameters. Observationally, we now know, in natural units, 
\begin{equation}
 \rho_{\rm eq} = 
 \begin{cases}
  \rho_{\rm QG}& \lesssim \ ( 10^{19}\ \text{GeV})^4 \\
\displaystyle{ \frac{\rho_m^4}{\rho_R^3} }&= \ [(0.86\pm 0.09) \ \text{eV} ]^4 \\
 \rho_\Lambda &= \ [(2.26\pm 0.05)\times 10^{-3}  \text{eV}]^4
 \end{cases}
\end{equation}
Of these, the classical era of the universe --- which is directly accessible to many different cosmological probes --- is described by the two densities $\rho_{\rm eq}$ and $\rho_\Lambda$. Using natural units $\hbar =c=1$ and the Planck length $L_P= (G\hbar/c^3)^{1/2} = G^{1/2}$, we can construct two dimensionless numbers from these densities. For the \cc, we get 
\begin{equation}
 \Lambda \left(\frac{G\hbar}{c^3}\right)=8\pi \rho_\Lambda L_P^4\approx 2.8 \times 10^{-122}
\end{equation}
which has led to the so called cosmological constant ``problem'' for $\Lambda$ since this extremely tiny value for a dimensionless number is supposed to indicate fine-tuning. On the other hand, the second density $\rho_{\rm eq}$, leads to the corresponding dimensionless number \cite{hptp2,lwf}
\begin{equation}
 (\rho_{\rm eq} L_P^4)\approx 2.4 \times 10^{-113}
\end{equation}
which is hardly commented upon in the literature!
(In fact, it comes as a bit of surprise to many cosmologists that $\rho_{\rm eq} L_P^4$ is indeed such a tiny number when they first hear it!)
As fine-tuning goes,  one would think that $10^{-113}$ does not fare much better than $10^{-122}$ for a dimensionless number. Therefore, if we think the value of $\rho_\Lambda L_P^4$ is a ``problem'', then we could also be concerned with the value of $\rho_{\rm eq} L_P^4$.

Most people are worried about $\rho_\Lambda L_P^4$ but not about $\rho_{\rm eq} L_P^4$ because of two reasons: 
(1) There is a feeling that the cosmological constant is ``in some way'' related to quantum gravity and hence the value $\rho_\Lambda L_P^4$ has some physical significance.
(2) There is a hope that the numerical value of $\rho_{\rm eq}$ can  be determined entirely by high energy physics. From its definition, we can relate $\rho_{\rm eq}$ to the ratio between the number density of the photons and the number density of  matter particles:
\begin{equation}
\rho_{\rm eq}=  \frac{\rho_m^4}{\rho_R^3}
=C \frac{(n_{\rm DM} m_{\rm DM}+n_{\rm B} m_{\rm B})^4}{n_\gamma^4}
=C\left[m_{\rm DM} \left(\frac{n_{\rm DM}}{n_\gamma}\right) + m_{\rm B} \left(\frac{n_{\rm B}}{n_\gamma}\right)\right]^{4}
\label{heprhoeq}
\end{equation} 
where $C =  15^3 (2 \zeta(3))^4  c^3 /\pi^{14} \hbar^3     \approx  2.845 \times 10^{110}$ (in cgs units) is a numerical constant, $n_{\rm DM},n_{\rm B},n_\gamma$ are the (current) number densities of dark matter particles, baryons and photons respectively and $m_{\rm DM},m_{\rm B}$ are the masses of the dark matter particle and baryon.
It is hoped that  the physics at (possibly) GUTs scale will  determine the ratios $(n_{\rm DM}/n_\gamma)$ and $(n_{\rm B}/n_\gamma)$ and specify $m_{\rm DM}$ and $m_{\rm B}$.
 Indeed, we do have a theoretical framework to calculate these numbers in different models of high energy  physics (for a review, see e.g. \cite{mazumdar}) though none of these models can be considered at present as compelling.  This leaves $\rho_\Lambda$ as the only constant to be determined. The small value of $\rho_{\rm eq} L_P^4$, or alternatively, the small value of 
 \begin{equation}
\Lambda L_P^2 \approx (2.85\pm 0.20) \times 10^{-122} \approx (1.1\pm 0.08)\times e^{-280}.                                                                     \end{equation} 
remains the key problem of which we will be concerned with in the later sections.

Incidentally, one can also rephrase the cosmological constant problem \textit{completely in classical terms} without introducing $\hbar $ or the Planck length $L_P$. There is, of course, \textit{no} fine-tuning problem for the cosmological constant in the classical, \textit{pure gravity} sector. The action for pure gravity in general relativity is given by
\begin{equation}
 A=\frac{c^4}{16\pi G}\int dt \ d^3 \mathbf{x} \g[R-2\Lambda]
\end{equation} 
which contains three constants, $G, c$ and $\Lambda$. It is not possible to form a dimensionless number from these three constants and hence it is meaningless to talk about the fine-tuning in the context of classical, matter-free gravity. The situation changes when we add matter to the action. Since the matter sector in our universe is characterized by $\rho_{\rm eq}$, we now have the dimensionless constant:
\begin{eqnarray}
 \sigma^4 &\equiv&\frac{\rho_\Lambda}{\rho_{\rm eq}}=  
 \frac{\rho_R^3(a)}{\rho_m^4(a)} \, \rho_\Lambda \nonumber\\
&=&[(2.62\pm 0.18) \times 10^{-3}]^4\simeq 10^{-10}\nonumber
\end{eqnarray} 
 The \textit{classical} cosmological constant problem can be stated as the fine tuning of the ratio \cite{hptp2,lwf}
\begin{equation}
\frac{\rho_\Lambda}{ \rho_{\rm eq}} \approx 10^{-10} 
\end{equation}
which governs the standard cosmological evolution, structure formation, etc. All the standard lore in cosmology depends only on this ratio because the classical evolution of our universe is completely determined by this number. This fine tuning is purely classical, in the sense that it does not require $\hbar$ or $L_P$, but e do not have any direct explanation for the smallness of this number.

 \subsection{Aside: Cosmological constant and the vacuum fluctuations}
 
 There is a tenacious myth in the literature that the sum of the zero point energies $((1/2) \hbar \omega_k)$ of the quantum modes contribute to the cosmological constant. The usual argument computes  this energy  with a  momentum scale cut-off at $k_{\rm max}$ to obtain
 \begin{equation}
 \rho_\Lambda=\int \frac{d^3\text{k}}{(2\pi)^3}\frac{1}{2}\hbar \omega_\text{k} \simeq k_{max}^4\simeq \frac{1}{L_{min}^4}
  \label{threenine}
 \end{equation} 
 If $k_{\rm max}$ is taken to be of the order of Planck energy, this will give $\rho_\Lambda L_P^4 \simeq \mathcal{O} (1)$ leading to confusion and  despair. \textit{This argument is completely wrong and the zero point energy used in \eq{threenine} does not contribute to the cosmological constant.} But given the prevalence of this erroneous argument, it is worth taking a moment to set it right (see \cite{cc1} for a clear discussion).
 
 To begin with, the zero point energy, $(1/2)\hbar \omega_k$,  of the quantum modes can be eliminated by straightforward normal ordering which is equivalent to  a redefinition of the Hamiltonian. But even if you do not resort to such a subtraction, what you get by the above computation is \textit{not} a cosmological constant but an ideal fluid with an equation of state $p \approx (1/3) \rho$ corresponding to radiation when the momentum cut-off scale, $k_{\rm max}$, is much larger than the mass of the field quanta. That is, the naive calculation (see, Appendix \ref{appen:vaclambda}) will lead to the result:
 \begin{equation}
 \langle \rho\rangle = \frac{1}{\left(2\pi\right)^3}
\frac12 \int {\rm d}^3{\text k}\, \omega (k); \qquad \langle p\rangle = \frac{1}{\left(2\pi\right)^3}
\frac16 \int {\rm d}^3{\text k}\, \frac{k^2}{\omega (k)}.
  \label{fournine}
 \end{equation} 
 This contribution is divergent  --- and will be large when computed with a Planck scale cut-off --- but it is not cosmological constant.  
 
 This fact also tells us that there is something fundamentally wrong with the above calculation.  The only Lorentz invariant energy momentum tensor $T^i_k$ which is acceptable for a Lorentz invariant vacuum is the one of the form $T^i_k = \rho_\Lambda \delta ^i_k$, which, of course, has the correct equation of state $p_\Lambda = - \rho_\Lambda$. The result in \eq{fournine} is not Lorentz invariant because the procedure of introducing a \textit{spatial} momentum cut-off breaks Lorentz invariance. 
 
 To get a sensible, Lorentz invariant result, one needs to compute the vacuum energy using a regularization procedure which preserves the symmetries of the theory. One possible choice is the dimensional regularization which will lead to the correct equation of state with 
 \begin{equation}
 \langle p \rangle  = -  \langle \rho \rangle; \quad  \langle \rho \rangle = \frac{m^4}{64\pi^2} \ln \left( \frac{m^2}{\mu^2}\right) 
  \label{forty}
 \end{equation} 
 Here $m$ is the mass of the scalar field contributing to the vacuum energy and $\mu$ is an arbitrary mass scale introduced by the regularization procedure. One can see from \eq{forty} that: (a) Massless fields do not contribute to this vacuum energy and (b) the sign of energy density depends on the relative values of $m$ and $\mu$. 
 The renormalization group argument would then suggest\footnote{This approach treats cosmological constant as a  coupling constant in the low energy effective action for gravity. There are reasons to believe that such an approach is fundamentally incorrect and hence  \eq{forty} and \eq{fortyone} will not have any relevance in the correct theory of quantum gravity. The numerical value of the cosmological constant, as we shall see later (see Sec. \ref{sec:cicc}), can be determined by a completely different procedure.} that  the effective cosmological constant is given by
 \begin{equation}
 \Lambda_{\rm eff} = \Lambda_{\rm bare} + \frac{Gm^4}{8\pi} \ln \left( \frac{m^2}{\mu^2} \right)
 \label{fortyone}
 \end{equation}
 Of course, when several matter fields are present, one needs to include contributions from all of them --- taking into account the fact that fermionic and bosonic degrees of freedom contribute with opposite signs --- to arrive at the final expression for $\Lambda_{\rm eff}$. One can then adjust $\Lambda_{\rm bare}$ so as to obtain any observed value for $\Lambda_{\rm eff}$. We shall not discuss this approach further since  there are reasons to believe that it is a totally incorrect route to approach the numerical value of the cosmological constant.

 \subsection{The necessary and sufficient conditions to solve the \cc\ problem}\label{sec:necsuff}
 
 As mentioned before, the \cc\ has a bad press and several ``problems'' have been attributed to it in the literature from time to time. A careful scrutiny of these issues reveals that the \cc\ is blameless and the so called problems arise because we have misunderstood the nature of gravity. These \cc\ 
 problem essentially arises from our misrepresentation of the dynamics of gravity --- which we will now describe \cite{tpap,A19}.

\subsubsection{Gravity breaks a key symmetry} 
 
Consider any theory of gravity interacting with matter fields and described by a total Lagrangian, given by the sum  $L_{\rm tot} = L_{\rm grav} [g_{ab}] + L_m [g_{ab},\phi_A]$. Here, $g_{ab}$ is the metric tensor which is assumed to describe the gravitational degrees of freedom and $\phi_A $ symbolically denotes all other matter degrees of freedom.  Varying the corresponding action  with respect to the matter degrees of freedom $\phi_A$ will lead to the equations of motion for matter in the presence of a given gravitational field. Similarly, varying the metric tensor is expected to lead to gravitational field equations of the form $\mathcal{G}^a_b = T^a_b$ where $\mathcal{G}_{ab}$ is a geometric variable obtained from the variation of $L_{\rm grav} $ (e.g., it will be the Einstein tensor $G_{ab}$ in general relativity, but could be a more complicated tensor in a general theory of gravity like, e.g., in \LL\ models \cite{dktpll}); our comments are applicable to a very general class of theories. Here $T_{ab}$ is the energy-momentum tensor of matter obtained from the variation of $L_m$ with respect to $g_{ab}$. We also note that the scalar nature of the Lagrangians lead to the generalized  Bianchi identity
($\nabla_a \mathcal{G}^a_b =0$)
and the conservation of the energy-momentum tensor $(\nabla_a T^a_b =0)$ in \textit{any} theory.

These variational principles leading to the equations of motion for matter and gravity, contain a peculiarity.  The equations of motion for the matter  are invariant under the addition of a constant to the matter Lagrangian. That is, the transformation\footnote{We will assume that the matter Lagrangian is \textit{not} supersymmetric invariant in the 
energy range we are interested in; supersymmetry  is the only symmetry that prevents the addition of a constant to the Lagrangian but since there is no evidence for this symmetry in nature, we shall not worry about it.}
\begin{equation}
 L_m\to L_m + ({\rm constant})
 \label{lmsym}
\end{equation} 
is a symmetry of the matter sector. The matter equations of motion remain invariant under the transformation in \eq{lmsym}. 
Though the addition of a constant to a Lagrangian is usually not stated as a ``symmetry'' principle, it definitely should be, because this mathematical operation does leave the equations of motion invariant; in that spirit, it is no different from any other mathematical operation, like for e.g., the Lorentz transformations, which leave the equations of motion invariant. Usually, one constructs the  theory  in such a way that the action itself remains manifestly invariant under all the relevant symmetry transformations (like, for e.g., the  Lorentz transformation). But, in the case of \eq{lmsym}, the Lagrangian and action \textit{do} change under \eq{lmsym} but the matter equations of motion 
do not. 
In fact, if we adopt the principle that Lagrangians should always be written in a manner making  the relevant symmetries apparent, then the matter sector Lagrangian should always be expressed in the form $(L_m + C)$ 
where $C$ is left as an unspecified constant. (This is analogous to not choosing a specific Lorentz frame while writing a Lagrangian, thereby exhibiting manifest Lorentz invariance.)\footnote{In the context of non-gravitational physics (classical mechanics, quantum mechanics,.....) we  know that it is only the differences in the energy which matter, and not the absolute zero of the energy. Nevertheless, it very surprising that \textit{no} formalism of non-gravitational physics, say, elementary classical or quantum mechanics, exists, which deals \textit{directly} with the energy differences! The natural development of these theories uses energy itself --- not energy differences. This is somewhat analogous to the natural description of gauge field theory, e.g., electrodynamics, which uses the gauge potential ($A_k$) though the observables are built from the field strengths $F_{ik}$ (in the classical context) and line integrals over gauge potentials (in the quantum theory). There is an attempt, usually called relational dynamics, which tries to study interacting particles using only relative velocities, differences in co-ordinates, etc. Even in this framework, I am not aware of a formulation which uses only the differences in energy rather than the energy itself. Given this background, it is indeed interesting that there \textit{does} exist a formulation of gravitational theories --- as we shall see later --- which is immune to the absolute value of the energy!}
Keeping such an arbitrary constant $C$ in the matter sector allows us to make manifest that the equations of motion for matter are indeed invariant under \eq{lmsym} because any additional constant merely changes the value of $C$.

On the other hand, the gravitational field equations, in contrast to the matter field equations, are \textit{not} invariant under the addition of a constant to the matter Lagrangian. The addition of a constant changes
the energy-momentum tensor of the matter: 
\begin{equation}
 T^a_b \to T^a_b + ({\rm constant}) \ \delta^a_b
 \label{tsym}
\end{equation} 
so that the gravitational field equations now become $\mathcal{G}^a_b = T^a_b + ({\rm constant}) \ \delta^a_b$. This is  equivalent to the introducing a \cc, if one was not present originally, or changing in its numerical value, if a \cc\ was present in the gravitational Lagrangian. 

The fact that any shift in the matter Lagrangian is physically equivalent to the introduction of a \cc\ has important (observable) consequences. When, for example,  the universe cools through the energy scale of, say, the electro-weak phase transition, the Higgs potential picks up a large shift in its energy and thus introduces a \cc. The numerical value of this \cc\ is very large compared to what we observe in the universe today; so we need a a careful fine-tuning (possibly with another \cc\ in the gravity sector) to cancel most of it. This is one of the key problems in understanding the nature of the \cc. In the literature, one often comes across discussion of the divergent contribution to \cc\ from quantum fluctuations leading to the running of the \cc, etc. But it needs to be appreciated that  there is a difficulty \textit{even at the tree-level} of the theory  due to the Higgs mechanism operating at, say, the electro-weak state. 

This is \textit{the} crucial problem related to the \cc\ viz., that its numerical value (either zero or non-zero) can be altered by the shift in \eq{lmsym}, which, however, leaves the matter equations unchanged. For example, particle physicists interested in the standard model may choose the overall constant in the matter Lagrangian arbitrarily since the standard model is unaffected by this constant. However, each such choice for the constant leads to a different value for the cosmological constant, and, hence, a different geometry for the universe. 

\subsubsection{What is needed to solve the \cc\ problem?}

We may restate the above issue as follows: If a fundamental principle, which enables the determination of the numerical value of the cosmological constant (either zero or non-zero) is discovered, it will not useful if the gravitational field equations are not invariant under the transformations in \eq{lmsym} or \eq{tsym}. We should  note that:

\begin{itemize}
 \item 
This problem  is  fundamental and is independent of the \textit{actual} value of the cosmological constant determined by observations (either zero or non-zero). The problem of the \cc\ existed even before observations showed that it has a non-zero value! It will persist even if the current observations about dark energy  turns out to be incorrect or if dark energy is found to be due to quintessence rather than \cc.
\item
The problem is also unrelated  to the energy densities of vacuum \textit{fluctuations}, regularization of zero-point energies, etc. If the Higgs mechanism operated during the evolution of the universe, causing the zero level of the energy densities to change by a large factor, then we will face a \cc\ problem in the form of extreme fine-tuning already \textit{at the tree-level} of quantum field theory.
\end{itemize}

Stated in this manner, we can immediately identify three ingredients which are \textit{necessary} to solve the \cc\ problem: 
\begin{itemize}
 \item[(a)] The field equations of gravity \textit{must be} made invariant under the transformations in \eq{lmsym} and \eq{tsym} so that gravity remains ``immune'' to the shift in the zero level of the energy densities. \label{page:conditions}
 
 \item[(b)] At the same time, we need to ensure that the \textit{solutions} to the field equations \textit{do} allow the cosmological constant to influence the geometry of the universe because, without this, the observed accelerated expansion of the universe  cannot  be explained. 
 
 \item[(c)] The \cc\ cannot be introduced as a low energy parameter in the Lagrangian if the theory is invariant under the transformation in \eq{lmsym}; so we need \cite{tphpplb} a fundamental physical principle to determine its numerical value.
\end{itemize}

The above demands, however, turns out to be extremely strong, and it has important consequences which are overlooked in attempts to ``solve'' the \cc\
problem. To appreciate this fact, consider any theory of gravity interacting with matter, which satisfies the following three conditions: 
\begin{enumerate}
 \item The theory is is based on a local, generally covariant, Lagrangian and  the matter action is constructed by integrating  a scalar Lagrangian $L_m(g_{ab},\phi_A)$ over the measure $\sqrt{-g} d^4x$. 
 \item The equations of motion for matter are invariant under the shift $L\to L + C$, where $C$ is a scalar constant. \label{page:conditions1}
 \item The field equations for gravity are obtained by the unrestricted variation of the metric tensor $g_{ab}$ in the total action.
\end{enumerate}
It is easy to see that the \cc\ problem  \textit{cannot} be solved in such any theory which satisfies the above three requirements. (This was first emphasized  in Section IV of ref. \cite{tpap}).
So even though all the three criteria stated above seem very reasonable, together, they prevent us from solving the \cc\ problem. Hence, (at least) one of them needs to be given up; if we do not give up general covariance and locality of the theory, or the freedom to add a constant to the matter Lagrangian, we must modify the third requirement.\footnote{One way of bypassing this issue is to postulate that the gravitational field equations are obtained by varying the metric, but keeping $\sqrt{-g} = $ constant. Such theories, known as unimodular theories of gravity, have been studied in the literature in the past \cite{unimod}. Unfortunately, there is, at best, weak  motivation to keep $\sqrt{-g} = $ constant. Since we will see that a more comprehensive approach exists, we will not discuss unimodular theories here.}  

At first sight, one might think that the requirements (a) and (b) in page  \pageref{page:conditions} are  impossible to satisfy simultaneously. However, it can be achieved by constructing  gravitational \textit{field equations} which are invariant under the transformation in  \eq{tsym}  but still allow the inclusion of a \cc\
as an \textit{integration constant} in their solutions. I will now describe how this can be achieved.

\subsubsection{What Einstein could have done: Avatars of field equations}
\label{sec:classgrav}

The above arguments show that, to solve the \cc\ problem we need a different perspective on gravitational field equations. The conventional approach of deriving it by varying the metric in an action is a wrong approach. I will now describe this aspect in somewhat greater detail to pave way for future discussion \cite{tp}.

The field equation in Einstein's theory is usually expressed, in standard textbook approach, in terms of $G_{ab} \equiv R_{ab} - (1/2)g_{ab}R$ in the form
$
 G_{ab} = \kappa_g T_{ab} 
$.
But, as I argued above, this is wrong because it is not invariant under the transformation in \eq{tsym}. There are, however,  two other ---  and as shall see,  nicer --- ways of writing the gravitational field equation.

The first alternative is to introduce a timelike, normalized vector field $u^i$ (which one could think of as the four-velocity of a fiducial observer) and demand that the equation 
\begin{equation}
 G_{ab}u^au^b = \kappa_g T_{ab}u^au^b 
\label{ee2}
\end{equation} 
holds for \textit{all} observers. This demand, of course, can be met only if  $G_{ab} = \kappa_g T_{ab}$  and we recover the standard result. 
In fact, Einstein could have obtained this equation instead of $G_{ab} = \kappa_g T_{ab}$ in a rather nice fashion (thereby saving us a lot of trouble!). Let me explain how.

Starting from the principle of equivalence and general covariance, you can conclude that: (a) Gravity is best described as the effect of curvature of spacetime by using a nontrivial line interval $ds^2=g_{ab}dx^adx^b$. (b) The influence of gravity on other systems can be obtained by demanding the validity of special relativity in the local inertial frames and the validity of general covariance. (c) In the Newtonian limit, the only nontrivial metric coefficient will be $g_{00} = - (1+2\phi)$.
What Einstein was looking for next was a generalization of the field equation for gravity in the Newtonian theory, viz.
the Poisson equation $\nabla^2 (2\phi) =  \kappa_g \rho$. The principle of equivalence identifies the gravitational potential with a component of the metric tensor through $g_{00} = - (1+2\phi)$ so that the Poisson equation can be formally written as  $-\nabla^2 g_{00} = \kappa_g T_{00}$ where $\rho$ is identified with the time-time component $T_{00}$ of the divergence-free, second rank symmetric energy momentum tensor $T_{ab}$. 
Since $\nabla^2$ is not Lorentz invariant, you might think  that it is preferable to  ``generalize'' the $\nabla^2$ to $\square^2$ so that the left hand side has second derivatives in \textit{both space and time}.  
The second derivatives of the metric tensor occurs in a  covariant combination  in the curvature tensor, which led Einstein to look for a   divergence-free, second rank symmetric tensor to replace $\nabla^2 g_{00}$ in the left hand side. After several false starts, he came up with  $G_{ab} = \kappa_g T_{ab}$ and postulated it to be the field equation.

But Einstein could have taken a  better, route! 
A relativistic generalization of Newton's law of gravity $\nabla^2\phi\propto\rho$, can be obtained \textit{retaining the right hand side as it is and without introducing second time derivatives in the left hand side}. 
To do this, we first note that: 
(i) The energy density  $\rho=T_{ab}u^au^b$, which appears in the right hand side, is foliation/observer dependent and involves the vector field  $u^i$. There is no way you can keep $u^i$ out of the definition of $\rho$ and hence you should accept it as a fact of life. 
(ii) Since $g_{ab}$ plays the role of $\phi/c^2$, a covariant scalar which generalizes  the left hand side, $\nabla^2\phi$, can indeed be constructed from the curvature tensor --- which contains the second derivatives of the metric. 
But, you must find a generalization  \textit{which depends on the four-velocity $u^i$ of the observer} because the right hand side does. A purely geometrical object (like e.g. $R$), simply won't do.
(iii) It is perfectly acceptable for the left hand side \textit{not} to have second \textit{time} derivatives of the metric, in the rest frame of the observer, since time derivatives do not occur in $\nabla^2\phi$. 

The main task is to obtain a scalar with  \textit{spatial} second derivatives which depends on $u^i$ to replace $\nabla^2\phi$. For this,  we first project the indices of $R_{abcd}$ to the space orthogonal to $u^i$, using the projection tensor $P^i_j=\delta^i_j+u^iu_j$, thereby obtaining the tensor
$\mathcal{R}_{ijkl}\equiv P^a_iP^b_jP^c_kP^d_l R_{abcd}$. The only scalar which can be constructed from $\mathcal{R}_{ijkl}$ is $\mathcal{R}^{-2}\equiv\mathcal{R}_{ij}^{ij}$ where $\mathcal{R}$ can be thought of as the radius of curvature of the space.\footnote{The $\mathcal{R}_{ijkl}$ and $\mathcal{R}$ should \textit{not} to be confused with the curvature tensor $^3R_{ijkl}$ and the Ricci scalar $^3R$ of the 3-dimensional space orthogonal to $u^i$.} The natural generalization of $\nabla^2\phi\propto\rho$ is then provided by $\mathcal{R}^{-2}\propto\rho=T_{ab}u^au^b$. Working out the left hand side (see e.g., p. 259 of 
Ref.~\cite{gravitation}) explicitly, one obtains
 $
G_{ab}u^au^b=\kappa_g T_{ab}u^au^b                                  
$
which is exactly \eq{ee2}!
So, \eq{ee2} tells you that the square of the radius of curvature of space is proportional to the reciprocal of the energy density, thereby giving a geometrical meaning to the left hand side.\footnote{As is well-known, the combination $G_{ab} u^au^b$ is also closely related to the ADM Hamiltonian in the conventional approach. But this is a \textit{dynamical} interpretation and not a purely \textit{geometrical} one. These ideas generalize in a simple manner to all \LL\ models of gravity \cite{dktpll,ll} and are not limited to Einstein's theory.}

The second alternative is to introduce a \textit{null} vector field $\ell^a$ (which we could think  of as a normal to  a null surface in the spacetime) and demand that the equation 
\begin{equation}
 G_{ab}\ell^a \ell^b = R_{ab}\ell^a \ell^b = \kappa_g T_{ab}\ell^a \ell^b
\label{ee3}
\end{equation}
holds for all null vectors $\ell^a$.  This leads to the result
\begin{equation}
 G_{ab} = \kappa_g T_{ab}  + \Lambda g_{ab}
\label{ee4}
\end{equation}
where $\Lambda$ is a constant.\footnote{\label{fn}Equation~(\ref{ee3}) implies that $R^{a}_{b} - \kappa_g T^a_{b} = f(x) \delta^a_{b}$. Taking the divergence and using the results, $\nabla_a T^a_b =0$ and 
$\nabla_a R^a_b = (1/2)\partial_b R$, we find  that $f(x) = (1/2)R +$ a constant, leading to \eq{ee4}. It is sometimes claimed in the literature that the Bianchi identity $\nabla_aG^a_b =0$ \textit{implies} $\nabla_aT^a_b =0$. But  $T^a_b $ can also be defined through the variation of the matter Lagrangian with respect to arbitrary coordinate transformations $x^a\to x^a+\xi^a(x)$. Its conservation, $\partial_a T^a_b =0$, expressed in Cartesian coordinates in local inertial frames, becomes $\nabla_aT^a_b =0$ in curvilinear coordinates in local inertial frames. The principle of equivalence  demands the validity of this condition in an arbitrary curved spacetime. 
That is, you can actually \textit{derive}  $\nabla_a T^a_b =0$ without using the Bianchi identity or the gravitational field equations.
It is, therefore, more appropriate to think of the Bianchi identity as \textit{being consistent} with $\nabla_aT^a_b =0$ rather than think of it as \textit{implying} it.} 
So, \eq{ee3} also leads to Einstein equation but with a crucial difference: It allows for a cosmological constant $\Lambda$ to arise as an integration constant to the field equation.
(Incidentally, one can associate with a non-geodesic congruence of observers a space-like vector field $\hat{a}^i$, as well as a null vector $\ell^i$, both derived from the time-like four velocity $u^i(x) $ defining the congruence. The $\hat a^i$ is constructed by normalizing the acceleration field $a^i = u^j \nabla_j u^i$ to get $\hat{a}^i \equiv a^i/a$ which is unit normalized. The null vector field $\ell^i$ can be defined as proportional to $u^i \pm \hat{a}^i$.)

These  formulations are  conceptually very different from the usual one.  To begin with   \eq{ee2} and \eq{ee3} are scalar equations but  involve additional vector fields. They have the same information content as the ten tensor components of standard Einstein equation because we demand them to hold for \textit{all} $u^i$ or all $\ell^i$. If you think of $u^i$ as a four velocity of an observer, then \eq{ee2} demands the equality of two quantities which this observer  measures in the matter sector and the geometrical sector. Such a demand, invoking a class of observers, is similar in spirit to the way we obtain the kinematics of gravity (``how gravity makes matter move'') by introducing special relativity in the coordinate frames adapted to the freely falling observers.

Second, nobody has provided a physical meaning for the text book field equation $G^a_b = \kappa_g T^a_b$. The right hand side, of course, is  the energy momentum tensor but not the left hand side has no simple meaning. 
In fact, in the conventional approach,  we  do not have an actual \textit{mechanism} which tells us how $T^a_b$ ends up curving the spacetime; in this sense, the  relation  $G^a_b = \kappa_g T^a_b$  equates apples and oranges; the left hand side is purely geometrical while the right hand side is made of  a large number of discrete (quantum) degrees of freedom of matter.\footnote{The usual `fix' is to use the quantum expectation value $\langle T^a_b\rangle$ in the right hand side but that is hardly appropriate as a fundamental description and provides us with no useful insights. We just do not know how to describe \textit{matter} close to Planck scales.} An equation like $G^a_bn_a n^b = \kappa_g  T^a_bn_a n^b$ (where $n_a = u_a$ or $\ell_a$), on the other hand, is conceptually better in this regard.  We have a better chance of interpreting both sides independently  and think of this equation as a balancing act performed by spacetime. 
Some of the later sections (esp. Sec. \ref{sec:zero-diss})  will be devoted to  providing the physical meaning for the two sides of \eq{ee3}.

So we now have three formulations --- based on $G^a_b = \kappa_g T^a_b$, \eq{ee2} and  \eq{ee3} --- all of which will lead to the same algebraic consequences for classical gravity.  That is, if  you specify $T_{ab}$ (and $\Lambda$ in case of \eq{ee4}) and solve the resulting differential equations, you will end up with the same spacetime geometry and same observable consequences. We need a 
a physical principle which will allow us to select one of them as the correct approach. This is precisely where the \cc\ provides a strong clue. As we said before, to solve the \cc\ problem we need to ensure that gravitational field equations remain invariant under \eq{tsym}.  We now raise this to the status of a  postulate \cite{A19,A11}:

\begin{itemize}
 \item[$\blacktriangleright$] The extremum principle  used to determine the spacetime dynamics (and hence the field equations) must remain invariant under the change $T^a_b \to T^a_b + $ (constant) $\delta^a_b$.
\end{itemize}

This guiding principle immediately rules out $G^a_b = \kappa_g T^a_b$ and \eq{ee2} as possible choices for the field equation and selects \eq{ee3} as the correct choice; of course, \eq{ee3} remains invariant under the shift $T^a_b \to T^a_b + $ (constant) $\delta^a_b$ because $\ell^2 =0$ for a null vector.  The postulate stated above will turn out to be as powerful in determining the  gravitational dynamics as the principle of equivalence was in determining the gravitational kinematics.

\section{Gravity from a Thermodynamic Variational Principle}
\label{sec:vpg}

The guiding principle for dynamics, introduced above rules out the possibility of varying the metric tensor $g_{ab}$ in any covariant, local action principle to obtain the field equations.  Instead, we are looking for a new variational principle which will give us \eq{ee3}.
A key constraint, on any such variational principle leading to \eq{ee3}, is the following: 
Since you cannot now bring in $T_{ab}$  by varying $g_{ab}$ in a matter action, the $T_{ab}$ must be present in the functional we vary \textit{in some form which does not violate our guiding principle}.\footnote{You can get \eq{ee3} by the following cheap trick: You restrict the variations of the metric in the conventional action principle to $\delta g^{ab}$ of the form $\delta g^{ab}=\ell^a\ell^b$ where $\ell^a$ is a null vector. But this is as bad as unimodular gravity and it is difficult to motivate such a restriction.} 
The most natural structure, built from $T^a_b$, which maintains the invariance we need, viz. under $T^a_b \to T^a_b + $ (constant) $\delta^a_b$, is given by
\begin{equation}
 \mathcal{H}_m \equiv T_{ab} \ell^a \ell^b
\label{Qtot}
\end{equation} 
where $\ell_a$ is a null vector.\footnote{We want to introduce a minimum number of extra variables to implement the required symmetry. In  $d$-dimensional spacetime, a null vector with $(d-1)$ degrees of freedom is the minimum additional structure one needs. For comparison, suppose you introduce, say, a combination like $T^{ab}V_{ab}$ with a symmetric traceless tensor $V_{ab}$, in order to maintain the invariance under $T^a_b \to T^a_b + $ (constant) $\delta^a_b$. This will  introduce $(1/2)d(d+1)-1$ extra degrees of freedom; in $d=4$, this introduces nine degrees of freedom, which is equivalent to  introducing three null vectors rather than one.}
This is exactly the combination that appears in the right  hand side of \eq{ee3}. 

The fact that you cannot vary the metric to get the equations of motion can come as a bit of a surprise, and this  constraint can indeed lead  to trouble if you want to obtain $G_{ab}=\kappa_g T_{ab}$. But  our guiding principle selected  out \eq{ee3} as the correct one. In this equation we have the auxiliary variable $\ell^a$ and so we can indeed construct variational principles in which we vary $\ell^a$ and obtain \eq{ee3} and thus \eq{ee4}. So everything is completely consistent within the spirit of our formalism.

Since our principle has selected out the combination \eq{Qtot},
it is worth asking whether it has  a physical interpretation. I will first show that this combination does have a clear thermodynamic interpretation\cite{tp}.  This fact is very nontrivial: The guiding principle was based on the nature of \cc\ problem and in conventional view point nobody has thought of linking the \cc\ to thermodynamics of horizons! All the same, this is precisely what  we are \textit{forced} into.

\subsection{Heat density of matter}
\label{sec:hdm}

That the combination $\mathcal{H}_m \equiv T_{ab} \ell^a \ell^b$ for any null normal $\ell_a$ can be thought of as the heat density  contributed by matter crossing  a null surface. 
To acquire some preliminary insight of the result, consider first the case of an ideal fluid, with $T^a_b = (\rho+p) u^au_b + p\delta^a_b$.  The combination $T^a_b \ell_a\ell^b$ is now precisely the \textit{heat density} $\rho +p = Ts$ where $T$ is the temperature and $s$ is the entropy density of the fluid. (The last equality arises from the Gibbs-Duhem relation. We have chosen the null vector such that $(\ell.u)^2=1$ for simplicity.)  
But  $T^a_bu^bu_a$ gives the energy density for \textit{any} kind of $T^a_b$, not just for that of an ideal fluid. How can we then interpret $T^a_b \ell_a\ell^b$ as the \textit{heat} density in a  \textit{general} context when $T^a_b$ could describe any  source --- not necessarily a fluid --- for which concepts like temperature or  entropy do not exist intrinsically?  \textit{Surprisingly enough,  you can do this!}. In any spacetime, around any event, you can introduce the  local Rindler observers who will indeed interpret $T^a_b \ell_a\ell^b$ as the heat density contributed by the matter to a null surface which they perceive as a horizon.  Let me describe how this comes about: 

Let us  begin by introducing a freely falling frame (FFF) with coordinates $(T, \mathbf{X})$ in a region around some fiducial event $\mathcal{P}$ in the spacetime.  Next,  transform from the FFF to a local Rindler frame (LRF) --- with coordinates $(t,\mathbf{x})$ --- using the transformations: $\kappa X=\sqrt{2\kappa x}\cosh(\kappa t), \kappa T=\sqrt{2\kappa x}\sinh (\kappa t)$ constructed using some acceleration $\kappa $. (This transformation is valid for  $X>|T|$ with  similar ones for other wedges.) One of the null surfaces passing though $\mathcal{P}$, will now get mapped to the $X=T$ null surface of the FFF and will act as a patch of horizon to the local Rindler observers with the trajectories $x=$ constant. The local vacuum state, appropriate for the freely-falling observers around $\mathcal{P}$, will now appear to be a thermal state to the local Rindler observers, with a temperature proportional to their acceleration $\kappa $.

Consider now the flow of energy associated with the matter --- with some \textit{arbitrary} $T_{ab}$ --- that crosses the null surface. Nothing peculiar happens when this phenomenon is viewed in the FFF by the locally inertial observer. But the local Rindler observer, who attributes a temperature $T$ to the horizon,  perceives it as a hot surface. Therefore, this observer will interpret the energy $\Delta E$, dumped on the horizon (by the matter that crosses the null surface in the FFF),  as  energy deposited on a \textit{hot} surface, thereby contributing a \textit{heat} content $\Delta Q=\Delta E$ to the surface. (Recall that, in the case of a \textit{black hole} horizon, an outside observer will find that  matter takes an infinite amount of time to cross the horizon,  thereby allowing for thermalization to take place. Similarly, a local Rindler observer will find that  matter takes a very long time to cross the local Rindler horizon.) 

It is fairly easy to compute $\Delta E$ in terms of $T^a_b$. The LRF has an approximate Killing vector field, $\xi ^{a}$, generating the Lorentz boosts in the FFF, which will coincide with a suitably defined%
\footnote{Since the null vectors have zero norm, there is an overall scaling ambiguity in some expressions involving them. This can be eliminated, in this case,  by considering a family of hyperboloids $\sigma^2\equiv X^2-T^2 =2\kappa x =$ constant and treating the light cone as the (degenerate) limit $\sigma\to0$ of these hyperboloids. We set $\ell_a= \nabla_a\sigma^2\propto\nabla_a x$ and then take its corresponding limit.} 
null normal $\ell ^{a}$ of the null surface in the appropriate limit. Using the heat current that arises from the  energy current $T_{ab}\xi ^{b}$, we find that the total heat energy dumped on 
the null surface is:
\begin{align}\label{Paper06_New_11}
 Q_{m}=\int \left(T_{ab}\xi ^{b}\right)d\Sigma ^{a}=\int T_{ab}\xi ^{b}\ell ^{a}\sqrt{\gamma}\ d^{2}x d\lambda
=\int T_{ab}\ell ^{b}\ell ^{a}\sqrt{\gamma}\ d^{2}x d\lambda
\end{align}
where we have used the fact that $\xi ^{a} \to \ell ^{a}$ on the null surface. 
So, the combination
\begin{equation}
 \mathcal{H}_m\equiv \frac{dQ_{m}}{\sqrt{\gamma}d^{2}xd\lambda}=T_{ab} \ell^a\ell^b
\label{hmatter} 
\end{equation}
is indeed  the heat density (energy per unit area per unit affine time) of the null surface, contributed by matter crossing a local Rindler horizon. This interpretation remains valid  for \textit{any} kind of $T^a_b$.  

\subsection{Einstein's equation as a zero-dissipation-principle}\label{sec:zero-diss}
 
 Given the interpretation that $T_{ab} \ell^a\ell^b$ is the heat density of matter, it is natural to ask what the field equation $R_{ab} \ell^a\ell^b=\kappa_g T_{ab} \ell^a\ell^b$ means. Unlike in the standard approach to Einstein's theory \textit{we now have a simple physical interpretation for the field equations!} Remarkably enough, the combination \cite{A19}
\begin{equation}
\mathcal{H}_g\equiv - \frac{1}{8\pi L_P^2}R_{ab}\ell^a\ell^b
\label{defhg}
\end{equation}
also
has an interpretation as the 
gravitational heat density (i.e., heating rate per unit area) of the null surface to which $\ell_a$ is the normal. Its integral, taken over the null surface, $Q_g$, can be interpreted  as the gravitational contribution to the heat content of the null surface. 
These results arise because  $R_{ab}\ell^a\ell^b$ is related to the concept of ``dissipation without dissipation'' \cite{A24,A25} of the null surfaces. Let me briefly describe how this result arises.

Let us start with  the standard description of a null surface by introducing the complementary null vector $k^a$ (with $k^a\ell_a=-1$) and defining the 2-metric on the cross-section of the null surface by $q_{ab}=g_{ab}+\ell_ak_b+k_a\ell_b$. We define the expansion $\theta\equiv\nabla_a\ell^a$ and shear $\sigma_{ab}\equiv \theta_{ab}-(1/2)q_{ab}\theta$ of the null surface where $\theta_{ab}=q^i_aq^j_b\nabla_i\ell_j$. (It is convenient to assume that the null congruence  to be affinely parametrized.) One can then show that \cite{A19}:
\begin{equation}
-\frac{1}{8\pi L_P^2}R_{ab}\ell^a\ell^b\equiv\mathcal{D}+\frac{1}{8\pi L_P^2}\frac{1}{\sqrt{\gamma}}\frac{d}{d\lambda}(\sqrt{\gamma}\theta)
\label{rai}
\end{equation} 
where $\gamma $ is the determinant of $q_{ab}$ and 
\begin{equation}
 \mathcal{D}\equiv\left[2\eta \sigma_{ab}\sigma^{ab}+\zeta\theta^2\right]
\end{equation} 
is the standard expression for the viscous heat generation rate of a fluid with shear and bulk viscous coefficients \cite{A26,A27,membrane}
defined%
\footnote{The fact that the null congruence  has negative bulk viscosity coefficient is well-known in the literature \cite{A26,A27,membrane}, especially in the context of the black hole membrane paradigm.} 
as $\eta=1/16\pi L_P^2,\zeta=-1/16\pi L_P^2$. Ignoring the total divergence term in \eq{rai}, we can identify the integral of $R_{ab}\ell^a\ell^b$ with: 
\begin{equation}
Q_{g}=-\frac{1}{8\pi L_P^2}\int \sqrt{\gamma}\, d^2x \, d\lambda\, R^a_b \ell_a\ell^b
=\int \sqrt{\gamma}\, d^2x \, d\lambda\, \mathcal{D}
=\int \sqrt{\gamma}\, d^2x \, d\lambda\, \left[2\eta \sigma_{ab}\sigma^{ab}+\zeta\theta^2\right]
\label{Qtoty}
\end{equation}
which represents the heat content of the null surface due to the gravitational degrees of freedom; the integrand $\mathcal{D}$ is 
the rate of heating of the null surface contributed by the gravitational degrees of freedom. 

Of course, you do not want the null surfaces in spacetime to exhibit either heating or dissipation! This is ensured by the presence of matter which is now needed when $R_{ab}\neq0$. The contribution to the heating from the microscopic degrees of freedom of the spacetime precisely cancels out the heating of  any null surface by the matter, if we demand: 
\begin{equation}
-\frac{1}{8\pi L_P^2}R_{ab}\ell^a\ell^b +  T_{ab}\ell^a\ell^b=\mathcal{H}_g+\mathcal{H}_m= 0  
\label{zerodisp}
\end{equation} 
This allows us to reinterpret the field equation
as a ``zero heat dissipation'' principle.

\subsection{Form of the variational principle}\label{sec:formvarprin}

Having ascertained the thermodynamic meaning of Einstein's equation, let us see how it can be obtained from a variational principle.
Let me begin with a simple variational principle which satisfies our dynamical principle and leads to \eq{ee3}. Since we cannot vary the metric, let us consider an action principle \cite{aseemtp} in which we vary a null vector field $\ell^a$. We take the variational principle to be based on the functional:
\begin{equation}
A[\ell, \nabla \ell]= \int \frac{d^4x}{L_P^4} \sqrt{-g}\, \left( L_P^4 T_{ab} \ell^a \ell^b + P^{ab}_{cd} \nabla_a \ell^c\, \nabla_b \ell^d\right)
\label{tpa}
\end{equation} 
where $P^{ab}_{cd}$ is a tensor with the algebraic symmetries of the curvature tensor and is divergence-free in all the indices. To obtain Einstein gravity, we take it to be\footnote{These ideas generalize to all \LL\ models of gravity with corresponding $P^{ab}_{cd}$.} 
\begin{equation}
P^{ab}_{cd} = \frac{L_P^2}{8\pi} \left( \delta^a_c \delta^b_d - \delta^b_c \delta^a_d \right)
\end{equation} 
It is now straightforward to show that varying $\ell^a$  after introducing a Lagrange multiplier to ensure $\ell^2=0$ will lead to the equation $R^i_j - \kappa_g T^i_j = f(x) \delta^i_j$ which --- in turn --- leads to \eq{ee4}. So, the action in \eq{tpa} ---
which seems to describe a garden variety null vector field with quadratic coupling --- leads to the result we want\footnote{Normally, when we vary a quantity $q_A$ in an extremum principle, we get an evolution equation for $q_A$. Here we vary $\ell_i$ in \eq{tpa} but get an equation constraining the background spacetime metric $g_{ab}$! This is because, after varying $\ell_i$, we demand that the equation must hold for all $\ell_i$.  While this makes our extremum principle conceptually different from the usual ones, it is perfectly well-defined --- and will make physical sense later on when we probe it deeper.} as long as the kinetic energy term has a specific structure! 

One can demystify the result  by noticing that the kinetic energy term in \eq{tpa} can be rewritten in the form 
\begin{equation}
P^{ab}_{cd} \nabla_a \ell^c\, \nabla_b \ell^d =  \nabla_a w^a +  \frac{L_P^2}{8\pi} R_{ij}\, \ell^c \ell^j
\end{equation} 
where $w^a = P^{ab}_{cd} \ell^c\nabla_b\ell^d$. So, except for an ignorable  total divergence, we are actually working with an action that is proportional to $(R_{ab} - \kappa_g T_{ab}) \ell^a\ell^b$.  The action, in fact, does not contain any kinetic energy term for $\ell^a$ at all once you discard the total divergence! Nevertheless, \eq{tpa} is also a perfectly legitimate action in which you can vary $\ell^a$ and get the equations we want.\footnote{The full action for matter plus gravity is obtained by adding to $A$ in \eq{tpa} the matter action; i.e., 
$
A_{tot}= A + A_{\rm matter}(\psi_A,g_{ab})
$
where $\psi_A$ denotes the matter variables.
This action works with the following extra prescription: You first vary $\ell_a$ \textit{first} to get the field equations for gravity, use the on-shell values in the first term $A$ in $A_{tot}$ and extremize  the resulting functional with respect to the matter variables $\psi_A$ to determine the matter equations of motion. In a path integral you should integrate over $\ell_a$ first. The reason why you need to vary $\ell_a$ first will become clearer, when we identify $\ell_a$ with internal variables describing the spacetime microstructure in Sec. \ref{sec:grfratoms}.}

One can actually construct a more general form of the action principle leading to the same result. To do this, 
define 
\begin{equation}
 q[x; \ell_a(x)] \equiv \left(T^a_b (x) - \frac{1}{\kappa_g} R^a_b (x)\right) \ell_a\ell^b \equiv E^a_b \ell_a\ell^b
\label{tpone}
\end{equation} 
which is a function of $x^i$ through $T^a_b$ and $R^a_b$ and a quadratic \textit{functional} of the null vector field $\ell^a(x)$. Consider now a variational principle based on the functional
\begin{equation}
Q [\ell_a(x)] = \int dV F(q[x;\ell_a])
\label{tptwo}
\end{equation} 
where $F(q)$ is a function of $q$ --- which is, at present, arbitrary --- and $Q$ is treated as a functional of $\ell^a$. 
The $F$ is a scalar and the integration in  \eq{tptwo} can be  over any (sub)domain of the spacetime with a covariant measure $dV$. (Most of the time, it  will be natural to do an integration over a null surface.)
Consider a variational principle of the form $\delta Q/\delta \ell =0$ subject to the additional constraint that  $\ell^2(x) =0$. 
Incorporating this constraint by a Lagrange multiplier $\lambda(x)$ changes
$F(q) \to F(q) + \lambda(x) \delta^a_b \ell_a\ell^b$.
Varying $\ell^b$ in this expression and demanding that $\delta Q=0$ for arbitrary $\delta \ell^b$ leads to the condition 
\begin{equation}
\left[ F'(q) E^a_b + \lambda(x) \delta^a_b \right] \ell_a =0; \qquad F'(q) \equiv \frac{dF}{dq}
\label{tpfour}
\end{equation} 
We expect the field equations to arise from the demand that the extremum condition $\delta Q =0$ should hold for all $\ell^a$.
This will work if: (i) the expression within the square bracket in \eq{tpfour}  vanishes and (ii) $q$, which appears in $F'(q)$,  becomes independent of $\ell^a$ on-shell. The  condition (ii), in turn, requires 
\begin{equation}
 E^a_b = f(x) \delta^a_b\, ,
\label{tpfive}
\end{equation} 
for some $f(x)$, 
so that $q=0$
on-shell. 
Substituting $E^a_b = f(x)\delta^a_b$ in the square bracket in \eq{tpfour} determines
 the Lagrange multiplier function $\lambda(x)$ to be 
$ \lambda(x)  = - F'(0) f(x)$
but is otherwise of no consequence.\footnote{Except that we need $F'(0)$ should be finite and non-zero. In fact,  the choice that is important to us is just $F(q) =q$.} 
Taking the divergence of $E^a_b = f(x)\delta^a_b$ and using the Bianchi identity and the result $\nabla_a T^a_b =0$ determines $f(x)$ to be 
$ f(x) = - (1/\kappa_g) \left( \Lambda +(1/2) R\right)$ 
where $\Lambda $ is a constant.
Plugging this back into \eq{tpfive}, we get the field equation to be 
\begin{equation}
G^a_b = \kappa_g\, T^a_b + \Lambda \, \delta^a_b
\end{equation} 
which, of course, is the same as  our \eq{ee4}. 
Thus,  a variational principle with an arbitrary function $F(q)$ --- where $q$ is defined by \eq{tpone} --- which will lead to our field equation in \eq{ee3} or \eq{ee4}. We do not  vary the metric in this approach. The variational principle and the resulting field equation remain invariant under the transformation $T^a_b \to T^a_b + $ (constant)$\delta^a_b$.

Incidentally, the above approach  also leads to a quantum theory based on the path integral 
\begin{equation}
Z\equiv \int \mathcal{D} \ell_a \, \delta(\ell^2)   \exp \int dV  F[L_P^4 q]
\label{tp16}
\end{equation} 
where we have used the dimensionless variable $L_P^4 q$.
  The path integral $Z$ in \eq{tp16} has to  restricted to null vectors which satisfy the condition $\ell^2=0$. (So the path integral is nontrivial even for $F\propto q$ which makes it a Gaussian in $\ell_a$.) 
As we have seen, the  classical field equations will arise from this expression when we evaluate it in the saddle point approximation and  demand that the result should hold for all $\ell^a$. 
But note that $Z = Z[g_{ab}, T_{ab}]$ is a nonlocal functional of $g_{ab}$ and $T_{ab}$. Varying $g_{ab}$ in $\ln Z$ will now lead to a nonlocal field equation relating $g_{ab}$ to $T_{ab}$.
  But since the path integral defining $Z$ is invariant under $T^a_b \to T^a_b + $ (constant)$\delta^a_b$ the extremization of $\ln Z$  will  lead to equations of motion which respects this symmetry.\footnote{To avoid possible misunderstanding, let me stress that this does not contradict our earlier claim, viz. you cannot vary the metric and get equations of motion which are invariant under $T^a_b \to T^a_b + $ (constant)$\delta^a_b$.
That claim,  as specifically  stated, is valid only for local  actions. The $Z$ in \eq{tp16} is  a   non-local  functional of the metric tensor and $T_{ab}$. Therefore the  equations resulting from an extremum principle based on $\ln Z$ will obey our guiding principle; but this is not a local variational principle obtained by integrating a scalar Lagrangian over the measure $\sqrt{-g}d^4x$.}

\section{Spacetime Evolution in  Thermodynamic Language}\label{sec:geotherm}

Since any spacetime (just like normal matter) will be perceived to be hot by a class of observers, the Boltzmann principle suggests that we should interpret the continuum physics of the spacetime as a thermodynamic description of unknown microscopic degrees of freedom (``atoms of spacetime''). The emergent gravity paradigm,   is such an attempt to obtain and interpret the \textit{field equations} of gravitational theories in a thermodynamic context.
The crucial thermodynamic inputs, viz. the temperature and entropy density of spacetime, are provided by the  temperature  (and a corresponding entropy) attributed to the null surfaces which are perceived as horizons by local Rindler observers.
\textit{It turns out that, using this single quantum input, we can rephrase and re-derive the entire description of classical gravity in a novel language}. I will now describe several aspects of this paradigm.

\subsection{The  Avogadro  number of the spacetime}

A crucial relation which arises in the study of, say, gases, is the equipartition law $E=(1/2)Nk_BT$ which should be more appropriately written as:
\begin{equation}
N=\frac{E}{(1/2)k_BT} 
\end{equation} 
The two variables in the right hand side, $E$ and $T$, have valid interpretations in the continuum, thermodynamic limit, but the left hand 
side has absolutely no meaning in the same, continuum, limit. The $N$  counts the microscopic degrees of freedom or --- more figuratively --- the number of atoms, the very existence of which is not recognized in continuum thermodynamics! A equation like this directly relates the macroscopic and microscopic descriptions of the system. Can we obtain a corresponding relation for spacetime? That is, an we count the number of spacetime degrees of freedom (`atoms of spacetime')? 

Remarkably enough, we can \cite{A17,A18}! Consider  a section of a spacelike surface $\mathcal{V}$ with boundary $\partial\mathcal{ V}$ corresponding to $\sqrt{-g_{00}}=$ constant. In any static spacetime, one can show that the gravitating (Komar) energy $E_{\rm Komar}$, located in the bulk,   is equal to the equipartition heat energy of the surface, leading to:
\begin{equation}
 E_{\rm Komar}\equiv\int_\mathcal{V} d^{3}x\sqrt{h} \ 2N\bar{T}_{ab}u^{a}u^{b}=\int_{\partial \mathcal{V}} \frac{\sqrt{q} \, d^2x}{L_P^2} \left( \frac{1}{2}k_B T_{\rm loc}\right) 
=\frac{1}{2} N_{\rm sur} (k_BT_{\rm avg})
\label{hequi1}
\end{equation} 
where $\sqrt{q} d^3x$ is the area measure on $\partial \mathcal{V}$.
Since there is a correspondence between the bulk and boundary energies as well as the concept of equipartition, we will call this relation  `holographic equipartition'.

The situation actually gets better. When we consider a general spacetime (rather than static spacetimes) we would expect the above relation to break down and the difference between the two energies to drive the evolution of the spacetime. This is exactly what happens and the result can be stated in terms of number of degrees of freedom in the bulk and the boundary. One can associate with the bulk energy $E_{\rm Komar}$ a dimensionless number  $N_{\rm bulk}$,  defined as the number of degrees of freedom in a bulk volume \textit{if} the (Komar) energy $E_{\rm Komar}$ contained in the bulk, is at equipartition at the temperature  $T_{\rm avg}$. That is, we define:
\begin{align}\label{Papper06_NewFin01}
 N_{\rm bulk}\equiv\frac{1}{(1/2)k_BT_{\rm avg}}\int d^{3}x\sqrt{h} \ 2N\bar{T}_{ab}u^{a}u^{b}=\frac{|E_{\rm Komar}|}{(1/2)k_BT_{\rm avg}}
\end{align}
(Of course, we do \textit{not}   assume that the equipartition is actually realized; this is just a dimensionless measure of the Komar energy in terms of the average boundary temperature.) 
One can then show \cite{A19} that the time evolution of spacetime geometry in a bulk region $\mathcal{V}$, bounded by the $\sqrt{-g_{00}}=\textrm{constant}$ surface, is sourced by the  bulk and boundary degrees of freedom. Specifically, we can show that:
\begin{align}\label{Paper06_NewFin02}
\frac{1}{8\pi}\int_\mathcal{V} d^{3}x\sqrt{h}u_a g^{ij}\pounds_\xi N^a_{ij}  =\frac{1}{2}k_BT_{\rm avg}\left(N_{\rm sur}-N_{\rm bulk}\right)
\end{align}
with $\xi_a=Nu_a$ being the time evolution vector, where $u_a$ is the velocity of the observers moving normal to the foliation.\footnote{The Lie variation term in \eq{Paper06_NewFin02} is also closely connected with  the canonical structure \cite{A19} of general relativity in the conventional approach, though the relation
$ \sqrt{h}u_{a}g^{ij}\pounds _{\xi}N^{a}_{ij}=-h_{ab}\pounds _{\xi}p^{ab}$,
where $p^{ab}=\sqrt{h}(Kh^{ab}-K^{ab})$ is the momentum conjugate to $h_{ab}$ in the standard approach.} This result  shows that \textit{it is the difference between the surface and the bulk degrees of freedom  which sources the time evolution of the spacetime!}
(A very similar result holds \cite{SCTPnull} for a null surface as well.)
A simple but elegant corollary is that in all static \cite{A17,A18} spacetimes, we have holographic equipartition, leading to the equality of the number of degrees of freedom in the bulk and boundary:
\begin{equation}
 N_{\rm sur}=N_{\rm bulk},
 \label{nsureqn}
\end{equation} 
which, of course, is a restatement of \eq{hequi1}.

In fact it is possible obtain an \textit{on-shell} relation similar to  \eq{nsureqn}, characterizing bulk-boundary relation for \textit{any} spacetime. To do this, 
we define a  
 vector field $P^a[v]$, which can be thought of  as the gravitational momentum attributed to spacetime \cite{A36}, by an observer with  velocity $v^a$. This is defined as:
\begin{equation}
 (16\pi) P^a[v] \equiv  -Rv^{a}-g^{ij}\pounds _{v}N^{a}_{ij}
 \label{deffourvel}
\end{equation}
 The physical meaning of $P^a[v]$ arises from the following result: The conservation $\nabla_a ( P^a + M^a) =0$ of total momentum $(P^a + M^a)$ of matter plus gravity for all observers will lead to \cite{A36} the field equations of general relativity; here $M^a=-T^a_bv^b$ is the momentum attributed  to matter by an observer with four-velocity $v^a$. So, the introduction of $P^a(v)$ restores the conservation of momentum in the presence of gravity!
(For a more detailed discussion of these results, and properties of $P^a$, see Ref. \cite{A36}).
Using this vector, one can  obtain a very general result \textit{valid for any spacetime (and foliation)}, which could be time-dependent. We can show that, when the equations of motion hold, i.e., on-shell, the total energy contained in a region $\mathcal{R}$ bounded by an equipotential surface $\partial\mathcal{R}$,
is   equal to the  heat content of the surface:
  \begin{equation}
 \int_\mathcal{R} d^3x\sqrt{h}u_a[ P^a(\xi)+ T^a_b\xi^b]=\int_{\partial\mathcal{R}} d^2x \ Ts; \qquad (\xi^b = N u^b)
\label{energytotal}
\end{equation}
This result generalizes \eq{hequi1} to an arbitrary spacetime.

\subsection{The  fluid mechanics of the null surfaces}

Given the thermodynamic properties of the null surfaces, we would expect the flow of gravitational momentum --- introduced in \eq{deffourvel} above ---  to be of primary importance vis-a-vis the thermodynamics of  null surfaces. This is indeed true. The Gaussian null coordinates (GNC)  generalizes the notion of the local Rindler frame  associated with an arbitrary null surface (see Refs. \cite{A41,A42,A43} for more details). We define
the time development vector as $\xi^a=Nu^a$ where $u^a$ is the four-velocity of observers at rest in GNC. When $P^a[v]$ evaluated for the  time evolution vector in for the Gaussian Null Coordinates (GNC)\footnote{This vector $\xi^a$ will reduce to the timelike Killing vector corresponding to the  Rindler time coordinate if we rewrite  the standard Rindler metric in the GNC form. So  $\xi^a$ provides  a natural generalization of the time evolution vector, corresponding to the local Rindler-like observers in the GNC, though, of course,  $\xi^a$  will not be a Killing vector in a general spacetime.} associated with a given null surface, 
 \cite{SCTPnull} its projection along $\ell_a,k_a$ and $q_{ab}$, associated with the given null surface. leads to three sets of equations all of which have direct thermodynamic interpretation.

 To do this, we construct the GNC associated with the given null surface and 
the $P^a(\xi)$ using the corresponding time evolution vector. The natural  basis vectors associated with the null surface are given by the set   $(\ell ^{a},k^{a},e^{a}_{A})$ where $e^{a}_{A}$ spans the two transverse directions. The gravitational momentum can be decomposed in this basis as: $P^{a}=A\ell ^{a}+Bk^{a}+C^{A}e^{a}_{A}$ and the  components $A,B$ and $C^{A}$ can be recovered from the projections of $P^a$  by the relations $A=-P^{a}(\xi)k_{a}$, $B=-P^{a}(\xi)\ell _{a}$ and $C^{A}=P^{a}(\xi)e^{A}_{a}$. So, the following combinations: $q^{a}_{b}P^{b}(\xi), k_{a}P^{a}(\xi)$ and $\ell_{a}P^{a}(\xi)$ contain complete  information about the flow of gravitational momentum with respect to  the given null surface. Each of these components lead to interesting thermodynamic interpretations. The calculations are somewhat involved (which can be found in Ref. \cite{SCTPnull}) but the final results are easy to state:

\begin{itemize}

\item The component $q^{b}_{a}P^{a}(\xi)$ allows us to rewrite the relevant component of the gravitational field equations in a form identical to  the Navier-Stokes equation for fluid dynamics \cite{A24,A25} (for a variable which can be interpreted as drift velocity on the horizons). This is  the most direct link between the field equations and the fluid mechanics of the degrees of freedom hosted by the null surface. This is a generalization of the corresponding result, known previously for black hole spacetimes \cite{A26,A27}, to \textit{any} null surface in \textit{any} spacetime. 

\item The projection $k_{a}P^{a}(\xi)$,  when evaluated on an arbitrary null surface can be \cite{A33} rewritten  in the form: $TdS=dE+PdV$, i.e., as a thermodynamic identity. All the variables in this relation have the conventional meanings and differentials are interpreted as changes in the relevant variables when we make an infinitesimal virtual displacement of the null surface in the direction of $k^a$. This again generalizes the corresponding results, previously known for spacetimes with some symmetry (see e.g., \cite{A30,A31,A32,A28,A29}) and the null surface in question is a horizon. This relation  also allows  us to associate a notion of energy  with an arbitrary null surface \cite{A44,A45}.

\item Finally, the component $\ell _{a}P^{a}(\xi)$ gives \cite{SCTPnull} an equation for the evolution of null surface, in terms of its heating rate involving both  $ds/d\lambda$ and $dT/d\lambda$, where $s$ is the entropy density, $T$ is the temperature associated with the  null surface and $\lambda$ is the parameter along the null generator $\ell _{a}$. 
\end{itemize}

\noindent To summarize, the evolution of spacetime can be expressed completely in thermodynamic/fluid mechanical language. This description, of course, is not unique and depends on the context just as in the case of macroscopic material bodies. 

I will now turn to the more challenging task of establishing a deeper layer of support for such a description. As in the case of ordinary fluid mechanics, such a layer is provided by description in terms of kinetic theory and distribution function. We will see that something similar happens in the case of the description of spacetime.

\section{Gravity at Mesoscopic Scales}
\label{sec:grfratoms}

It is possible to synthesis the above ideas together and obtain the gravitational field equations from an extremum principle  based on the density of states of the quantum geometry. I will motivate a suitable definition of the \dqg\ ---  which is the key new concept in this approach --- that leads to deeper insights into the \md\ and,  as a bonus, provides \cite{tp}  a fresh perspective on the numerical value of the \cc. 

\subsection{Overview  of the paradigm}

In this section, I will summarize the results
and highlight the key concepts. Later subsections provide some mathematical details and I refer the reader to two previous reviews \cite{tp,lwf} for a more extensive discussion.

\subsubsection{Distribution function: Counting   the continuum}

 Let us begin by recalling   that the physics of the fluids can be presented at two different levels, one more fundamental than the other. The first level is the continuum description in which we completely ignore the fact that there are microscopic degrees of freedom in the fluid in the form of atoms/molecules. This level of description uses  variables like density $\rho(x^i)$, pressure $P(x^i)$, temperature $T(x^i)$, fluid momentum $P^\mu(x^i)$, etc. and the dynamics  is governed by the continuity equation and, say, the Navier-Stokes equation.    Of course, the existence of temperature (which we know is proportional to the random kinetic energy of the atoms) as well as the  transport coefficients tells us that such a description is fundamentally incomplete. 
 
 The second, deeper, level of description uses  the  distribution function $f(x^i, p_j)$ for the system, which \textit{counts}  the number of  atoms $dN = f(x^i, p_j)d^3xd^3p$   per unit phase space volume $d^3xd^3p$ (with the constraint $p^2=m^2$ making the phase space  six-dimensional).
 This description, in terms of a distribution function, is indeed remarkable because it allows us to use the continuum language and --- at the same time --- to recognize the discrete nature of the fluid. (One could equivalently think of $f$ as the number of \textit{degrees of freedom} per unit phase space volume. I will use the terminology `atoms' to describe these degrees of freedom.)
 The new variable that is required at this level of description is the  ``internal'' variable $p^\mu$ which allows us to  describe atoms with different \textit{microscopic} momenta co-existing at the same event $x^i$. The \textit{macroscopic} momentum  of the fluid,  $P^\mu(x^i)$, used at the first level, is given by the average value $\langle p^\mu\rangle =P^\mu(x^i)$. Similarly, we have the relation $\langle p^\mu p^\nu\rangle =P^\mu(x^i)P^\nu(x^i)+\Sigma^{\mu\nu}$ where the second term arises from the dispersion of momentum.
 
 In a similar manner, we want to introduce a function $\rho_g (x^i,\phi_A)$ to describe the \dqg. Here, $\phi_A$ (with $A=1,2,3,...$) denotes the internal degrees of freedom (analogous to the momentum $p_i$ for the distribution function for the molecules of a fluid) which  exist as fluctuating \textit{internal} variables at \textit{each} event $x^i$. Their behaviour,  at any given event $x^i$, is determined by a probability distribution $P(\phi_A,x^i)$, the form of which depends on the microscopic quantum state of the spacetime. Of course, we cannot determine this function at this stage, without knowing more about the quantum structure of spacetime, but --- fortunately, as we will see --- we only need some basic properties of the  average values like $\langle \phi_A\rangle, \langle \phi_A\phi_B\rangle$ etc., for the purpose of obtaining and interpreting the classical field equations of gravity.
 The dependence of $\rho_g (x^i,\phi_A)$ on the event $x^i$ arises only indirectly through   the geometrical variables like the metric tensor, curvature tensor etc., (which I will collectively denote as $\mathcal{G}_N(x)$ with $N=1,2,3,...$), so that $\rho_g (x,\phi_A)=\rho_g (\mathcal{G}_N(x), \phi_A)$. Such a description will not be exact but will be valid at some mesoscopic scales larger than $L_P$ so that the variables  $\mathcal{G}_N(x)$ can be defined. At these scales, the Planck length plays a role   analogous to that of the mean-free-path in the kinetic theory.
 
It turns out that there is a natural way of defining $\rho_g (x^i,\phi_A)$, once we introduce discreteness into the spacetime through a zero-point length. Remarkably enough, this procedure \textit{also}  identifies for us the internal variable $\phi_A$ as a  four-vector $n^a$ with constant norm,  which can be thought of  as a microscopic, fluctuating, quantum variable at each event $x^i$. 
 Once we determine  the form of $\rho_g (x^i,n_a)\equiv\exp S_g$ --- and the corresponding density of states for matter, $\rho_m \equiv\exp S_m$ ---  the equilibrium state of spacetime plus matter is determined by extremum of 
 $\rho_g \rho_m =\exp[S_g+S_m]$. This leads to Einstein's equations, along with a rather elegant interpretation.
  
\subsubsection{Events in quantum spacetime have finite area but zero volume} 

Let us begin by determining $\rho_g$. The two, primitive, geometrical constructs one can think of in any spacetime  are the area and the volume. It is, therefore,  natural to assume that the \dqg, $\rho_g(\mathcal{P})$,  at a given event $\mathcal{P}$, should be some function $F$ of either the area $A(\mathcal{P})$ or the volume $V(\mathcal{P})$  that we can ``associate with'' the event $\mathcal{P}$. Further, the total degrees of freedom, $\rho_g(\mathcal{P})\rho_g(\mathcal{Q})$, associated with two separate events $\mathcal{P}$ and $\mathcal{Q}$ is multiplicative
 while the primitive area/volume elements are additive. This suggests that the function $F$ should be an exponential. In terms of area, for example, we then have:
\begin{equation}
 \ln 
 \left\{ 
 \begin{array}{c}
  \text{density of states of the}\\
  \text{quantum spacetime at $\mathcal{P}$}
 \end{array}
 \right\}\quad
 \propto\quad
\left\{ 
 \begin{array}{c}
  \text{area ``associated with''}\\
  \text{the event $\mathcal{P}$}
 \end{array}
 \right\}
\end{equation}
That is, we take
$\ln\rho_g(\mathcal{P})\propto A(\mathcal{P})$.

Next, we need to give a precise meaning to the phrase, area (or volume) ``associated with'' the event $\mathcal{P}$. 
To do this, let us consider the Euclidean extension of a local neighborhood around $\mathcal{P}$ and  all possible geodesics emanating from  $\mathcal{P}$. The surface $\mathcal{S}(\mathcal{P},\sigma)$ formed by all the events, which are at a geodesic distance $\sigma$, forms a \textit{equi-geodesic surface} around 
$\mathcal{P}$. Let $A(\mathcal{S})$ be its area and let $V(\mathcal{S})$ be the volume enclosed by this surface. The limiting values of  $A(\mathcal{S})$ and  $V(\mathcal{S})$ in the limit of $\sigma\to0$ provide a natural definition of the area/volume 
``associated with'' the event $\mathcal{P}$. 

In standard Riemannian geometry --- which, of course, knows nothing about the discreteness of microscopic spacetime --- both the area and volume will vanish in the limit of $\sigma\to0$; events have zero area and zero volume associated with them,  as to be expected.  But when we introduce the discreteness of the spacetime in terms of a zero-point length, we  find that \cite{paperD} the area associated with an event  becomes nonzero but the volume  still remains zero!. (Some key results related to this approach are summarized in Appendix \ref{appen:qmetric}.)  What is more, this approach will introduce an arbitrary, constant norm, vector $n_a$ into the fray. (Its norm is unity in the Euclidean sector and --- on analytic continuation --- it will map to a  null vector with zero norm in the Lorentzian sector.). This variable is analogous to the momentum variable $p^a$ (of constant norm) which is needed to define the distribution function. So, just like $p^a$, this quantum degree of freedom $n^a$ --- as well as the area ``associated with'' an event --- will  be a fluctuating, indeterminate quantity. 
It turns out that, in terms of this internal, vector degree of freedom, the $\rho_g(x^i,n_a)$ is given by
\begin{equation}
\ln \rho_g \propto \left[1- \frac{L_0^2}{6} R_{ab}(x) n^an^b\right] =\frac{\mu}{4}\left[1-\frac{L_0^2}{6}R_{ab}n^an^b\right]\, ; \qquad L_0^2 = \frac{3}{\pi} L_P^2
\label{rhogresult}
\end{equation}
where $\mu$ is a dimensionless proportionality constant.
Eventually, we can set $\mu = 1$ by suitable choice of measure. We have separated out a factor $(1/4)$ so that, with this normalization, the on-shell entropy density of null surfaces will be $(1/4)$ per unit area.\footnote{Equation \ref{rhogresult} is valid to leading order in $L_P^2$; one can obtain higher order terms of $\mathcal{O}(L_P^4R^2)$ etc but the geometric description may not be valid at higher orders. So it is probably not consistent to proceed to higher orders. It should be kept in mind that the entire description is valid only when $(L_P^2R)$ is small compared to unity so that the right hand side remains positive. A more exact expression for $\rho_g$, involving the Van-Vleck determinant, is given in \eq{vv-dos} of Appendix \ref{appen:qmetric}.} 
The fact that the term involving $R_{ab}$  comes with a \textit{minus sign} in \eq{rhogresult} is very crucial for the success of our programme and we have \textit{no} control over it! 

 The fluctuations of $n^a$ will be  governed by some  probability functional $P[n^i(x),x]$, which is the probability that the quantum geometry is leads to a vector field  $n_a(x)$ at every $x$. As I said before, the form of $P$  is not known at present; but, fortunately, we only need  two properties of this  probability distribution $P[n^i(x),x]$ which can be derived: (i) It preserves the norm of $n^i$; i.e  $P[n^i(x),x]$ will have the form $F[n(x),x]\delta(n^2-\epsilon)$ with $\epsilon=1$ in the Euclidean sector and $\epsilon=0$ in the Lorentzian spacetime. (ii) The   average of $n^a$ over the fluctuations (evaluated in a given quantum state of the geometry) gives,%
 \footnote{This is analogous to the average value of the microscopic momenta of fluid particles $\langle p^\mu\rangle =P^\mu(x^i)$ being equal to the macroscopic momentum of the fluid. The key difference is that, in standard fluid mechanics, the distribution function itself is used to do the averaging while here the fluctuations are governed by some other probability distribution $P(x,n)$. Recall that the null normal $\ell_a$ \textit{of} a null surface also defines the tangent vector to the null geodesic congruence \textit{on} the null surface; in this sense, it is actually the momentum of the ``photons'' traveling along the null geodesics.} 
 in the Lorentzian spacetime, a null normal $\ell^a(x^i)$ to any patch of null surface; i.e.,  $\langle n^a\rangle = \ell^a(x^i)$. Such  averages are defined through the functional integral
\begin{equation}
 \langle n_a\rangle=\int\mathcal{D}n\; n_a(x)P[n^a(x),x]=\ell_a(x)
\end{equation} 
where $P[n^a(x),x]$ is  parametrized by some null vector field $\ell_a(x)$. (For example, $P$ could be a narrow Gaussian functional in the variable $[n^a(x)-\ell^a(x)]$.) 
 Different quantum states of the spacetime geometry will lead to  different $P$ with different null normals $\ell_a (x^i)$ as their mean values; so, in fact,  the expectation value $\langle n_a\rangle $ actually leads to the \textit{set of all null normals} $\{\ell_a(x^i)\}$ at a given  event $x^i$ when we take into account all quantum states which have sensible classical limit, described by some metric. (Of course, not all quantum states of \md\ will have a classical limit.) 
 Similarly, we will have $\langle n_in_j\rangle =\ell_i\ell_j+\sigma_{ij}$ where the second term $\sigma_{ij}$ represents quantum gravitational corrections to the mean value, etc. Thus, the mean value $\langle \ln\rho_g(x^i,n_a)\rangle $,
 in the continuum limit, is given by:
\begin{equation}
\langle \ln\rho_g(x^i,n_a)\rangle =\frac{\mu}{4}\left[ 
 1-\frac{L_0^2}{6}  R_{ab}\ell^a\ell^b +....\right]= \frac{\mu}{4} \left[ 1 -\frac{L_P^2}{2\pi}  R_{ab}\ell^a\ell^b \right]
\label{denast2}
\end{equation}
where we have not displayed terms proportional to $R_{ab}\sigma^{ab}$ which are of higher order and independent of $\ell_a(x)$. (It is possible to describe the null normals as a limiting case of timelike/spacelike normals and obtain this result. It is also possible to derive it, more rigorously, from first principles \cite{pesci}.) 
The combination 
\begin{equation}
\mathcal{H}_g \propto  - R_{ab}\ell^a\ell^b
\label{defhg1}
\end{equation}
which occurs here can be rewritten in terms of an expression quadratic in the \textit{derivatives}  $\nabla_a\ell_b$ (plus an ignorable total divergence) and
has a physical interpretation as the 
gravitational contribution to the heating rate (per unit area) of the null surface to which $\ell_a$ is the normal (see Sec. \ref{sec:zero-diss}). Its integral, $Q_g$, taken over the null surface,  can be interpreted  as the gravitational contribution to the heat content of the null surface. 

To complete the picture, we also need the corresponding expression for  matter (see Sec. \ref{sec:hdm}). In the continuum limit, it is straightforward to show --- using the concept of local Rindler horizons --- that  this is given by:
\begin{equation}
\langle \ln \rho_m\rangle  \propto  L_P^4 T_{ab} \ell^a \ell^b=L_P^4\mathcal{H}_m
\label{rhomresult}
\end{equation}
where - $\mathcal{H}_m$ can be interpreted as the heat density contributed by matter crossing a null patch.
Taking into account both matter and spacetime, the total number of degrees of freedom, in the continuum limit --- in a state characterized by the vector field $\ell_a(x)$ --- is given by:
\begin{equation}
 \langle \Omega_{\rm tot}\rangle_\ell  =\ \prod_{x}\, \langle \rho_g\rangle  \langle \rho_m\rangle  
 = \exp \sum_x \left( \langle \ln \rho_g\rangle  + \langle \ln \rho_m\rangle \right)\equiv \exp [S_{\rm grav}(\ell)+S_{\rm m}(\ell)]
 \label{omtot}
\end{equation}
to the leading order (when we ignore the fluctuations, so that $\ln\langle \rho\rangle \approx\langle \ln\rho\rangle $).
The proportionality constants appearing in \eq{rhogresult} and \eq{rhomresult} can be taken to be the same factor by choosing the measure in this sum appropriately. 
 The  $\ell_a$ dependent part of the configurational entropy $S_{\rm tot}=S_{\rm grav}+S_{\rm m}$ is then given by the functional
\begin{equation}
S_{\rm tot}[\ell(x)]=
\int_{\mathcal{S}} d^3V_x\; \mu E^a_b \langle n_an^b\rangle =\int_{\mathcal{S}} d^3V_x\;\mu\left(T^a_b (x) - \frac{1}{\kappa} R^a_b (x)\right) \ell_a(x)\ell^b(x) + ....
\label{av1}
\end{equation}
where, in the continuum limit, the sum over $x$ is replaced by integration over the null surface $\mathcal{S}$ for which $\ell^a(x)$ is the normal, with the measure
$d^3 V_x=(d\lambda d^2 x \sqrt{\gamma}/L_P^3)$ and the proportionality constant $\mu$ is introduced. 

The  gravitational field equations can now be obtained by extremizing the expression for $\langle \Omega_{\rm tot}\rangle_\ell$ or, equivalently, the configurational entropy $S_{\rm tot}=S_{\rm grav}+S_{\rm m}$ over $\ell$ and demanding that the extremum condition holds for all $\ell_a$.
(This is precisely the variational principle we discussed in Sec. \ref{sec:formvarprin}; we now have a deeper level of support for the same.)
Because different quantum states of geometry will lead to different $\ell_a(x)$ at the same event $x^i$, this is equivalent to demanding the validity of the extremum condition for all quantum states of the geometry, which are relevant in the classical limit.
The extremum of \eq{av1} with respect   to the variation $\ell^a \to \ell^a + \delta \ell^a$, 
subject to the constraint $\ell^2=$ constant, will then lead to Einstein's equations, with a cosmological constant arising as an integration constant. (See footnote \ref{fn})
Moreover,  the integrand of \eq{av1} \textit{itself} can be interpreted as the heating rate of the null surface. So, as to be expected, the classical limit makes perfect thermodynamic sense.
I emphasize that we have obtained the field equations \textit{without} treating the metric as a dynamical variable to be varied in an extremum principle.

I will now describe, in some detail,  how the expression for $\rho_g$ is obtained and how the internal degree of freedom $n_a$ arises.

\subsection{Degrees of freedom of geometry and matter}
\label{sec:zpl}

I will now describe how the \md\ in \eq{rhogresult} arises in slightly more detail. 
We first introduce the notion of an equi-geodesic surface,  
which can be done either in 
the Euclidean sector or in the Lorentzian sector;  let us work in the Euclidean sector. 
An equi-geodesic surface $\mathcal{S}$ is made of the set of all points located at the same geodesic distance $\sigma$ from some specific point $P$, which we take to be the origin \cite{D1,D4,D5,D6}.  
We next  ``associate'' an area element with a point $P$ in a fairly natural way  by the following limiting procedure:
(i) Construct an equi-geodesic surface $\mathcal{S}$ around a point $P$ at some geodesic distance $\sigma$. (ii) Calculate the area element $\sqrt{h}$ from the induced metric $h_{ab}$ on this $\mathcal{S}$. (iii) Use the limit $\sigma\to0$ to define and compute the area element associated with the point $P$.
 In standard differential geometry, one can show \cite{D8} that, in the limit of $\sigma \to 0$,  the area element, normalized to the flat spacetime value (to the leading order in $\sigma^2$), is given by: 
\begin{align}
\frac{\sqrt{h}}{\sqrt{h_{flat}}}=\left(1-\frac{1}{6}\sigma ^{2}R^a_bn_an^b\right)
\label{gh}
\end{align}
where $n_a=\nabla_a\sigma$ is the normal to $\mathcal{S}$.  The second term containing  $\mathcal{E}\equiv R^a_bn_an^b$ gives the curvature correction to the area element of  an equi-geodesic surface. In fact, \eq{gh} is a  textbook result in differential geometry and is often presented as a measure of the curvature at any event. 

As can be readily seen from \eq{gh}, $\sqrt{h}\to0$  when $\sigma\to 0$ since, in this limit $\sqrt{h_{flat}}\propto\sigma^3$. Even though  the normal $n_a=\nabla_a\sigma$ becomes ill-defined in this limit, this ambiguity in the second term is irrelevant when $\sigma\to 0$ because  the $\sigma^2$ factor kills this terms.
This is, of course,  an expected result. The existence of non-zero \md\ requires some sort of discrete structure in the spacetime. They do not arise if  the spacetime is treated as a continuum all the way. (This is similar to the fact that you can't associate a finite number  of molecules of a fluid with an event $P$ if the fluid is treated as a continuum all the way.)  Classical differential geometry, which leads to \eq{gh}, knows nothing about any discrete spacetime structure and hence cannot provide us with a nonzero $\rho_g$. To obtain a nonzero $\rho_g$, we need to know how the geodesic interval and the  metric  in the quantum description of spacetime. In particular, we  expect to have a $\sqrt{h}$  which does not vanish in the coincidence limit in  such a quantum description. I will now  describe how this issue can be addressed without knowing the exact theory of quantum gravity.

There is a large amount of evidence (see e.g., \cite{D2a,D2b,D2c,D2d,D2e,D2f}) which indicates that a primary effect of quantum gravity will be to introduce into the spacetime  a zero-point length, by modifying the geodesic interval $\sigma^2(x,x')$ between any two events $x$ and $x'$  to a form like $\sigma^2 \to \sigma^2 + L_0^2$ where $L_0$ is a length scale of the order of the Planck length.\footnote{A more general modification will be of the form $\sigma^2 \to S(\sigma^2)$  where the function $S(\sigma^2)$ satisfies the constraint $S(0) = L_0^2$ with $S'(0)$ finite. Our results     are actually  insensitive to the explicit functional form of  $S(\sigma^2)$. For the sake of  illustration, I will use the simple function $S(\sigma^2) = \sigma^2 + L_0^2$. The introduction of zero-point-length is also likely to eliminate spacetime singularities \cite{pesci-sc-dk} but I will not discuss this aspect here.} This feature allows us to determine a quantum corrected, effective metric
along the following lines:
Just as the original $\sigma^2$ is obtained from the original metric $g_{ab}$, we demand that the correct geodesic interval $S(\sigma^2)$ --- which incorporates the effects of quantum gravity --- arises from a corresponding, quantum gravity-corrected metric \cite{D1}, which we will call the qmetric $q_{ab}$. 
(Some key results, related to this approach, are summarized in Appendix \ref{appen:qmetric}.)
Of course, no such local, non-singular $q_{ab}$ can exist;  for any such $q_{ab}$, the resulting geodesic interval will vanish in the coincidence limit,  by definition of the integral. The $q_{ab}(x,x')$ will  be a bitensor, which must be singular at all events in the coincidence~limit $x\to x'$. It turns out that 
one can determine \cite{D4,D5,D7} the form of $q_{ab}$ just from the demand that
the pair $(q_{ab},S(\sigma^2))$ should satisfy the same relationships as $(g_{ab},\sigma^2)$.  Once we have determined $q_{ab}$ we can  
compute the area element ($\sqrt{h}\, d^3 x$) of an equi-geodesic surface 
using the  qmetric.  

The computation is 
straightforward  and --- for $S(\sigma^2)=\sigma^2+L_0^2$ in $D=4$, though similar results \cite{paperD,D7} hold in the more general case and in $D$ dimensions ---  we get the result:\footnote{One might think that the result in \eq{hfinal}  arises from the standard result in \eq{gh}, just by the simple replacement of $\sigma^2\to(\sigma^2+L_{0}^{2})$. This just happens to be true in this case; but it turns out that this replacement  does \textit{not} work for the volume element $\sqrt{q}$ which actually vanishes \cite{D4,D5,D7} when $\sigma\to0$. As a result,  each event has zero volume, but a finite area, associated with it! A further insight into this rather curious feature is provided by the following fact:
The leading order dependence of $\sqrt{q}d\sigma\approx\sigma d\sigma$ leads to the volumes scaling universally as $\sigma^2$ while the area measure remains finite. This, in turn, leads to the result \cite{paperD} that \textit{the effective dimension of the quantum spacetime tends to $D=2$ close to Planck scales,} independent of the original $D$. Similar `dimensional reduction' has been noticed earlier by several people \cite{deq2}
in different, but specific, models of quantum gravity. The procedure described here leads to this result in a fairly \textit{model-independent} manner.}
\begin{align}
\frac{\sqrt{h}}{\sqrt{h_{flat}}}
=\left[1-\frac{1}{6}(R_{ab}n^an^b)\left(\sigma ^{2}+L_{0}^{2}\right)\right];\qquad L_0^2=\frac{3L_P^2}{\pi}
\label{hfinal}
\end{align}
(The correct Newtonian limit of the theory requires $L_0^2=3L_P^2/4\pi$ which is the value we will use.)
When $L_{0}^{2}\to0$, we recover the standard result in \eq{gh}, as we should. Our interest, however, is in the coincidence limit $\sigma^2\to0$ computed with finite $L_0$.
Something nice happens when we take this limit and we get a non-zero result:
\begin{align}
\lim_{\sigma\to 0} \frac{\sqrt{h}}{\sqrt{h_{flat}}}
= \left[1-\frac{L_0^2}{6}R_{ab}n^an^b\right]
\label{hlimit}
\end{align}
That is, the quantum corrected metric 
attributes to every point in the spacetime a finite area measure (but a zero volume measure)! 
So we define \cite{tpentropy,tpreviews1} the dimensionless \dqg, as:\footnote{It is gratifying that the term involving $R_{ab}$  comes with a \textit{minus sign} in \eq{denast} which is crucial for the success of our programme;  we have \textit{no} control over it!} 
\begin{equation}
\ln\rho_g(x^i,n_a)\propto\lim_{\sigma\to 0}  \frac{\sqrt{h(x,\sigma)}}{\sqrt{h_{flat}(x,\sigma)}}
\propto \left[1-\frac{L_0^2}{6}R_{ab}n^an^b\right]=\frac{\mu}{4}\left[1-\frac{L_P^2}{2\pi}R_{ab}n^an^b\right]
\label{denast}
\end{equation}
where $\mu$ is a dimensionless proportionality constant. This is related to our choice of the measure;  alternatively, we can set $\mu=1$ in \eq{denast} and redefine  the measure. All these results in \eq{hfinal},\eq{hlimit} and \eq{denast} are valid to leading order in $L_P^2$, when the second term is small compared to unity,  so that the expression for $\sqrt{h}$ remains positive definite. (It is possible to show that, when higher order terms are retained, we get $\ln \rho_g=(1/4\Delta)$ where $\Delta$ is the Van-Vleck determinant defined in \eq{vv-dos} in Appendix \ref{appen:qmetric}. There is probably no point in computing higher order terms since the geometric description itself might break down when these terms are relevant.)

We also see that $\rho_g$, defined by the first equality in \eq{denast}, depends on the extra, internal degree of freedom, $n_a$ which could take all possible values (at a given $x^i$) except for the constraint that it has  unit norm in the Euclidean space. \textit{This quantity arises as a relic of the discrete nature of the spacetime.} This $n_a$ is analogous to the $p_j$ which appears in the fluid distribution function $f(x^i,p_j)$ --- as a relic of the discrete nature of the fluid ---  which can also take all possible values at a given $x^i$ except for the constraint that it has a constant norm $p^2=m^2$. In the case of the fluid, we have $\langle p^\mu\rangle =P^\mu(x^i)$ where $P^\mu(x^i)$ is the average momentum of the fluid in the continuum description and
 $\langle p^\mu p^\nu\rangle =P^\mu(x^i)P^\nu(x^i)+\Sigma^{\mu\nu}$, where the second term originates from the dispersion of momentum. Similarly  $n^a$ is a microscopic, fluctuating  variable such that its average over  fluctuations gives some vector field  $\langle n^a\rangle = \ell_E^a(x^i)$ of unit norm in the Euclidean continuum limit. Similarly, $\langle n^in^j\rangle =\ell^i_E\ell^j_E+\sigma^{ij}$ where the second term $\sigma_{ij}$ represents higher order corrections to the mean value.

We have been working so far in the Euclidean sector with $n_a \propto \nabla_a\sigma$ being the unit normal to the equi-geodesic surface, $\sigma=$ constant. In the limit of $\sigma\to0$ in the Euclidean sector,  the equi-geodesic surface shrinks  to the origin. \textit{However, in the Lorentzian sector, the same limit leads to the null surface which acts as the local Rindler horizon around the chosen event.} So, in this limit, we need to identify $n^a$ and its mean value $\ell^a_E$ with Lorentzian null vectors. This is how an internal degree of freedom  enters into the \dqg, $\langle \ln\rho_g\rangle $, as a null vector $\ell^a$.

We can summarize the net result of this exercise as follows: Take any event $x^i$ and a local patch of any null surface ---  with normal $\ell_a(x)$ and passing through $x^i$ --- that can act as a horizon to the local Rindler observers. Identify the mean value $\langle n_a\rangle $ with $\ell_a(x)$ in the continuum limit. There exists, of course, an infinite number of null patches (and normals) at any event. The quantum state of the microscopic geometry determines the expectation value $\langle n_a\rangle $ in a more fundamental description with  different quantum states  leading to  different null normals $\ell_a (x^i)$. Thus, the expectation value $\langle n_a\rangle $ actually corresponds to the \textit{set of} all possible null normals $\{\ell_a(x^i)\}$ at an event $x^i$ when we consider all possible quantum states. Then the \md\ associated with any event in the continuum spacetime is given by expression: 
\begin{equation}
\langle \ln\rho_g(x^i,n_a)\rangle = \frac{\mu}{4}\left[ 
 1-\frac{L_P^2}{2\pi}R_{ab}\ell^a\ell^b\right] +....
\label{denast1}
\end{equation}
where we have not displayed terms proportional to $R_{ab}\sigma^{ab}$ which are of higher order and independent of $\ell_a(x)$.

Let me also comment on the corresponding degrees of freedom for matter, determined through its contribution to the heat density.
The entropy $S_m$ associated with the heat $Q_m$ in \eq{Paper06_New_11} is given by
$S_m = Q_m/T_H$ where $T_H$ is a temperature introduced essentially for dimensional purposes. (A natural choice will be to take it to be the temperature associated with the acceleration of the Rindler observers very close to the horizon, say, at a distance of one Planck length. But, as to be expected, none of the results will depend on its numerical value when we choose the measure of integration appropriately.) We then have
\begin{equation}
S_m(\ell) = \frac{1}{T_H} \int d\lambda\, d^2 x \sqrt{\gamma}\, \mathcal{H}_m(x,\ell) 
= \mu \int \frac{d\lambda\, d^2 x \sqrt{\gamma}}{L_P^3} \, \left(L_P^4 \mathcal{H}_m(x,\ell)\right)
\label{tpnine}
\end{equation}  
where we have introduced suitable factors of $L_P$ to exhibit clearly the dimensionless nature of $S_m$ and defined $\mu \equiv (1/L_PT_H)$. 
Replacing the integration by a summation over a set of spacetime events for conceptual clarity, we can write 
\begin{equation}
S_m(\ell) = \sum_x L_P^4 \mathcal{H}_m (x,\ell) \equiv \sum_x \ln \rho_m(x,\ell) = \ln \prod_x  \rho_m(x,\ell)
\label{tpten}
\end{equation} 
so that $\ln\rho_m(x,\ell)=L_P^4 \mathcal{H}_m (x,\ell)$ with a proportionality constant set to unity if we absorb the factor $\mu$ in the integration measure. (Alternatively, we could have set the proportionality constants  in \eq{rhogresult} and \eq{rhomresult} to be $\mu$ and then used the measure without the factor $\mu$.) 
This connects up with our discussion in Sec. \ref{sec:grfratoms}   and, specifically, with the result  in \eq{rhomresult} in the continuum limit. 
The total number of degrees of freedom is now correctly given by
\begin{equation}
\Omega_m(\ell) = \exp S_m(\ell) = \prod_x \rho_m = \prod_x \exp(L_P^4 \mathcal{H}_m ) = \exp\left[\mu \int \frac{d\lambda\, d^2 x \sqrt{\gamma}}{L_P^3} \, \left(L_P^4 \mathcal{H}_m\right)\right]
\label{tpeleven}
\end{equation} 
Here, the first equality is the standard relation between the entropy and the degrees of freedom, the second expresses the result as a product over the degrees of freedom associated with each event and the third equality presents it in terms of the variable $\mathcal{H}_m = T^{ab}\ell_a\ell_b$.  

We see that the matter stress tensor appears only through the combination $T^a_b \ell_a \ell^b$ in the theory, making it  invariant  under the shift $T^a_b \to T^a_b + $ (constant)$\delta^a_b$ arising by the addition of a constant to matter Lagrangian. This takes care of our guiding principle, introduced in Sec. \ref{sec:classgrav}

\subsection{A deeper level of description: Two speculations}

The above ideas heavily relied on the existence of geometrical constructs, like an effective metric, at mesoscopic scales which we think of scales intermediate to Planck scales (requiring description in pregeometric variables) and macroscopic scales (with classical gravity is described in thermodynamic language). In this section I speculate about how these results might fit with a deeper level of description.

\subsubsection{Geodesic interval as the two-point correlator of pregeometric variables}

Our approach chooses the geodesic interval $\sigma^2(x,x')$ (rather than the metric) as the useful variable to describe spacetime geometry \cite{D7}. In the classical spacetime, both $\sigma^2(x,x')$ and $g_{ab}(x)$ contain the same amount of information and one is derivable from the other. But the geodesic interval $\sigma^2(x,x')$ is far better suited to take into account quantum gravitational effects in the mesoscopic scales.
Introducing a zero-point length in the spacetime by the modification $\sigma^2 \to S(\sigma^2)$ (with a finite $S(0)$) allows us to obtain all the results presented above.

At a more fundamental level, one could think of the correct geodesic interval $S[\sigma^2(x,y)]$, incorporating the zero-point-length, as a correlator of pregeometric variable, say $J(x)$, along the lines:
\begin{equation}
 S[\sigma^2(x,y)] = \langle J(x) J(y)\rangle
 \label{ssig}
\end{equation} 
where the average is taken using a probability distribution function, with the leading order behaviour:
\begin{equation}
 \mathcal{P}[J(x)] = \exp\left[ -\frac{1}{2} \int \frac{J(x)J(y)}{S[\sigma^2(x,y)] }\, dV_x \ dV_y+\cdots \right]
 \label{pgp}
\end{equation} 
At the pregeometric level, $x,y$ probably should be thought of as discrete variables in an abstract space which acquire a continuum meaning in the spacetime.  The details of this mapping from an abstract $x$-space to the physical spacetime is unclear at this stage but one can still proceed with algebraic manipulation as indicated above. (Since $S$ has the dimensions of  square of the length, it is natural to assume that $J$ has the dimensions of length; the integration measures $dV_x$ etc in \eq{pgp} can made dimensionless, by suitable factors of $L_P$, like e.g. $d^4x/L_P^4$.)

This relation suggests that $G(x,y) \equiv (1/ S[\sigma^2(x,y)])$ could be thought of as a propagator for a theory with $J(x)$ acting \textit{as the sources}. The coincidence limit of this propagator $G(x,x) = (1/L_0^2)$ is finite due to the existence of the zero-point-length.  One can also think of $G(x,y)$ as the VEV in a field theory with a highly nonlocal Lagrangian. [When $\sigma^2$ is only a function of $(x-y)$, the Lagrangian will be  $L\propto \phi(x)G^{-1}(k=i\nabla)\phi(x)$ where $G(k)$ is the Fourier transform of $G(x-y)$].

The detailed formalism for constructing such theories, with propagators having finite coincidence limit, has already been developed \cite{D2d}. 
 In the study of quantum fields, the notion of zero-point-length  was introduced into the propagator  earlier using the concept of path integral duality. The key idea was that the action for a particle propagating in a spacetime should remain invariant under the transformation $\sigma (x,y) \to L_0^2/\sigma (x,y)$. This suggests defining the (Euclidean) propagator for a particle of mass $m$, propagating in a given spacetime by the sum over paths:
 \begin{equation}
  G(x,y) = \sum_\sigma \exp\left[- m \left(\sigma + \frac{L_0^2}{\sigma}\right)\right]
  \label{modact}
 \end{equation}
 The resulting propagator has a very simple property:\footnote{The action for a relativistic particle of mass $m$ is $A = - m \sigma = - \sigma/\lambda_c$ where $\lambda_c = \hbar/mc$ is the Compton wavelength of the particle. When we bring in GR, it makes no sense to sum over paths with length $\sigma$ less than the Schwarzchild radius $R_g = Gm/c^2$ of the particle. This suggests suppressing the contribution from paths with $\sigma \lesssim R_g$. Assuming that this suppression preserves a duality symmetry under $\sigma \to 1/\sigma$, one would modify the relativistic action to the form $A_g = -(\sigma/\lambda_c) - (R_g/\sigma)$ which can be written in the form $A_g =  - (1/\lambda_c)[\sigma + (L_0^2/\sigma)]$ where $L_0$ is of the order of Planck length. This is one motivation for the modification of the action in \eq{modact} though, with rescaling, it can be applied even for a massless particle in the form of \eq{modker}.} 
 It can be obtained from a heat kernel in which $\sigma^2(x,y)$ is replaced by $S[\sigma^2(x,y)] = \sigma^2(x,y) + L_0^2$. That is, the modified propagator is given by:
 \begin{equation}
  G(x,y) = \int_0^\infty ds\ e^{-(L_0^2/4s)} \, K_{\rm std} (x,y; s)
  \label{modker}
 \end{equation} 
 where $K_{\rm std} (x,y; s)$ is the heat kernel in the standard spacetime without zero-point-length. So, in the case of $S[\sigma^2(x,y)] = \sigma^2(x,y) + L_0^2$, we can describe the probability distribution in \eq{pgp} in terms of a suitable heat kernel for pregeometric variables:
 \begin{equation}
  \mathcal{P}[J(x)] = \exp\left[ -\frac{1}{2} \int_0^\infty ds e^{-(L_0^2/4s)} \int dx \ dy\ J(x)J(y) K_{\rm std} (x,y; s)\,  +\cdots \right]
 \end{equation} 
 This relation, in which $J(x)$ can be thought of as smearing fields, along with \eq{ssig} expresses the geodesic interval --- and thus the qmetric and $g_{ab}$ --- in terms of a deeper layer of description.
 These ideas are under exploration and I hope to discuss them in a future publication.  

\subsubsection{Fluctuations around the equilibrium}

 In the kinetic theory of a normal fluid, the distribution function $f(x^a,p_i)$ --- which counts the microscopic degrees of freedom --- depends not only on $x^a$ but also on the internal variable $p_i$, a fluctuating four-vector of constant norm. This internal variable is a relic of the discrete nature of the fluid,   due to the existence of atoms/molecules of matter. 
 
 Similarly, when we develop the kinetic theory of the mesoscopic spacetime, and count the corresponding mesoscopic  degrees of freedom of geometry $\rho(x^i,n_a)$ (through a simple limiting procedure, associating an  area with each event), we \textit{discover} that it depends not only on $x^i$ but also on a internal variable $n_a$. This internal variable, again,  is  a fluctuating four-vector of constant norm and arises as  a relic of the discrete nature of the spacetime fluid.

This  \textit{discovery} of the mesoscopic spacetime degree of freedom $n_a$, in turn,  allows us to define the equilibrium state for matter and geometry, purely from combinatorics --- viz., by maximizing the total degrees of freedom. 
This extremum condition, is just the  Einstein's equations! 
  The concept of equilibrium also has a direct physical meaning in the classical limit, in which we identify the mean value $\langle n_a\rangle$ of the fluctuating internal variable with a null vector field $\ell_a(x)$. The equilibrium condition then reduces to the equation  $\mathcal{H}_g+\mathcal{H}_m=0$, where $\mathcal{H}_g$ is the dissipational heat density of gravity and $\mathcal{H}_m$ is the corresponding quantity for matter. Equilibrium tantamounts to zero-dissipation.

  There is, however, another possibility which is more exciting and, of course, speculative. This is based on the idea that in standard statistical thermodynamics one can postulate that the probability $P$ for the existence of a particular configuration is related to the entropy $S$ by $P=e^S$. The equilibrium configuration is determined by maximizing $S$; but more importantly, the relation $P=e^S$  allows us to study fluctuations around the equilibrium if we know the form of the entropy functional for different configurations. That is, equilibrium statistical mechanics also contains information about fluctuations around the equilibrium. In our case, the entropy functional for the degrees of freedom is given by
  \begin{equation}
   S_{\rm tot} (x,n) = L_P^4 T_{ab}(x) n^a n^b + \frac{1}{4\Delta (x,n)}
  \end{equation} 
  where $\Delta(x,n)$ is the Van Vleck determinant. (See \eq{vv-dos}.) If you expand $\Delta$ in powers of $L_P^2$, the probability $P=\exp S$ will acquire the form:
  \begin{equation}
P \propto \exp \left[ \frac{1}{4} - \left(\frac{L_P^2}{8\pi } R^a_b (x) - L_P^4T^a_b (x) \right) n_an^b + \cdots \right]
\label{boltz}
\end{equation} 
 The higher order terms denoted by ..... in this equation are important to ensure convergence.  In the  context of using \eq{boltz}, the sign of the first term can be arbitrarily large or small which needs to be regularized by higher order corrections. 
 Ideally, one would like to interpret this as the joint probability for the existence of a given geometry as well as the internal variable $n^a$. However, we only have knowledge about spacetime geometry at very long wavelengths so that it is probably more appropriate to interpret the above expression as the probability for $n^a$ at a given event $x$. This probability is determined by the long wavelength description of background spacetime and the matter stress tensor. When Einstein's equations hold, the leading term vanishes and the sub-leading terms give us the probability distribution for $n^a$, suppressed by Planck scale effects. In this approach, Einstein's equations arise from a requirement that, to  leading order, the internal degree of freedom $n^a$ should not be excited in any spacetime event.\footnote{The probability for a given spacetime geometry can be determined by doing a functional integral over $n^a$. This is a complicated task even in the saddle-point limit.}

 The extremum condition, leading to the vanishing of the argument of the exponent in \eq{boltz} ensures that these degrees of freedom are not excited at the lowest order. This again translates, in the classical limit, to the zero-dissipation-principle.
 So,  we now have a strikingly simple physical meaning for the gravitational field equations. In the standard form, Einstein's equation   $G^a_b=(8\pi L_P^2) T^a_b$, equates apples to oranges, viz., geometry with matter. In our interpretation, we relate the geometrical heat of dissipation, $R^{ab}n_an_b$ (arising from the coupling of internal variable $n_a$ with geometry) to  heat of dissipation of matter, $T^{ab}n_an_b$ (arising from the coupling of internal variable $n_a$ with matter). 
 
 This also suggests a possible route to understanding the question: What is the \textit{actual mechanism} by which the energy momentum tensor $T^a_b$ generates the curvature $R^a_b$ ? In Einstein's theory, this  is just a hypothesis, in the form of the field equation. In our approach, both the  spacetime geometry and matter couples to the variable $n_a$ through the terms $R^a_b n_a n^b$ and $T^a_b n_a n^b$ respectively, thereby leading to an effective coupling between them. The Einstein's equation is then just an average, equilibrium condition and we will expect --- as in any statistical system involving large number of degrees of freedom --- fluctuations around this equilibrium.

\section{Cosmology: The Test-bed for the New Paradigm}\label{sec:altpara}

The conventional approach  to cosmology begins by assuming the validity of GR to describe the evolution of spacetime --- at even very large scales --- and then obtains a specific solution to the field equation to describe the evolution of the cosmos. But, in the paradigm described earlier,  the field equations of gravity themselves  have only the  conceptual status similar to  the equations describing an elastic solid or a fluid  \cite{A19,tp,tpreviews1} in thermodynamic equilibrium. In this alternative perspective, gravity is the thermodynamical limit of the statistical mechanics of the underlying \md\ (the `atoms of space'). The field equations are obtained from a thermodynamic variational principle which is similar to extremizing a thermodynamic potential to obtain the equilibrium state of the normal matter. The validity of the GR field equations then correspond to a maximum entropy configuration of the \md. 
The evolution of spacetime itself is then described in a \textit{purely thermodynamic language} in terms of suitably defined degrees of freedom in the bulk and boundary of a 3-volume; see \eq{Paper06_NewFin02}.
I stress that, even though \eq{Paper06_NewFin02} describes a time evolution, it is obtained from an extremum condition for a thermodynamic variational principle and represents the thermodynamic equilibrium between matter degrees of freedom and \md. 
 
In the specific context of cosmology, one can write a similar, but even simpler (and elegant), equation of the form \cite{tpemeuniv}:
 \begin{equation}
 \frac{dV_H}{dt}=L_P^2(N_{sur}-N_{bulk})
 \label{evl1}
\end{equation}
where $V_H= (4\pi/3)H^{-3}$ is the volume of the Hubble sphere, $N_{\rm sur}= A_H/(L_P^2)=4\pi H^{-2}/L_P^2$ is the number of \md\ on the Hubble sphere, $N_{\rm bulk}=-E/[(1/2)k_BT]$ is the equipartition value for the bulk degrees of freedom corresponding to the Komar energy $E$ contained in the Hubble sphere, and $T=(H/2\pi)$ is the Hubble temperature. (I have assumed $E<0$, which will describe the current accelerated phase of the universe; otherwise one needs to flip  signs to keep $N_{bulk}>0.$) This equation is equivalent to the space-space component of the Einstein equation describing the FRW spacetime. 

One can also rewrite the time-time component of the Einstein equation in thermodynamic language as an energy balance relation:
\begin{equation}
 \rho V_H=TS 
 \label{evl2}
\end{equation}
where $S=A_H/(4L_P^2) = \pi H^{-2}/L_P^2$ is the entropy associated with area of the Hubble sphere making $TS$ in the right hand side the heat energy of the boundary surface. This equation tells us that the total energy within the Hubble sphere is equal to the heat energy of the boundary surface.\footnote{It is again quite gratifying that the entire cosmological evolution can be concisely described in terms of two equations \eq{evl1} and \eq{evl2} with direct thermodynamic interpretation. Notice that all the numerical factors work out appropriately to ensure the description to be simple and elegant.}

However, the field equations of GR, representing some kind of thermodynamic equilibrium between matter and the \md, cannot be universally valid.  
 Recall that, in the case of standard fluid mechanics, we  have to abandon the thermodynamic description in \textit{two} different contexts. First id the well-known situation, which arises when we probe the fluid at scales comparable to the mean free path; we then need to take into account the discreteness of molecules etc., and the fluid description breaks down. Second, somewhat less appreciated case, arises when a fluid simply has not reached local thermodynamic equilibrium at the scales (which can be large compared to the mean free path) we are interested in. In the first case, the fluid description itself breaks down; in the second case, we do have a continuum description of the fluid, but it needs to be explored using non-equilibrium kinetic theory. 

Something very analogous happens in the description of gravity. 
The \md\ could have reached the maximum entropy configuration at sub-cosmic scales, so that the standard field equations of gravity remain valid at these scales (say at scales $10^6 L_P\lesssim x\lesssim H^{-1}$). Equation~(\ref{evl1}) and \eq{evl2}  hold at these scales and standard cosmology retains its validity. For scales close to $L_P$, of course, the discrete nature of spacetime has to be taken into account and this is similar to probing a fluid at scales comparable to the mean free path. But it is also possible that the \md\ have not evolved into the maximum entropy configuration at very large scales comparable to the \textit{horizon} scale (which is much larger than the scale of the \textit{Hubble radius} in the RD and MD phases). At these large scales we again expect \eq{evl1} and \eq{evl2} to be modified because the \md\ are not in the maximum entropy configuration. This is similar to the situation  for normal fluids, when we have to use  non-equilibrium thermodynamics. 

In our approach, the symmetry of Einstein's equations, viz. general covariance, emerges when the \md\ reach the maximum entropy configuration at the intermediate scales. At very large scales, this `equilibrium' has not yet been achieved and 
 the universe, at very large scales,  picks out a cosmic frame of rest viz. the one in which CMBR is isotropic. (Of course,  we also do not know how to introduce the idea of general covariance in a meaningful way close to Planck scales; but that is a different --- and more well-known --- story.) 
 
 A  figurative analogy of this situation is provided by a large chunk of ice containing a point source of heat inside \cite{lwf}. The heat source melts the ice around it, creating a  region containing  water, which 
 expands, maintaining thermodynamic equilibrium. The microscopic degrees of freedom in the form of water have a higher degree of symmetry (rotational invariance), compared to the microscopic degrees of freedom locked up in the ice-lattice. In the case of the universe, the expansion  actually leads to the emergence of space \cite{tpemeuniv} with the \md\ reaching the maximum entropy configuration. This region exhibits a higher degree of symmetry (via., the general covariance of \eq{evl1} and \eq{evl2}, (which  are just components of $G^a_b=\kappa_g T^a_b$), compared to the larger scales of the universe.

What could possibly be the additional ingredient we need to introduce into the standard GR to describe this situation? \textit{It is the concept of information stored in the spacetime and its accessibility  to  different observers \cite{tpreviews1,tp,tpreviews2}.}
A key feature of gravity is indeed its ability to control the amount of information accessible to any given observer.  
The lack of access to spacetime regions leads to a configurational entropy related to the \md. Over decades, we have come to realize \cite{info} that information is a very physical entity and that anything which affects the flow and accessibility of information will have direct physical significance. This will make matter and geometry to be more closely tied together (through the information content) than in the conventional approach.

It turns out that, by using this   idea, we can solve \cite{tphpplb} \textit{the} deepest mystery about our universe, viz., the small numerical value ($\Lambda L_P^2 \approx 10^{-122}$) of the \cc, $\Lambda$. It turns out that the numerical value of the \cc\ is directly related to the 
 amount of information $I$ accessible to an eternal observer in our universe. If $\Lambda =0$, such an observer can access all of spacetime and can acquire an infinite amount of information. But when $\Lambda \neq 0$, the information accessible to the observer turns out to be  finite and is related to the numerical value of $\Lambda$. So, if we have an independent way of fixing $I$, then you can determine the value of  $\Lambda$ in terms of the other observable parameters. 
 
 The value of $I$, in turn,  is fixed by the following fact: As I mentioned before, \textit{the spacetime becomes effectively 2-dimensional close to Planck scales},  irrespective of the dimension exhibited by the spacetime at large scales \cite{paperD,deq2}. This, in turn, implies that the basic unit of information stored in the \md\ is just $A_P/L_P^2= 4\pi$ where $A_P=4\pi L_P^2$ is the area of the 2-sphere with radius $L_P$. This unit of quantum gravitational information  allows us to determine the numerical value of the cosmological constant.

 \subsection{Cosmic information and  the value of the cosmological constant}\label{sec:cicc}
 
 The key issue in formulating the connection between \cc\ and the cosmic information lies in quantifying the amount of spacetime information. This is indeed difficult for a \textit{general} spacetime;  but it is possible to introduce a natural definition of information content  in the context of \textit{cosmological} spacetimes (`CosmIn')  and use it to link the quantum and classical phases of the universe. Moreover,  this information paradigm allows us to determine both, (i) the numerical value of the cosmological constant and (ii) the amplitude of the primordial, scale invariant, power spectrum of  perturbations, thereby providing a holistic description of cosmology.

In any Friedmann model, the \textit{proper} length-scales (e.g., the wavelengths of the modes of a field) scale as $\lambda(a) \propto a$ and can cross the \textit{proper} Hubble radius $H^{-1}(a)=(\dot a/a)^{-1}$ as the universe expands. 
The number of modes  $dN$  located within the comoving Hubble volume $V_H(a) = (4\pi/3) (aH)^{-3}$, which have comoving wave numbers in the range  $d^3k$, is given by $dN = V_H(a) d^3k/(2\pi)^3\equiv V_H(a) dV_k/(2\pi)^3$ where $dV_k=4\pi k^2 dk$. 
A mode with a comoving wave number $k$ will cross the Hubble radius when $k=k(a)\equiv a H (a)$. So, modes with wave numbers between  $k$ and $k+dk$, where $dk= [d(aH)/da]\, da$, will cross the Hubble radius 
during the interval ($a, a+da$). We will \textit{define} \cite{tphpplb,hptp1,hptp2} the information associated with  modes which cross the Hubble radius during any interval $a_1<a<a_2$ by 
\begin{equation}
N(a_2,a_1) = \pm\int_{a_1}^{a_2} \frac{V_H(a)}{(2\pi)^3} \, \frac{dV_k[k(a)]}{da} \, da = \pm\frac{2}{3\pi}\ln \left( \frac{h_1}{h_2}\right)
\label{defN}
\end{equation} 
where $h(a)\equiv H^{-1}(a)/a$ is the \textit{comoving} Hubble radius and $h_1=h(a_1),h_2=h(a_2)$. The sign is chosen so as to keep $N$ positive, by definition. 

If we do not introduce any untested physics from the matter sector (like e.g., inflationary scalar fields, which we will \textit{not} need), the universe would have been radiation dominated at early epochs and, classically,  has a singularity at $a=0$.  The classical description, of course,  breaks down when quantum gravitational effects set in; we will 
assume that the universe makes a transition from a quantum, pre-geometric phase to the classical, geometric phase at some epoch $a=a_{\rm QG}$ when the radiation energy density  is $\rho_R=\rho_{\rm QG}$ where $(8\pi/3)\rho_{\rm QG} \equiv  E_{\rm QG}^4$. We express  this energy scale as $E_{\rm QG}\equiv\nu^{-1} E_{\rm Pl}$ 
where $E_{\rm Pl}\equiv \hbar c/L_{\rm P}=1/L_{\rm P}$  and $L_P\equiv(G\hbar/c^3)^{1/2}=G^{1/2}$ is the Planck length; here $\nu$ is a numerical factor which, as we shall see, can be determined from observations.\footnote{We do \textit{not} assume that $\nu$ is of order unity, thereby allowing the possibility that quantum gravitational effects can have a long tail.  The transition, from pregeometric to classical phase, could be sudden (e.g. like a phase transition) or gradual, and in the latter case $\nu^{-1} E_{\rm Pl}$ is equivalent to the \textit{effective} scale at which the transition can be approximated as a sudden occurrence.} The Hubble radius at the transition epoch $a=a_{\rm QG}$ is $H_{\rm QG}^{-1}\equiv \nu^2 L_P$.  

If the universe was populated by sources with  $(\rho+3p)>0$ for all $a>a_{\rm QG}$, then the function $N(a,a_{\rm QG})$, defined by \eq{defN}, is a monotonically increasing function of $a$ and --- more importantly --- \textit{diverges} as $a\to\infty$. 
It seems reasonable to demand that \textit{$N(a,a_{\rm QG})$ should be finite and its finite value should be determined by purely quantum gravitational (QG) considerations.} 
This will be  a natural consequence of the discreteness of space and the existence of a minimal length --- which is a generic feature of quantum gravity models --- leading to a finite reservoir of information in the QG phase.
This  requires the comoving Hubble radius $H^{-1}(a)$ to reach a maximum value  at some epoch, say, $a=a_\Lambda$ and then decrease. 
So, the number of modes $N(a_\Lambda,a_{\rm QG})$ which \textit{enter} the Hubble radius during the entire history of the universe --- which we call `CosmIn' --- will be a finite constant, say $N(a_\Lambda,a_{\rm QG})\equiv I_c$. This, in turn, requires that $\rho+3p=0$ at $a=a_\Lambda$ with $\rho+3p<0$ for $a>a_\Lambda$. \textit{Interestingly finiteness of CosmIn thus demands that we must have an accelerating phase in the universe.} 

The simplest way to ensure that $(\rho+3p)<0$ at late times, (again without invoking untested physics like e.g., quintessence) is to introduce a non-zero cosmological constant, with an energy density $\rho_\Lambda$.  The expansion of such a universe, in the classical phase $a>a_{\rm QG}$,  is driven by the energy density of matter $\rho_m \propto a^{-3}$, radiation $\rho_R \propto a^{-4}$ and the cosmological constant $\rho_\Lambda$. As described in Sec. \ref{sec:cosmicconstants}, we can  define the density $\rho_{\rm eq} \equiv \rho_m^4(a)/\rho_R^3(a)$ and parametrize the   universe as a dynamical system described by three densities: ($\rho_{\rm QG},\rho_{\rm eq},\rho_\Lambda$). 
The challenge is to understand the value of $\rho_\Lambda$.
Of course, such an attempt, in which the numerical value of the cosmological constant is determined using a physical principle, makes sense 
only if the
 field equation is  invariant under the addition of a constant to the matter Lagrangian. As we stressed in Sec. \ref{sec:necsuff},
the cosmological constant problem can be clearly posed (and solved) \textit{only if} the gravitational field \textit{equations} are made invariant under the addition of a constant to the matter Lagrangian, but their \textit{solutions} permit an inclusion of the cosmological constant.
This is accomplished naturally in the emergent gravity paradigm which is the backdrop in which we proceed further.

The idea is to relate the value of $\rho_\Lambda$ to the value of  $N(a_\Lambda,a_{\rm QG})\equiv I_c$. 
The calculation of $I_c$ is completely straightforward (see Appendix C of \cite{hptp2} for details.) and the result is given by:
\begin{equation}
 I_c = -\frac{2}{3\pi}\ln\left[\frac{k_1 (\rho_\Lambda^2\rho_{\rm eq})^{1/12}}{E_{\rm QG}} \right]
 \label{ic}
\end{equation}
where $k_1=(3^{1/2}/2^{1/3})(8\pi/3)^{1/4}\approx 2.34$.
Inverting this equation, the cosmological constant can be expressed in terms of $I_c, \nu, \rho_{\rm eq}$ as:
\begin{equation}
\rho_\Lambda L_P^4 = \frac{4}{27} \ \left(\frac{3}{8\pi} \right)^{3/2} \frac{1}{\nu^6 (\rho_{\rm eq}L_P^4)^{1/2}} \ \exp \left( - 9 \pi I_c\right)
\label{four}
\end{equation}
As mentioned before, the \textit{non-zero} value of the cosmological constant is related to the \textit{finite} value of $I_c$. The fact that even an eternal observer can only access a finite amount of information (quantified here in terms of the number of modes which cross the Hubble radius) implies that the cosmological constant is non-zero; clearly, $\rho_\Lambda\to0$ when $I_c\to\infty$ and vice-versa. 
So if $I_c$ is known  from an independent consideration, \eq{four} will determine the numerical value of the cosmological constant in terms of $(\rho_{\rm eq}, \rho_{\rm QG})$. 

To have an independent handle on $I_c$, we recall the   result  that \textit{the effective dimension of the quantum-corrected spacetime becomes $D=2$ close to Planck scales,} independent of the original $D$. This result was obtained, in a reasonably model-independent manner  in Ref. \cite{paperD}; similar results have been established earlier by several authors (for a sample, see e.g., \cite{deq2}) in a number of approaches to quantum gravity. This result, in turn, implies that \cite{paperD,tp} the unit of information associated with a quantum gravitational 2-sphere of radius $L_P$ can be taken to be $I_{\rm QG}=4\pi L_P^2/L_P^2=4\pi$. We shall therefore  \textit{postulate} that:
\begin{equation}
I_c= N(a_\Lambda, a_{\rm QG}) = 4\pi 
\label{onlypost}
\end{equation} 
This relation should be viewed as a relic of the pre-geometric phase described by quantum gravitational considerations. While it suggests the notion of a ``single Planckian sphere'' from which the cosmogenesis occurred, it is \textit{not} possible to model such an idea rigorously, given our current ignorance of quantum gravity and cosmogenesis. What we actually  postulate --- \textit{and  is sufficient for our purpose, independent of the details of the model} --- is that the information content, as measured by CosmIn, $N(a_\Lambda, a_{\rm QG})$, is equal to the information contained in the two dimensional surface of a Planckian sphere, viz. $4\pi$. As long as the relevant cosmogenesis model maintains this equality of information, our results will follow.
Using, $I_c= 4\pi$ in \eq{four} we obtain 
\begin{equation}
\rho_\Lambda L_P^4 = \frac{4}{27} \  \left(\frac{3}{8\pi} \right)^{3/2}\frac{1}{\nu^6 (\rho_{\rm eq}L_P^4)^{1/2}} \ \exp \left( - 36\pi^2\right)
\label{fourpi}
\end{equation} 
Given the scale $E_{\rm QG}=\nu^{-1} E_P$, at which the transition to classical geometry  from quantum pre-geometry occurs, the above equation determines $\rho_\Lambda$. 
(We can, of course, reverse the argument and use the observed value of $\rho_\Lambda$ to determine the factor $\nu$. Using  $\rho_\Lambda L_P^4 = (1.14 \pm 0.09) \times 10^{-123}$ and $\rho_{\rm eq}L_P^4 = (2.41 \pm 1.01) \times 10^{-113} $, we find that $\nu = (6.2 \pm 0.3) \times 10^{3} $ making $E_{\rm QG}$ close to the GUTs scale. It suggests that cosmogenesis was completed and the universe acquired a classical geometry only close to GUT scale.)

Fortunately, there is an \textit{independent} way of estimating $\nu$ by calculating the amplitude of primordial  perturbations  in terms of $\nu$, and comparing it with the observations. In our model, the matter fields inherit the primordial, pre-geometric quantum fluctuations at $a=a_{\rm QG}$. 
There are two ways of estimating the resulting amplitude and spectral characteristics of the density fluctuations generated in this process:  One (conservative) procedure is to quantize a field  in the Friedmann universe, by the standard procedure of decomposing it into different Fourier modes,  each  labeled by the comoving wave number $k$.  Any given oscillator starts in its ground state when the  quantum of (proper) energy associated with this mode, $\hbar k/a$, is equal to $E_{\rm QG}$. (This initial condition is different from choosing the Bunch-Davies vacuum for the field, as is often done in inflationary models; see e.g., \cite{wald} for a discussion). The calculation of quantum fluctuations is then completely straightforward. (See, for e.g. \cite{wald,tp-sriram}). The final result is: 
\begin{equation}
 \mathcal{A} = \left[\frac{k^3P(k)}{2\pi^2}\right]^{1/2}  = \frac{c_1}{\nu} \sqrt\frac{4}{3 \pi} \left[\frac{3 w^{1/2} (6 w + 5)}{4 (3w + 5)^2}\right]^{1/2} = \frac{0.19 c_1} {\nu}
 \label{eqn6}
\end{equation} 
for $w=1/3$, where $c_1$ is a numerical factor of order unity whose exact value can be determined by more detailed analysis.\footnote{This can be obtained, for example, from eq.(16) of Ref. \cite{wald}, making note of the fact that $l_p^2$ in Ref. \cite{wald} is $(8\pi/3)L_P^2$ and $l_0=\nu L_P$.} Using the value of $\nu$ obtained from \eq{fourpi}, we find that 
 $\mathcal{A}_{\rm theory} = 3.05 c_1 \times 10^{-5}$  which has to be compared with the observed value $\mathcal{A}_{\rm obs} \approx 4.69  \times 10^{-5}$. We see that the results are remarkably consistent with $c_1=1.54=\mathcal{O}(1)$. 
 
 A more speculative -- but exciting -- possibility is to generate the perturbations directly from the quantum pre-geometric phase \cite{lee}. This is based on the fact that if the pre-geometric phase obeys holographic equipartition \cite{tp}, it can be modeled as a thermal system with energy $E\propto A T_c$ where $T_c\approx E_{\rm QG}=E_{\rm Pl}/\nu$ is the critical temperature at which the quantum to classical transition occurs and $A\propto R^2$ is the area of the boundary. Such a system will have a specific heat  $C\propto A \propto R^2$ leading to energy fluctuations $\sigma_E^2 = CT^2 \propto A\propto R^2$.
This, in turn, will lead to perturbations in the energy density $\delta \rho  = \delta E/V $ such that $\sigma^2_\rho = \sigma^2_E/V^2\propto\sigma^2_E/R^6$. This will give rise to (see Ref. \cite{lee} for details; for similar ideas, see e.g., Ref. \cite{related}) to a scale invariant  spectrum with $\mathcal{A} \approx T_c/E_{\rm Pl}\approx\nu^{-1}$. The observed result for $\mathcal{A}$ is again obtained when  $\nu \approx \mathcal{O}(1) \times 10^4$. In this model, we  have a clear identification of a transition from the pre-geometric phase to geometric phase occurring at the energy scale $\nu^{-1} E_{\rm Pl}$, with consistent results.

\begin{figure}
\begin{center}
\scalebox{0.22}{\input{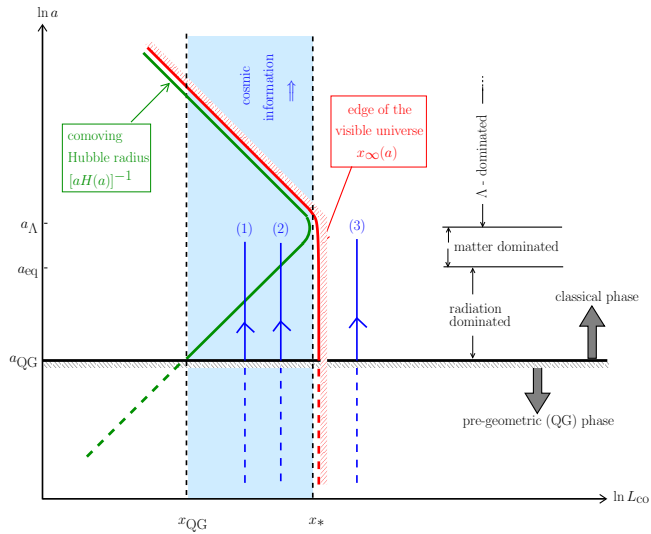}}
\end{center}
 \caption{Various length scales of interest in cosmological evolution. See text for the description.}
\label{fig:lengthscales}
\end{figure}

This discussion can also be presented in terms of  the notion of an \textit{information horizon}, which is  summarized in Figure~\ref{fig:lengthscales}. In the figure,
 the comoving Hubble radius (green line) increases during the radiation dominated ($h\propto a$) and matter dominated ($h\propto a^{1/2}$) phases and decreases ($h\propto a^{-1}$) in the $\rho_\Lambda$ dominated phase. The turn-around occurs when $a=a_\Lambda$. The classical description loses its meaning  at $a=a_{\rm QG}$; this limit is indicated as a horizontal (black) line at $a=a_{\rm QG}$. 
An eternal observer (viz., an observer at the origin who is making observations at arbitrarily late times) will be able to receive signals emitted at the epoch $a$, from a maximum comoving distance 
\begin{equation}
 x_\infty(a) = \int_{t}^{\infty} \frac{dt}{a(t)} = \int_a^\infty \frac{d\bar a}{\bar{a}^2 H(\bar a)}
\label{xinfin}
\end{equation} 
which is also shown in Figure~\ref{fig:lengthscales} by a  red line.  During the phase dominated by the cosmological constant, $x_\infty(a)$ decreases as $1/a$. But at earlier times, 
$x_\infty(a)$
stays very nearly constant (changing only by a factor 3    
when  $a$ changes by nearly a factor $3000$). 
The signals travel a \textit{finite} comoving distance $x_*\equiv x_\infty(0)$ during the \textit{entire cosmic history},\footnote{This quantity  $x_\infty(0)\approx x_\infty(a_{\rm QG})$
can be expressed in terms of an elliptic integral. 
For $(\rho_\Lambda/\rho_{eq})^{1/4}\approx 2.62 \times 10^{-3} $, we find that $a_{eq} H_\Lambda x_* = 9.99 \times 10^{-4}$ which is the maximum comoving distance we can ever probe. We have already probed a fraction 0.74 of this today.} $0<t<\infty$  . The $x_\infty(a)$ will be divergent for \textit{all} $a$ if  $(\rho+3p)>0$ \textit{asymptotically} (i.e., when $a\to\infty$). On the other hand, an accelerated phase for all $a>a_\Lambda$, due to $(\rho+3p)<0$, 
ensures that $x_\infty(a)$ is finite. In particular 
 the maximum comoving distance $x_\infty(a_{\rm QG})$ an eternal observer can probe on the spatial hypersurface $a=a_{\rm QG}$ --- which indicates to the birth of the classical spacetime --- is also finite. 
One can, in fact, rewrite the expression for $I_c$ more suggestively in terms of the proper information horizon $r_*\equiv a_{\rm QG}x_*$ at $a=a_{\rm QG}$ as:
\begin{equation}
 I_c = \frac{2}{3\pi}\ln \left[k_2 \frac{r_*}{H_{\rm QG}^{-1}}\right]
 \label{ic1}
\end{equation}
where  $k_2 = 2^{1/3}/3^{3/2}\approx 0.24$. 
We see from \eq{ic1} that except for a numerical factor $k_2=\mathcal{O}(1)$, the argument of the logarithm occurring in $I_c$ is the ratio $r_*/H_{\rm QG}^{-1}$, relating the finite value of the proper size of the information horizon, $r_*$, to the finiteness of $I_c$. The region of visibility on the $a=a_{\rm QG}$ surface,  $r_*$, is finite but large (compared to $H_{\rm QG}^{-1}$) when  $\exp(3\pi I_c/2)$ is finite but large.

\subsection{Comments on the result}

Finally, let me emphasize the underlying logical structure of the framework, and elaborate on some of the ingredients which have gone into obtaining the results.

We study a universe which 
makes a transition from a quantum, pre-geometric phase to the classical geometric description at $a=a_{\rm QG}$ when the characteristic energy scale is $E_{\rm QG}\equiv \nu^{-1} E_{\rm Pl}$. The matter fields reside in their natural vacuum state at the beginning. 
 All our results will then follow from a single postulate (see \eq{onlypost})
$
N(a_\Lambda, a_{\rm QG}) = 4\pi                                
$ 
where $N(a_\Lambda, a_{\rm QG})$,  is the total  number of modes which enter the Hubble radius from the epoch of cosmogenesis (i.e the time the universe made a transition to classicality), ($a_{\rm QG}$) up to the epoch $a_\Lambda$, until when the modes continue to enter the Hubble radius.  
Given this single postulate, it follows that $N(a_\Lambda, a_{\rm QG})$  as well as $a_\Lambda$ have to be finite. This, in turn, \textit{demands} a turn around in the Hubble radius and leads to a late time acceleration phase.  
Computing $N(a_\Lambda, a_{\rm QG})$ for a universe with radiation, matter and the cosmological constant, and using $N(a_\Lambda, a_{\rm QG}) = 4\pi$, we obtain the result in \eq{fourpi}. Previous work, on the other hand, \cite{wald,lee} leads to \eq{eqn6} of the paper. We can satisfy both \eq{fourpi} and \eq{eqn6} with a single value of $\nu$, which is a nontrivial test of consistency.   These results  also bring to center-stage the notion of spacetime information and its role in gravitational dynamics, already seen in several other contexts \cite{tp}. It also strengthens the viewpoint \cite{tp,lwf}, that the universe should not be treated as a particular solution to the gravitational field equations but instead, be approached as a special dynamical system.

The assumption that cosmogenesis was preceded by a quantum gravitational phase, is a natural one. What is new, important and intriguing is the validity of the postulate in \eq{onlypost}, which arises as a  relic of the pre-geometric phase. This postulate equates --- independent of the details of the model ---  the information content, as measured by CosmIn, $N(a_\Lambda, a_{\rm QG})$, with the information contained in the two dimensional surface of a Planckian sphere, viz. $4\pi$. Given our ignorance of quantum gravity,  this result cannot be `derived' rigorously; but we have motivated it based on the occurrence of dimensional reduction to $D=2$ in a large class of QG models, for which there is considerable evidence. \eq{onlypost} is a postulate, but that is the only  postulate we need.

There is, however, another aspect to this postulate which is on very firm ground. From the observed values of $\rho_\Lambda,\rho_{\rm eq}$ and amplitude of the primordial  spectrum,
we can actually determine the numerical value of $I_c$. We then find that:
\begin{equation}
 I_c = -\frac{2}{3\pi}\ln\left[\frac{k_1 (\rho_\Lambda^2\rho_{\rm eq})^{1/12}}{E_{\rm QG}}\right] = 4\pi [1+\mathcal{O}(10^{-3})]
 \label{eqn7}
\end{equation} 
 directly from cosmological observations. The fact that this peculiar combination of parameters defining $I_c$ has a simple value equal to $4\pi$ (to the accuracy of one part in a thousand!) definitely cries out for an explanation. Such an explanation can be provided  by identifying $I_c$ with the information accessible to the eternal observer and $4\pi$ with the quantum gravitational unit of information.  So the postulate in \eq{onlypost} is certainly true in our universe, even though we do not have a first-principle, quantum gravitational, explanation for it.
Obviously, we need to understand better how the postulate $N(a_\Lambda, a_{\rm QG})=4\pi$ using the definition of $N(a_\Lambda, a_{\rm QG})$, based on counting the modes by $d^3x\ d^3k/(2\pi)^3$,  relates to other notions of information used in quantum gravity. 

Let us next examine the assumptions related to the computation of the amplitude of primordial perturbations. 
It is well-known from standard inflationary calculations that $\mathcal{A} \sim E_{\rm inf}/E_{\rm Pl}$ where $E_{\rm inf}$ is the energy scale of inflation. So, it is  expected that $\nu^{-1} \approx 10^{-4}$ gives the correct amplitude for the perturbations. But the key new feature is that the \textit{same} value of $\nu$ leads to the precise, observed value of the cosmological constant.  
That is, we can determine \textit{two} quantities $\mathcal{A}$ and $\rho_\Lambda L_P^4$ ---  neither of which can be computed from first principles in conventional cosmology --- from a \textit{single} parameter $\nu$ (\textit{and} our postulate in \eq{onlypost}). There is no a priori reason why a specific value for $\nu$ should give the correct, observed values for both $\mathcal{A}$ and $\rho_\Lambda L_P^4$. \textit{This is clearly a strong argument in favour of this scenario.}

Our approach did not invoke inflation in the standard manner with inflaton fields.
Conventional cosmology actually requires the  inflationary paradigm \textit{only} to produce a scale invariant primordial spectrum.\footnote{Incidentally, the scale invariant spectrum goes under the name ``Harrison-Zeldovich spectrum''. Harrison derived this spectrum in an often cited \cite{harrison}--- but rarely read --- paper using \textit{quantum gravitational} considerations decades before inflation was invented. Clearly, observational support for a scale invariant spectrum does not prove the existence of a conventional inflationary phase.}
 As argued in, for e.g., Ref. \cite{wald, tpsesh}, the other ``problems'' which inflation is claimed to ``solve''  are not sufficient motivation for inflation. For example, consider the horizon problem in non-inflationary models
which is  tied to the  existence of the singularity at $t=0$ with the scale factor varying as $t^n$ with $n<1$ near $t=0$, thereby rendering the integral of $dt/a(t)$ finite near $t=0$. Any sensible quantum gravity model will eliminate the singularity (through the existence of a minimum length). Of course,   the usual notion of the horizon distance itself ceases to exist in the pre-geometric, QG phase. But to the extent  an (effective) expansion factor can be introduced to describe this phase, $a_{\rm eff}(t)$ will be   nonsingular for all $t$, eliminating the horizon problem.  The elimination of the singularity will generically solve\cite{models} the horizon problem as well.   
Further, the generation of the primordial spectrum in the models  mentioned above \cite{wald,lee} uses a single parameter to predict the spectrum --- which is  conceptually superior to the plethora of models with various fine-tuned potentials $V(\phi)$ for the inflaton fields.

 \section{Conclusions}
 
 I started by identifying two vital clues --- the thermality of horizons and the existence of the cosmological constant --- about the quantum microstructure of spacetime. These are the two clues which are totally ignored in the conventional approaches to quantum gravity. In the first part of the review I described several features of horizon thermality which leads us to the following conclusions: 
 
\begin{enumerate}
 
  \item The  origin of the Euclidean manifold maps to the horizons $\mathcal{H}^\pm$ in the Lorentzian spacetime. This transformation of a \textit{point} in Euclidean space to a null \textit{surface} in the Lorentzian spacetime is a result of crucial importance. Such a mapping works even when both the Euclidean space and the spacetime are not flat, when we use locally flat/inertial coordinates. 
  
  \item The rotational invariance around the origin in a Euclidean plane translates to invariance under Lorentz boost in spacetime. This  leads to the result that  the vacuum functional of the quantum field theory will induce a thermal density matrix for the description of physics in the right wedge. 
  
  \item The null surfaces, which act as local Rindler horizons, possess a heat density $H_{\rm sur} = Ts$. Through the Boltzmann principle, this allows us to conclude that the spacetime possess microscopic degrees of freedom.
  
  \item This heat density of the null surfaces is given by the surface term in Hilbert action thereby providing the first direct link between horizon thermodynamics and gravitational dynamics.
 
 \end{enumerate}
The null surfaces play a very special role in magnifying the \md\ thereby allowing a thermodynamic interpretation of geometry. This is related to the fact that, null \textit{surfaces} originate --- in analytic continuation --- from an infinitesimally localized \textit{points} in the Euclidean sector which are susceptible to quantum fluctuations.
  
In the second part of the review, I highlighted the lesson we could learn from the existence of a non-zero \cc.

 \begin{enumerate}
 
  \item The so called cosmological constant problem is a problem about the description of dynamics of gravity. In the conventional approach, gravity breaks a key symmetry present in the matter sector which is the root cause of this problem.
  
  \item To solve the cosmological constant problem it is necessary to formulate the dynamics of gravity such that the field equations remain invariant under the shift $T^a_b \to T^a_b + (\text{constant})\, \delta^a_b$; at the same time, these equations must allow the introduction of the \cc\ as an integration constant in the solution.
  
  \item It is not possible to satisfy the above demand if the dynamical equations are derived from a local, generally covariant, Lagrangian by an unrestricted variation of the metric tensor as a dynamical variable. 
  
  \item The simplest alternative is to construct and interpret an extremum principle, leading to the equations $G^a_b \ell_a \ell^b = \kappa_g T^a_b \ell_a \ell^b$, and demanding that this equation holds for all null vectors $\ell_a$ at a given event. this gives Einstein's equations with the \cc\ as an integration constant.
 
 \end{enumerate}

  To implement this program, we needed to achieve three goals: 
  (a) We needed to provide an interpretation for an extremum principle which leads to the equation $G^a_b \ell_a \ell^b = \kappa_g T^a_b \ell_a \ell^b$. In particular, we need to understand the origin of the null vector $\ell_a$ and the reason to demand that this equation holds for all $\ell_a$ at a given event. 
  (b) We needed a physical principle which will determine the numerical value of the observed \cc\ in the universe. 
 The last part of the review described how these aspects merge together nicely to provide an alternate description of spacetime evolution. The main highlights of this approach are as follows:
 
 \begin{itemize}
 
 \renewcommand{\labelitemi}{$\blacktriangleright$}
 
  \item One can attribute an observer/foliation dependent heat density to every event in spacetime by constructing local Rindler horizons. This, through the Boltzmann principle, suggests that there exists microscopic degrees of freedom and the field equations of gravity has the same conceptual status as, say, the equations of fluid dynamics. 
  
  \item It is indeed possible to reformulate gravitational field equations in a thermodynamic language in terms of heating/cooling of spacetime.  This goes well beyond the exercise of deriving Einstein's equations from some thermodynamic arguments but finally ending up with just $G^a_b = \kappa_g T^a_b$. What one must have is a complete reformulation of geometry in terms thermodynamic variables. This is what has been achieved in this approach.
  
  \item  The  thermodynamic interpretation must be foliation/observer dependent. We can, however, achieve ``general covariance without general covariance'' by demanding, say, an equation like $G^a_b \ell_a \ell^b = \kappa_g T^a_b\ell_a \ell^b $ should hold for all null vectors $\ell_a$. The major advantage of this approach is that both sides of this equation render themselves to simple thermodynamic interpretation, which we miss if we insist on equations like $G^a_b = \kappa_g T^a_b$.

  \item In the case of normal matter, the equipartition relation $N=[E/(1/2)k_BT]$ offers a key link between macroscopic, thermodynamic variables in the right hand side and the number of microscopic degrees of freedom of the system in the left hand side. An identical result holds for gravity but with the link connecting bulk degrees of freedom  ($N_{\rm bulk}$) in a volume to surface degrees of freedom ($N_{\rm surface}$) in the boundary of the volume. The evolution of spacetime is sourced by the difference $(N_{\rm bulk} - N_{\rm surface})$ and all static spacetimes obey holographic equipartition $N_{\rm surface} = N_{\rm bulk}$. So you can actually count the atoms of spacetime though you don't know what they are. 
  
  \item At the mesoscopic level, one can introduce a distribution function to count the \md. This requires introduction of a zero point length in the spacetime by postulating a modification of the geodesic interval $\sigma^2$ to a form $S(\sigma^2)$ such that $S(0) $ is non-zero. At a deeper level, this allows us to think of the spacetime interval as a correlator of underlying pregeometric degrees of freedom in an effective theory with $S[\sigma^2(x,y)]=\langle J(x)J(y) \rangle$.
  
  \item This modification associates a finite area (but zero volume) with every spacetime event and also introduces a fluctuating null vector 
  at every event as a relic of Planck scale physics.  The field equations can now be obtained by extremizing the configurational entropy of \md\ plus the degrees of freedom of matter.

  \item The field equation is invariant under the shift  $T^a_b \to T^a_b + (\text{constant})\, \delta^a_b$, thereby solving the \cc\ problem. But they allow the introduction of the \cc\ as an integration constant which needs to be determined by an extra physical principle.
  
  \item The zero point length of spacetime also leads to the following result: The spacetime behaves as a 2-dimensional system close to Planck scales irrespective of its dimension at macroscopic scales. This, in turn, gives
  a basic unit of information ($4\pi$) for a quantum spacetime. 
  
  \item This paradigm suggests a very different perspective on the evolution of the universe, viz., that cosmic evolution cannot be described by a specific solution to the gravitational field equation. This is because the gravitational field equation arises in the thermodynamic equilibrium limit of the underlying \md; such an equilibrium will not hold at very large scales and one needs to study cosmic evolution from a different perspective. 
  
  \item General covariance will hold only when thermodynamic limit for the underlying \md\ hold and it can break down at very large, cosmic, scales. This could  lead to the existence of a rest frame for the universe in which CMBR is isotropic.
  
  \item This approach also suggests that one can relate the numerical value of the \cc\ to the information accessible to an eternal observer in the universe. This relation, along with the fact that the unit of quantum information is $4\pi$, is sufficient to predict the numerical value of the \cc. The result agrees extremely well with observations even though the emergent gravity paradigm was not designed with this goal in mind. 
  
 \end{itemize}

%%%%%%%%%%%%%%%%%%%%  Appendices   %%%%%%%%%%%%%%%%%%%%%%%%%%%%%%%%%%%%
 
   \renewcommand{\theequation}{\thesection\arabic{equation}}  %%%  Equation numbers to appear as A.., B.. etc.
 
   \setcounter{equation}{0}  % reset counter for appendix A
   
\appendix
 
 \section*{Appendix}

\section{Calculational Details}\label{appen:calc}

  \subsection{The  Euclidean continuation}\label{appen:apenA}

  In this Appendix, I will briefly outline the steps involved in obtaining \eq{rt13}, \eq{rt14}, \eq{more1}, \eq{more2}, \eq{rt20} and some related results. (I will  use mostly positive signature so that the analytic continuation of the Lorentzian time coordinate leads to a positive definite metric.)
  
  One can, of course, obtain  \eq{rt13} and  \eq{rt14} by doing the remaining integral in \eq{rt9} (and the analogous one for RR case) but this requires somewhat complicated manipulation of known integrals over Bessel functions. Since I also want to describe how to do the analytic continuation from the Euclidean sector to get all the four wedges (R, F, L, P), I will follow an alternative route. I will start from the \textit{Euclidean} propagator and obtain all the  results we need by careful analytic continuation. 
  
  The Euclidean (inertial) propagator  can be expressed in the polar coordinates (with $x=\rho \cos \theta, \ t_E = \rho \sin\theta$)  in the  form:
  \begin{equation}
 G_{\rm Eu} (\bm{k}_\perp ; \, \rho_1, \rho_2, \theta) = \frac{1}{2\pi^2} \int_{-\infty}^\infty d\nu\, e^{\pi \nu} \, K_{i\nu}(\mu \rho_2) \, K_{i\nu}(\mu\rho_1)\ e^{-\nu|\theta|}
   \label{rt22}
  \end{equation}
  In  presenting this result, I have already Fourier transformed both sides with respect to the transverse coordinate difference  ($\bm{x}^\perp_1-\bm{x}^\perp_2$) thereby introducing the conjugate variable $\bm{k_\perp}$ and defined  $\mu^2 = k_\perp^2 + m^2$. 
  This result is well known in literature and is simple to obtain. When you Fourier transform the transverse coordinates in the Euclidean version of the propagator in \eq{rt2} you just get the reduced (two-dimensional) propagator, viz. $K_0 (\mu\ell)/2\pi$ where 
  $\ell = |\bm{\rho}_1 - \bm{\rho}_2|$.
  One can then use the  standard identity
  \begin{equation}
   \frac{1}{2\pi} K_0(\mu \ell) = \frac{1}{\pi^2}\int_0^\infty d\nu\ K_{i\nu} (\mu \rho_1)\ K_{i\nu} (\mu \rho_2)\, \cosh[\nu(\pi - |\theta|)]
   \label{stdid}
  \end{equation}
  to express it as an integral over the range $0<\nu<\infty$.
  Extending the integration range  to  $(-\infty < \nu < \infty)$ we can obtain \eq{rt22}. 
  
  To proceed from \eq{stdid} (which contains $|\theta_1-
  \theta_2|$) to \eq{rt13} or \eq{rt14} (both of which have $(\theta_1-
  \theta_2)$), one needs to do the analytic continuation of the variables in an unusual way. (This simple procedure is explained in \cite{tppropagator}.) Alternatively, 
 one can get this result from published tables of integrals. You  recall that, when you Fourier transform with respect to transverse coordinates in the \textit{Lorentzian} propagator, you  get the two-dimensional result $G_{Min}=iK_0(\mu\ell)/2\pi$ with 
 $
 \ell^2=\rho_{<}^2+\rho_{>}^2-2\rho_{<}\rho_{>}\cosh(\tau_2-\tau_1)
 $
 where we have ordered the $\rho$-s as $\rho_>\ >\ \rho_<$ for future convenience. (The  $\tau$ ordering is irrelevant in this expression; note that,  interchanging $\tau$ and $\tau'$ corresponds to reversing the sign of $\nu$ which makes no difference because $K_{i\nu}$ is an even function of $\nu$.) Next, 
 look up the integral 6.792 (2) of \cite{gr} which gives, as a special case, the result:
 \begin{align}
	\int_{-\infty}^{\infty}\frac{d\omega}{\pi}e^{-i\omega \tau}K_{i\omega}(a)K_{i\omega}(b)= K_{0}(\sqrt{a^2+b^2+2ab\cosh\tau}); \quad(|\arg[a]|+|\arg[b]|+|\textrm{Im}[\tau]|<\pi)
	\label{spe1}
\end{align}
The left hand side is almost  what we want but in the right hand side, the argument of $K_0$ has a term with $(+\cosh\tau)$ while our $\ell^2$ has $(-\cosh\tau)$. We need to take care of this while also ensuring that $\sigma^2$ comes up as the limit of $\sigma^2+i\epsilon$ in the Lorentzian sector (i.e, Im$(\sigma^2)>0$) with our mostly negative signature.  To this end, we can make
the following identification in \eq{spe1}:
\begin{align}
	a=\mu\rho_{<}e^{i(\pi-\epsilon)};\qquad b=\mu\rho_{>}
\end{align}
with real $\tau$. Then  we have $|\arg[a]|+|\arg[b]|+|\textrm{Im}[\tau]|=\pi-\epsilon<\pi$ taking care of the constraint in \eq{spe1}. 
Further, one can verify that the ordering $\rho_>\ >\ \rho_<$ also ensures that  Im$(\ell^2)>0$ leading to the correct $i\epsilon$ prescription in the Lorentzian sector. (The sign of imaginary part is decided by the sign of $(\rho_{>}\cosh(\tau)-\rho_{<})$ which remains positive due to the ordering of $\rho$-s.)
This leads to our advertised result:
\begin{align}\label{rt24}
	\frac{i}{2\pi^2}\int_{-\infty}^{\infty}d\omega\ e^{-i\omega \tau}K_{i\omega}\left(-\mu\rho_{<}\right)K_{i\omega}(\rho_{>})= \frac{i}{2\pi}K_{0}\left(\mu\ell\right)=G_{Min}
\end{align}

 To obtain the result in \eq{rt14} we need to know how to proceed from the Euclidean sector to the wedge F. This is nontrivial because,  in the usual  analytic continuation ($\theta\to i\tau$) you go from $(\rho \sin\theta, \rho \cos\theta)$ to 
 $(i\rho \sinh\tau, \rho \cosh\tau)$ which \textit{only} covers the right wedge! But one can actually get all the four wedges from the Euclidean sector by using the following (four) sets of analytic continuations \cite{rk-tp}:
 \begin{align}
  &R: \ \rho \to \rho,\ \theta\to i\tau\,; && x=\rho \cosh \tau, \ t=\rho \sinh\tau\label{start}\\
  &F: \ \rho \to i\rho,\ \theta\to i\tau + \frac{\pi}{2}\,; && x=\rho \sinh \tau, \ t=\rho \cosh\tau\\
  &L: \ \rho \to \rho, \ \theta\to i\tau -\pi\,; && x=-\rho \cosh \tau, \ t=-\rho \sinh\tau\\
  &P: \ \rho \to i\rho, \theta =  i\tau - \frac{\pi}{2}\,; && x=-\rho \sinh \tau, \ t=-\rho \cosh\tau\label{end}
 \end{align} 
 Now using   $(\rho,\theta)\to (\rho, i\tau)$ in R and using $(\rho,\theta)\to (i\rho, i\tau+\pi/2)$ in F,
  along with the identity
 \begin{equation}
  K_{i\nu} (iz) = -\frac{i\pi}{2}\ e^{-\pi \nu/2} \ H_{-i\nu}^{(2)} (z) =-\frac{i\pi}{2}\ e^{\pi \nu/2} H_{i\nu}^{(2)} (z)
 \end{equation} 
 one obtains a result similar to \eq{rt24} with a Hankel function replacing one McDonald function. This leads to \eq{rt14}. 
 
 The analytic continuations in \eq{start} to \eq{end} allow us to obtain the propagator for any pair of points located in any two wedges directly --- and  easily --- from the Euclidean propagator. The results are as follows: You will get a $K_{i\nu}K_{i\nu}$ structure in RR, LL, RL and LR. (The notation AB implies that the first event is  in wedge A and second is in wedge B.) In FF, PP, FP and PF the McDonald functions will be  replaced by the Hankel functions. In PR, FL, RF, LP, RP and LF one obtains a product of Hankel and McDonald functions. 
 The interchange of F with P (or R with L) just reverses the sign of $\nu$; so does the interchange of the two events. The similarity  with Minkowski-Bessel modes \cite{gerlach} is obvious. (These results agree with the ones obtained by more complicated procedure, in \cite{boulware}  except for some inadvertent typos in \cite{boulware}). More details of this procedure and results can be found in Ref. \cite{rk-tp}.
 
 You can now obtain \eq{rt20}, working in the Lorentzian sector, by some further  manipulations.
 One starts with \eq{rt24} and converts it to an integral in the range $(0<\nu<\infty )$. Then using the results 
 \begin{equation}
 n_\nu = \frac{e^{-\pi \nu}}{2\, \sinh \pi\nu}\, ; \qquad 1+n_\nu = \frac{e^{\pi \nu}}{2\, \sinh \pi\nu}
   \label{rt25}
  \end{equation}
  you can rewrite the propagator as
  \begin{eqnarray}
  G^{(RR)} &=& \frac{i}{\pi^2} \int_0^\infty d\nu \,  K_{i\nu}(\mu\rho_>)\, K_{i\nu} ( -\mu\rho_<)
  \sinh \pi\nu \left[ e^{-\pi \nu} \, (n_\nu+1)\ e^{-i\nu \tau} + n_\nu\, e^{\pi \nu} \, e^{i\nu \tau} \right]\nonumber\\
  &=& \frac{i}{\pi^2} \int_0^\infty d\nu \,  K_{i\nu}(\mu\rho_>)\, K_{i\nu} ( -\mu\rho_<)
  \sinh \pi\nu \left[ (n_\nu+1) e^{-i\nu(\tau-i\pi)} + n_\nu e^{i\nu(\tau-i\pi)} \right]
   \label{rt26}
  \end{eqnarray}
   The pre-factors (outside the square bracket) will lead to the product of wave functions in \eq{rt20} and the shift $(\tau-i\pi)$ will lead to the reflected coordinate.

 The thermal factor in \eq{rt20}, however, finds a more natural home in the Euclidean sector. Let us see how this comes about --- using again  a set of identities related to Bessel functions ---  when we work in the Euclidean sector. First, we note that the Euclidean propagator $K_0 (\mu\ell)/2\pi$ (obtained after transverse coordinates are removed by a Fourier transform) satisfies a Bessel function addition theorem (see page 351 (8) of \cite{watson})
 given by:
 \begin{equation}
G_E= \frac{1}{2\pi} K_0 (\mu \ell) = \frac{1}{2\pi} \sum_{m=-\infty}^\infty K_m (\mu \rho_{_>})\, I_m (\mu\rho_{_<}) \, \cos m (\theta - \theta')
 \end{equation} 
 The $K_mI_m$ product in the above result can be rewritten using another identity 
 you can look up (see 6.794(10) of \cite{gr}):
 \begin{equation}
 \frac{2}{\pi^2} \int_0^\infty d\omega \ \omega \, \sinh \pi \omega \, \frac{K_{i\omega}(\mu \rho) \, K_{i\omega}(\mu \rho')}{\omega^2+m^2} 
 = K_m(\mu \rho_{_>})\, I_m(\mu \rho_{_<})
 \label{tpeq2}
 \end{equation}
thereby leading to:
 \begin{equation}
G_E = \frac{1}{\pi^3} \sum_{m=-\infty}^\infty  \int_0^\infty d\omega \ \omega \, \sinh \pi \omega \, \frac{K_{i\omega}(\mu \rho) \, K_{i\omega}(\mu \rho')}{\omega^2+m^2}\ \cos m (\theta - \theta')
 \label{tpeq3}
 \end{equation}
The sum occurring in the above expression can again be looked up (see 1.445 (2) of \cite{gr}); it is precisely the thermal factor in \eq{rt20} written in Euclidean sector:
\begin{eqnarray}
\mathcal{T}_\omega(\theta-\theta')&\equiv& \sum_{m=-\infty}^\infty\frac{1}{\pi}\frac{\omega}{\omega^2+m^2}\ \cos m (\theta - \theta')
=\frac{\cosh\omega(\pi-|\theta - \theta'|)}{\sinh\pi\omega}\nonumber\\
&=&(n_\omega+1)e^{-\omega|\theta - \theta'|}+n_\omega e^{\omega|\theta - \theta'|}
\label{tpeq3new}
 \end{eqnarray}
This will lead to the Euclidean version of \eq{rt20}:
\begin{equation}
G_E = \frac{1}{\pi^2} \int_0^\infty d\omega\ (\sinh \pi \omega) \, K_{i\omega} (\mu \rho)\, K_{i\omega}(\mu\rho')\, \mathcal{T}_\omega(\theta-\theta')
\end{equation} 
The thermal factor in the Euclidean sector
  can  also be expressed as a periodic sum in the Euclidean angle; that is, we can easily show that:
 \begin{equation}
 \mathcal{T}_\omega(\theta-\theta') =  \sum_{n=-\infty}^\infty e^{-\omega|\theta -\theta' +2\pi n|}
  \end{equation}
  thereby making the periodicity in the Euclidean, Rindler time obvious. This is yet another hidden thermal feature of the standard inertial propagator!
So we can write the Euclidean, inertial, propagator as a thermal sum:
\begin{equation}
G_E =  \sum_{n=-\infty}^\infty\frac{1}{\pi^2} \int_0^\infty d\omega\ (\sinh \pi \omega) \, K_{i\omega} (\mu \rho)\, K_{i\omega}(\mu\rho')\, e^{-\omega|\theta -\theta' +2\pi n|}
\end{equation} 
This equation has a simple interpretation (which is explored extensively in Ref. \cite{rk-tp}): In the right hand side the $n=0$ term is just the Euclidean propagator for \textit{the Rindler vacuum}. The periodic, infinite, sum `thermalises' it thereby producing the  propagator for the inertial vacuum.

  \subsection{Boost-mode expansion}\label{appen:boost}
 
Some of the details of the boost-mode expansion are given in this Appendix.
The standard expansion in plane wave modes is given by  $\phi(x)$ as $\phi(x) = \mathcal{A}(x) + \mathcal{A}^\dagger (x)$ where 
\begin{equation}
  \mathcal{A} 
  = \int\frac{d^3 k}{(2\pi)^{3/2}} \, a_{\bm{k}}\, \frac{e^{-i\omega_{\bm{k}} t + i \bm{k} \cdot \bm{x}}}{\sqrt{2\omega_{\bm{k}}}}
  \equiv\int\frac{d^3 k}{(2\pi)^{3/2}} \frac{ a_{\bm{k}} f_{\bm{k}}}{ \sqrt{2\omega_{\bm{k}}} }  
  = \int \frac{d^2 k_\perp}{(2\pi)} \frac{dk_x}{\sqrt{2\pi}} \frac{1}{\sqrt{2\omega_{\bm{k}}}}\ a_{\bm{k}} f_{\bm{k}}
  \label{elevena}
\end{equation}
Here $\omega_{\bm{k}}^2\equiv m^2+\mathbf{k}^2$ and, in the last expression,  we have separated the $d^3k$ integration into $dk_x$ and transverse integrations. 
This expansion is parameterized by the three components of the momentum $\bm{k}$. We now rewrite the  momentum vector $\bm k$ as                                                                                                                                                    
\begin{equation}  
\omega_{\bm k} \equiv \mu \cosh \theta; \qquad k_x \equiv \mu \sinh \theta
\label{krapida} 
\end{equation}
 with $\mu^2 = m^2 + \bkp^2$. This allows us to use the parameterizations in terms of the variables $(\bm{k}_\perp, \theta)$ instead of $(\bkp,  k_x)$. 
 The integration measure in \eq{elevena} transforms as 
 \begin{equation}
 \frac{dk_x}{\sqrt{\omega_{\bm{k}}}} \, a_{\bm{k}} = \frac{dk_x}{\omega_{\bm{k}}} \, \left(\sqrt{\omega}_{\bm{k}} \, a_{\bm{k}}\right) \equiv d\theta\ a_{\bkp \theta};\qquad a_{\bkp \theta}\equiv 
\sqrt{\omega}_{\bm{k}} \, a_{\bm{k}}
 \label{twelvea}
\end{equation} 
 where we have used the result $d\theta= {dk_x}/{\omega_{\bm{k}}}$. This allows us to write the last expression in \eq{elevena} as:
  \begin{equation}
  \mathcal{A} = \int \frac{d^2k_\perp \ d\theta}{(2\pi)^{3/2}} \ a_{\bkp \theta}  \frab{f_{\bm{k}}}{\sqrt{2}} \equiv\int \frac{d^2k_\perp \ d\theta}{(2\pi)^{3/2}} \,a_{\bkp \theta} \, f_{\bkp \theta};\qquad  f_{\bkp \theta}\equiv \frab{f_{\bm{k}}}{\sqrt{2}}
 \label{thirteena}
\end{equation} 
The modes $f_{\bkp \theta} (\tau, \rho, \bm x_\perp)$  are the just the original mode functions $f_{\bm k}(t, x, \bm x_\perp)$
expressed in terms of the variables $(\bkp, \theta)$  and $(\tau, \rho)$ using \eq{krapida} and the coordinate transformations in R
given by \eq{paramet11}. (For the moment we will concentrate on the R wedge; we will comment about the other wedges towards the end.)
The combination $(k_x x - \omega_k t)$ occurring in the original modes  $f_{\bm k}$ transforms to $\mu \rho \sinh (\theta - \tau)$
so that
\begin{equation}
f_{\bkp \theta} (\tau, \rho, \bm x_\perp)=\frac{1}{\sqrt{2}}e^{-i\omega_{\bm{k}} t + i \bm{k} \cdot \bm{x}}
=\frac{1}{\sqrt{2}} \,e^{i\bkp \cdot \bm{x}_\perp}\ e^{i\mu \rho \sinh (\theta - \tau)}
\label{tp2371a}
\end{equation} 
We will next  express the modes $f_{\bkp \theta}$ as a Fourier transform in the variable $\theta$ through the equation 
\begin{equation}
  f_{\bkp \theta} \equiv \frac{1}{\sqrt{2}} \,e^{i\bkp \cdot \bm{x}_\perp}\ e^{i\mu \rho \sinh (\theta - \tau)} \equiv \int_{-\infty}^\infty \frac{d\omega}{\sqrt{2\pi}} \, e^{+i\omega \theta} f_{\bkp \omega}
\end{equation} 
 These modes $f_{\bkp \omega}$ are given by the inverse Fourier transform:
 \begin{align}
  f_{\bkp \omega} &=  \int_{-\infty}^\infty \frac{d\theta}{\sqrt{2\pi}} \,  f_{\bkp \theta}  \ e^{-i\omega \theta} = \frac{e^{i\bkp \cdot \bm{x}_\perp}}{\sqrt{2\pi}} \, \frac{1}{\sqrt{2}} \int_{-\infty}^\infty  d\theta\ e^{-i\omega \theta + i \mu \rho \sinh (\theta - \tau)}\nonumber\\
  &  = \frac{1}{\sqrt{\pi}} K_{i\omega}(\mu\rho) \, e^{\pi\omega/2} \, e^{-i\omega \tau + i \bkp \cdot \bm{x}_\perp}
  \label{MB1a}
\end{align} 
 where $K_{i\omega}(z)$ is the modified Bessel function 
 with the property, $K_{i\omega}(x) = K_{-i\omega}(x)$. We see from \eq{MB1a} that the modes $f_{\bkp \omega}$ have a simple behaviour with respect to translation in $\tau$ coordinate; viz. they get multiplied by a pure phase. In terms of these modes we can write the mode expansion of the scalar field as $\phi=\mathcal{A}+\mathcal{A}^\dagger$ with:
 \begin{align}
  \mathcal{A}(\tau,\rho,\bp{x}) \equiv  \int_{-\infty}^\infty \frac{d^2 k_\perp d\theta}{(2\pi)^{3/2}}a_{\bkp \theta} \,f_{\bkp \theta} 
 & =  \int_{-\infty}^\infty \frac{d^2 k_\perp d\theta}{(2\pi)^{3/2}}  \int_{-\infty}^\infty \frac{d\omega}{\sqrt{2\pi}}\, a_{\bkp \theta} \,f_{\bkp \omega} \ e^{i\omega \theta} \nonumber\\
  &\equiv  \int_{-\infty}^\infty \frac{d^2 k_\perp d\omega}{(2\pi)^{3/2}} a_{\bkp \omega} \,f_{\bkp \omega} 
 \label{threea}
\end{align} 
 where all the integrals range over the entire real line and we have defined a new set of annihilation operators by a simple Fourier transform:
 \begin{equation}
  a_{\bkp \omega} \equiv  \int_{-\infty}^\infty \frac{d\theta}{\sqrt{2\pi}} \, e^{i\omega \theta}\ a_{\bkp \theta}
 \end{equation}
 As it stands, \eq{threea} involves integration over $\omega$ in the range $(-\infty < \omega < +\infty)$. We would next like to re-express this result with the integration range limited to positive frequencies $(0<\omega<\infty)$ with respect to the $\tau$ coordinate. This can be easily done using the fact that $K_{i\omega}(x) = K_{-i\omega}(x)$ and
 leads to the expression 
 \begin{align}
 \mathcal{A}(\tau,\rho,\bp{x}) &=  \int_{-\infty}^\infty \frac{d^2 k_\perp d\omega}{(2\pi)^{3/2}}\, \frac{1}{\sqrt{\pi}} \, K_{i\omega}(\mu\rho)\, e^{\pi\omega/2}\, a_{\bkp \omega}\, e^{-i\omega \tau + i \bkp \cdot \bm{x}_\perp}\nonumber\\
&= \int_{-\infty}^\infty \frac{d^2 k_\perp}{\sqrt{2}(2\pi^2)} \int_0^\infty d\omega K_{i\omega}(\mu\rho) \left[ a_{\bkp \omega} e^{\pi\omega/2}  e^{-i\omega \tau} + a_{\bkp - \omega}\ e^{i\omega \tau} \, e^{-\pi\omega/2}\right] e^{i\bkp \cdot\bm{x}_\perp}
 \label{fivea}
\end{align} 
 This expression can be re-written in a nicer form  by re-labeling $\bkp$ as $-\bkp$ in the second term. This gives
 \begin{align}
  \mathcal{A}(\tau,\rho,\bp{x}) =  \int \frac{d^2 k_\perp}{(2\pi^2)}\int_0^\infty \frac{d\omega}{\sqrt{2}} \, K_{i\omega}(\mu\rho)
    \left\{  a_{\bkp \omega} e^{\pi\omega/2}  e^{-i\omega \tau + i\bkp \cdot\bm{x}_\perp} + a_{-\bkp-\omega}\ e^{-\pi\omega/2}\, e^{i\omega \tau - i\bkp \cdot\bm{x}_\perp} \right\}
 \label{sixa}
\end{align} 
 Adding up $\mathcal{A}$ and its Hermitian conjugate, we find that the scalar field $\phi(x)$ has the expansion 
 \begin{align}
 \phi(x) &= \mathcal{A} +  \mathcal{A}^\dagger  \nonumber\\
    &=\int \frac{d^2 k_\perp}{(2\pi^2)}\int_0^\infty \frac{d\omega}{\sqrt{2}} \, K_{i\omega}(\mu\rho) \left\{e^{-i\omega \tau + i \bkp \cdot \bm{x}_\perp} \left[  a_{\bkp \omega} e^{\pi\omega/2} +  a^\dagger_{-\bkp-\omega}\ e^{-\pi\omega/2}\right] + \text{h.c.}\right\}
 \label{sevena}
\end{align}

 \subsection{Wightman function and CFT}\label{appen:wightman}
 
 The Wightman function for a massless scalar field in $D = d+1$ dimensions, in $(U,V,\bp{x})$ coordinates, in flat spacetime is given by 
 \begin{equation}
  G_M(x_1,x_2) = \frac{A_d}{(\bp{x}^2 - UV)^{\frac{1}{2} (d-1)}} \equiv \frac{A_d}{(\sigma^2)^{\frac{1}{2} (d-1)}}; \qquad A_d = \frac{\Gamma\left(\frac{1}{2} (d-1)\right)}{4\pi^{\frac{1}{2}(d+1)}}
  \label{gmx1}
 \end{equation} 
 where $U,V, \bp{x}$ stand for  the coordinate differences like $U \equiv U_2-U_1$ etc. 
 This, of course, depends on the dimension $d$. However, the second derivative of the Wightman function $Q_M (x_1,x_2) \equiv \partial_{V_1} \partial_{V_2} G_M(x_1,x_2)$ has a universal form, independent of the dimension when evaluated on the null plane $U \to 0$, keeping $UV = -\epsilon$. It is given by 
 \begin{equation}
  Q_M^{U\to0} (x_1,x_2) = -\frac{1}{4\pi}\frac{\delta(\bp{x})}{V^2}
  \label{qmx1}
 \end{equation} 
 This result arises because of the conformal dimension of the fields $\partial_V(\phi)$ in CFT, but, of course, can be derived directly along the following lines. 
 From the expression for $G_M(x_1,x_2)$ in \eq{gmx1}, we find that
 \begin{equation}
  Q_M(U,V,\bp{x}) = - A_d \frac{(d^2-1)}{4} \frac{U^2}{(\sigma^2)^{\frac{1}{2}(d+3)}}
  \label{tpqm}
 \end{equation} 
 We consider the limit of this expression when $U\to 0$ keeping $UV$ infinitesimally negative. Obviously the expression 
$ {U^2}/{(\sigma^2)^{\frac{1}{2}(d+3)}}$
 vanishes unless $\bp{x}^2=0$, that is, except in the coincidence limit of all the transverse coordinates. We can therefore write 
 this expression, in the $U\to0$ limit,   as ${U^2}/{(\sigma^2)^{\frac{1}{2}(d+3)}}=f(V)\delta(\bp{x})$. To determine the function $f(V)$, we integrate both sides over 
 all $\bp{x}$.
 This leads to
 \begin{equation}
 f(V)= \int d^{d-1} x_\perp\, \frac{U^2}{(\sigma^2)^{\frac{1}{2}(d+3)}}= \int d^{d-1} x_\perp\, \frac{U^2}{(x_\perp^2 - UV)^{\frac{1}{2}(d+3)}}
 =\frac{1}{V^2} \int d^{d-1} q_\perp \, \frac{1}{(q_\perp^2 + 1 )^{\frac{1}{2}(d+3)}}
 \end{equation} 
 where $q^2 \equiv \bp{x}^2/|UV|$. The integral has the value
 \begin{equation}
\Omega_d\equiv \int d^{d-1} q_\perp \, \frac{1}{(q_\perp^2 + 1 )^{\frac{1}{2}(d+3)}}= \frac{\pi^{\frac{1}{2}(d-1)}}{\Gamma\left( \frac{1}{2}(d+3)\right) } 
 \end{equation} 
 so that $Q_M$ in \eq{tpqm}, in the limit of $U\to0$, reduces to the form
 \begin{equation}
 Q_M^{U\to0} = - \frac{1}{4}  (d^2-1)\ \frac{A_d \Omega_d}{V^2} \ \delta (\bp{x})
  \label{fourfive}
 \end{equation} 
 While both $\Omega_d$ and $A_d$ depends on the dimension, the combination which occurs in the limiting form of $Q_M$, viz.,
 \begin{equation}
-\frac{1}{4} (d^2-1) \frac{\Gamma\left( \frac{1}{2}(d-1)\right)}{4\pi^{\frac{1}{2}(d+1)}} \frac{\pi^{\frac{1}{2}(d-1)}}{\Gamma\left( \frac{1}{2}(d+3)\right)}
= - \frac{1}{16} \frac{(d^2-1)}{\pi}\, \frac{1}{\frac{1}{2}(d+1)}\, \frac{1}{\frac{1}{2}(d-1)} = - \frac{1}{4\pi}
 \end{equation} 
 does not. This shows that, in the null limit, $Q_M$ is given by the expression in \eq{qmx1}. This is the result used in the text.
 
 \subsection{Exponential redshift  and the Planck spectrum}\label{appen:planck}
 
 A monochromatic, complex, plane  wave traveling along positive $x$-direction is described by a function $\exp (-i\Omega u)$.  A similar wave undergoing exponential redshift of the frequency can be described by the function $\exp i\theta (u) $ with $\theta(u) \equiv (\Omega/\kappa) e^{-\kappa u}$; in that case, the instantaneous frequency will be $\Omega_{\rm ins} \equiv  -\partial \theta/\partial u = \Omega e^{-\kappa u}$ which  clearly signals  exponential redshift. The power spectrum 
 $|F(\Omega,\nu)|^2$ of this signal $\exp i\theta (u)$
 can be obtained by evaluating its Fourier transform $F(\Omega,\nu)$ with respect to $u$ and computing its squared modulus. The Fourier transform is given by
 \begin{equation}
  F(\Omega,\nu) \equiv \int_{-\infty}^{+\infty} \frac{du}{2\pi} \exp\left( i\nu u + i \frac{\Omega}{\kappa } e^{-\kappa u}\right)
= \frac{1}{2\pi \kappa } \int_0^{+\infty} dx \, x^{-(i\nu/\kappa ) - 1} \, e^{(i\Omega/\kappa )x}  
  \label{eqns1}
 \end{equation} 
 In fact it is essentially this integral which appears in several computations of horizon thermality. To evaluate it, we start with the standard integral
 \begin{equation}
  \int_0^{+\infty} x^{s-1} e^{-bx} dx  = \exp(-s\, \ln b ) \, \Gamma(s)
 \end{equation} 
 and identify $\ln b$ as the limit obtained through 
 \begin{equation}
 \ln b = \lim_{\epsilon\to +0} \ln \left(- \frac{i\Omega}{\kappa }+ \epsilon\right)  = \ln \left|\frac{\Omega}{\kappa }\right| - i \frac{\pi}{2} \, \mathrm{sign}\left(\frac{\Omega}{\kappa }\right)
 \end{equation} 
 This immediately leads to the result
 \begin{equation}
 F(\Omega,\nu) = \frac{1}{2\pi \kappa }  \exp\left[ \frac{i\nu}{\kappa } \ln\left|\frac{\Omega}{\kappa }\right| + 
 \frac{\pi\nu}{2\kappa } \, \mathrm{{sign}}\left(\frac{\Omega}{\kappa }\right)\right] \Gamma \left( - \frac{i\nu}{\kappa }\right)
 \end{equation} 
 We will assume here that $\Omega>0$ and $\nu>0$.
  The power spectrum $|F(\Omega,-\nu)|^2$ at negative frequencies (usually corresponding to the Bogoliubov coefficient $\beta_{\Omega\nu}$) can be computed   using the identity:
 \begin{equation}
 |\Gamma(ix)|^2=\frac{\pi}{x\sinh(\pi x)}
\end{equation} 
 This leads to the thermal spectrum per unit logarithmic range of frequencies:
 \begin{equation}
 \nu |F(\Omega,-\nu)|^2=\frac{1}{2\pi\kappa}\frac{1}{e^{2\pi\nu/\kappa}-1}
 \end{equation}
 corresponding to the temperature $T=\kappa/2\pi$.

 \subsection{The response of detectors in different spatial dimensions}\label{appen:detectorresp}
 
 The standard lore on computing detector response proceeds as follows. One starts with a system having the Hamiltonian
 $
{H} = {H}_{\rm field} (\Phi) + H_0 + H_{\rm int}
 $
 where $H_{\rm field}(\Phi)$ is the Hamiltonian of a free scalar field, $H_0$ is the internal Hamiltonian of the detector with discrete energy levels. We will concentrate on the ground state ($\ket{E_0}$) and the first excited state ($\ket{E_1}$) with the energy difference $E \equiv E_1-E_0>0$. The interaction part of the Hamiltonian
 $
  H_{\rm int} = \mu(\tau) \, \Phi[x(\tau)]
 $
 represents the coupling between the field at the location of the detector $x(\tau)$ and the monopole moment of the detector $\mu(\tau)$. Here $\tau$ is the proper time measured along the trajectory of the detector which will correspond to the rapidity parameter if the detector is on the integral curve of the boost Killing vector field. We will also assume that the monopole moment $\mu(\tau)$ evolves as 
 $
 \mu(\tau)=e^{i\tau H_0}\mu(0)e^{-i\tau H_0}
 $.
 
 We consider the transitions induced between the initial state 
 $
\ket{i} = \ket{0} \otimes \ket{E_0}   
 $
 and the final state 
 $
\ket{f} = \ket{\psi} \otimes \ket{E_1}  
 $
  under the action of $H_{\rm ind}$ where $\ket{0}$ is the initial, inertial, vacuum state of the field and $\ket{\psi}$ is the final state of the field which we are not particularly concerned with. Straightforward application of perturbation theory will lead to the result that the rate of transitions for any stationary trajectory (that is, for  motion along the integral curve of  any time-like Killing vector field) is given by $|\mu|^2 \mathcal{R}(E)$ where 
  \begin{equation}
  \mathcal{R}(E) 
 =   \int_{-\infty}^\infty ds \ e^{-iEs} \ G^+[s];\qquad s\equiv \tau -\tau'
  \end{equation} 
  Here $G^+(x,x')$ is the Wightman function, which, for (1+3) dimensions is given by: 
  \begin{equation}
  G^+(x, x')
 = - \frac{1}{4\pi^2}\ \frac{1}{[(t-t'-i\epsilon)^2 - |\bm x - \bm x'|^2]}
   \label{eqns20}
  \end{equation} 
 evaluated along the trajectory $t(\tau), \bm{x}(\tau)$. The Killing nature of the trajectory ensures that $G^+$ is only a function of the proper time difference $s=\tau - \tau'$ between the two events. It is trivial to show that along the inertial trajectory we will get $\mathcal{R}(E) \propto \theta(-E)$ which vanishes for $E>0$. So, quite gratifyingly, the detector on an inertial trajectory does not get excited in the inertial vacuum. 
 
 Let us next consider the response along the integral curves of the boost parameter which we will now write as
 $
 \kappa x=\cosh \kappa\tau$, $\kappa t=\sinh \kappa\tau
 $
 where we have introduced  $\kappa$ by rescaling the coordinates. The $\sigma^2$ in the denominator of \eq{eqns20} now becomes $\sigma^2(s) = (2/\kappa)^2 \sinh^2 [(\kappa/2)(s-i\epsilon)]$.   Therefore the rate of transition is given by the integral 
 \begin{equation}
 \mathcal{R}(E) =  -\frac{1}{4\pi^2}\int_{-\infty}^\infty ds\ e^{-iEs}  \, \frac{(\kappa/2)^2}{\sinh^2 \left[ \frac{\kappa}{2} (s - i \epsilon)\right]}
 \label{4dint}
 \end{equation} 
 The integral can be evaluated by standard application of residue theorem. We will display it in gory detail for future reference:
 \begin{align}
 &  \int_{-\infty}^\infty dx \  \frac{e^{-2i  \Omega x}}{\sinh^{2} (x-i\epsilon)}
   = -\frac{2\pi i}{\Gamma(2)} \sum_{n=1}^\infty\lim_{x\to - i\pi n} \, \frac{d}{dx} \, \frac{(x+ i\pi n)^{2} e^{-2i \Omega x}}{\sinh^{2} x}\nonumber\\
  &  = - \frac{2\pi i}{\Gamma(2)} \sum_{n=1}^\infty e^{-2\pi n \Omega }\lim_{z\to 0}\frac{d}{dz} \, \frac{z^{2} e^{-2i  \Omega z}}{\sinh^{2} (z-i\pi n)}
   = -  \frac{2\pi i}{\Gamma(2)} 
 \left[ \lim_{z\to 0}\frac{d}{dz} \frac{z^{2} e^{-2i  \Omega z}}{\sinh^{2}z}\right]
  \sum_{n=1}^\infty
  e^{-2\pi n   \Omega}
  \label{eqns29}
\end{align} 
 Both factors can be easily computed thereby leading to the result 
 \begin{equation}
  \mathcal{R}(E)=\frac{E}{2\pi}\ \sum_{n=1}^\infty e^{-\frac{2\pi}{\kappa} En}=\frac{1}{2\pi} \frac{E}{e^{\frac{2\pi E}{\kappa}} - 1}
 \end{equation} 
 This is thermal with temperature $T=\kappa/2\pi$ leading to the illusion that our detector is ``seeing'' the Rindler particles in the inertial vacuum determined by standard QFT and Bogoliubov coefficients.
 
 To dispel this illusion, let us next perform the corresponding computation in $1+D$ dimensions. The Wightman function in \eq{eqns20} now gets replaced by
 \begin{equation}
   G^+ (x,x') = \frac{\Gamma((D-1)/2)}{4e^{i\pi(D-1)/2}\, \pi^{(D+1)/2}} 
 \left[ (t- t' - i \epsilon)^2 - |\bm{x} - \bm{x}'|^2\right]^{-(D-1)/2}
 \end{equation} 
 This means we now have to evaluate the integral in 
 \begin{equation}
  F(\Omega)= \frac{e^{i\pi(1-D)/2}\Gamma((D-1)/2)}{2^{D} \pi^{(D+1)/2}}
 \ \int_{-\infty}^\infty dx \ \frac{e^{-2i  \Omega x}}{\sinh^{D-1} (x-i\epsilon)}
 \end{equation} 
 which should be compared with \eq{4dint}.
 This integral can again be computed as before and here are the details (which should be compared with \eq{eqns29}):
  \begin{align}
 & \int_{-\infty}^\infty dx \  \frac{e^{-2i  \Omega x}}{\sinh^{D-1} (x-i\epsilon)}
  = -\frac{2\pi i}{\Gamma(D-1)} \sum_{n=1}^\infty\lim_{x\to - i\pi n} \, \frac{d^{D-2}}{dx^{D-2}} \, \frac{(x+ i\pi n)^{D-1} e^{-2i \Omega x}}{\sinh^{D-1} x}\nonumber\\
  & = - \frac{2\pi i}{\Gamma(D-1)} \sum_{n=1}^\infty e^{-2\pi n \Omega }\lim_{z\to 0}\frac{d^{D-2}}{dz^{D-2}} \, \frac{z^{D-1} e^{-2i  \Omega z}}{\sinh^{D-1} (z-i\pi n)}\nonumber\\
  &  = -  \frac{2\pi i}{\Gamma(D-1)} 
 \left[ \lim_{z\to 0}\frac{d^{D-2}}{dz^{D-2}} \frac{z^{D-1} e^{-2i  \Omega z}}{\sinh^{D-1}z}\right]
  \sum_{n=1}^\infty
 (-1)^{n(D-1)}
  e^{-2\pi n   \Omega}
  \label{eqns34}
\end{align} 
 Comparing \eq{eqns34} with \eq{eqns29}, we see two differences. The pre-factor in front of the sum has become more complicated but this is something you can live with. What is crucial is the appearance of the factor $(-1)^{n(D-1)}$ within the sum which, of course, will be absent whenever $D-1$ is an even number like in (1+3) and (1+1) spacetime dimensions --- which are the most popular ``examples''. Using 
 \begin{equation}
 \sum_{n=1}^\infty(-1)^{n(D-1)}\, e^{-2\pi n   \Omega} = \frac{(-1)^{D-1}}{e^{2\pi   \Omega} + (-1)^D}
 \end{equation} 
 and defining 
 \begin{equation}
  Q_D(\Omega) \equiv -e^{i\pi D/2}\, \lim_{z\to 0}\frac{d^{D-2}}{dz^{D-2}} \,\frac{z^{D-1} e^{-2i\Omega z}}{\sinh^{D-1}z}
 \end{equation} 
 we now get the detector response to be 
 \begin{equation}
  F(\Omega) = \frac{2^{3-2D}\, \pi^{1-D/2}}{\Gamma(D/2)} \ \frac{Q_D( \Omega)}{e^{2\pi  \Omega} + (-1)^D}
  \label{eqns38}
 \end{equation} 
 The function $Q_D$ has the form like, for e.g., 
 \begin{equation}
 Q_3(\Omega) = 2\Omega\, ,  Q_4(\Omega) = 1+ 4\Omega^2\, ,  Q_5(\Omega) = 8\Omega\, (1+\Omega^2)
 \end{equation} 
 and can possibly be attributed to some kind of density of states. However, the factor $(-1)^D$ in the denominator in \eq{eqns38} shows a ``reversal of statistics''; a bosonic field will exhibit a thermal spectrum corresponding to fermionic degrees of freedom when $D$ is even. This is one of the reasons why one cannot blindly assume that the detectors detect excitations of the underlying quantum field. 
 
 \subsection{Computation of the vacuum contribution to $\Lambda$}\label{appen:vaclambda}
 
 Consider a massive scalar field satisfying the standard field equation
 \begin{equation}
 -\ddot{\Phi}+\delta ^{\alpha\beta}\partial _\alpha\partial_\beta\Phi-m^2\Phi=0.
 \end{equation} 
 The canonical quantization of this field is achieved by expanding it in terms of the creation and annihilation operators in the form
 \begin{equation}
\Phi\left(t,{\bm x}\right)=\frac{1}{\left(2\pi\right)^{3/2}}
\int \frac{{\rm d}^3{\bm k}}{\sqrt{2\omega(k)}}\biggl(a_{\bm k}
{\rm e}^{-i\omega t +i{\bm k}\cdot {\bm x}}
+
a_{\bm k}^{\dagger}{\rm e}^{i\omega t -i{\bm k}\cdot {\bm x}}
\biggr)
 \end{equation} 
 where 
 $
 \omega (\bm{k})\equiv \sqrt{\bm{k}^2+m^2}
 $
 and 
 $
 [a_{\bm k},a^{\dagger}_{\bm k'}]=\delta ^{(3)}\left(
{\bm k}-{\bm k'}\right)
 $. The energy density of the scalar field is given by 
 the expectation value $\bk{0}{T_{00}}{0}$. Similarly, the contribution to the pressure is determined by the VEV of $(1/3) (\perp^{\mu\nu} T_{\mu\nu})$. An elementary calculation gives \cite{cc1} 
\begin{equation}
 \bk{0}{T_{00}}{0} = \bk{0}{\frac12 \dot{\Phi}^2+\frac12 
\delta ^{\alpha\beta}\partial _\alpha\Phi\partial _\beta\Phi
+\frac12 m^2\Phi^2}{0}
 \end{equation}  
 and 
 \begin{equation}
\langle p\rangle = \left \langle 0\left\vert 
\frac13 \perp ^{\mu \nu}T_{\mu \nu} \right \vert 0\right \rangle
= \left \langle 0\left\vert 
\frac12 \dot{\Phi}^2-\frac16 \delta ^{\alpha\beta}\partial _\alpha\Phi\partial _\beta\Phi
-\frac12 m^2\Phi^2\right \vert 0\right \rangle
 \end{equation} 
 Both these can be computed using the results 
\begin{align}
\langle 0\vert \dot{\Phi}^2\vert 0\rangle
&=\frac{1}{(2\pi)^3}\int 
\frac{{\rm d}^3{\bm k}}{2\omega(k)}\, \omega ^2(k),
\\
\langle 0\vert \delta ^{\alpha\beta}\partial_\alpha\Phi\partial _\beta\Phi \vert 0\rangle
&=\frac{1}{(2\pi)^3}\int \frac{{\rm d}^3{\bm k}}{2\omega(k)}\, {\bm k}^2,
\\
\langle 0\vert \Phi^2\vert 0\rangle
&=\frac{1}{(2\pi)^3}\int \frac{{\rm d}^3{\bm k}}{2\omega(k)}.
\end{align}
 We then get 
 the results quoted in the main text
 \begin{equation}
\langle \rho\rangle =\langle 0\vert  T_{0 0}\vert 0\rangle
= \frac{1}{\left(2\pi\right)^3}
\frac12 \int {\rm d}^3{\bm k}\, \omega (k); \qquad \langle p\rangle = \frac{1}{\left(2\pi\right)^3}
\frac16 \int {\rm d}^3{\bm k}\, \frac{k^2}{\omega (k)}
  \label{newfornine}
 \end{equation} 
 Both these integrals, of course, are divergent but formally the equation of state is $p=(1/3)\rho$ which represents radiation rather than cosmological constant. This can be made more precise by explicitly evaluating the integrals using a cut-off. We get
 \begin{align}
 \langle \rho \rangle &= \frac{1}{4\pi ^2}\int _0^{M} {\rm d}k 
k^2 \sqrt{k^2+m^2}\label{54}\\
&= \frac{M^4}{16 \pi^2}\Biggl[\sqrt{1+\frac{m^2}{M^2}}
\left(1+\frac12\frac{m^2}{M^2}\right)
 -\frac12\frac{m^4}{M^4}\ln \left(
\frac{M}{m}+
\frac{M}{m}\sqrt{1+\frac{m^2}{M^2}}\right)
\Biggr]\label{55}
\\
&= \frac{M^4}{16 \pi^2}\left(1+\frac{m^2}{M^2}
+\cdots \right)
 \end{align} 
 and 
 \begin{align}
 \langle p \rangle &= \frac13 \frac{1}{4\pi ^2}\int _0^{M} {\rm d}k 
\frac{k^4}{\sqrt{k^2+m^2}}\label{57}\\
&= \frac13\frac{M^4}{16 \pi^2}
\Biggl[\sqrt{1+\frac{m^2}{M^2}}
\left(1-\frac32\frac{m^2}{M^2}\right)
+\frac32\frac{m^4}{M^4}\ln \left(
\frac{M}{m}+
\frac{M}{m}\sqrt{1+\frac{m^2}{M^2}}\right)
\Biggr]\label{58}\\
&= \frac13 \frac{M^4}{16 \pi^2}\left(1-\frac{m^2}{M^2}
+\cdots \right).
 \end{align} 
 The leading order terms do give $p= (1/3)\rho$.
 
 As mentioned in the main text Lorentz invariance requires that $\bk{0}{T^i_k}{0}$ should be proportional to $\delta^i_k$. The results above do not confirm to this requirement, because introducing a cut-off in the 3-momentum breaks Lorentz invariance.  To do this properly, we can use any Lorentz invariant regularization procedure, like for e.g.,
 dimensional regularization. The basic idea behind dimensional regularization is to compute the relevant quantities in 
 $d\equiv 4-\epsilon$ dimensions, take $\epsilon \to 0$ limit, isolate and discard the divergences and thus obtain a finite result. We will now see what this procedure gives \cite{cc1}. The expansion of the scalar field in $d$ spacetime dimensions is given by
 \begin{equation}
\Phi\left(t,{\bm x}\right)=\frac{1}{\left(2\pi\right)^{(d-1)/2}}
\int \frac{{\rm d}^{d-1}{\bm k}}{\sqrt{2\omega(k)}}\biggl(a_{\bm k}
{\rm e}^{-i\omega t +i{\bm k}\cdot {\bm x}}
 +
a_{\bm k}^{\dagger}{\rm e}^{i\omega t -i{\bm k}\cdot {\bm x}}
\biggr).
 \end{equation} 
 The vacuum energy density now has the form
 \begin{equation}
 \langle \rho\rangle 
= \frac{\mu^{4-d}}{2}
 \int \frac{{\rm d}^{d-1}{\bm k}}{\left(2\pi\right)^{(d-1)}}\, \omega (k)
=
\frac{\mu^{4-d}}{2}
 \int \frac{{\rm d}^{d-1}{\bm k}}{\left(2\pi\right)^{(d-1)}}\, (k^2+m^2)^{1/2}
 \end{equation} 
 where $\mu$ is a parameter introduced to maintain correct dimensionality. Integrals of this kind are usually evaluated using the result 
 \begin{equation}
 \int\frac{d^\alpha k}{(2\pi)^\alpha}\frac{1}{(k^2+m^2)^\beta}
 =\frac{1}{(4\pi)^{\alpha/2}}\frac{\Gamma[\beta-(\alpha/2)]}{\Gamma(\beta)}
 \left(\frac{1}{m^2}\right)^{\beta-(\alpha/2)}
 \end{equation} 
 and analytically continuing to all the relevant values of $\alpha$ and $\beta$. A simple calculation then gives, for the energy density, the result:
 \begin{equation}
\langle \rho\rangle 
= \frac{\mu^4}{2\left(4\pi\right)^{(d-1)/2}}
\frac{\Gamma(-d/2)}{\Gamma(-1/2)}\left(\frac{m}{\mu}\right)^d
 \end{equation} 
 Similarly, the computation of pressure leads to 
 \begin{equation}
 \langle p \rangle =
\frac{\mu^{4-d}}{\left(2\pi\right)^{(d-1)}}
\frac{1}{2(d-1)} \int {\rm d}^{d-1}{\bm k}\, \frac{k^2}{\omega (k)}
= -\frac{\mu^4}{2\left(4\pi\right)^{(d-1)/2}}
\frac{\Gamma(-d/2)}{\Gamma(-1/2)}\left(\frac{m}{\mu}\right)^d
 \end{equation} 
 so that we have a manifestly Lorentz invariant result $\langle p \rangle = - \langle \rho \rangle$. To isolate the divergences and extract the finite quantity, we use the expansions 
 \begin{equation}
 \Gamma\left(-2+\frac{\epsilon}{2}\right)=\frac{1}{(-2+\epsilon/2)}
\frac{1}{(-1+\epsilon/2)}\frac{1}{(\epsilon/2)}\Gamma\left(1+\frac{\epsilon}{2}
\right),
 \end{equation} 
 as well as 
 \begin{equation}
 \left(4\pi\right)^{-3/2+\epsilon/2} \simeq \frac{1}{(4\pi)^{3/2}}
\left[1+\frac{\epsilon}{2}\ln \left(4 \pi\right)\right];\qquad
\left(\frac{m}{\mu}\right)^{4-\epsilon}\simeq 
\left(\frac{m}{\mu}\right)^4\left(1-\epsilon \ln\frac{m}{\mu}\right)
 \end{equation} 
 This leads to the result
 \begin{equation}
 \langle \rho\rangle \simeq -\frac{m^4}{64 \pi^2}
\left[\frac{2}{\epsilon}+\frac32-\gamma -\ln 
\left(\frac{m^2}{4\pi \mu^2}\right)\right]+\cdots
 \end{equation} 
 The first term within the square bracket is divergent and hence the finite term $(3/2) -\gamma $ as well as $\ln 4\pi$ have no intrinsic meaning. Discarding these divergent contribution we get the finite part to be
 \begin{equation}
\langle \rho\rangle =\frac{m^4}{64 \pi^2}
\ln \left(\frac{m^2}{\mu^2}\right)
 \end{equation} 
 which was the result quoted in the text. Essentially the dimensional regularization has killed the power law divergences in \eq{54} and \eq{57} and has retained the logarithmic terms. As one can see  from \eq{55} and \eq{58}, the logarithmic contributions do lead to the correct equation of state $p =-\rho$ which is what we have obtained.
 
 \subsection{Zero-point-length and the quantum metric}\label{appen:qmetric}
 
At mesoscopic scales --- intermediate to Planck scale and macroscopic scales --- it should be possible to explore the spacetime structure in terms of an effective, renormalized, quantum metric, or qmetric, for short. In the absence of a complete theory of quantum gravity, the form of the qmetric needs to be determined by some physical considerations. 

Recall that the spacetime geometry can be completely determined in therms of the geodesic interval (also called the world function), $\sigma^2(x,y)$ which is a biscalar related to the metric by two equations. The first is:
\begin{equation}
\sigma =\int_{x'}^{x} \sqrt{g_{ab}dx^{a}dx^{b}}=\int _{\lambda _{0}}^{\lambda} \sqrt{ g_{ab}n^{a}n^{b}}d\lambda
\label{int-rel}
\end{equation}
where $n^{a}=dx^{a}/d\lambda$ is the tangent vector to the geodesic. This equation
 tells you how the metric determines $\sigma^2(x,y)$. 
The second one is the differential version of the same, given by:
\begin{equation}
\frac{1}{2}\,\lim_{x\to x'} \left[\nabla _{a}\nabla _{b}\sigma ^{2}(x,x')\right]=g_{ab}(x)
\label{diff-rel}
\end{equation} 
This allows you to determine the metric, from $\sigma^2(x,y)$. These two equations, together, imply that $g_{ab}(x)$ and 
$\sigma^2(x,y)$ contain the same amount of information. Classical gravity can be described entirely in terms of the single \textit{biscalar}
function $\sigma^2(x,y)$ instead of the ten functions in the metric $g_{ab}(x)$!. 

There is sufficient evidence which suggests that, at mesoscopic scales, the primary effect of quantum gravity is to change $\sigma^2$ to another function $S(\sigma^2)$ such that $S(0)\equiv L_0^2$ is finite and non-zero, signifying a zero-point-length to the spacetime. (While most of the ideas will work for arbitrary $S(\sigma^2)$, I will illustrate the results for the simple choice 
$S(\sigma^2)=\sigma^2(x,y)+L_0^2$; with a mostly positive signature, the zero -point-length adds to the spatial distances.) The qmetric is defined to be that ``metric`` which will lead to $S(\sigma^2)$ as the interval, just as the classical metric $g_{ab}(x)$ led to $\sigma^2(x,y)$ as the interval. That is the pair $(q_{ab}, S(\sigma^2))$ is related to each other through \eq{int-rel} and \eq{diff-rel}, just like the pair $(g_{ab}, \sigma^2)$. 

It is trivial to see that no such, local, metric exists because --- if it does --- \eq{int-rel} implies that the interval will vanish at coincidence limit. So qmetric is actually a bitensor $q_{ab}(x,x')$ and will diverge at \textit{all} events in the coincidence limit. 
It is indeed possible to determine such a bitensor $q_{ab}(x,x')$ just using the relations, analogous to \eq{int-rel} and \eq{diff-rel}, for the pair $(q_{ab}(x,x'), S(\sigma^2))$. The form of the qmeteic is given by:
\begin{equation}
 q_{ab}=Ah_{ab}+B n_{a}n_{b};\qquad q^{ab}=\frac{1}{A}h^{ab}+\frac{1}{B}n^{a}n^{b}
 \label{qmetric-def}
\end{equation} 
with
\begin{equation}
 B=\frac{\sigma ^{2}}{\sigma ^{2}+L_{0}^{2}};\qquad A=\left(\frac{\Delta}{\Delta _{S}}\right)^{2/(D-1)}\frac{\sigma ^{2}+L_{0}^{2}}{\sigma ^{2}}
 \label{qmetric-ab}
\end{equation}
where 
\begin{equation}
 \Delta (x,y)=\frac{1}{\sqrt{g(x)g(y)}}\textrm{det}\left\lbrace \nabla _{a}^{x}\nabla _{b}^{y}\frac{1}{2}\sigma ^{2}(x,y) \right\rbrace
 \label{qmmetric-vv}
\end{equation}
is the Van-Vleck determinant. The $\Delta_S$ is the Van-Vleck determinant with $\sigma^2$ replaced by $S(\sigma^2)$. All the effects of zero-point-length quoted in the text (area element, volume element, dimensionality at Planck scales ....) can be obtained from the qmetric.

The 
Van-Vleck determinant also plays an important role in the density of states. it can be shown that the exact expression for the density of states for geometry is given by
\begin{equation}
 \ln \rho_g(x,n)=\frac{1}{\Delta (x,n)}
 \label{vv-dos}
\end{equation} 
The Van-Vleck determinant has the expansion 
\begin{equation}
 \Delta (x,n)=1+\frac{L_0^2}{6}R_{ab}n^an^b+\mathcal{O}(L_0^4)
\end{equation} 
so that the density of states takes the approximate form 
\begin{equation}
 \ln \rho_g(x,n)=\frac{1}{\Delta (x,n)}\approx 1-\frac{L_0^2}{6}R_{ab}n^an^b
\end{equation} 
to the leading order.
 
\section*{Acknowledgement}

I thank Sumanta Chakraborty, Dawood Kothawala and 
Karthik Rajeev for extensive discussions. I thank Sumanta Chakraborty and Dawood Kothawala for comments on an earlier draft. My research  is partially supported by the J.C.Bose Fellowship of Department of Science and Technology, Government of India.


\begin{thebibliography}{000}

\bibitem{leesmolin} Lee Smolin, (1979), \textit{What is the problem of quantum gravity?}, preprint (based on Chap 1 of Ph.D dissertation, Harvard University).\footnote{The full sentence reads as: ``While there has been a lot of very interesting and imaginative work done on both sides in the past five or so years, it is safe to say that nothing which could be definitively called progress has been accomplished in this time.''}

\bibitem{tpcourses}
Part of this review is based on a set of lectures I have given recently; links to videos/slides of these lectures are available at this URL: https://www.iucaa.in/$\sim$paddy/transients/downloads.htm


  \bibitem{davies-unruh} 
P C W Davies,  (1975), \textit{J. Phys.} \textbf{A 8} 609 -- 616;
W. G. Unruh, \textit{Phys. Rev.} \textbf{D 14}(1976), 870.


\bibitem{A19}
Padmanabhan T., 2014, {\em Gen.~Rel.~Grav.}  {\bf 46}, 1673.  [arXiv:1312.3253]



\bibitem{ll}
D.~Lovelock,  1971, {{\em J. Math. Phys.} {\bfseries 12}, 498};
C.~Lanczos,  1932, {{\em Rev. Mod. Phys.}  {\bfseries 39} 716};
C.~Lanczos,  1938, {{\em Annals Math.} {\bfseries 39}, 842}.

 \bibitem{dktpll} T. Padmanabhan, Dawood Kothawala (2013),  \textit{Phys. Repts.} \textbf{531}, 115  [arXiv:1302.2151]. 


\bibitem{llwork} 
 T. Padmanabhan,  \textit{Int.J.Mod.Phys.},\textbf{ D 15}, 1659-1675 (2006) [gr-qc/0606061];
 Sumanta Chakraborty, T. Padmanabhan, \textit{Phys. Rev.}, \textbf{D 90}, 124017 (2014) [arXiv:1408.4679];
 Sumanta Chakraborty, T. Padmanabhan,  \textit{Phys. Rev.}, \textbf{D 90}, 084021 (2014) [arXiv:1408.4791]; 
 Sumanta Chakraborty, Krishnamohan Parattu, \textit{Gen. Rel. Grav.} \textbf{51}, 23 (2019) [arXiv:1806.08823];
 Sumanta Chakraborty, \textit{Adv. High Energy Phys.}, \textbf{2018}, 6509045 (2018) [arXiv:1704.07366];
\bibitem{scbook}
 Sumanta Chakraborty, (2017), \textit{Classical and Quantum Aspects of Gravity in Relation to the Emergent Paradigm (Springer Theses)}, Springer (Switzerland).


 \bibitem{ayan} A. Mukhopadhyay, T. Padmanabhan, \textit{Phys.Rev.,} \textbf{D 74}, 124023 (2006) [hep-th/0608120] 

\bibitem{ccreviews} 
Edmund J. Copeland, M. Sami, Shinji Tsujikawa (2006), \textit{Int.J.Mod.Phys.} \textbf{D15}, 1753;
Varun Sahni, Alexei Starobinsky (2000), \textit{Int.J.Mod.Phys.} \textbf{D9}, 373;
T. Padmanabhan (2003), \textit{Phys.Repts.} \textbf{380}, 235 [arXiv:hep-th/0212290]. 

\bibitem{cc1}
J. Martin (2012),  \textit{C. R. Physique} \textbf{13}, 566 [arXiv:1205.3365]


 \bibitem{inherentqm} T. Padmanabhan, \textit{Gen.Rel.Grav.,} \textbf{34} 2029- (2002) [gr-qc/0205090];  
 Mod.Phys.Letts. \textbf{A 17}, 1147 (2002) [hep-th/0205278]. 
 

\bibitem{tpap}
T. Padmanabhan (2009),  \textit{Adv. Sci. Lett.}, \textbf{2}, 174 [arXiv:0807.2356] 


 
 \bibitem{tphpplb} T.  Padmanabhan, Hamsa Padmanabhan, \textit{Phys. Letts. B} \textbf{773}, 81-85 (2017) [arXiv:1703.06144]
 
 \bibitem{leeNL} Lee T D,  \textit{Nucl. Phys.} \textbf{B 264}, 437 (1986).
 
 

 \bibitem{tppropagator} 
 T.   Padmanabhan, (2019),  \textit{Phys. Rev.},\textbf{ D 100}, 045024 (2019) [arXiv:1905.08263]
 
 

 \bibitem{tpqft}
  J. Schwinger, (1998), \textit{Particles, Sources, and Fields: Vol. 1}, Perseus Books, USA.  For a textbook discussion see, e.g., 
  T. Padmanabhan,  (2016), \textit{Quantum Field Theory: Why, What and How}, Springer International Publishing, Switzerland.

  


\bibitem{tpreviews2}
T. Padmanabhan (2010), \textit{ Reports in Progress of Physics}, \textbf{73}, 046901 [arXiv:0911.5004]



\bibitem{Keski-Vakkuri:1997xp}
Keski-Vakkuri E and Kraus P  1997 \textit{Nucl.  Phys.} B \textbf{491} 249--262;
also see Peltola A 2009 \textit{Class.Quant.Grav} \textbf{26} 035014 [arXiv:0807.3309].



\bibitem{matsasrev}  L. C. B. Crispino, A. Higuchi, and G. E. A. Matsas, \textit{Rev. Mod. Phys.} \textbf{80}, 787 (2008)




 \bibitem{kkvp} 
 Kinjalk Lochan, Karthik Rajeev, Amit Vikram and T. Padmanabhan,  \textit{Phys. Rev.,} \textbf{D 98,} 105015 (2018) [arXiv:1805.08800]
 
\bibitem{sc-trg} T.R. Govindarajan, Sumanta Chakraborty, \textit{Embedding into flat spacetime and black hole thermodynamics},
 [arXiv:1908.09074]

 
\bibitem{gerlach} U. H. Gerlach, (1988) \textit{Phys.Rev.} \textbf{D38}:514-521 [gr-qc/9910097].



 \bibitem{agullo} Ivan Agullo et al., Phys.Rev.\textbf{D77}, 124032 (2008).
 

 
 \bibitem{takagi} Takagi S (1986) \textit{Prog. Theor. Phys. Suppl.}, \textbf{88} 1–142.
 
 

  \bibitem{detector} 
  B. S. DeWitt, Quantum gravity: The new synthesis, in: S. Hawking, W. Israel
(Eds.), General Relativity: An Einstein Centenary Survey, Cambridge
University Press, Cambridge, 1979, pp. 680 - 745.

  
 
 
 
 \bibitem{vacprobe} L. Sriramkumar and T. Padmanabhan, \textit{Int. Jour. Mod. Phys.}, \textbf{D 11},1 (2002) [gr-qc-9903054] 
 

 
 \bibitem{aseem-tp-bh} Aseem Paranjape, T. Padmanabhan,  \textit{Phys.Rev.} \textbf{D 80}, 044011 (2009) [arXiv:0906.1768] 
 
 
 

 \bibitem{brout} Brout R et al 1995 \textit{Phys. Rept.} \textbf{260} 329–454
 

 
 
\bibitem{D2c} 
 Padmanabhan T 1985 \textit{Ann. Phys.} \textbf{165} 38 

\bibitem{D2d} 
 Padmanabhan T 1997 \textit{Phys. Rev. Lett.} \textbf{78} 1854  [hep-th/9608182];
T. Padmanabhan, \textit{Phys. Rev.}, \textbf{D 57}, 6206 (1998) 
 
\bibitem{tplimitations} 
Snyder H S 1947 \textit{Phys. Rev.} \textbf{71} 38 ;
DeWitt B S  1964 \textit{ Phys. Rev. Lett.} \textbf{13} 114;
Yoneya T  1976 \textit{Prog. Theor. Phys.} \textbf{56} 1310;
Padmanabhan T  1987  \textit{Class. Quantum Grav.} \textbf{4} L107;
For a review, see Garay L J  1995 \textit{ Int. J. Mod. Phys.} A \textbf{10} 145. 
 
 
 
 
 
\bibitem{Padmanabhan:1998jp}
Padmanabhan T 1998 \textit{Phys.  Rev. Lett.} \textbf{81}  4297--4300 [hep-th/9801015].

\bibitem{Padmanabhan:1998vr}
Padmanabhan T  1999 \textit{Phys. Rev.} D \textbf{59}  124012 [hep-th/9801138].



\bibitem{tpreviews3}
T. Padmanabhan  (2011),  \textit{J. Phys. Conf. Ser.} \textbf{306}, 012001 [arXiv:1012.4476] 


 
 
 \bibitem{horts} Bibhas Ranjan Majhi, T. Padmanabhan,  \textit{Eur. Phys. J.,} \textbf{C 73} 12651 (2013) [arXiv:1302.1206]
 


\bibitem{KBP} Krishnamohan Parattu, Bibhas Ranjan Majhi, T. Padmanabhan,    \textit{Phys.Rev.,} \textbf{D 87},  124011 (2013) [arXiv:1303.1535].

 
 \bibitem{36a} Sumanta Chakraborty and T. Padmanabhan, (2019), \textit{Boundary Term in the Gravitational Action is the Heat Content of the
  Null surfaces}, [arXiv:1909.00096].
 
\bibitem{tpcqg}
T. Padmanabhan,  \textit{Class.Quan.Grav.}, \textbf{21}, 4485 (2004) [gr-qc/0308070]





 \bibitem{krooth}
 T. Padmanabhan, \textit{Mod. Phys. Letts.} \textbf{A 29}, 1450037 (2014).



\bibitem{gravitation}
Padmanabhan T 2010 \textit{Gravitation: Foundations and Frontiers}, (Cambridge University Press UK).



\bibitem{tpstructure}
Padmanabhan T.,\textit{AIP Conf. Proc.,}  \textbf{1483},  212 (2012)  [arXiv:1208.1375]



\bibitem{tpnoeeng}
 T. Padmanabhan, \textit{Gen.Rel.Grav,}  \textbf{44}, 2681 (2012)  [arXiv:1205.5683]

 
 
\bibitem{surH}
S. Carlip and C. Teitelboim,  \textit{Class.Quant. Grav}, \textbf{12}, 1699 (1995);
S. Massar and R. Parentani,\textit{ Nucl.Phys.} \textbf{B575}, 333 (2000); [gr-qc/9903027].

 
 

\bibitem{bibhas-tp}  
Bibhas Ranjan Majhi,  T. Padmanabhan,  \textit{ Phys.Rev.}, \textbf{D 86}, , 101501 (2012)  [arXiv:1204.1422]






\bibitem{hptp1}
Hamsa Padmanabhan and  T. Padmanabhan,  \textit{Int.Jour.Mod.Phys,} \textbf{D22}, 1342001 (2013) [arXiv:1302.3226].

\bibitem{hptp2} 
T. Padmanabhan and Hamsa Padmanabhan, \textit{Int.Jour.Mod. Phys.}, \textbf{D 23}, (2014)  1430011 [arXiv:1404.2284]. 

 
 
 


\bibitem{lwf} T. Padmanabhan (2017), Comptes Rendus Physique, \textbf{28}, 275-291  [arXiv:1611.03505]


\bibitem{mazumdar} A. Mazumdar, \textit{The origin of dark matter, matter-anti-matter asymmetry, and inflation}, [arXiv:1106.5408]





 
 \bibitem{unimod}
See e.g., G. F. R.  Ellis (2014), \textit{Gen. Rel. Grav.} \textbf{46}, 1619 [arXiv:1306.3021];
James L. Anderson and David Finkelstein (1971), \textit{Am. J. Phys.}, \textbf{39}, 901;
David R. Finkelstein, Andrei A. Galiautdinov and James E. Baugh (2001), \textit{J. Math. Phys.} \textbf{42}, 340 [arXiv:gr-qc/0009099];
William G. Unruh (1989), \textit{Phys. Rev.} \textbf{D 40}, 1048-1052.





 
 \bibitem{tp} T. Padmanabhan,  \textit{Int. Jour. Mod. Phys.,} \textbf{D 25} 1630020 (2016) [arXiv:1603.08658] 

  






\bibitem{A11}
Padmanabhan, T., 2015, {\em Mod. Phys. Lett. A},
 {\bf 30}, 1540007.





   
  \bibitem{A24}  
  T.~Padmanabhan, 
  {{\em Phys.Rev.}
  {\bfseries D83} (2011) 044048}
  {{  [arXiv:1012.0119]}}.

   \bibitem{A25}  
  S.~Kolekar and T.~Padmanabhan, 
  {{\em Phys.Rev.}
  {\bfseries D85} (2012) 024004}
{{  [arXiv:1109.5353]}}. 
   



 

 \bibitem{A26} 
  Damour, T., \textit{Quelques Propri´et´es M´ecaniques, Electromagn´etiques, Thermodynamiques et Quantiques des Trous Noirs}, Ph.D. Thesis, Universit´e
Paris, Paris, France, August 1979.  
  
  
   
 \bibitem{A27}
 T.~Damour, ``{Surface effects in black hole physics},'' {\em Proceedings of the
  Second Marcel Grossmann Meeting on General Relativity} (1982).
  


\bibitem{membrane} Thorne K S,  Price R H and MacDonald D A 1986 \textit{Black Holes: The Membrane Paradigm} (Yale University Press)


\bibitem{aseemtp}
T. Padmanabhan, Aseem Paranjape (2007), \textit{Phys.Rev.}, \textbf{D75} 064004 [arXiv:gr-qc/0701003];  
T. Padmanabhan  (2008), \textit{ Gen.Rel.Grav.}, \textbf{40}, 529 [arXiv:0705.2533]







 \bibitem{A17}
 T.~Padmanabhan, 
  {{\em Mod.Phys.Lett.}
  {\bfseries A25} (2010) 1129}
{{  [arXiv:0912.3165]}}.
 
 \bibitem{A18}
  T.~Padmanabhan,
  {{\em Phys.Rev.}
  {\bfseries D81} (2010) 124040}
{{  [arXiv:1003.5665]}}. 





\bibitem{SCTPnull}
Chakraborty S. and T. Padmanabhan, (2015) \textit{Phys. Rev.}, \textbf{D 92}, 104011 [arXiv:1508.04060].




 \bibitem{A36}
 T.~Padmanabhan, (2016) \textit{Gen. Rel. Grav. (Letts),} \textbf{48:4}, 1-7
{{  [arXiv:1506.03814]}}.
  




  
  \bibitem{A41}
  V.~Moncrief and J.~Isenberg, 
  {{\em Communications in
  Mathematical Physics} {\bfseries 89}, (1983) 387}.
  
  
   \bibitem{A42}
   E.~M. Morales, ``{On a Second Law of Black Hole Mechanics in a Higher
  Derivative Theory of Gravity},''
{\em available at
  http://www.theorie.physik.uni-goettingen.de/forschung/qft/theses/dipl/Morfa-%
Morales.pdf} (2008).
   
  \bibitem{A43}  
  K.~Parattu, S.~Chakraborty, B.~R. Majhi, and T.~Padmanabhan, (2016), \textit{ Gen. Rel. Grav.}, \textbf{48}, 94
{{  [arXiv:1501.01053]}}. 
   
   


  \bibitem{A33}
  S.~Chakraborty, K.~Parattu, and T.~Padmanabhan, (2015),  JHEP, 10, 097
{{  [arXiv:1505.05297]}}.

 
  
 \bibitem{A30}
 D.~Kothawala, S.~Sarkar, and T.~Padmanabhan,
{{\em Phys.Lett.}
  {\bfseries B652} (2007) 338--342},
{{  [gr-qc/0701002]}}.
   
 \bibitem{A31} 
 A.~Paranjape, S.~Sarkar, and T.~Padmanabhan, 
 {{\em Phys.Rev.}
  {\bfseries D74} (2006) 104015},
{{  [hep-th/0607240]}}.
    

 
  
 \bibitem{A32}
  D.~Kothawala and T.~Padmanabhan, 
  {{\em Phys. Rev.}
  {\bfseries D79} (2009) 104020},
  {{  [arXiv:0904.0215]}}. 
         
 \bibitem{A28}
 R.-G. Cai and S.~P. Kim,
 {{\em JHEP} {\bfseries
  0502} (2005) 050},
{{  [hep-th/0501055]}}.
        
 \bibitem{A29}
  M.~Akbar and R.-G. Cai, 
  {{\em Phys.Lett.}
  {\bfseries B635} (2006) 7--10},
{{  [hep-th/0602156]}}. 

         
  \bibitem{A44} 
  S.~A. Hayward,  {{\em
  Class.Quant.Grav.} {\bfseries 15} (1998) 3147--3162},
{{  [gr-qc/9710089]}}.
  
   \bibitem{A45}
    D.~Kothawala, 
  {{\em Phys.Rev.}
  {\bfseries D83} (2011) 024026},
{{  [arXiv:1010.2207]}}.      
   

 

\bibitem{paperD} 
Padmanabhan T, Chakraborty S and Kothawala D 2016 \textit{ Gen. Rel. Grav.,} \textbf{48} 55  [arXiv:1507.05669].



\bibitem{D1} 
Kothawala D and Padmanabhan T 2014 {\em Phys. Rev.} D \textbf{90} 124060 [arXiv:1405.4967].

\bibitem{pesci}
Alessandro Pesci, \textit{Class. Quantum Grav.}, \textbf{36} (2019) 075009 [arXiv:1812.01275];
Alessandro Pesci, \textit{Looking at spacetime atoms from within the Lorentz sector}, [arXiv:1803.05726]

 
\bibitem{D4}
Kothawala D 2013 {\em Phys. Rev.} D \textbf{88} 104029. 
 
\bibitem{D5}
 Stargen D J and Kothawala D 2015  \textit{Phys. Rev.} D \textbf{92} 024046  [arXiv:1503.03793].

 
 
\bibitem{D6}
Kothawala D and Padmanabhan T 2015 {\em Phys.~Lett.} B \textbf{748} 67. 

\bibitem{pesci-sc-dk}
Sumanta Chakraborty, Dawood Kothawala, Alessandro Pesci, \textit{Raychaudhuri equation with zero point length}, \textit{Phys. Lett.} \textbf{B 797}, 134877 (2019) [arXiv:1904.09053].



\bibitem{D8} 
Gray A  1974 {\em Mich. Math. J.} \textbf{20} 329. 
 
 


\bibitem{D2a}  
 DeWitt B S 1964 \textit{Phys. Rev. Lett.} \textbf{13} 114.

\bibitem{D2b} 
 Padmanabhan T 1985 \textit{Gen. Rel. Grav.} \textbf{17} 215 


\bibitem{D2e}
 Garay L 1998 \textit{ Phys. Rev. Lett.} \textbf{80} 2508 [gr-qc/9801024]

\bibitem{D2f} 
 Garay L 1995 \textit{Int. J. Mod. Phys.} A \textbf{10} 145.  


 
\bibitem{D7} 
Chakraborty S and Padmanabhan T [work in progress].




\bibitem{deq2}  
For a sample of  earlier results, see e.g., 
S. Carlip, R. Mosna, J. Pitelli, Phys. Rev. Lett. \textbf{107}, 021303 (2011) arXiv:1103.5993; AIP Conf.Proc. \textbf{1196}, 72 (2009) arXiv:0909.3329;
G. Calcagni, Phys. Rev. Lett. \textbf{104}, 251301 (2010) arXiv:0912.3142;
J. Ambjorn, J. Jurkiewicz, R. Loll, Phys. Rev. Lett. \textbf{95}, 171301 (2005) arXiv:hep-th/0505113;
L. Modesto, P. Nicolini, 2010, Phys. Rev. D 81, 104040 [arXiv:0912.0220];
G. Calcagni, D. Oriti, J. Thurigen, Class. Quant. Grav. \textbf{30}, 125006 (2013) arXiv:1208.0354;
V. Husain, S.S. Seahra and E.J. Webster, Phys. Rev. D \textbf{88}, 024014 (2013) arXiv:1305.2814. 

 

  
 

\bibitem{tpentropy}  Padmanabhan  2015  \textit{ Entropy} \textbf{17}  7420-7452 [arXiv:1508.06286].

\bibitem{tpreviews1}
 T. Padmanabhan, \textit{Exploring the Nature of Gravity}, [arXiv:1602.01474]
 
 
 

\bibitem{tpemeuniv} 
T. Padmanabhan,  \textit{Res. Astro. Astrophys.,} \textbf{12}, 891 (2012) [arXiv:1207.0505] 





\bibitem{info} 
Landauer R., \textit{Information is Physical}, Proc. Workshop on Physics and Computation PhysComp '92 (IEEE Comp. Sci.
Press, Los Alamitos, 1993) pp. 1-4.

 
 

\bibitem{wald} S. Hollands and R. M. Wald (2002), \textit{Gen. Rel. Grav.} {\bf 34}, 2043


 
\bibitem{tp-sriram} L. Sriramkumar and T. Padmanabhan (2005), \textit{Phys. Rev.} {\bf D 71}, 103512



\bibitem{lee} J. Magueijo, L. Smolin and C. R. Contaldi (2007), \textit{Class. Quant. Grav.} {\bf 24}, 3691



\bibitem{related} Yun-Song Piao, \textit{Phys.Rev.,} \textbf{D76}:043509 (2007) [arXiv:gr-qc/0702071]; Phys.Rev. \textbf{D74} (2006) 043509  	[arXiv:gr-qc/0512161]; Joao Magueijo, \textit{Phys.Rev.,} \textbf{D76}:123502 (2007) [arXiv:astro-ph/0703781].

\bibitem{tpsesh} T.Padmanabhan and T.R. Seshadri (1988), \textit{Class. Quan. Grav.}, \textbf{5}, 221.

\bibitem{models}
Patrick Peter, Nelson Pinto-Neto (2008), \textit{Phys.Rev.}\textbf{ D78} 063506 [arXiv:0809.2022];
J.V. Narlikar and T.Padmanabhan (1983),  \textit{Ann. Phys. }, \textbf{150}, 289;
Cristiano Germani, Nicolas Grandi, Alex Kehagias (2008), AIP Conf.Proc. \textbf{1031},  172 [arXiv:0805.2073]. 

\bibitem{gr}  I. S. Gradshteyn, I. M. Ryzhik (2007), \textit{Table of Integrals, Series and Product}s (7th edition), Academic Press.
 
 \bibitem{rk-tp} Karthik Rajeev and T. Padmanabhan, (2019), \textit{Exploring the Rindler vacuum and the Euclidean Plane}, [arXiv:1906.09278]

\bibitem{boulware} David G. Boulware, (1975), \textit{Phys. Rev.}, \textbf{D 11}, 1404.
 
\bibitem{watson}  G. N. Watson, (1995), \textit{A Treatise on the Theory of Bessel Functions}, Cambridge University Press.
 
\bibitem{harrison} E. R. Harrison (1970), \textit{Phys. Rev.} {\bf D 1}, 2726



\end{thebibliography}
\end{document}